\documentclass[
11pt, % The default document font size, options: 10pt, 11pt, 12pt
oneside, % Two side (alternating margins) for binding by default, uncomment to switch to one side
english, % ngerman for German
singlespacing, % Single line spacing, alternatives: onehalfspacing or doublespacing
parskip, % Uncomment to add space between paragraphs
headsepline, % Uncomment to get a line under the header
]{MastersDoctoralThesis} % The class file specifying the document structure
\usepackage{mathtools}
\usepackage{amsmath}
\usepackage{amssymb}
\usepackage{amsthm}
\usepackage{amsfonts}
\usepackage{graphicx}
\usepackage{caption}
\usepackage{subcaption}

\usepackage{tikz}
\usepackage[compat=1.1.0]{tikz-feynhand}
\usepackage{ytableau}
\DeclareUnicodeCharacter{2212}{-}
\newtheorem{thm}{Theorem}
\newtheorem*{thm*}{Theorem}
\newtheorem{definition}{Definition}
\newtheorem*{definition*}{Definition}
\newtheorem{conj}{Conjecture} 

\usepackage[utf8]{inputenc} % Required for inputting international characters
\usepackage[T1]{fontenc} % Output font encoding for international characters

\usepackage{mathpazo} % Use the Palatino font by default

\usepackage[square,numbers]{natbib}
\usepackage[english, ku, titlepage, dropcaps]{ku-frontpage}

\assignment{Master thesis}
\author{Xin Qian}

\title{One-point functions in AdS/dCFT}
\subtitle{MPS and twisted Yangian} % No subtitle
\date{Handed in: May 20, 2022}
\advisor{Advisor: Prof. Charlotte Fl\o e Kristjansen}

\AtBeginDocument{
\hypersetup{pdftitle=thesis} % Set the PDF's title to your title
\hypersetup{pdfauthor=Xin} % Set the PDF's author to your name
\hypersetup{pdfkeywords=One-point functions} % Set the PDF's keywords to your keywords
}
\begin{document}
\frontmatter % Use roman page numbering style (i, ii, iii, iv...) for the pre-content pages

\pagestyle{plain} % Default to the plain heading style until the thesis style is called for the body content

%----------------------------------------------------------------------------------------
%	TITLE PAGE
%----------------------------------------------------------------------------------------
\maketitle

\begin{abstract}
\addchaptertocentry{\abstractname} % Add the abstract to the table of contents
I focus on the scalar one-point functions in $SO(6)$ sector of D5-D3 probe-brane set-up. Start with a general introduction of integrability, I explore both coordinate Bethe ansatz and algebraic Bethe ansatz, with possible generalization. I then shortly review how to use the Bethe ansatz in $\mathcal{N}=4$ super Yang-Mills theory, and then apply such procedure to the D5-D3 system. The dual field theory of such system corresponds to a defected version of  $\mathcal{N}=4$ super Yang-Mills theory, where the one-point functions of certain scalars are non-zero. The calculation of one-point functions is mapped to the overlap between matrix product states and Bethe states. The matrix product states are found to be solutions of the twisted Boundary Yang-Baxter equation, and equivalently the representations of extended twisted Yangian. By dressing procedure or coproduct property, we can connect the scalar matrix product state and higher dimension matrix product states. We have used the branching rules to find the connection with some detailed parameters needed to be fixed. Such method can not only be used for calculations of one-point functions in probe-branes system, but also shed some light on non-equilibrium system. 
\end{abstract}

%----------------------------------------------------------------------------------------
%	ACKNOWLEDGEMENTS
%----------------------------------------------------------------------------------------

\begin{acknowledgements}
\addchaptertocentry{\acknowledgementname} % Add the acknowledgements to the table of contents
I would like to thank my supervisor Charlotte F\o le Kristjansen, for her guidance in such elegant project, and also for keeping answer many questions from mathematical definition to general physical meaning. Further, I would like to thank Tamás Gombor, for his help on this project, with useful code and knowledge on integrability. I also want to thank Vasilis Moustakis for our many interesting discussion on this project. Finally, I would like to thank Xiaoming Chen and Shanzhong Han, for their constant support. 
\end{acknowledgements}

%----------------------------------------------------------------------------------------
%	LIST OF CONTENTS/FIGURES/TABLES PAGES
%----------------------------------------------------------------------------------------

\tableofcontents % Prints the main table of contents

%	THESIS CONTENT - CHAPTERS
%----------------------------------------------------------------------------------------

\mainmatter % Begin numeric (1,2,3...) page numbering

\pagestyle{thesis} % Return the page headers back to the "thesis" style

% Include the chapters of the thesis as separate files from the Chapters folder
% Uncomment the lines as you write the chapters
\chapter{Introduction} % Main chapter title

\label{Introduction} % For referencing the chapter elsewhere, use \ref{Introduction} 

%----------------------------------------------------------------------------------------

% Define some commands to keep the formatting separated from the content 
\newcommand{\keyword}[1]{\textbf{#1}}
\newcommand{\tabhead}[1]{\textbf{#1}}
\newcommand{\code}[1]{\texttt{#1}}
\newcommand{\file}[1]{\texttt{\bfseries#1}}
\newcommand{\option}[1]{\texttt{\itshape#1}}

Postulated in 1997 by Maldacena \cite{Maldacena_1999}, AdS/CFT correspondence is one of the major developments in theoretical physics over the last two decades (now generally known as gauge/gravity duality). The most studied version is the one has a duality between $\mathcal{N}=4$ super Yang-Mills (SYM)  theory with $U(N)$ gauge symmetry and type II B string theory with the near-horizon limit of a stack of D3-branes placed in flat space. 

Even though such theory is conjectured to be true at general setting, AdS/CFT is actually lack of rigorous proof.  It turns out that in the free/planar sector, where the gauge coupling $g_{YM}\rightarrow 0$ and colors $N\rightarrow \infty$ with 't Hooft coupling $\lambda=g^2_{YM}N$ kept fixed, one can solve the AdS/CFT completely. The miracle tool which leads to such a solution is called $integrability$ \cite{Beisert_2003,Beisert_2011}. Integrability techniques are originally introduced to solve $2D$ many-body systems, particularly spin chains, in condensed matter physics. How can we relate a $2D$ system to a $4D$ field theory? The answer lies on the state-operator correspondence \cite{EPFL_CFT}, which is a mapping of the Hilbert spaces between two systems, the $\mathcal{N}=4$ SYM and $2D$ spin chain system \cite{Minahan_2003}. 

Integrability,  a useful tool to solve physical system from a non-perturbatively way. In classical sense, there is a very clear definition of integrability due to Liouville \cite{intro_int_NB}. That the integrals of motions are many enough to fully determine the phase space. In quantum level, a system is said integrable if the scattering process is factorized. That is to say that higher point collision can be decomposed into two point collision. In an equivalent way, our R-matrix should satisfy the Yang-Baxter equations \cite{Yang1967,BAXTER1972}, and the conserved charges, classical analog of integrals of motions, can be generated by transfer matrix constructed from R-matrix, and they are commutative. The eigenstates of the transfer matrix are called the Bethe states, which encode different kind of excitations of our system. From the coordinate Bethe ansatz, by imposing the periodic boundary condition on the wave function in spin chain system, one can obtain Bethe equation on each level of excitation. These Bethe equations completely solve our system. On the other hand, in stead of constructing complicated wave function, one can get the eigenvalues of transfer matrix, Bethe equations and Bethe states using the RTT-relation as commutation relation in algebraic Bethe ansatz. This method can be generalized to arbitrary symmetry property of certain integrable system using fusion procedure \cite{fusion_procedure} and Bazhanov-Reshetikhin determinant formula \cite{BR_formula} (or Tableau sum \cite{T_system_Y_system}).

In the real world, all systems have impurities and different types of defects that break translational and rotational symmetries into subgroups. Hence, it is worth studying a defected generalization of the above $\mathcal{N}=4$ SYM. According to the correspondence, what we are interested in are those which have a gravity dual in the string theory side. An important gravity set-up is to add a stack of $N_f$ Dp-branes, the $flavor\ branes$, into the near-horizon geometry, which will give rise to fields that transform in the fundamental representation of the gauge group. This is quite useful for investigating quark fields in QCD. Anyway in this thesis, we will restrict to a single $N_f=1$ D5-branes codimension-one intersection (we will assume it sits at $x_3=0$) so that probe limit $N_f/N\rightarrow 0$ is satisfied. In the set-up we can neglect the backreaction of D5-branes, and treat it as a $probe\ brane$ \cite{Karch_2001}. In the field theory side, the associated theory is a $defect\ conformal\ field\ theory$ where (2+1) dimensional matter fields are coupled to a gauge theory in (3+1) dimensions. The brane intersection preserves 8 of 32 real supercharges, (2+1) dimensional $SO(3,2)$  conformal symmetry, and our global $SO(6)$ R-symmetry breaks down to a subgroup $SU(2)\times SU(2)$.

In our probe-brane system, the codimension-one defect will separate our space into $x_3<0$ and $x_3>0$. As a consequence, there are three scalar fields at $x_3>0$ obtain a $x_3$-dependent vacuum expectation value (vev) parametrized by $\mathfrak{su}(2)$ algebra, all the other fields have zero vev at both side. Due to the intersection of branes between these two region differs, one would expect another parameter $k$ as the intersection number identifies our system. The gauge group will break to $U(N-k)$ in $x_3<0$ and $U(N)$ in $x_3>0$. However, due to the vev, the gauge symmetry $U(N)$ in $x_3>0$ will further break down to $U(N-k)$ \cite{one_pt_dfc}. 

The non-vanishing vev gives us a non-zero $x_3$ dependent one-point functions even at tree level for scalar fields. Such data is quite important for bootstrapping general correlation functions. As show in \cite{de_Leeuw_2016}, we can use the integrability techniques to extract such information, and the so-called matrix product states (MPS) will come into play, as the role of "defect states". And correspondingly the computation of the one-point functions in dCFT maps to the computation of overlaps between the MPS and the Bethe eigenstates of the spin chain. Due to the codimension one property, our system will also show up in the dCFT applications in condensed matter system with space two dimensional defect. Moreover, the MPS turn out to coincide with the integrable initial state in a quantum quench of an integrable system, for calculation of time evolution after the quench. As showed in \cite{Pozsgay_2019}, the integrable initial states, can also be achieved by solving the twisted Boundary Yang-Baxter equation (BYB). From the commutativity of transfer matrix, one can get build initial integrable states, with higher auxiliary dimensions, out of the original one by definition. And we call it $dressing\ procedure$. Such property is captured by the coproduct of the twisted Yangian for $SU(N)$ spin chain, extended twisted Yangian for $SO(N)$ spin chain. More interestingly, the BYB takes a quite similar form as part of the definition of twisted Yangian or extended twisted Yangian. The MPS maps the representations of the twisted Yangian algebra or extended twisted Yangian algebra. The RTT relations are encoded in definition of Yangian and extended Yangian, of which the twisted Yangian and extended twisted Yangian are subalgebra. There is also an important property that Yangian and its descendances may satisfy, that a relation satisfied by classical group, it also satisfies in quantum group level. We will use this property to guide our calculation, like branching rules. 

So in conclusion, we calculate the simplest scalar representation of twisted Yangian, find the overlap between this solution with Bethe states using thermodynamic Bethe ansatz technique. Then do the dressing procedure, we can build the higher dimensional MPS, compare dressed solution with the complicated original MPS, then one can obtain the overlap formula. 

The remainder of the thesis is structured as follows: In Chapter \ref{Chapter1}, we will give a short introduction to both coordinate Bethe ansatz and algebraic Bethe ansatz. With the help of Appendix A, B, C, D, one can get a taste of how Bethe ansatz derived in general symmetry of integrable system. In Chapter \ref{Chapter2}, the famous mapping from $\mathcal{N}=4$ SYM to spin chain system is introduced, by calculating the one-loop anomalous dimension in scalar section, which corresponds to energy eigenvalue in spin chain system. Further we introduce the state-operator correspondence, which is a mapping from scalar operators to certain excitations states in spin chain. In Chapter \ref{Chapter3}, the set-up of D5-D3 probe brane system is introduced, and we use the state-operator correspondence to map our one-point functions to the overlap between MPS and Bethe states. The result for $SU(2)$ sector $SU(3)$ sector as well as $SO(6)$ are presented by direct calculation. In Chapter \ref{Chapter4}, follow \cite{Pozsgay_2019}, we connect our MPS used in Chapter \ref{Chapter4} to the solutions of BYB. We reproduce the derivation of $SU(3)$ sector \cite{de_Leeuw_2020} in Chapter \ref{Chapter5}. And guided by the same method, we present the unfinished work in full $SO(6)$ sector in Chapter \ref{Chapter6}.
% Chapter 1

\chapter{Integrability and Bethe ansatz} % Main chapter title

\label{Chapter1} % For referencing the chapter elsewhere, use \ref{Chapter1} 

%----------------------------------------------------------------------------------------
Integrability is an elegant tool to solve physical system from a non-perturbatively way. In classical sense, if a phase has $2n$ degrees of freedom, and there are number $n$ coordinates are in involution. We will call such system integrable due to Liouville. In quantum world, the Poisson bracket turns to commutator. Our integrability condition becomes the famous \textit{Yang-Baxter equations}. I will first present \textit{XXX} spin chain system to give a taste of what are dealing with, the possible generalizations are presented in Appendix. The coordinate Bethe ansatz (CBA) is quite physical, we are basically dealing with $2D$ scattering problem, and it turns out integrability in this case is just \textit{factorized scattering}. In algebraic Bethe anzatz (ABA), we need to construct the R-matrix for satisfying certain system and \textit{Yang-Baxter equations}. And the scattering is replaced by commutation relation, the \textit{RTT relation}. And in this way, the calculation is more compact and easy to generalize. 

\section{Coordinate Bethe ansatz for XXX Spin Chain}
\subsection{The XXX Hamiltonian and Scattering phase factor}
A  spin is described by an element of the vector space
\begin{equation}
\mathbb{V}=\mathbb{C}^2.
\label{eqn:eq1}
\end{equation}
Hence a spin chain of length $L$ is a $L$-fold tensor product
\begin{equation}
\mathbb{V}^{\otimes L}=\mathbb{V}_1\otimes\dots\otimes\mathbb{V}_L.
\label{eqn:eq2}
\end{equation}
The Hamiltonian operator $H:\mathbb{V}^{\otimes L}\rightarrow\mathbb{V}^{\otimes L}$ is homogenous and acts on nearest neighbours

\begin{equation}
\mathcal{H}=\sum_k H_{k,k+1},\quad H_{k,l}:\mathbb{V}_k\otimes\mathbb{V}_l \rightarrow \mathbb{V}_k\otimes\mathbb{V}_l
\label{eqn:eq3}
\end{equation}
The  pairwise kernel $H_{k,l}$ for the Heisenberg chain reads 
\begin{equation}
H_{k,l}=J_0(1_{k}\otimes 1_{l})+J_x(\sigma^x_{k}\otimes \sigma^x_{l})+J_y(\sigma^y_{k}\otimes \sigma^y_{l})+J_z(\sigma^z_{k}\otimes \sigma^z_{l}).
\label{eqn:eq4}
\end{equation}
For $XXX$ spin chain, we take the value
\begin{equation}
J_0=-J_x=-J_y=-J_z=\frac{1}{2}J.
\label{eqn:eq5}
\end{equation}
With this choice, the Hamiltonian kernel reads 
\begin{equation}
H_{k,l}=\frac{J}{2}(I_{k,l}-P_{k,l}),
\label{eqn:eq6}
\end{equation}
where $P_{k,l}$ is the permutation operator acting with rules
\begin{equation}
P_{k,l}(\mathbb{V}_k\otimes\mathbb{V}_l)=\mathbb{V}_l\otimes\mathbb{V}_k.
\label{eqn:eq7}
\end{equation}
We will focus on periodic boundary condition mainly, such that $\mathbb{V}_{L+1}=\mathbb{V}_l$. Using the commutation relations for $SU(2)$ generators, one can first prove $[\sigma_k^{\alpha}\otimes I+I\otimes\sigma_{k+1}^{\alpha},H_{k,k+1}]=0$, then it is easily to show that 
\begin{equation}
[Q^\alpha,\mathcal{H}]=0,
\label{eqn:eq8}
\end{equation}
with $Q^{\alpha}=\sum_{k=1}^L\frac{1}{2}\sigma_k^{\alpha}, \ \alpha=1,2,3$.
Prepared with this setup, we are now ready to calculate the spectrum of our $XXX$ spin chain system. Start with the vacuum state
\begin{equation}
\vert 0 \rangle=\lvert \uparrow \uparrow\dots\uparrow\rangle.
\label{eqn:eq9}
\end{equation}
This state has zero energy

\begin{equation}
H_{k,k+1}\lvert 0\rangle=\frac{J}{2}(I_{k,k+1}-P_{k,k+1})\lvert 0\rangle=J\lvert0\rangle-J\lvert0\rangle=0.
\label{eqn:eq10}
\end{equation}
Now we flip one spin at site $n$
\begin{equation}
\lvert n\rangle=\lvert \uparrow \uparrow\dots\uparrow \overset{n}\downarrow\uparrow\dots\uparrow\rangle
\label{eqn:eq11}
\end{equation}
Analogous with Bloch wave, one can find the following eigenstate with one spin down, also fixed momentum
\begin{equation}
\lvert p\rangle=\sum_n^L\mathrm{e}^{ipn}\lvert n\rangle.
\label{eqn:eq12}
\end{equation}
This state is called a magnon state, it can be viewed as one particle excitation of the vacuum state. Act with $\mathcal{H}$ on $\lvert p\rangle$ and obtain

\begin{equation}
\begin{split}
\mathcal{H}\lvert p\rangle&=\frac{J}{2}\sum_k\mathrm{e}^{ipn}\left(\overbrace{\lvert n\rangle-\lvert n-1\rangle}^{H_{n-1,n}}+\overbrace{\lvert n\rangle-\lvert n+1\rangle}^{H_{n,n+1}}\right)\\
&=J/2\sum_n\mathrm{e}^{ipn}(1-\mathrm{e}^{ip}+1-\mathrm{e}^{-ip})\lvert n\rangle\\
&=E(p)\lvert p\rangle
\end{split}
\label{eqn:eq13}
\end{equation}
with eigenenergy

\begin{equation}
E(p)=J(1-\cos p)=2J\sin^2(\frac{1}{2}p)
\label{eqn:eq14}
\end{equation}
For a closed chain, the momentum would further be quantized by the periodic boundary condition to

\begin{equation}
p=\frac{2\pi n}{L},\quad \mathrm{with}\ n=0,\dots,L-1.
\label{eqn:eq15}
\end{equation}
A shift of $2\pi$ corresponds to a change of "Brillouin zone" which leaves the eigenstate unchanged. We continue with states with two spin flipped
\begin{equation}
\lvert n_1<n_2\rangle=\lvert \uparrow \uparrow\dots\uparrow \overset{n_1}\downarrow\uparrow\dots\uparrow \overset{n_2}\downarrow\uparrow\dots\uparrow\rangle.
\label{eqn:eq16}
\end{equation}
here we assume $n_1<n_2$. Since our periodic boundary condition, it is just a way to indicate the relative position of $n_1$ and $n_2$. Here, analogous with one magnon state, we make a plane wave eigenstate ansatz

\begin{equation}
\lvert p<q\rangle=\sum_{n_1<n_2} \mathrm{e}^{ipn_1+iqn_2}\lvert n_1<n_2\rangle.
\label{eqn:eq17}
\end{equation}
Act with $\mathcal{H} $ on $\lvert p<q\rangle$,

\begin{equation}
\begin{split}
\mathcal{H}\lvert p<q\rangle=\sum_{n_1<n_2-1} \mathrm{e}^{ipn_1+iqn_2}&[\lvert n_1,n_2\rangle-\lvert n_1-1,n_2\rangle+\lvert n_1,n_2\rangle-\lvert n_1+1,n_2\rangle\\
&+\lvert n_1,n_2\rangle-\lvert n_1,n_2-1\rangle+\lvert n_1,n_2\rangle-\lvert n_1,n_2+1\rangle]\\
+\sum_{n_1=n_2-1} \mathrm{e}^{ipn_1+iqn_2}&[\lvert n_1,n_2\rangle-\lvert n_1-1,n_2\rangle+\lvert n_1,n_2\rangle-\lvert n_1,n_2\rangle\\
&+\lvert n_1,n_2\rangle-\lvert n_1,n_2+1\rangle]\\
=\sum_{n_1<n_2} \mathrm{e}^{ipn_1+iqn_2}4\lvert &n_1,n_2\rangle-\sum_{n_1=n_2-1} \mathrm{e}^{ipn_1+iqn_2}2\lvert n_1,n_2\rangle\\
-\mathrm{e}^{ip}\sum_{n_1<n_2} \mathrm{e}^{ipn_1+iqn_2}&\lvert n_1,n_2\rangle+\mathrm{e}^{ip}\sum_{n_1=n_2-1} \mathrm{e}^{ipn_1+iqn_2}\lvert n_1,n_2\rangle\\
-\mathrm{e}^{iq}\sum_{n_1<n_2} \mathrm{e}^{ipn_1+iqn_2}&\lvert n_1,n_2\rangle-\mathrm{e}^{-ip}\sum_{n_1<n_2} \mathrm{e}^{ipn_1+iqn_2}\lvert n_1,n_2\rangle\\
-\mathrm{e}^{-iq}\sum_{n_1<n_2} \mathrm{e}^{ipn_1+iqn_2}&\lvert n_1,n_2\rangle+\mathrm{e}^{-iq}\sum_{n_1=n_2-1} \mathrm{e}^{ipn_1+iqn_2}\lvert n_1,n_2\rangle,
\end{split}
\label{eqn:eq18}
\end{equation}
then
\begin{equation}
\begin{split}
\ [&\mathcal{H}-E(p)-E(q)]\lvert p<q\rangle \\=&-\sum_{n_1=n_2-1} \mathrm{e}^{ipn_1+iqn_2}2\lvert n_1,n_2\rangle+\mathrm{e}^{ip}\sum_{n_1=n_2-1} \mathrm{e}^{ipn_1+iqn_2}\lvert n_1,n_2\rangle\\
+&\mathrm{e}^{-iq}\sum_{n_1=n_2-1} \mathrm{e}^{ipn_1+iqn_2}\lvert n_1,n_2\rangle\\
=&(\mathrm{e}^{ip+iq}-2\mathrm{e}^{iq}+1)\sum_{n_1} \mathrm{e}^{i(p+q)n_1}\lvert n_1,n_1+1\rangle.
\end{split}
\label{eqn:eq19}
\end{equation}
In Eq.(\ref{eqn:eq18}) and Eq.(\ref{eqn:eq19}), I have replaced $\lvert n_1<n_2\rangle$ with $\lvert n_1,n_2\rangle$. Use exactly the same procedure for $\lvert q<p\rangle$, one would obtain
\begin{equation}
\begin{split}
\ [&\mathcal{H}-E(p)-E(q)]\lvert q<p\rangle \\
=&(\mathrm{e}^{ip+iq}-2\mathrm{e}^{ip}+1)\sum_{n_1} \mathrm{e}^{i(p+q)n_1}\lvert n_1,n_1+1\rangle.
\end{split}
\label{eqn:eq20}
\end{equation}
Combine Eq.(\ref{eqn:eq19}) with Eq.(\ref{eqn:eq20}), we can construct an eigenstate

\begin{equation}
\lvert p,q\rangle=\lvert p<q\rangle+S(p,q)\lvert q<p\rangle
\label{eqn:eq21}
\end{equation}
with a scattering phase factor
\begin{equation}
S(p,q)=-\frac{\mathrm{e}^{ip+iq}-\mathrm{e}^{iq}+1}{\mathrm{e}^{ip+iq}-\mathrm{e}^{ip}+1}.
\label{eqn:eq22}
\end{equation}

\subsection{Coordinate Bethe Ansatz}
\begin{figure}[h]
\centering
\includegraphics{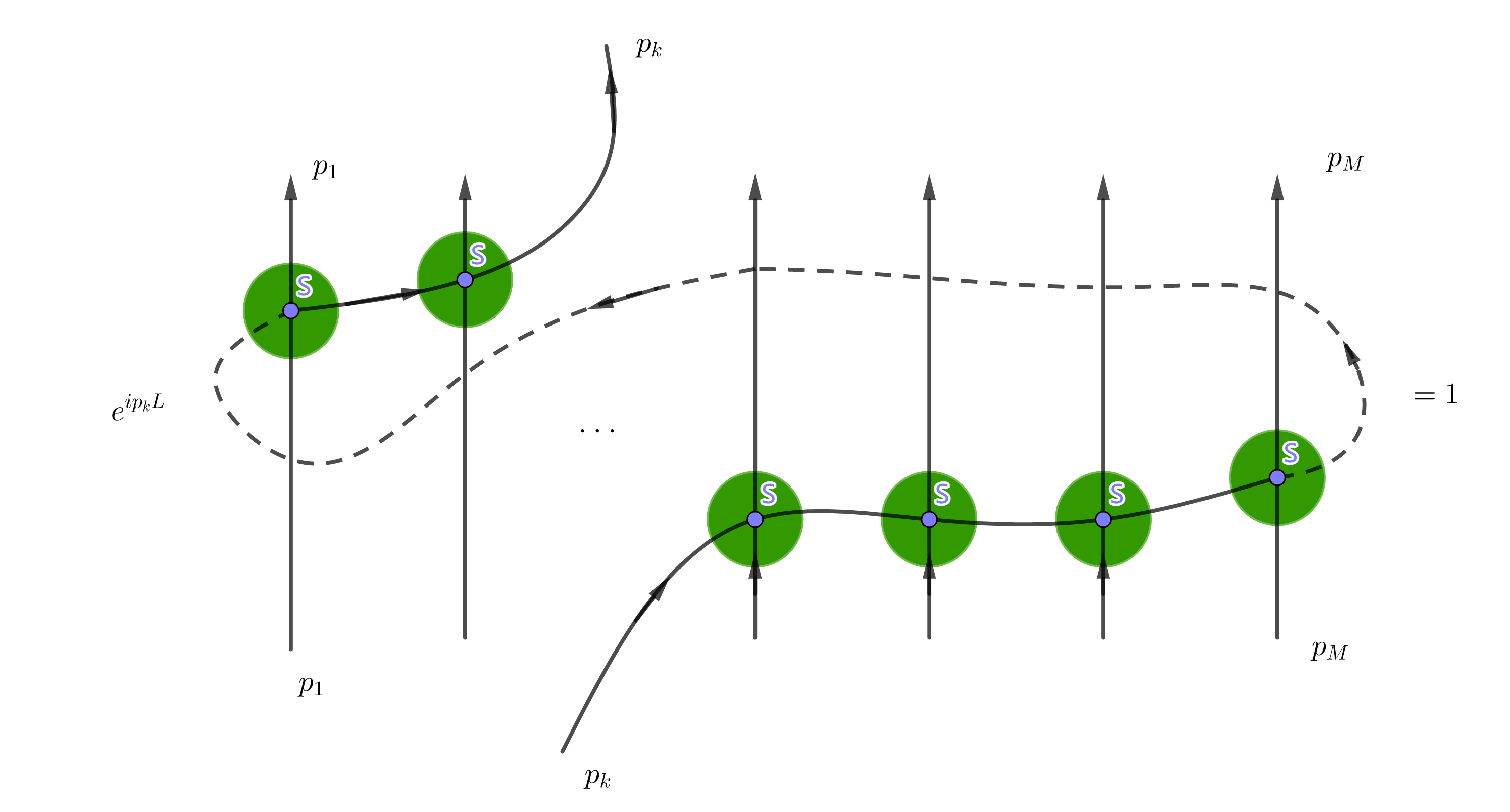}
\decoRule
\caption{Bethe equations in periodic spin chain \cite{intro_int_NB}}
\label{fig:fig001}
\end{figure}

Now let's have a look at three-magnon states, there are $6=3!$ different configurations for magnons which carry one momentum $p_n$ each. A useful ansatz for an eigenstate is the so-called $Bethe \ ansatz$
\begin{equation}
\begin{split}
\lvert p_1,p_2,p_3\rangle=&\lvert p_1<p_2<p_3\rangle+S_{12}\lvert p_2<p_1<p_3\rangle\\
&+S_{23}\lvert p_1<p_3<p_2\rangle+S_{13}S_{23}\lvert p_3<p_1<p_2\rangle\\
&+S_{12}S_{13}\lvert p_2<p_3<p_1\rangle+S_{12}S_{13}S_{23}\lvert p_3<p_2<p_1\rangle
\end{split}
\label{eqn:eq23}
\end{equation}
In this combination, all pairwise contact terms $n_1+1=n_2$ of the Hamiltonian action conserved by the construction due to the choice of appropriate pairwise scattering factors between any two partial wave function.
Normally, one should expect a triple contact term
\begin{equation}
(\mathcal{H}-E)\lvert p_1,p_2,p_3\rangle\sim \sum_k \mathrm{e}^{iPk}\lvert k<k+1<k+2\rangle.
\label{eqn:eq24}
\end{equation}
However, due to a miracle this contact term is absent without further fuss. This miracle is called $integrability$. It works for any number of magnons as we will discuss below. We only need the two-magnon scattering factor to construct all the states. This also give rise to the Yang-Baxter relation:
\begin{equation}
S_{12}S_{13}S_{23}=S_{23}S_{13}S_{12}.
\label{eqn:eq25}
\end{equation}
For closed spin chain with the periodic boundary condition, the M-magnon state must satisfy
\begin{equation}
\lvert p_1,p_2,\dots,p_M\rangle= \lvert p_2,\dots,p_M,p_1+L\rangle.
\label{eqn:eq26}
\end{equation}
This leads to the $Bethe\ equations$ for a closed chain
\begin{equation}
\mathrm{e}^{ip_kL}\prod_{j=1,j\ne k}^M S(p_k,p_j)=1, \quad \mathrm{for\ all\ k=1,\dots,M}.
\label{eqn:eq27}
\end{equation}
Graphically, the Bethe equation can be represented as Fig. ~\ref{fig:fig001}
The total energy and total momentum of the energy eigenstate can be read off from the set of magnon momenta $p_k$
\begin{equation}
E=\sum_{k=1}^{M}E(p_k),\quad P=\sum_{k=1}^{M}p_k.
\label{eqn:eq28}
\end{equation}
It's convenient to introduce a set of different variables $rapidities$ or $Bethe\ roots$ $u_k$ to replace the momenta $p_k$
\begin{equation}
Hp_k=2\,\mathrm{arccot}\, 2u_k,\quad u_k=\frac{1}{2}\cot\frac{1}{2}p_k,\quad\mathrm{e}^{ip_k}=\frac{u_k+\frac{i}{2}}{u_k-\frac{i}{2}}
\label{eqn:eq29}
\end{equation}
This transformation will kill the $2\pi$-ambiguity of $p_k$. The scattering phase factor reduces to 
\begin{equation}
S(u,v)=\frac{u-v-i}{u-v+i}.
\label{eqn:eq30}
\end{equation}
The Bethe equations take the form
\begin{equation}
\left(\frac{u_k+\frac{i}{2}}{u_k-\frac{i}{2}}\right)^L=\prod_{j=1,j\ne k}^M \frac{u_k-u_j+i}{u_k-u_j-i} \quad \mathrm{for}\ k=1,\dots,M.
\label{eqn:eq31}
\end{equation}
Each solution of the Bethe equations corresponds to an eigenstate, whose momentum and energy eigenvalues can be extracted easily via
\begin{equation}
\mathrm{e}^{iP}=\prod_{k=1}^M \frac{u_k+\frac{i}{2}}{u_k-\frac{i}{2}},\quad E=\sum_{j=1}^{M}\frac{2J}{4+u_j^2}.
\label{eqn:eq32}
\end{equation}
\section{The algebraic Bethe Ansatz}
\subsection{Lax operator and R-matrix}
In this section we will use a different approach to study the XXX Heisenberg spin chain \cite{intro_int_ETH}. In this way, our formalism is easy to generalize to other spin chain model.

The basic tool of the algebraic Bethe Ansatz\cite{Faddeev_aba} is the so-called Lax operator $L$. Consider a spin chain with $N$ sites and the corresponding Hilbert space $\mathcal{H}=\bigotimes_{n=1}^Nh_n$. $h_n$ are called local quantum space, in this model it will be $\mathbb{C}^2$. We also need to add an additional auxiliary space $V_a$, which again is $\mathbb{C} ^2$. The Lax operator is an operator $L:h_n\otimes V_a\rightarrow h_n\otimes V_a$ and for us it takes the form
\begin{equation}
L_{n,a}(u)=u\mathrm{I}_n\otimes \mathrm{I}_a +i\sum_{\alpha} S_n^{\alpha}\otimes \sigma^{\alpha},
\label{eqn:eq01}
\end{equation}
where $\sigma^{\alpha}$ are Pauli matrices
\begin{equation}
\sigma^x=\begin{pmatrix}0&1\\1&0\end{pmatrix},\quad \sigma^y=\begin{pmatrix}0&-i\\i&0\end{pmatrix},\quad\sigma^x=\begin{pmatrix}1&0\\0&-1\end{pmatrix}.
\label{eqn:eq02}
\end{equation}
For spin-$\frac{1}{2}$ $S^{\alpha}=\frac{1}{2}\sigma^{\alpha}$, $u$ is a complex parameter, called the spectral parameter. Similar to classical Lax operator, we can express Eq.(\ref{eqn:eq01}) as a matrix in auxiliary space with entries as the operator acting on the physical space
\begin{equation}
L_{n,a}(u)=\begin{pmatrix}u+iS_n^z&iS_n^-\\iS_n^+&u-iS_n^z\end{pmatrix}.
\label{eqn:eq03}
\end{equation}
Let us introduce the permutation operator
\begin{equation}
P=\frac{1}{2}(\mathrm{I}\otimes\mathrm{I}+\sum_{\alpha}\sigma^{\alpha}\otimes \sigma^{\alpha})=\begin{pmatrix}1&0&0&0\\0&0&1&0\\0&1&0&0\\0&0&0&1\end{pmatrix},
\label{eqn:eq04}
\end{equation}
and it acts in $\mathbb{C}^2 \otimes \mathbb{C}^2$
\begin{equation}
P(x\otimes y)=y\otimes x
\label{eqn:eq05}
\end{equation}
Then Lax operator yields
\begin{equation}
L_{n,a}(u)=(u-\frac{i}{2})\mathrm{I}_{n,a}+iP_{n,a}.
\label{eqn:eq06}
\end{equation}
Now, since we know the commutation relation of spin operators
\begin{equation}
[S^a,S^b]=i\epsilon^{abc}S^c,
\label{eqn:eq07}
\end{equation}
we can figure out the commutation between two Lax operators $L_{n,a_1}(u_1)$ and $L_{n,a_2}(u_2)$, with the same quantum space but different auxiliary space $V_1$ and $V_2$ respectively. Written in a compact form
\begin{equation}
R_{a,b}(u_1-u_2)L_{n,a}(u_1)L_{n,b}(u_2)=L_{n,b}(u_2)L_{n,a}(u_1)R_{a,b}(u_1-u_2),
\label{eqn:eq08}
\end{equation}
where $R$ is called the quantum R-matrix with
\begin{equation}
R_{a,b}(u)=u\mathrm{I}_{a,b}+iP_{a,b}.
\label{eqn:eq09}
\end{equation}
Relation Eq. (\ref{eqn:eq08}) is called the fundamental commutation relation, it is easily checked by using permutation operators
\begin{equation}
\begin{split}
\left((u_1-u_2)\mathrm{I}_{a,b}+iP_{a,b}\right) \left((u_1-\frac{i}{2})\mathrm{I}_{n,a}+iP_{n,a}\right)\left((u_2-\frac{i}{2})\mathrm{I}_{n,b}+iP_{n,b}\right)=\\
\left((u_2-\frac{i}{2})\mathrm{I}_{n,b}+iP_{n,b}\right)\left((u_1-\frac{i}{2})\mathrm{I}_{n,a}+iP_{n,a}\right)\left((u_1-u_2)\mathrm{I}_{a,b}+iP_{a,b}\right),
\end{split}
\label{eqn:eq010}
\end{equation}
and also relation $P_{n,a_1}P_{n,a_2}=P_{a_1,a_2}P_{n,a_1}=P_{n,a_2}P_{a_2,a_1}$.\footnote{One can directly check this by acting on $X\otimes Y\otimes Z$ in $n\otimes a_1\otimes a_2$, in the end all three get $Y\otimes Z\otimes X$. $P_{x,y}=P_{y,x}$ is trivial.} 

From the fundamental commutation relation (FCR) one can show that the R-matrix needs to satisfy the Yang-Baxter equation. Consider the following product of L-operators (omit the common $n$ index)
\begin{equation}
L_1L_2L_3=R_{12}^{-1}L_2L_1L_3R_{12}=R_{12}^{-1}R_{13}^{-1}L_2L_3L_1R_{13}R_{12}=R_{12}^{-1}R_{13}^{-1}R_{23}^{-1}L_3L_2L_1R_{23}R_{13}R_{12},
\label{eqn:eq011}
\end{equation}
on the other hand
\begin{equation}
L_1L_2L_3=R_{23}^{-1}L_1L_3L_2R_{23}=R_{23}^{-1}R_{13}^{-1}L_3L_1L_2R_{13}R_{23}=R_{23}^{-1}R_{13}^{-1}R_{12}^{-1}L_3L_2L_1R_{23}R_{13}R_{12}.
\label{eqn:eq012}
\end{equation}
Therefore, for both these two equations to be satisfied, we need to impose that our R-matrix 
\begin{equation}
R_{12}R_{13}R_{23}=R_{23}R_{13}R_{12}.
\label{eqn:eq013}
\end{equation}
This is the so-called Yang Baxter equation, and it tells us that our model is indeed integrable.
\subsection{Monodromy and transfer matrix}
The spin system has $N$ sites. Then, using the Lax operator to define the monodromy matrix
\begin{equation}
T_a(u)=L_{N,a}(u)\dots L_{1,a}(u). 
\label{eqn:eq014}
\end{equation}
It can be seen as a Wilson loop on auxiliary space whose entries are operators act on the physical Hilbert space. Write it as 
\begin{equation}
T(u)=\begin{pmatrix}A(u)&B(u)\\ C(u)& D(u)\end{pmatrix}.
\label{eqn:eq015}
\end{equation}
From this monodromy matrix we can generate the tower of the conserved charges that characterize integrable systems.
First we need to find the commutation relations between the different entries of the monodromy matrix. It can be easily derived by repeated using Eq. (\ref{eqn:eq08}) for Lax operator. We find
\begin{equation}
R_{ab}(u_1-u_2)T_a(u_1)T_b(u_2)=T_b(u_2)T_a(u_1)R_{ab}(u_1-u_2)
\label{eqn:eq016}
\end{equation}
It's convenient to introduce the transfer matrix
\begin{equation}
t(u)=\mathrm{Tr}_aT_a(u)=A(u)+D(u)
\label{eqn:eq017}
\end{equation}
Rewrite the Eq. (16) as follows 
\begin{equation}
T_{a_1}(u_1)T_b(u_2)=R_{ab}(u_1-u_2)^{-1}T_b(u_2)T_a(u_1)R_{ab}(u_1-u_2),
\label{eqn:eq018}
\end{equation}
and then take the trace over the auxiliary space, by the cyclicity of the trace one find
\begin{equation}
[t(u_1),t(u_2)]=0.
\label{eqn:eq019}
\end{equation}
Since the transfer matrix depends on a spectral parameter, we can make a Taylor expansion in $u_1$ and $u_2$, generating a set of commutating operators for different powers. Moreover from classical counterpart, the Hamiltonian will be one of them.
In order to see it's indeed so, we expand the monodromy matrix around $u=\frac{i}{2}$. To the zeroth order, 
\begin{equation}
T_a(\frac{i}{2})=i^N P_{N,a}P_{N-1,a}\dots P_{1,a}=i^NP_{12}P_{23}\dots P_{N-1,N}P_{N,a},
\label{eqn:eq020}
\end{equation}
where we have used the properties of permutation operator listed in below Eq. (~\ref{eqn:eq010}). Now take the trace and use $\mathrm{tr}_a P_{N,a}=\mathrm{I}_N$, one can achieve 
\begin{equation}
i^{-N}t(\frac{i}{2})=P_{12}P_{23}\dots P_{N-1,N}=\mathcal{U},
\label{eqn:eq021}
\end{equation}
which is simply the shift operator. By definition, momentum $P$ produces an infinitesimal shift and one can identify
\begin{equation}
\mathrm{e}^{iP}=\mathcal{U}
\label{eqn:eq022}
\end{equation}
Proceed with taking the next order expansion

\begin{equation}
\begin{split}
    \frac{dT_a}{du}\biggr\rvert_{u=\frac{i}{2}}&=i^{N-1}\sum_{n=1}^N P_{N,a}\dots\hat{P}_{n,a}\dots P_{1,a}\\
&=i^{N-1}\sum_{n=1}^N P_{12}P_{23}\dots\hat{P}_{n-1,n+1}\dots P_{N-1,N}P_{N,a},
\end{split}
\label{eqn:eq023}
\end{equation}
where $\hat{P}$ means that the corresponding term is absent. Use the result in Eq. (\ref{eqn:eq021}), take the trace
\begin{equation}
\left(t(u)^{-1}\frac{dt(u)}{du}\right)\biggr\rvert_{u=\frac{i}{2}}=\frac{d}{du}\ln t(u)\biggr\rvert_{u=\frac{i}{2}}=\frac{1}{i}\sum_n P_{n,n+1}.
\label{eqn:eq024}
\end{equation}
Then the Hamiltonian can be rewritten in terms of the above quantity,
\begin{equation}
H=\frac{NJ}{2}-\frac{J}{2}\sum_nP_{n,n+1},
\label{eqn:eq025}
\end{equation}
So as promised,we can indeed generate the Hamiltonian from the transfer matrix.Actually it will spit out $N-1$ commuting operators, plus the total spin component $Q_{\alpha}$, completes the N commuting operator. In classical language, there are $N$ operator in involution, so as a classical counterpart of this system. 
\subsection{Algebraic Bethe Ansatz}
What remains is to find the spectrum of the Hamiltonian. From the transfer matrix, we figure out a family of commuting operators, now we are going to diagonalize them simultaneously. In particular, we would like to find the eigenvalue of the transfer matrix directly. Since Hamiltonian can be constructed from transfer matrix, as a matter of fact we have worked out the spectrum of the Hamiltonian. Even more we have found all the eigenvalues of all the commuting conserved charges.
Again we assume that the state where all spin are flipped up
\begin{equation}
\lvert 0\rangle=\lvert \uparrow \uparrow\dots\uparrow\uparrow\rangle,
\label{eqn:eq026}
\end{equation}
it is easy to show directly by Eq. (\ref{eqn:eq03}) and Eq. (\ref{eqn:eq014}) that
\begin{equation}
T(u)\lvert 0 \rangle=\begin{pmatrix}\alpha^N(u)&*\\0&\delta^N(u)\end{pmatrix}\lvert 0 \rangle,
\label{eqn:eq027}
\end{equation}
where
\begin{equation}
\alpha(u)=u+\frac{i}{2},\quad \delta(u)=u-\frac{i}{2}.
\label{eqn:eq028}
\end{equation}
In other words we have
\begin{equation}
\begin{split}
A\lvert 0 \rangle&=\prod_n(u+iS_n^z)\lvert 0\rangle+0=\alpha^N(u)\lvert 0\rangle,\\D\lvert 0 \rangle&=\prod_n(u-iS_n^z)\lvert 0\rangle+0=\delta^N(u)\lvert 0\rangle,\\C\lvert 0 \rangle&=0.
\end{split}
\label{eqn:eq029}
\end{equation}
Hence, $\lvert 0 \rangle$ is an eigenstate of the transfer matrix with eigenvalue 
\begin{equation}
t\lvert 0 \rangle=(A+D)\lvert 0 \rangle=[\alpha^N(u)+\delta^N(u)]\lvert 0 \rangle.
\label{eqn:eq030}
\end{equation}
We have found the reference state $\lvert 0 \rangle $, and now we need to a way to flip down the spin. It turns out to be $B$ operator. 
To prove $B$ play the role as flipping spin, first we need to know the commutation relation between different entries of monodromy matrix. In space $V_1\otimes V_2$ (again we drop the quantum space), according to Eq. (\ref{eqn:eq015}), the monodromy matrix yields

\begin{equation}
\begin{split}
T_1(u)&=T_{n,1}\otimes\mathrm{I}_2=\begin{pmatrix}A(u)&0&B(u)&0\\0&A(u)&0&B(u)\\C(u)&0&D(u)&0\\0&C(u)&0&D(u)\end{pmatrix},\\ T_2(u)&=\mathrm{I}_1\otimes T_{n,2}=\begin{pmatrix}A(u)&B(u)&0&0\\C(u)&D(u)&0&0\\0&0&A(u)&B(u)\\0&0&C(u)&D(u)\end{pmatrix}.
\end{split}
\label{eqn:eq031}
\end{equation}
And the R-matrix takes the form
\begin{equation}
R_{1,2}(u-v)=(u-v)\mathrm{I}_{1,2}+iP_{1,2}\\=\begin{pmatrix}u-v+i&0&0&0\\0&u-v&i&0&\\0&i&u-v&0\\0&0&0&u-v+i\end{pmatrix}.
\label{eqn:eq032}
\end{equation}
The fundamental relation Eq. (\ref{eqn:eq016}) implies the following relations between the operators 
\begin{equation}
\begin{split}
\ &[B(u),B(v)]=0,\\
&A(u)B(v)=f(u-v)B(v)A(u)+g(u-v)B(u)A(v),\\
&D(u)B(v)=h(u-v)B(v)D(u)+k(u-v)B(u)D(v).
\end{split}
\label{eqn:eq033}
\end{equation}
where
\begin{equation}
f(u)=\frac{u-i}{u},\quad g(u)=\frac{i}{u}\\
h(u)=\frac{u+i}{u},\quad k(u)=-\frac{i}{u}
\label{eqn:eq034}
\end{equation}
From this commutation relation, we could interpret $B$ as a creation operators that create magnons. To confirm our guess, let us first make an ansatz 
\begin{equation}
\lvert u_1,u_2,\dots,u_M\rangle=B(u_1)B(u_2)\dots B(u_M)\lvert 0\rangle.
\label{eqn:eq035}
\end{equation}
The condition that the above state is an eigenstate of transfer matrix will impose a set of constraints on the spectral parameter $u_i$, and the constraints are the Bethe equations we have found in coordinate Bethe ansatz.
Now we need to verify that Eq. (\ref{eqn:eq035}) is indeed an eigenstate of the transfer matrix. Under relation of Eq. (\ref{eqn:eq033}) we get
\begin{equation}
\begin{split}
A(u)\lvert u_1,\dots,u_M\rangle &= \alpha(u)^N\prod_{n=1}^{M}f(u-u_n)\lvert u_1,\dots,u_M\rangle\\&+\sum_n^M W_{n}^{A}(u,u_n)\lvert \dots u_{n-1},u,u_{n+1}\dots\rangle,
\end{split}
\label{eqn:eq036}
\end{equation}
for some coefficient $W_n^A$, the first element comes from moving $A(u)$ to right without exchange spectral parameter, the second one corresponds to a combination of $2^M-1$ terms. But with the commutation relation of operator $B$ and $g(u)$ is an odd function, i.e. 

\begin{equation}
\begin{split}
A(u)B(v)B(w)&=\left(f(u-v)B(v)A(u)+g(u-v)B(u)A(v)\right)B(w)\\
            &=f(u-v)f(u-w)B(v)B(w)A(u)+f(u-v)g(u-w)B(v)B(u)A(w)\\
            &+g(u-v)f(v-w)B(u)B(w)A(v)+g(u-v)g(v-w)B(u)B(v)A(w)\\
            &=f(u-v)f(u-w)B(v)B(w)A(u)+f(u-v)g(u-w)B(v)B(u)A(w)\\
            &+g(u-v)f(v-w)B(u)B(w)A(v)+0,
\end{split}
\label{eqn:eq037}
\end{equation}
one can easily observe that we have only $M$ terms left in the second term of R.H.S. of Eq. (\ref{eqn:eq036}). That is
\begin{equation}
W_{n}^A(u,u_n)=\alpha^N(u_n) g(u-u_n)\prod_{k\ne n}^{M}f(u_n-u_k)
\label{eqn:eq038}
\end{equation}
The discussion of $D$ operator is completely the same, and we obtain
\begin{equation}
\begin{split}
D(u)\lvert u_1,\dots,u_M\rangle &= \delta(u)^N\prod_{n=1}^{M}h(u-u_n)\lvert u_1,\dots,u_M\rangle\\
&+\sum_n^M W_{n}^{D}(u,u_n)\lvert \dots u_{n-1},u,u_{n+1}\dots\rangle,
\end{split}
\label{eqn:eq039}
\end{equation}
where
\begin{equation}
W_{n}^A(u,u_n)=\delta^N(u_n) k(u-u_n)\prod_{k\ne n}^{M}h(u_n-u_k)
\label{eqn:eq040}
\end{equation}
Combine Eq. (\ref{eqn:eq038}) and Eq. (\ref{eqn:eq039}), one would get
\begin{equation}
(A(u)+D(u))\lvert u_1,\dots u_M\rangle=\Lambda(u)\lvert u_1,\dots,u_M\rangle,
\label{eqn:eq041}
\end{equation}
with
\begin{equation}
\Lambda(u)=\alpha(u)^N\prod_{n=1}^{M}f(u-u_n)+\delta(u)^N\prod_{n=1}^{M}h(u-u_n),
\label{eqn:eq042}
\end{equation}
provided $W_{n}^A(u,u_n)+W_{n}^D(u,u_n)$ term vanishing
\begin{equation}
\prod_{k\ne n}^M f(u_n-u_k)\alpha^N(u_n)=\prod_{k\ne n}^M h(u_n-u_k)\delta^N(u_n)
\label{eqn:eq043}
\end{equation}
for $j=1,\dots,M$. Substituting Eq. (\ref{eqn:eq028}) and Eq. (\ref{eqn:eq034}) into Eq. (\ref{eqn:eq043})
\begin{equation}
\left(\frac{u_n+\frac{i}{2}}{u_n-\frac{i}{2}}\right)^N=\prod_{j\ne n}^{M}\frac{u_n-u_j+i}{u_n-u_j-i},\quad\mathrm{with}\quad n=1,\dots,M,
\label{eqn:eq044}
\end{equation}
 which is exactly the same Bethe equation we get from coordinate Bethe Ansatz. 
Last but not least, we need to find the spectrum. We have already found the explicit expression for the momentum and energy in Eq. (\ref{eqn:eq022}) and Eq. (\ref{eqn:eq025})
\begin{equation}
\mathrm{e}^{iP}=U=i^{-N}t(\frac{i}{2})=i^{-N}\Lambda(\frac{i}{2}). 
\label{eqn:eq045}
\end{equation}
Plugging the eigenvalue in Eq. (\ref{eqn:eq042}), the above equation yields
\begin{equation}
P=\sum_jp_j,\quad \mathrm{e}^{ip_j}=\frac{u_j+\frac{i}{2}}{u_j-\frac{i}{2}}\quad\Leftrightarrow\quad u_j=\frac{1}{2}\cot \frac{p_j}{2}
\label{eqn:eq046}
\end{equation}
This is the same relation we have encounter in coordinate Bethe Ansatz. The energy is same procedure
\begin{equation}
E=\frac{NJ}{2}-\frac{J}{2}\left(i\frac{d}{du}\ln \Lambda(u)\right)\biggr\rvert_{u=\frac{i}{2}}=\sum_{j=1}^{M}\frac{2J}{4+u_j^2}
\label{eqn:eq047}
\end{equation}

\section{The norm and Gaudin matrix}
\label{sec2.3}
In this section we will prove a useful expression for the norm of Bethe states \cite{Korepin1982CalculationON,Slavnov1989CalculationOS}, and with possible generalization. The results are prepared for Chapter \ref{Chapter3}.

\subsection{Inner product between off-shell Bethe state and Bethe state}

As we have already show in Algebraic Bethe ansatz, one of the key ingredients is the monodromy matrix $T(\lambda)$ Eq. (\ref{eqn:eq015})
\begin{equation}
T(\lambda)=\begin{pmatrix}A(\lambda)&B(\lambda)\\C(\lambda)&D(\lambda)
\end{pmatrix}.
\label{eqn:eq301}
\end{equation}
And the commutation relations between the entry of the monodromy matrix are specified by FCR Eq. (\ref{eqn:eq016})

\begin{equation}
R_{a,b}(\lambda,\mu)T_a(\lambda)T_b(\mu)=T_b(\mu)T_a(\lambda)R_{a,b}(\lambda,\mu)
\label{eqn:eq302}
\end{equation}

For this moment, we focus on the R-matrix
\begin{equation}
R_{a,b}(\lambda,\mu)=\begin{pmatrix}f(\mu,\lambda)&0&0&0\\0&1&g(\mu,\lambda)&0\\0&g(\mu,\lambda)&1&0\\0&0&0&f(\mu,\lambda)\end{pmatrix},
\label{eqn:eq303}
\end{equation}
where $f(\mu,\lambda)=\frac{\mu-\lambda + i}{\mu-\lambda}$, $g(\mu,\lambda)=\frac{i}{\mu-\lambda}$ \footnote{Note that this R-matrix is actually the same as the one we define in last section, $R_{new}(u)=-\frac{1}{u}R_{old}(-u)$, an overall scalar function gives us the same Hamiltonian, and the eigenvalues in Eq. (\ref{eqn:eq306}) become $a(u)=(-\frac{1}{u})^L(-u+\frac{i}{2})^L$, $d(u)=(-\frac{1}{u})^L(-u-\frac{i}{2})^L$}.  It is convenient to define
\begin{equation}
h(\mu,\lambda)=\frac{f(\mu,\lambda)}{g(\mu,\lambda)}=\frac{\mu-\lambda+i}{i}.
\label{eqn:eq304}
\end{equation}
From FCR Eq. (\ref{eqn:eq302}), one can get the part of commutation relations
\begin{equation}
\begin{split}
&[B(\lambda),B(\mu)]=[C(\lambda),C(\mu)]=0,\\&[C(\mu),B(\lambda)]=g(\mu,\lambda)(A(\mu)D(\lambda)-A(\lambda)D(\mu)).
\end{split}
\label{eqn:eq305}
\end{equation}
Moreover, $B(\lambda )$ is the Hermitian conjugate of  $C(\lambda)$.One also need to specify how the operators in monodromy matrix acting on the vacuum state
\begin{equation}
\begin{split}
    A(\lambda)\lvert 0\rangle=a(\lambda)\lvert 0 \rangle,\quad D(\lambda)\lvert 0\rangle=d(\lambda)\lvert 0 \rangle,\quad C(\lambda)\lvert 0\rangle= 0,\\
\langle 0\rvert A(\lambda)=a(\lambda)\langle 0\rvert,\quad \langle 0\rvert D(\lambda)=d(\lambda)\langle 0\rvert,\quad \langle 0\rvert B(\lambda)= 0,
\end{split}
\label{eqn:eq306}
\end{equation}
where $a(\lambda)$ and $d(\lambda)$ are complex function. From now on, if there are further restriction on $\lambda$, we shall regard $\lambda $ and $r(\lambda)=\frac{a(\lambda)}{d(\lambda)}=\Big(\frac{-\lambda+\frac{i}{2}}{-\lambda-\frac{i}{2}}\Big)^L$ as independent variables.
The eigenfunctions of the Hamiltonian are constructed by means of the operators $B(\lambda)$:
\begin{equation}
\Psi_{N}(\{\lambda_j\})=\prod_{j=1}^{N}B(\lambda_j)\lvert 0 \rangle,
\label{eqn:eq307}
\end{equation}
and the parameters $\lambda_j$ constraint on shell (or Bethe equations)
\begin{equation}
r(\lambda_j)\prod_{k=1,k\ne j}\frac{f(\lambda_j,\lambda_k)}{f(\lambda_k,\lambda_j)}=1,\quad j=1,\dots,N
\label{eqn:eq308}
\end{equation}
Also the dual eigenfunction have the same form by means of the operator $C(\lambda )$:
\begin{equation}
\bar{\Psi}_N(\{\lambda_i\})=\langle 0\rvert\prod_{j=1}^{N}C(\lambda_j),
\label{eqn:eq309}
\end{equation}
where $\lambda_j $ should be also on shell as Eq. (\ref{eqn:eq308}).

Now we consider the scalar product of two general states, which are not necessary the eigenstates of the Hamiltonian. Or in other words, the spectral parameters doesn't need to satisfy on shell condition Eq. (\ref{eqn:eq308}).  The scalar product is defined by
\begin{equation}
S_N=\langle 0\rvert\prod_{j=1}^{N}\mathbf{C}(\lambda_j^C)\prod_{j=1}^N\mathbf{B}(\lambda_j^B)\lvert0\rangle,
\label{eqn:eq310}
\end{equation}
with $\mathbf{C}(\lambda)=C(\lambda)/d(\lambda)$, $\mathbf{B}(\lambda)=B(\lambda)/d(\lambda)$. It is believed that $S_N$ is a rational function of $4N$ independent variables $S_N=S_N(\{\lambda_j^C\},\{\lambda_j^B\},\{r_j^C\},\{r_j^B\})$, where $r_j^{B,C}=r(\lambda_j^{B,C})$.The points $\lambda_k^C=\lambda_m^B,\quad k,m=1,\dots,N$ are simple poles of $S_N$, when approaching to these points, we would surprisingly get a recursion relation
\begin{equation}
\begin{split}
    S_N(\{\lambda_j^C\},\{\lambda_j^B\},&\{r_j^C\},\{r_j^B\})\bigg\rvert_{\lambda_k^C\rightarrow\lambda_m^B}=S_{N-1}(\{\lambda_j^C\}_{j\ne k},\{\lambda_j^B\}_{j\ne m},\{\tilde{r}_j^C\}_{j\ne k},\{\tilde{r}_j^B\}_{j\ne m})\\&\times g(\lambda_k^C,\lambda_m^B)(r_k^C-r_m^B)\prod_{j=1, j\ne k}f(\lambda_k^C,\lambda_j^C)\prod_{j=1, j\ne k}f(\lambda_m^B,\lambda_j^B)
\end{split}
\label{eqn:eq311}
\end{equation}
where
\begin{equation}
\tilde{r}_j^{B,C}=r_j^{B,C}\frac{f(\lambda_j^{B,C},\lambda_m^B)}{f(\lambda_m^B,\lambda_j^{B,C})}.
\label{eqn:eq312}
\end{equation}
As indicated by $g(\lambda_k^C,\lambda_m^B)$ in Eq. (\ref{eqn:eq311}), that $S_N\sim\frac{1}{\lambda_{k}^{C}}$ as $\lambda_k^{C}\rightarrow \infty$ (holds same for $\lambda_m^B$). To get the norm of two on shell states, first let us pay attention to a special case with the dual wave function $\langle 0\rvert \prod_{j=1}^N \mathbf{C}(\lambda_j^C)$ on shell as in Eq. (\ref{eqn:eq308}). 
\begin{equation}
r^C(\lambda_j)\prod_{k=1,k\ne j}\frac{f(\lambda^C_j,\lambda^C_k)}{f(\lambda^C_k,\lambda^C_j)}=1,\quad j=1,\dots,N
\label{eqn:eq313}
\end{equation}
The scalar product $\tilde{S}_N$ depends on $3N$ independent variables instead of $4N$, yields
\begin{equation}
\tilde{S}_N(\{\lambda_j^C\},\{\lambda_j^B\},\{r_j^B\})=\langle 0\rvert\prod_{j=1}^{N}\mathbf{C}(\lambda_j^C)\prod_{j=1}^N\mathbf{B}(\lambda_j^B)\lvert0\rangle.
\label{eqn:eq314}
\end{equation}
We are ready to prove the key point of this section, the $\tilde{S}_N $ is given by
\begin{equation}
\tilde{S}_N=G_N(\{\lambda_j^C\},\{\lambda_j^B\})\det_N M_{lk}(\{\lambda_j^C\},\{\lambda_j^B\},\{r_j^B\}),
\label{eqn:eq315}
\end{equation}
where
\begin{equation}
G_N(\{\lambda_j^C\},\{\lambda_j^B\})=\prod_{j>k}^N g(\lambda_j^B,\lambda_k^B)g(\lambda_k^C,\lambda_j^C)\prod_{j=1}^N\prod_{k=1}^N h(\lambda_j^C,\lambda_k^B),
\label{eqn:eq316}
\end{equation}
when $N=1$ we set the first product term equal to one and
\begin{equation}
M_{lk}(\{\lambda_j^C\},\{\lambda_j^B\},\{r_j^B\})=\frac{g(\lambda_k^C,\lambda_l^B)}{h(\lambda_k^C,\lambda_l^B)}-r_l^B\frac{g(\lambda_l^B,\lambda_k^C)}{h(\lambda_l^B,\lambda_k^C)}\prod_{m=1}^{N}\frac{f(\lambda_l^B,\lambda_l^C)}{f(\lambda_k^C,\lambda_l^B)}.
\label{eqn:eq317}
\end{equation}
$Proof$: Now we prove this relation by induction. For future convenience, denote $\Theta_N=\tilde{S}_N\bigg\rvert_{\lambda_N^C\rightarrow\lambda_m^B}$. We consider $\Theta_N$ as the function of $\lambda_N^C$. It is obvious that $\Theta_N $ is a rational function of $\lambda_N^C$, behaves like $\Theta_N\sim\frac{1}{\lambda_{N}^{C}}$ as $\lambda_N^{C}\rightarrow \infty$. For $N=1$, $\tilde {S}_N $ can be calculated by relation Eq. (\ref{eqn:eq305}):
\begin{equation}
\begin{split}
    \tilde{S}_1&=\langle 0\rvert\mathbf{C}(\lambda^C)\mathbf{B}(\lambda^B)\lvert0\rangle=\frac{1}{d(\lambda^C)d(\lambda^B)}\langle 0\rvert [C(\lambda^C),B(\lambda^B)]\lvert0\rangle\\
&=\frac{g(\lambda^C,\lambda^B)}{d(\lambda^C)d(\lambda^B)}\langle 0\rvert A(\lambda^C)D(\lambda^B)-A(\lambda^B)D(\lambda^C)\lvert0\rangle\\
&=\frac{g(\lambda^C,\lambda^B)}{d(\lambda^C)d(\lambda^B)}\left(a(\lambda^C)d(\lambda^B)-a(\lambda^B)d(\lambda^C)\right)\\
&=g(\lambda^C,\lambda^B)(r^C-r^B)\\
&=g(\lambda^C,\lambda^B)(1-r^B),
\end{split}
\label{eqn:eq318}
\end{equation}
where the last line we used the periodic boundary condition. For one magnon, there is no scattering phase factor, the Bethe equation reduces to
\begin{equation}
r^C=\mathrm{e}^{ipL}=1.
\label{eqn:eq319}
\end{equation}
Where on the other hand, $\Theta_1$ can be calculated as in formula Eq. (\ref{eqn:eq315}). First, set the limit $\lambda^B=\lambda^C$, then
\begin{equation}
\begin{split}
    M_{11}(\{\lambda^C\},\{\lambda^B\},\{r^B\})\big\rvert_{\lambda^C\rightarrow\lambda^B}= \frac{g(\lambda^C,\lambda^B)}{h(\lambda^C,\lambda^B)}-r^B\frac{g(\lambda^B,\lambda^C)}{h(\lambda^B,\lambda^C)}\times(-1)+\mathrm{small\ terms}.
\end{split}
\label{eqn:eq320}
\end{equation}
Then the norm at $N=1$ becomes 
\begin{equation}
\begin{split}
    \Theta_1&=[G_1M_{11}]\big\rvert_{\lambda^C\rightarrow\lambda^B}\\
&=g(\lambda^C,\lambda^B)+r^B g(\lambda^B,\lambda^C)\frac{h(\lambda^C,\lambda^B)}{h(\lambda^B,\lambda^C)}+\mathrm{small\ terms}\\
&=g(\lambda^C,\lambda^B)(1-r^B)+\mathrm{small\ terms},
\end{split}
\label{eqn:eq321}
\end{equation}
where the last step we use the fact $g(\lambda)$ is an odd function and $\frac{h(\lambda^C,\lambda^B)}{h(\lambda^B,\lambda^C)}\bigg\rvert_{\lambda^C\rightarrow\lambda^B}=1$. From Eq. (\ref{eqn:eq321}) and Eq. (\ref{eqn:eq318}), we can set $\Theta_1=\tilde{S}_1+\Delta$. Since both $\Theta_1\rightarrow\frac{1}{\lambda^C}$ and $\tilde{S}_1\rightarrow\frac{1}{\lambda^C}$ as $\lambda^C \rightarrow \infty$, we can conclude $\Delta = 0$, hence $\Theta_1=\tilde{S}_1$. 

Since the form of $G_1$ doesn't not fall into the form of $G_i,\ i=2,\dots,N$, we need to prove it indeed gives us the right recursion relation. First let us calculate the recursion relation for $G_N,\ N>2$ with the limit $\lambda_N^C=\lambda_m^B$
\begin{equation}
G_{N-1}(\{\lambda_j^C\}_{j\ne N},\{\lambda_j^B\}_{j\ne m})=\prod_{j>k;k,j\ne m}^{N-1}g(\lambda_j^B,\lambda_k^B)g(\lambda_k^C,\lambda_j^C)\prod_{j=1}^{N-1}\prod_{k=1,k\ne m}h(\lambda_j^C,\lambda_k^B)
\label{eqn:eq322}
\end{equation}
then
\begin{equation}
\begin{split}
    G_N(\{\lambda_j^C\},\{\lambda_j^B\})\big\rvert_{\lambda_N^C\rightarrow\lambda_m^B}
&=\prod_{j=1,j\ne m}^{N}g(\lambda_m^B,\lambda_j^B)(-1)^{N-m}\prod_{j=1}^{N-1}g(\lambda_j^C,\lambda_N^C)\\
&\times\prod_{k=1,k\ne m}^{N}h(\lambda_N^C,\lambda_k^B)\prod_{j=1}^{N-1}h(\lambda_j^C,\lambda_m^B)h(\lambda_N^C,\lambda_m^B)\\
&\times G_{N-1}(\{\lambda_j^C\}_{j\ne N},\{\lambda_j^B\}_{j\ne m}).
\end{split}
\label{eqn:eq323}
\end{equation}
We consider $N=2,m=2$ case, above equation becomes
\begin{equation}
G_2=g(\lambda_2^B,\lambda_1^B)g(\lambda_1^C,\lambda_2^C)(-1)^0h(\lambda_2^C,\lambda_1^B)h(\lambda_1^C,\lambda_2^B)h(\lambda_2^C,\lambda_2^B)G_1(\lambda_1^C,\lambda_1^B),
\label{eqn:eq324}
\end{equation}
on the other hand, from Eq. (\ref{eqn:eq316}) we can also get an expression for $G_2$. For these two to coincide, we must set
\begin{equation}
G_1=h(\lambda_1^C,\lambda_1^B)
\label{eqn:eq325}
\end{equation}
For $N=2$, $m=1$, $G_1$ take similar form as in Eq. (16). Consequently, $N=1$ indeed can be the starting point for our induction procedure. 
Let $\Theta_{N-1}=S_{N-1}$. Now we are going to prove that $\Theta_N =\tilde{S}_N$, begin with
\begin{equation}
\begin{split}
    \det_N M_{lk}(\{\lambda_j^C\},\{\lambda_j^B\},\{r_j^B\})\big\rvert_{\lambda_N^C\rightarrow\lambda_m^B}&=\det_{N-1}M_{lk}(\{\lambda_j^C\}_{j\ne m},\{\lambda_j^B\}_{j\ne N},\{\tilde{r}_j^B\}_{j\ne m})\\
&\times(-1)^{N+m}M_{mN}+\mathrm{small\ terms}.
\end{split}
\label{eqn:eq326}
\end{equation}
then 
\begin{equation}
\begin{split}
    \Theta_N\big\rvert_{\lambda_N^C\rightarrow\lambda_m^B}&=\left[G_N(\{\lambda_j^C\},\{\lambda_j^B\})\times\det_N M_{lk}(\{\lambda_j^C\},\{\lambda_j^B\},\{r_j^B\})\right]\big\rvert_{\lambda_N^C\rightarrow\lambda_m^B}\\
&=A\Theta_{N-1}+\mathrm{small\ terms},
\end{split}
\label{eqn:eq327}
\end{equation}
with
\begin{equation}
\begin{split}
    A&=(-1)^{2N}h(\lambda_N^C,\lambda_m^B)\prod_{j=1,j\ne m}^{N}f(\lambda_m^B,\lambda_j^B)\prod_{j=1}^{N-1}f(\lambda_N^C,\lambda_m^B)\\&\times\left\{ \frac{g(\lambda_N^C,\lambda_m^B)}{h(\lambda_m^B,\lambda_N^C)}-r_m^B\frac{g(\lambda_m^B,\lambda_N^C)}{h(\lambda_m^B,\lambda_N^C)}\prod_{j=1}^{N}\frac{f(\lambda_m^B,\lambda_j^C)}{f(\lambda_j^C,\lambda_m^B)}\right\}\\&=\prod_{j=1,j\ne m}^{N}f(\lambda_m^B,\lambda_j^B)\prod_{j=1}^{N-1}f(\lambda_j^C,\lambda_N^C)g(\lambda_N^C,\lambda_m^B)\times\left(1+r_m^B\prod_{j=1}^N\frac{f(\lambda_m^B,\lambda_j^C)}{f(\lambda_j^C,\lambda_m^B)}\right)\\&=\prod_{j=1,j\ne m}^{N}f(\lambda_m^B,\lambda_j^B)\prod_{j=1}^{N-1}f(\lambda_N^C,\lambda_j^C)g(\lambda_N^C,\lambda_m^B)\\&\times\left(\prod_{j=1}^{N-1}\frac{f(\lambda_j^C,\lambda_N^C)}{f(\lambda_N^C,\lambda_j^C)}+r_m^B\frac{f(\lambda_m^B,\lambda_N^C)}{f(\lambda_N^C,\lambda_m^B)}\prod_{j=1}^{N-1}\frac{f(\lambda_m^B,\lambda_j^C)}{f(\lambda_j^C,\lambda_m^B)}\prod_{j=1}^{N-1}\frac{f(\lambda_j^C,\lambda_N^C)}{f(\lambda_N^C,\lambda_j^C)}\right)\\&=\prod_{j=1,j\ne m}^{N}f(\lambda_m^B,\lambda_j^B)\prod_{j=1}^{N-1}f(\lambda_N^C,\lambda_j^C)g(\lambda_N^C,\lambda_m^B)\times(r_N^C-r_m^B)
\end{split}
\label{eqn:eq328}
\end{equation}
Combine Eq. (\ref{eqn:eq327}) and Eq. (\ref{eqn:eq328}), 
\begin{equation}
\begin{split}
    \Theta_N\big\rvert_{\lambda_N^C\rightarrow\lambda_m^B}&=
g(\lambda_N^C,\lambda_m^B)\prod_{j=1,j\ne m}^{N}f(\lambda_m^B,\lambda_j^B)\prod_{j=1}^{N-1}f(\lambda_N^C,\lambda_j^C)\times(r_N^C-r_m^B)\\
&\times\Theta_{N-1}(\{\lambda_j^C\}_{j\ne N},\{\lambda_j^B\}_{j\ne m},\{\tilde{r}_j^B\}_{j\ne m})+\mathrm{small\ terms},
\end{split}
\label{eqn:eq329}
\end{equation}
compare it with Eq. (\ref{eqn:eq311}), using the same argument as $\tilde{S}_1 $. Since both $\Theta_N\rightarrow\frac{1}{\lambda_N^C}$ and $\tilde{S}_N\rightarrow\frac{1}{\lambda_N^C}$ as $\lambda_N^C \rightarrow \infty$, that the difference $\Delta=\tilde{S}_N-\Theta_N$ as a function of $\lambda_N^C $, is equal to zero identically. This completes the proof.

\subsection{Norm in on-shell limit}
From the formula for the scalar product $\tilde{S}_N$ we can also readily construct the square of the norm of an eigenfunction of the Hamiltonian. Set $\lambda_j^B=\lambda_j^C+\epsilon,\ \epsilon\rightarrow 0$. Now let us calculate $G_N$ and $M_{lk}$ separately in this limit.

The following limit in the leading order turns out to be useful for our later calculation:
\begin{equation}
\begin{split}
&f(\lambda,\lambda+\epsilon)=1-\frac{i}{\epsilon},\quad g(\lambda,\lambda+\epsilon)=-\frac{i}{\epsilon},\quad h(\lambda,\lambda+\epsilon)=1+i\epsilon,\\
&f(\lambda^B_{l},\lambda^C_{m})=f(\lambda^C_{l},\lambda^C_{m})\times[1+\frac{\epsilon}{\lambda^C_{l}-\lambda^C_{m}+i}],
\end{split}
\label{eqn:eq330}
\end{equation}
 as well as 
\begin{equation}
\begin{split}
\frac{f(\lambda^B_{l},\lambda^C_{m})}{f(\lambda^C_{m},\lambda^B_{l})}&=\frac{f(\lambda^C_{l},\lambda^C_{m})}{f(\lambda^C_{m},\lambda^C_{l})}\Big(1+\epsilon[\frac{1}{\lambda^C_{l}-\lambda^C_{m}+i}-\frac{1}{\lambda^C_{l}-\lambda^C_{m}-i}]+O(\epsilon^2)\Big),\\
r^B_{j}&=r^{C}_j\Big(1+\epsilon\frac{\partial}{\partial\lambda^C_{j}}\ln r(\lambda^C_j)\Big)+O(\epsilon^2)\\&=\prod_{k=1,k\ne j}^N\frac{f(\lambda^C_{k},\lambda^C_{j})}{f(\lambda^C_{j},\lambda^C_{k})}\Big(1+\epsilon\frac{\partial}{\partial\lambda^C_{j}}\ln r(\lambda^C_j)\Big)+O(\epsilon^2)
\end{split}
\label{eqn:eq331}
\end{equation}

$G_N$ in limit:
\begin{equation}
\begin{split}
G_N(\{\lambda_j^C\},\{\lambda_j^B\})&=\prod_{j>k}^N g(\lambda_j^B,\lambda_k^B)g(\lambda_k^C,\lambda_j^C)\prod_{j=1}^N\prod_{k=1}^N \frac{f(\lambda^{C}_{j},\lambda^{C}_{k})}{g(\lambda^{C}_{j},\lambda^{C}_{k})}\\
&=\prod_{j=1,j\ne k}^N\prod_{k=1}^N g(\lambda^C_{j},\lambda^C_{k})\prod_{j=1,j\ne k}^N\prod_{k=1}^N \frac{f(\lambda^C_{j},\lambda^C_{k})}{g(\lambda^C_{j},\lambda^C_{k})}\prod_{j=1}^{N}\frac{f(\lambda^C_{j},\lambda^C_{j})}{g(\lambda^C_{j},\lambda^C_{j})}\\
&=\prod_{j=1,j\ne k}^N\prod_{k=1}^Nf(\lambda^C_{j},\lambda^C_{k})\times 1.
\end{split}
\label{eqn:eq332}
\end{equation}

$M_{kl}$ in the limit:
Let us first calculate the diagonal terms, 
\begin{equation}
\begin{split}
M_{ll}(\dots)&=\frac{g(\lambda_l^C,\lambda_l^B)}{h(\lambda_l^C,\lambda_l^B)}-r_l^B\frac{g(\lambda_l^B,\lambda_l^C)}{h(\lambda_l^B,\lambda_l^C)}\frac{f(\lambda_l^B,\lambda_l^C)}{f(\lambda_l^C,\lambda_l^B)}\prod_{m\ne l}^{N}\frac{f(\lambda_l^B,\lambda_m^C)}{f(\lambda_m^C,\lambda_l^B)}\\
&=g(\lambda_l^C,\lambda_l^B)\Big[\frac{1}{h(\lambda^C_{l},\lambda^B_{l})}+r^B_{l}\frac{1}{h(\lambda^B_{l},\lambda^C_{l})}\frac{f(\lambda_l^B,\lambda_l^C)}{f(\lambda_l^C,\lambda_l^B)}\times A\Big]\\
&=\frac{-i}{\epsilon}\Big[(1-i\epsilon)+r^{B}_{l}(1+i\epsilon)(-1+2i\epsilon)\times A\Big]\\
&=i(\frac{\partial}{\partial\lambda^C_{l}}\ln r^C_{l})+\sum_{m\ne l}^{N}\frac{2}{(\lambda^C_{l}-\lambda^C_{m})^2+1},
\end{split}
\label{eqn:eq333}
\end{equation}
where have use Eq. (\ref{eqn:eq330}) and Eq. (\ref{eqn:eq331}), and $A=\prod_{m\ne l}^{N}\frac{f(\lambda_l^B,\lambda_m^C)}{f(\lambda_m^C,\lambda_l^B)}$. The off-diagonal terms can be calculate in the same way
\begin{equation}
M_{lk}=\frac{-2}{(\lambda^C_{k}-\lambda^C_{l})^2+1}+ O(\epsilon).
\label{eqn:eq334}
\end{equation}
Therefore generally speaking, the $Gaudin\ matrix$ is 
\begin{equation}
M_{lk}=\delta_{lk}\Big[i(\frac{\partial}{\partial\lambda^C_{l}}\ln r^C_{l})+\sum_{m}^{N}\frac{2}{(\lambda^C_{l}-\lambda^C_{m})^2+1}\Big]-\frac{2}{(\lambda^C_{k}-\lambda^C_{l})^2+1}.
\label{eqn:eq335}
\end{equation}
To simplify our result, let us introduce a set of new variables 
\begin{equation}
\begin{split}
\phi_{k}&=i\ln r_k +i\sum_{j=1,j\ne k}^N\ln\frac{f(\lambda_{k},\lambda_{j})}{f(\lambda_{j},\lambda_{k})}\\
&=i\ln \Bigg(\Big(\frac{\lambda_k-\frac{i}{2}}{\lambda_k+\frac{i}{2}}\Big)^L\prod_{j\ne k}\frac{\lambda_k-\lambda_j+i}{\lambda_k-\lambda_j-i}\Bigg),\quad k=1,\dots,N
\end{split}
\label{eqn:eq336}
\end{equation}
notice that expression in the big parentheses is the right hand side of Bethe Equation (8). Then the Gaudin matrix can be written as
\begin{equation}
M_{lk}=\frac{\partial\phi_{k}}{\partial\lambda_{l}},
\label{eqn:eq337}
\end{equation}
Then the square of the norm in the on-shell limit yields
\begin{equation}
S_N=\prod_{j=1,j\ne k}^N\prod_{k=1}^Nf(\lambda_{j},\lambda_{k})\det \frac{\partial \phi_{p}}{\partial\lambda_{q}}.
\label{eqn:eq338}
\end{equation}
We will use this result in Chapter \ref{Chapter3} for the calculation of one-point functions, and the Gaudin matrix. For relevant calculation of $SU(3)$ model and $SO(6)$ model, one can refer to \cite{su3_norm}, \cite{phd_Escobed_Jorge} respectively.
% Chapter 2

\chapter{Bethe ansatz in $\mathcal{N}=4$ SYM}
\label{Chapter2} % For referencing the chapter elsewhere, use \ref{Chapter1} 

%----------------------------------------------------------------------------------------

%----------------------------------------------------------------------------------------

\section{$\mathcal{N}=4$ SYM}
\subsection{Action}
Symmetries usually introduce more constraints in a theory, since they give more conserved quantities and invariance that need to be satisfied by physical quantities. Likewise, the supersymmetry(SUSY) gives rise to non-renormalization theorems: the correlation functions are better behaved and the interacting terms in the action are not renormalized at any order both in perturbative level and non-perturbative level. The situation becomes better when we consider $\mathcal{N}=4$ SUSY: the system contains so many constraints that it actually conformal invariant. That is to say not only the interaction term in the action stays same at all order, but also there is no renormalization of coupling constants, i.e. $\beta$ functions vanishes at all loops. The action of $N=4$ SYM theory reads \cite{action_SYM}
\begin{equation}
\begin{split}
S=\frac{2}{g^2_{YM}}\int d^4x\, \mathrm{tr}\, \Big[-&\frac{1}{4}F_{\mu\nu}F^{\mu\nu}-\frac{1}{2}D_{\mu}\phi^{i}D^{\mu}\phi^{i}\\
&+\frac{i}{2}\bar{\psi}\Gamma^{\mu}D_{\mu}\psi +\frac{1}{2}\bar\psi \Gamma^{i}[\phi_{i},\psi]+\frac{1}{4}[\phi_i,\phi_j][\phi_i,\phi_j]\Big]
\end{split}
\label{eqn:eq001}
\end{equation}
where $\Gamma$ denotes the ten-dimensional gamma matrices, which is a reminiscent of Kaluza Klein reduction from ten dimensional $\mathcal{N}=1$ SYM to four dimensional $\mathcal{N}=4$ SYM. And the field strength $F_{\mu\nu}$ and the covariant derivative $D_{\mu}$ are defined via 
\begin{equation}
\begin{split}
 F_{\mu\nu}&=\partial_{\mu}A_{\nu}-\partial_{\nu}A_{\mu}-i[A_{\mu},A_{\nu}],\\
 D_{\mu}\phi_{i}&=\partial_{\mu}\phi_{i}-i[A_{\mu},\phi_{i}],\quad D_{\mu}\psi=\partial_{\mu}\psi-i[A_{\mu},\psi].
\end{split}
\label{eqn:eq002}
\end{equation}
The field content of the theory corresponds to $\mathcal{N}=4$ massless vector multiplet, can be fully determined by supersymmetry. It contains six real scalar fields $\phi^{i}$, transforms in the anti-symmetric $\mathbf{6}$ representation of global R-symmetry group $SO(6)\simeq SU(4)$, one gauge field $A_{\mu}$ which is a singlet $\mathbf{1}$ under the $R$-symmetry, and four Weyl fermions $\psi^A$ transforms in the fundamental representation $\mathbf{4}$ of $SU(4)$. In addition, the local gauge symmetry is $SU(N)$. All the fields transforms in the adjoint representation of the gauge group. We will denote the color components of, say, the scalar $\phi^i$ as $[\phi^{i}]_{ab}$, where $a,b=1,\dots,N$ are color indices. 
\subsection{Gauge invariant operator in $\mathcal{N}=4$ SYM}
In our theory, the gauge invariant operators are made up of the trace of the various fields that belong to the $\mathcal{N}=4$ multiplet. The fields $\chi$ (scalar field $\phi$, Weyl spinor field $\psi$ and the field strength $F$) as mentioned before transform under adjoint representation of $SU(N)$, the infinitesimal gauge transformation takes
\begin{equation}
\chi \rightarrow \chi +[\varepsilon,\chi]
\label{eqn:eq003}
\end{equation}
where $\varepsilon$ is the generator of gauge transformation, a field that transforms like this is said to transform covariantly. And the gauge field $A_{\mu}$ transforms as
\begin{equation}
A_{\mu} \rightarrow A_{\mu}+\partial_{\mu}\varepsilon+[\varepsilon, A_{\mu}].
\label{eqn:eq004}
\end{equation}
It can be shown that $D_{\mu} \chi$ also transforms covariantly. 
We can make gauge invariant local composite operators by taking traces of combinations of different field that transforms covariantly under gauge group, of course these fields need to be valued at same space-time point. A single trace composite gauge-invariant operator $\mathcal{O}$ is given by
\begin{equation}
\mathcal{O}(x)=\mathrm{Tr}[\chi_1(x)\chi_2(x)\dots\chi_L(x)],
\label{eqn:eq005}
\end{equation}
 where $L$ is the number of combined fields. Obviously the product of single trace operators is also gauge invariant.
 Since we are only interested in the planar limit of our theory, where $g_{YM}\rightarrow 0$, $N\rightarrow \infty$ with 't Hooft coupling
\begin{equation}
\lambda=g^2_{YM}N
\label{eqn:eq006}
\end{equation}
kept fixed. If we restrict ourselves to colors $N\rightarrow \infty$ limit, one can show that only the single trace operators will contribute, the operators composed of single trace operators will be suppressed by $O(\frac{1}{N})$ when considering the correlation functions.

\section{One-loop anomalous dimensions}
In conformal world, the correlation functions are highly restrict. One point functions can be chosen to be vanished, two-point functions are completely fixed by their scalar dimension $\Delta$, and three-point functions can be fixed up to a structure constant $\lambda$. With the knowledge of conformal data $(\Delta,\lambda)$, higher-point correlation functions can be completely determined by conformal bootstrap. The two-point functions take the form as \cite{Minahan_2006_intro}
\begin{equation}
\langle \mathcal{O}^i(x)\bar{\mathcal{O}}^j(y)\rangle\sim\frac{c_{ij}}{|x-y|^{2\Delta_{i}}}, 
\label{eqn:eq007}
\end{equation}
with $c_{ij}=0$ for $\Delta_{i}\ne \Delta_j$. One can also do a rotation of fields and a normalization to set $c_{ij}=\delta_{ij}$ given the composite operators has the same matter content. 

To keep our store short, we will only focus on one-loop order of the planar limit. And the single trace operators are only made up of the six scalar fields, the so-called SO(6) sector since these scalars transform in the fundamental representation of it. We define our un-renormalized single trace operator as 
\begin{equation}
\mathcal{O}_I^{bare}=\mathrm{Tr}[\phi_{i_1}\dots\phi_{i_L}].
\label{eqn:eq008}
\end{equation}
Where $i_k$ runs $\{1,\dots,6\}$, and bare dimension $\Delta^{(0)}_I=L$. Due to the fact that there is no conformal anomaly in our theory, so the conformal symmetry is valid at all loop level. Then the two-point function of a operator like Eq. (8) with itself reads
\begin{equation}
\langle\mathcal{O}^{bare}_I(x)\bar{\mathcal{O}}^{bare}_I(y)\rangle\sim c_I N^L \frac{1}{|x-y|^{2\Delta}},
\label{eqn:eq009}
\end{equation}
where the dimension $\Delta=\Delta^{(0)}+\gamma$, with $\Delta^{(0)}$ being the bare dimension, $\gamma$ being the anomalous dimension because of the quantum correction, and $c_I$ is the number of cyclic permutation that keep $I=\{i_1,i_2,\dots,i_L\}$ invariant \cite{one_pt_dfc}. For our case the single trace operators are made up only of scalar field, hence $\Delta^{(0)}=L$, and if we consider, say, operators with $\{121212\}$, then $c_I=3$.

Consider the small gauge coupling, then the corresponding quantum correction is also way more small than classical quantity, $\gamma\ll \Delta^{(0)}$. Expand Eq. (\ref{eqn:eq009}) around classical value
\begin{equation}
\langle\mathcal{O}^{bare}_I(x)\bar{\mathcal{O}}^{bare}_I(y)\rangle\sim c_I N^L\frac{1}{|x-y|^{2\Delta^{(0)}}}\Big(1-\gamma \ln \Lambda^2(x-y)^2\Big),
\label{eqn:eq0010}
\end{equation}
where $\Lambda$ is an energy cut-off to keep the dimension correct. (About this approximation, one should specify it use the diagram to calculate the number of cyclic permutations that leaves $I=\{i_1,i_2,\dots,i_L\}$ invariant.) Instead of using a cut-off, we will use dimensional regularization in later calculation, which is
\begin{equation}
d\rightarrow d-2\varepsilon,\quad g_{YM}\rightarrow g_{YM}\mu^{\varepsilon},
\label{eqn:eq0011}
\end{equation}
where $d=4$ is the dimension, and $\mu$ is a parameter with dimension of mass.

\subsection{Two-point functions}
Since we are only interested in the scalar part of our theory, then we could write the free bosonic part of of our Lagrangian to get the propagator
\begin{equation}
\mathcal{L}_{free}=\frac{2}{g^2_{YM}}\mathrm{Tr}\,[\frac{1}{2}A_{\mu}(\eta^{\mu\nu}\Box-\partial^{\mu}\partial^{\nu})A_{\nu}+\frac{1}{2}\phi_i\Box\phi_{i}],
\label{eqn:eq0012}
\end{equation}
one can read the following free propagators
\begin{equation}
\begin{split}
\langle \phi^i_{ab}(x)\phi^j_{cd}(y)\rangle_0&=\frac{g^2_{YM}}{2}\delta_{ij}\delta_{ad}\delta_{bc}\int\frac{d^4p}{(2\pi)^{4}}\frac{-i}{p^2}e^{ip(x-y)}\\
\langle A^{\mu}_{ab}(x) A^{\nu}_{cd}(y)\rangle_0&=\frac{g^2_{YM}}{2}\eta^{\mu\nu}\delta_{ad}\delta_{bc}\int\frac{d^4p}{(2\pi)^{4}}\frac{-i}{p^2}e^{ip(x-y)}.
\end{split}
\label{eqn:eq0013}
\end{equation}
with $0$ denote vacuum in free theory. The above integral can be evaluated using
\begin{equation}
\begin{split}
\int&\frac{d^dp}{(2\pi)^{d}}\frac{-i}{(p^2)^n}e^{ipx}\\
&=\frac{-i}{i^n}\frac{1}{(n-1)!}\int_{0}^{\infty}ds s^{n-1}\int \frac{d^dp}{(2\pi)^d}e^{isp^2+ip\cdot x}\\
&=\frac{-i}{i^n}\frac{1}{(n-1)!}\frac{(\pi)^{d/2}}{(2\pi)^d}i^{(2-d)/2}\int_{0}^{\infty}ds\, e^{-i\frac{x^2}{4s}} s^{n-1-d/2}\\
&=(i)^{-d}\frac{1}{4^n \pi^{d/2}}\frac{\Gamma(\frac{d}{2}-n)}{\Gamma(n)}\frac{1}{(x^2)^{d/2-n}}.
\end{split}
\label{eqn:eq0014}
\end{equation}
Where the second line we used Schwinger parametrization and third line Gauss integral. We denote 
\begin{equation}
\begin{split}
K_{\varepsilon}(x,y)&=\frac{(g_{YM}\mu^{\varepsilon})^2}{2}\int\frac{d^{4-2\varepsilon}p}{(2\pi)^{4-2\varepsilon}}\frac{-i}{p^2}e^{ip(x-y)}=\frac{(g_{YM}\mu^{\varepsilon})^2}{2}\frac{\Gamma(1-\varepsilon)}{4\pi^{2-\varepsilon}}\frac{1}{[(x-y)^2]^{1-\varepsilon}}\\
&=\frac{g^2_{YM}}{8\pi^2(x-y)^2}\Big(1+\varepsilon(\gamma_E+\log\pi(x-y)^2+2\log\mu)+O(\varepsilon^2)\Big),
\end{split}
\label{eqn:eq0015}
\end{equation}
as the scalar propagator after the dimensional regularization. 

In path integral formalism the n-point correlation function are given
\begin{equation}
\langle\Omega\lvert T\{\hat{\phi}(x_1)\dots\hat{\phi}(x_n)\}\lvert\Omega \rangle=\frac{\int\mathcal{D}\phi\, \phi(x_1)\dots\phi(x_n)e^{iS}}{\int\mathcal{D}e^{iS}},
\label{eqn:eq0016}
\end{equation}
where $T$ is the time-ordering operator, and $\lvert\Omega\rangle$ is the vacuum of the full action. The generating functions can be written in the presence of a classical external source term $J(x)$ for $\phi(x)$

\begin{equation}
Z[J]=\int \mathcal{D}\phi\, \exp\Big(iS+i\int d^4x\, J(x)\phi(x)\Big),
\label{eqn:eq0017}
\end{equation}
then Eq. (\ref{eqn:eq0016}) can be rewritten as

\begin{equation}
\begin{split}
\langle&\Omega\lvert T\{\hat{\phi}(x_1)\dots\hat{\phi}(x_n)\}\lvert\Omega \rangle=(-i)^n\frac{1}{Z[0]}\frac{\delta Z(J)}{\delta J(x_1)\dots\delta J(x_n)}\Big\vert_{J=0}\\
&=\frac{1}{Z[0]}\int\mathcal{D}\phi\, \phi(x_1)\dots\phi(x_n)e^{iS}\\
&=\frac{1}{Z[0]}\int\mathcal{D}\phi\, \phi(x_1)\dots\phi(x_n)e^{i\int d^4x \mathcal{L}_{free}+\mathcal{L}_{int}}\\
&=\frac{1}{Z[0]}\int\mathcal{D}\phi\, \phi(x_1)\dots\phi(x_n)\Big(1+i\int d^4x\,\mathcal{L}_{int}(x)\\&\qquad-\frac{1}{2}\int d^4x\,\mathcal{L}_{int}(x)\int d^4y\,\mathcal{L}_{int}(y)+\dots\Big)e^{i\int d^4x \mathcal{L}_{free}}\\
&=\langle\phi(x_1)\dots\phi(x_n)\Big(1+i\int d^4x\,\mathcal{L}_{int}(x)-\frac{1}{2}\int d^4x\,\mathcal{L}_{int}(x)\int d^4y\,\mathcal{L}_{int}(y)+\dots\rangle_0,
\end{split}
\label{eqn:eq0018}
\end{equation}
where we have choose the normalization $Z[0]=0$. Then one can evaluate the correlation functions using the Feynman rules.

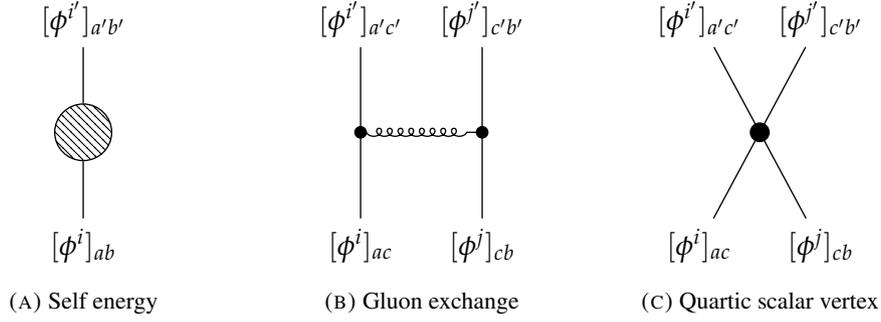
\begin{figure}[t!]
    \centering
    \begin{subfigure}[t]{.3\textwidth}
     \centering
     \begin{tikzpicture}
\begin{feynhand}
    \vertex (a) at (0,1.5) {$[\phi^{i^{\prime}}]_{a^{\prime}b^{\prime}}$}; \vertex (b) at (0,-1.5) {$[\phi^{i}]_{ab}$};
\vertex [NWblob] (o) at (0,0) {}; \propag [plain] (a) to (o);
\propag [plain] (b) to (o);
  \end{feynhand}
\end{tikzpicture}
  \caption{Self energy}
  \label{fig:feynman001b}
\end{subfigure}%
~
    \begin{subfigure}[t]{.3\textwidth}
  \centering
  \begin{tikzpicture}
\begin{feynhand}
    \vertex (a) at (-0.8,1.5) {$[\phi^{i^{\prime}}]_{a^{\prime}c^{\prime}}$}; \vertex (b) at (-0.8,-1.5) {$[\phi^{i}]_{ac}$};
    \vertex (c) at (0.8,1.5) {$[\phi^{j^{\prime}}]_{c^{\prime}b^{\prime}}$};
    \vertex (d) at (0.8,-1.5) {$[\phi^{j}]_{cb}$};
    \vertex [dot] (o1)  at (-0.8,0) {}; \vertex [dot] (o2) at (0.8,0) {};
\propag [plain] (a) to (o1); \propag [plain] (b) to (o1);\propag[gluon] (o1) to (o2);\propag[plain] (c) to (o2); \propag[plain] (d) to (o2);
  \end{feynhand}
  
\end{tikzpicture}
  \caption{Gluon exchange}
  \label{fig:feynman001a}
\end{subfigure}%
~
\begin{subfigure}[t]{.3\textwidth}
  \centering
  \begin{tikzpicture}
\begin{feynhand}
    \vertex (a) at (-0.8,1.5) {$[\phi^{i^{\prime}}]_{a^{\prime}c^{\prime}}$}; \vertex (b) at (-0.8,-1.5) {$[\phi^{i}]_{ac}$};
    \vertex (c) at (0.8,1.5) {$[\phi^{j^{\prime}}]_{c^{\prime}b^{\prime}}$};
    \vertex (d) at (0.8,-1.5) {$[\phi^{j}]_{cb}$};
    \vertex [dot] (o) at (0,0) {z};
\propag [plain] (a) to (o); \propag [plain] (b) to (o);\propag[plain] (c) to (o); \propag[plain] (d) to (o);
  \end{feynhand}
  
\end{tikzpicture}
  \caption{Quartic scalar vertex}
  \label{fig:feynman001c}
\end{subfigure}%

    \caption{one-loop Feynman diagrams \cite{one_pt_dfc}}
    \label{fig:feynman001}
\end{figure}

At one-loop order, there are only three type of diagrams \cite{Minahan_2003} that contribute to the two-point functions, see Fig.~\ref{fig:feynman001}

The color structure is completely fixed since we want planar Feynman diagram, the flavor structure is what we are interested in. We could compute the first two diagrams, but since gluon fields $A_{\mu}$ have no $SO(6)$ R-charges and the flavor of a scalar must be same in the self energy diagram. Therefore, these two diagram will only give us terms like $\delta_{i,i^{\prime}}$ and $\delta_{ii^{\prime}}\delta_{jj^{\prime}}$. And later we will fixed it by choose a special correlation function. 

Now we can calculate the only no-trivial diagram quartic interaction, it turns out quiet easy.

\begin{figure}
\centering
\begin{subfigure}[t]{.4\textwidth}
  \centering
  \includegraphics[width=1.0\linewidth]{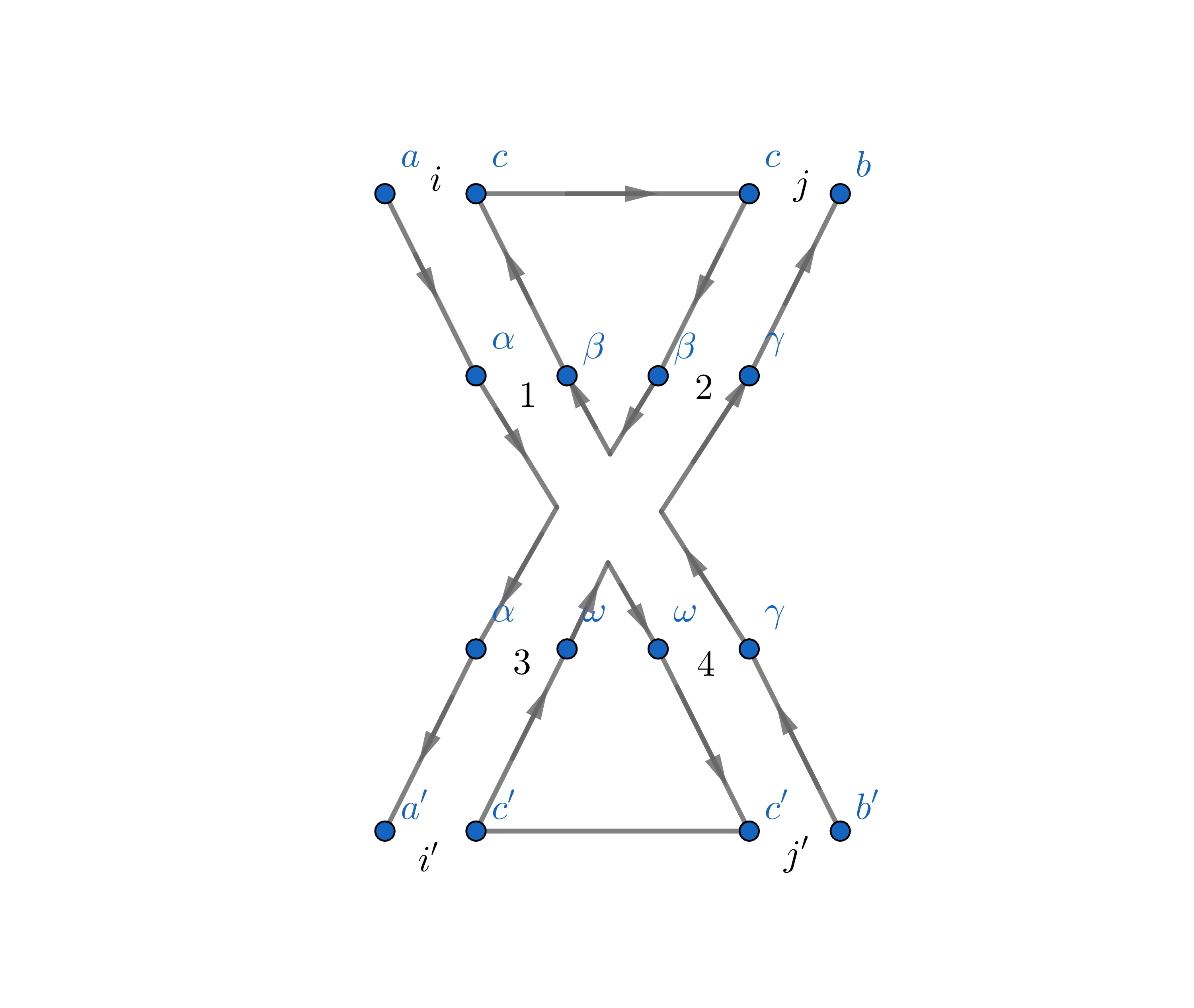}
  \caption{4-pt quartic scalar interaction}
  \label{fig:feynman002}
\end{subfigure}%
\begin{subfigure}[t]{.45\textwidth}
  \centering
  \includegraphics[width=1.1\linewidth]{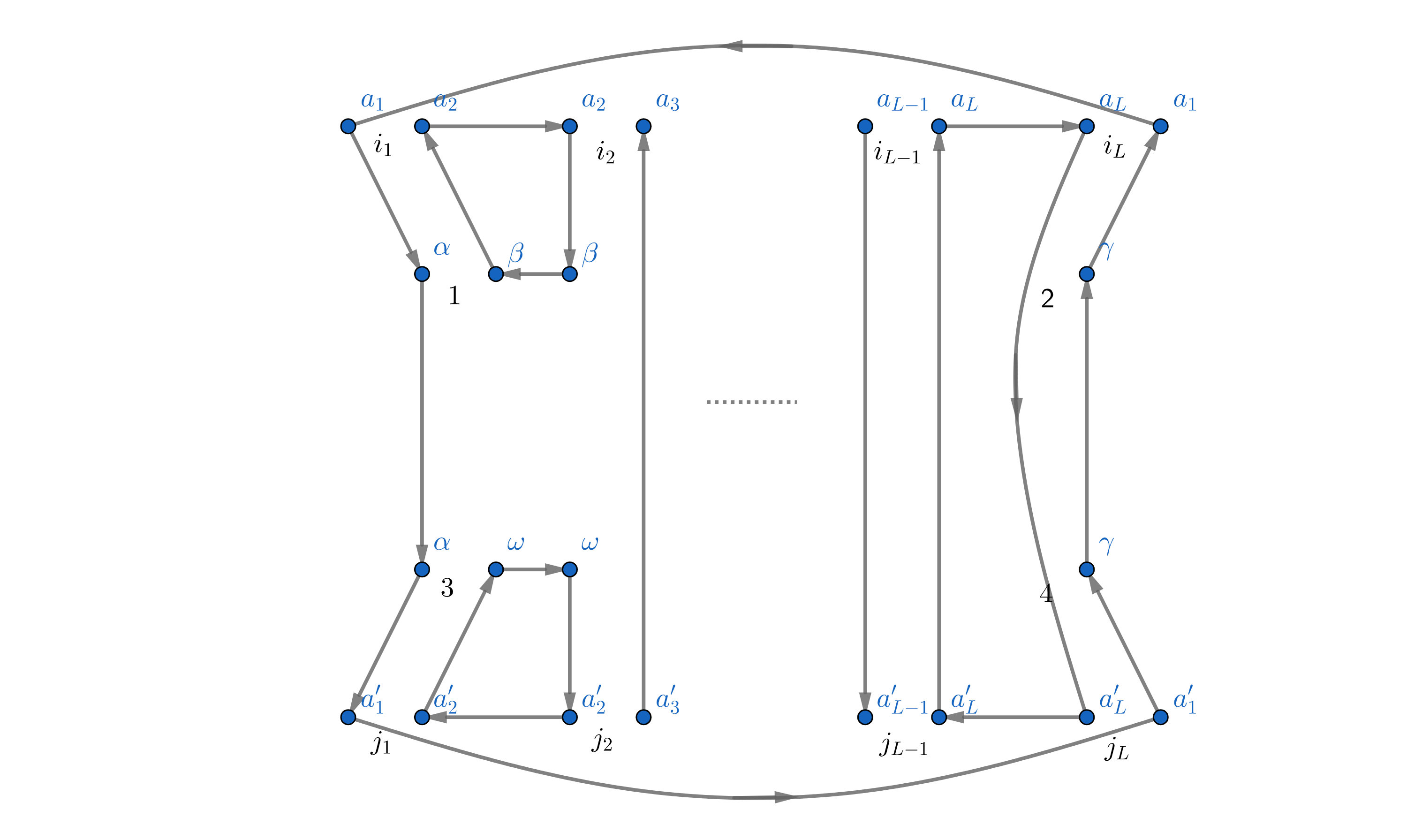}
  \caption{2L-pt quartic scalar vertex in one-loop level}
  \label{fig:feynman003}
\end{subfigure}
\caption{Quartic scalar vertex interaction}
\label{fig:test}
\end{figure}

The tree level calculation is quite easy
\begin{equation}
\langle [\phi_i\phi_j]_{ab}(x)[\phi_{j^{\prime}}\phi_{i^{\prime}}]_{b^{\prime}a^{\prime}}(y)\rangle_{0}=NK^2_{\varepsilon}(x,y).
\label{eqn:eq0019}
\end{equation}
 One loop level would be
\begin{equation}
\begin{split}
&\langle [\phi_i\phi_j]_{ab}(x)[\phi_{j^{\prime}}\phi_{i^{\prime}}]_{b^{\prime}a^{\prime}}(y)\rangle\\&=\frac{i}{2(g_{YM}\mu^{\varepsilon})^2}\langle [\phi_i\phi_j]_{ab}(x)(\int d^{4-2\varepsilon}z\, [\phi_k(z),\phi_l(z)][\phi_k(z),\phi_l(z)])[\phi_{j^{\prime}}\phi_{i^{\prime}}]_{b^{\prime}a^{\prime}}(y)\rangle_{0}\\
&=\frac{i}{2(g_{YM}\mu^{\varepsilon})^2}\int d^{4-2\varepsilon}zK^2_{\varepsilon}(x,z)K^2_{\varepsilon}(z,y)2N^2\delta_{a,a^{\prime}}\delta_{b,b^{\prime}}(4\delta_{i,j^{\prime}}\delta_{j,i^{\prime}}-2\delta_{i,i^{\prime}}\delta_{ij^{\prime}}-2\delta_{i,j}\delta_{i^{\prime},j^{\prime}}).
\end{split}
\label{eqn:eq0020}
\end{equation}
To get the flavor indices $4\delta_{i,j^{\prime}}\delta_{j,i^{\prime}}$ and $2\delta_{i,i^{\prime}}\delta_{j,j^{\prime}}+2\delta_{i,j}\delta_{i^{\prime},j^{\prime}}$, we can set $1,2,3,4=[k,l,k,l]$ or $1,2,3,4=[k,k,l,l]$ in Fig.~\ref{fig:feynman002} respectively , with $[]$ denote as circling permutation. So we are left with the integral
\begin{equation}
\begin{split}
&\int d^{4-2\varepsilon}z\,K_{\varepsilon}^2(x,z)K_{\varepsilon}^2(z,y)\\
&=[\frac{(g_{YM}\mu^{\varepsilon})^2}{2}\frac{\Gamma(1-\varepsilon)}{4\pi^{2-\varepsilon}}]^4\int d^{4-2\varepsilon}z\,\frac{1}{[(x-z)^2]^{2-2\varepsilon}[(z-y)^2]^{2-2\varepsilon}}\\
&=i(g_{YM}\mu^{\varepsilon})^2\Big(\frac{\Gamma(1-\varepsilon)}{8\pi^{2-\varepsilon}}\Big)^4\pi^{2-\varepsilon}G(2-2\varepsilon,2-2\varepsilon),
\end{split}
\label{eqn:eq0021}
\end{equation}
where we have used the formula
\begin{equation}
\int d^{4-2\varepsilon}p\frac{1}{[p^2]^{\lambda_1}[(p-q)^2]^{\lambda_2}}=i\pi^{2-\varepsilon}G(\lambda_{1},\lambda_{2})\frac{1}{[q^2]^{\lambda_{1}+\lambda_{2}+\varepsilon-2}},
\label{eqn:eq0022}
\end{equation}
with
\begin{equation}
G(\lambda_1,\lambda_2)=\frac{\Gamma(\lambda_{1}+\lambda_{2}+\varepsilon-2)\Gamma(2-\lambda_{1}-\varepsilon)\Gamma(2-\lambda_{2}-\varepsilon)}{\Gamma(\lambda_{1})\Gamma(\lambda_{2})\Gamma(4-\lambda_{1}-\lambda_{2}-2\varepsilon)}.
\label{eqn:eq0023}
\end{equation}

So Eq.~\ref{eqn:eq0020} is equivalent to 
\begin{equation}
\begin{split}
\langle [\phi_i\phi_j]_{ab}(x)[\phi_{j^{\prime}}\phi_{i^{\prime}}]_{b^{\prime}a^{\prime}}(y)\rangle_{\lambda}&=-\frac{\lambda}{16\pi^2}NK^2_{\varepsilon}(x,y)\delta_{aa^{\prime}}\delta_{bb^{\prime}}(2\delta_{i,j^{\prime}}\delta_{j,i^{\prime}}-\delta_{i,i^{\prime}}\delta_{j,j^{\prime}}-\delta_{i,j}\delta_{i^{\prime}j^{\prime}})\\
&\quad \times\Big(\frac{1}{\varepsilon}+1+\gamma_E+\log (\pi(x-y)^2\Big)+O(\varepsilon).
\end{split}
\label{eqn:eq0024}
\end{equation}
Combine Eq. (\ref{eqn:eq0024}) and Eq. (\ref{eqn:eq0019})
\begin{equation}
\begin{split}
\langle [\phi_i\phi_j]_{ab}(x)[\phi_{j^{\prime}}\phi_{i^{\prime}}]_{b^{\prime}a^{\prime}}(y)\rangle_{\lambda}=NK^2_{\varepsilon}(x,y)\Big[&1-\frac{\lambda}{16\pi^2}\Big(\frac{1}{\varepsilon}+1+\gamma_E+\log (\pi(x-y)^2\Big)\\ 
&\times\delta_{aa^{\prime}}\delta_{bb^{\prime}}(2\delta_{i,j^{\prime}}\delta_{j,i^{\prime}}-\delta_{i,i^{\prime}}\delta_{j,j^{\prime}}-\delta_{i,j}\delta_{i^{\prime}j^{\prime}})+O(\varepsilon)\Big].
\end{split}
\label{eqn:eq0025}
\end{equation}
If we promote this result to the operator like the type in Eq. (\ref{eqn:eq008}), the corresponding two-line diagram will be Fig.~\ref{fig:feynman003}. 
The corresponding one-loop contribution to the correlator is 
\begin{equation}
\begin{split}
\langle\mathcal{O}^{bare}_{I}(x)\bar{\mathcal{O}}^{bare}_{I}(y)\rangle_{\lambda}&=-\frac{\lambda}{16\pi^2}N^{L}K_{\varepsilon}^L(x,y)\Big(\frac{1}{\varepsilon}+1+\gamma_E+\log (\pi(x-y)^2\Big)\\
&\quad\times\sum_{\ell=1}^{L}(C-2\mathbb{P}_{l+l+1}+\mathbb{K}_{l,l+1})(\delta_{i_1,j_1}\delta_{i_2,j_2}\dots\delta_{i_L,j_L}+\mathrm{cycles\ perm.}).
\end{split}
\label{eqn:eq0026}
\end{equation}
Above equation, to simplify our notion for too many Kronecker deltas, we use a permutation operator and a trace operator
\begin{equation}
\begin{split}
\mathbb{P}_{\ell,\ell+1}\dots\delta_{i_{\ell}j_{k}}\delta_{i_{\ell+1}j_{k+1}}\dots=\dots\delta_{i_{\ell+1}j_{k}}\delta_{i_{\ell}j_{k+1}}\dots,\\
\mathbb{K}_{\ell,\ell+1}\dots\delta_{i_{\ell}j_{k}}\delta_{i_{\ell+1}j_{k+1}}\dots=\dots\delta_{i_{\ell}i_{\ell+1}}\delta_{j_{k}j_{k+1}}\dots,
\end{split}
\label{eqn:eq0027}
\end{equation}
to keep track of the flavor structure. So in conclusion, one would get the following correlation function for operators Eq. (\ref{eqn:eq008})  
\begin{equation}
\langle \mathcal{O}^{bare}_I(x)\bar{\mathcal{O}}^{bare}_I(y)\rangle_{\varepsilon}=c_IN^{\Delta^{(0)}}K^{\Delta^{(0)}}_{\varepsilon}(x,y)[1-\frac{1}{c_I}\Big(\frac{1}{\varepsilon}+1+\gamma_E+\log (\pi(x-y)^2\Big)\hat{D}]
\label{eqn:eq0028}
\end{equation}
where 
\begin{equation}
\begin{split}\hat{D}&=\hat\Gamma(\delta_{i_1,j_1}\delta_{i_2,j_2}\dots\delta_{i_L,j_L}+\mathrm{cycles\ perm.}),\\
\mathrm{with}\quad \hat{\Gamma}&=\frac{\lambda}{16\pi^2}\sum_{\ell=1}^{L}(C-2\mathbb{P}_{l,l+1}+\mathbb{K}_{l,l+1}).
\end{split}
\label{eqn:eq0029}
\end{equation}
Compare with Eq. (\ref{eqn:eq0010}), one could say that the one-loop anomalous dimension is actually the operator $\Gamma$ whose eigenvalue is $\gamma$. And the rest variables are independent of the specific form $\mathcal{O}^{bare}$, except $c_I$ which is trivial. 

To find $C$, we notice that an operator in the symmetric representation of $SO(6)$ is a chiral primary operator, i.e. $\mathcal{O}=\mathrm{Tr}[Z^L]$ with $Z=\phi_3+i\phi_6$, which means its quantum number is protected, hence anomalous dimension is zero. The symmetric representations are traceless, hence $\mathbb{P}$ and $\mathbb{K}$ will give us 1 and $0$ retrospectively, so $C=2$.
\begin{equation}
\hat\Gamma=\frac{\lambda}{16\pi^2}\sum_{\ell}^{L}(2-2\mathbb{P}_{l,l+1}+\mathbb{K}_{l,l+1})
\label{eqn:eq0030}
\end{equation}

\section{SO(6) spin chain and state-operator correspondence}
To make our life simple, we make a combination of real scalar fields into complex ones as follows

\begin{equation}
\begin{split}
X=\phi_1+i\phi_4,\quad Y=\phi_2+i\phi_5, \quad Z=\phi_3+i\phi_6,\\\bar{X}=\phi_1-i\phi_4,\quad \bar{Y}=\phi_2-i\phi_5, \quad \bar{Z}=\phi_3-i\phi_6.
\end{split}
\label{eqn:eq0031}
\end{equation}
Then we can identify the $SU(2)$ sub-sector with $X$ and $Y$ or other two non-conjugated fields, the $SU(3)$ sub-sector is form by $X$, $Y$, $Z$. To make contact with later periodic spin chain, we will consider a general operator of the form
\begin{equation}
    \mathcal{O}=\Psi^{S}\mathcal{O}_S^{bare}, 
\label{eqn:eq0031a}    
\end{equation}
where the coefficient $\Psi^{S}$ is invariant under cyclic permutation
\begin{equation}
\Psi^{(s_1,s_2,\dots,s_L)}=\Psi^{(s_L,s_1,s_2,\dots,s_{L-1})}.
\label{eqn:eq0031b}
\end{equation}

The $SO(6)$ sector is proved to be closed at one-loop level \cite{so6_one_loop}, so we can make the following identification

\begin{equation}
\mathcal{O}=\Big(\frac{4\pi^2}{\lambda}\Big)^{\frac{L}{2}}\frac{\mathcal{Z}}{\sqrt{L}}\frac{\mathrm{Tr}\,\prod_{\ell=1}^{L}\big(\langle 1_{\ell}\lvert\otimes X+\langle 2_{\ell}\lvert\otimes Y+\langle 3_{\ell}\lvert\otimes Z+\langle 4_{\ell}\lvert\otimes \bar{X}+\langle 5_{\ell}\lvert\otimes \bar{Y}+\langle 6_{\ell}\lvert\otimes \bar{Z}\big)\lvert\mathbf{u}\rangle}{\sqrt{\langle \mathbf{u}\lvert\mathbf{u}\rangle}}
\label{eqn:eq0032}
\end{equation}
where the Bethe state $\lvert \mathbf{u}\rangle$ is the eigenstate of Hamiltonian 
\begin{equation}
H=\sum_{\ell=1}^{L}(2-2\mathbb{P}_{l,l+1}+\mathbb{K}_{l,l+1}).
\label{eqn:eq0033}
\end{equation}
By comparing Eq.(\ref{eqn:eq0032}) with Eq.(\ref{eqn:eq0028}) (where $c_I=L$ due to Eq.(\ref{eqn:eq0031b}), and there is an extra factor of 2 in $(2N)^{\Delta^{(0)}}$ from Eq. (\ref{eqn:eq0031})), we know that $\mathcal{O}$ is a normalized operator.
And the states $\lvert \#\rangle$ are basis states in the defining representation of $SO(6)$, we have calculated 
\begin{equation}
\mathcal{Z}=1+\frac{\lambda}{16\pi^2}\frac{\Delta^{(1)}}{2}\Big(\frac{1}{\varepsilon}+1+\gamma_E+\log \pi\Big)
\label{eqn:eq0034}
\end{equation}
with $\Delta^{(1)}$ is the eigenvalue of the Hamiltonian on Bethe state. 
% Chapter 3

\chapter{dCFT and One-point functions} % Main chapter title

\label{Chapter3} % For referencing the chapter elsewhere, use \ref{Chapter1} 
%----------------------------------------------------------------------------------------

\section{D3-D5 defect in AdS/CFT}
As we stated before, the one-point function in the conformal theory should be zero due to scaling invariance
\begin{equation}
\langle \mathcal{O}\rangle=0.
\label{eqn:eq0001}
\end{equation}
But when the conformal symmetry is broken by a defect, one can speculate that the one-point will not vanish.  Indeed, when we put a codimension $1$ defect in our four-dimensional SYM, consider we put it at the $x_3=0$ plane, and the one-point function of the operator $\mathcal{O}_i(x)$ is fixed up to a constant
\begin{equation}
\langle \mathcal{O}_{i}(x)\rangle=\frac{a_i}{x_3^{\Delta_i}},
\label{eqn:eq0002}
\end{equation}
 where $x_3$ is the distance from the operator to the defect.
 \begin{figure}
     \centering
     \includegraphics[scale=1.2]{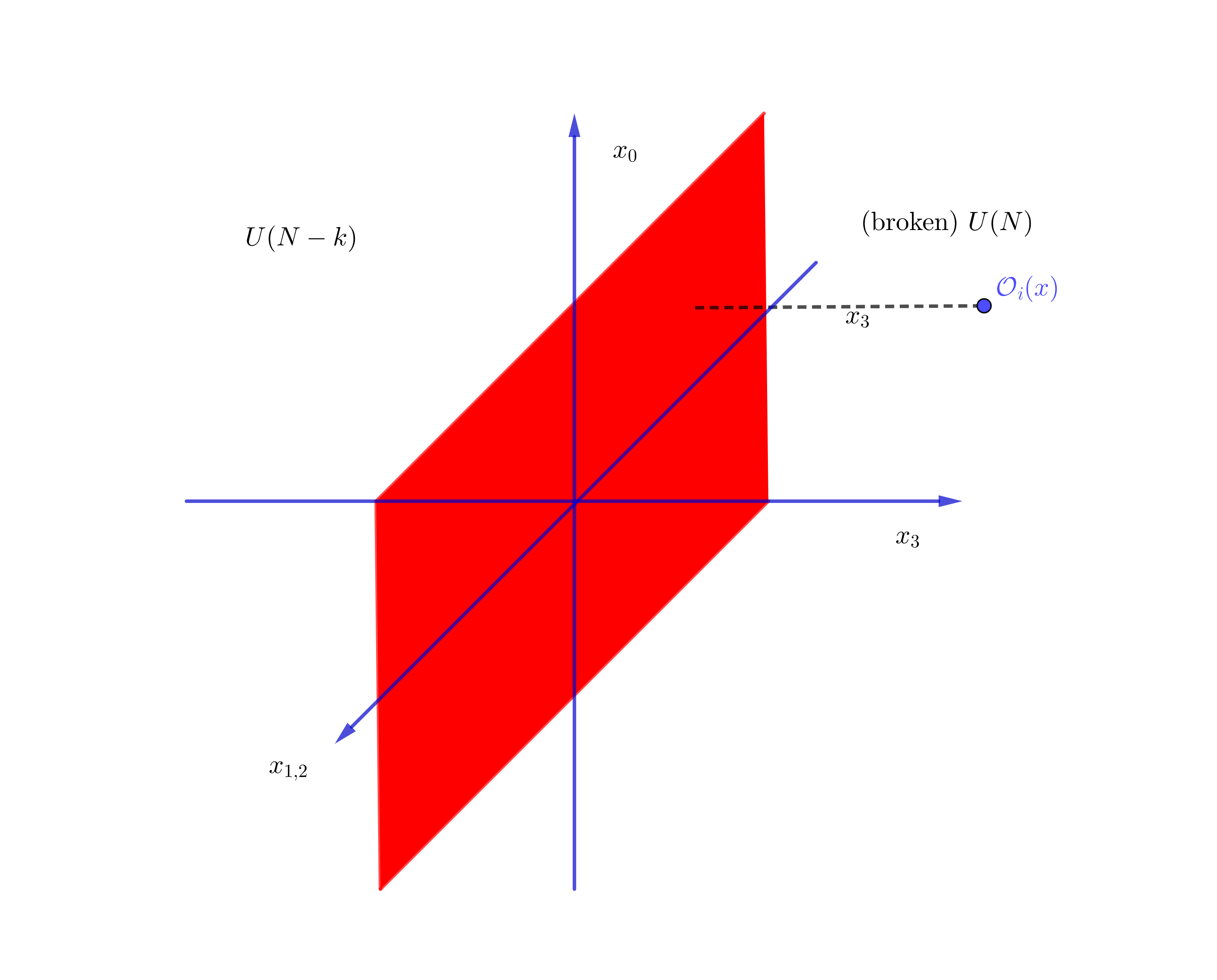}
     \caption{A codimension one defect sit at $x_3=0$ \cite{one_pt_dfc}}
     \label{fig:dcft001}
 \end{figure}
 By the $AdS/CFT$ correspondence, such defected theory has a gravitational dual on the type $IIB$ superstring theory side in $AdS_5\times S^5$. The defects role is played by insertion of $D5$ or $D7$ probe brane in the $AdS_5\times S^5$ space. We will stick to the $D5$ probe brane case in this thesis only. These brane will intersect with the usual stacks of $D3$ branes \cite{Karch_2001},\cite{DeWolfe_dcft_probe_brane}. The number of $D3$ branes on the sides of probe brane can differ, resulting in a new parameter $k$, which indicates non-zero background flux.  
 
 The gauge group for our $\mathcal{N}=4$ SYM theory will also be different on sides of the defects, $U(N-k)$ for $x_3<0$ while $U(N)$ for $x_3>0$. However, the $U(N)$ symmetry is also broken for $x_3>0$ by assigning a non-zero vacuum expectation value. From the $\mathcal{N}=4$ SYM, one can derive the classical equations of motion for scalar fields \cite{Myers_noncommutative}
\begin{equation}
\frac{d^2\phi_i^{cl}}{dx^2_3}=[\phi_j^{cl},[\phi_j^{cl},\phi_i^{cl}]],\quad i=1,\dots,6,
\label{eqn:eq0003}
\end{equation}
where we have setting fermions $\psi$ and gauge fields $A_{\mu}$ to zero classically, and only concerning time independent solution. We will get the following solutions.

The D3-D5 defect solution with geometry $AdS_4\times S^2$ for $x_3>0$ is solved by $\mathfrak{su}_2$ 
\begin{equation}
\phi^{cl}_{i}=-\frac{(t_i)_{k\times k}\oplus 0_{(N-k)\times(N-k)}}{x_3},\quad i=1,2,3,\quad \phi^{cl}_{j}=0,\quad j=4,5,6,
\label{eqn:eq0004}
\end{equation}
where the $k\times k$ matrices $t_{1,2,3}$ are $k$ dimensional representations of $\mathfrak{su}_2$ satisfying
\begin{equation}
[t_i,t_j]=i\epsilon_{ijk}t_k.
\label{eqn:eq0005}
\end{equation}
For $x_3<0$, all classical fields are vanishing.
\subsection{$\mathfrak{su}_2$ sector}
Consider a general composite single trace operator in $SO(6)$ sector
\begin{equation}
\mathcal{O}=\Phi^{i_1\dots i_L}\mathrm{Tr}\, \phi_{i_1}\dots\phi_{i_L},
\label{eqn:eq0006}
\end{equation}
with coefficient $\Phi^{i_1\dots i_L}$ is invariant under cycli permutations. It turns out that at tree level, the one-point functions are given by inserting the above classical solutions,
\begin{equation}
\langle \mathcal{O}\rangle^{cl}=(-1)^L\Phi^{i_1\dots i_L}\frac{\mathrm{Tr}(t_{i_1}\dots t_{i_L})}{x_3^L},
\label{eqn:eq0007}
\end{equation}
recall that the trace is taken over the color space, and we will leave the $k$ dependence implicit. Actually, a systematical way of calculating one-point function via spin chain can be achieved by the so called matrix product state (MPS), for $\mathfrak{su}_2$ sector MPS takes the form 
\begin{equation}
\lvert \mathrm{MPS}\rangle_{k}=\mathrm{Tr}\prod_{\ell=1}^{L}\Big[t_1\otimes\lvert\uparrow\rangle_{\ell}+t_2\otimes\lvert\downarrow\rangle_{\ell}\Big]=\sum_{i_n=1}^2\mathrm{Tr}[t_{i_1}\dots t_{i_L}]\lvert i_1\dots i_L\rangle,
\label{eqn:eq0008}
\end{equation}
where we have used $\lvert 1\rangle$ and $\lvert 2\rangle$ stands for spin up $\lvert\uparrow\rangle$ and spin down $\lvert\downarrow\rangle$. Using the explicit mapping between the spin chain states and the field operators Eq. (\ref{eqn:eq0032}), the one-point function Eq. (\ref{eqn:eq0007}) reduces to 
\begin{equation}
\langle \mathcal{O}\rangle_k^{cl}=(-1)^{L}\Big(\frac{8\pi^2}{\lambda}\Big)^{L/2}\frac{1}{\sqrt{L}}\frac{C_k}{x_3^L},\quad C_k=\frac{\langle\mathrm{MPS}\lvert\mathbf{u}\rangle}{\sqrt{\langle\mathbf{u}\lvert\mathbf{u}\rangle}}.
\label{eqn:eq0009}
\end{equation}
Following \cite{Buhl_Mortensen_2016}, one can actually proof $C_k$ take the following form
\begin{equation}
C_k=i^LT_{k-1}(0)\sqrt{\frac{Q(\frac{i}{2})Q(0)}{Q^2(\frac{ik}{2})}}\sqrt{\frac{\det G_+}{\det G_-}},
\label{eqn:eq00010}
\end{equation}
where 
\begin{equation}
T_{n}(u)=\sum_{a=-n/2}^{n/2}(u+ia)^L\frac{Q(u+\frac{n+1}{2}i)Q(u-\frac{n+1}{2}i)}{Q(u+(a-\frac{1}{2})i)Q(u+(a+\frac{1}{2})i)}.
\label{eqn:eq00011}
\end{equation}
And we introduce the Baxter Q-function for $SU(2)$ spin chain
\begin{equation}
Q(v)=\prod_{i=1}^{M}(v-u_i),
\label{eqn:eq00012}
\end{equation}
and also the Gaudin determinant $\det G$ for the norm of Bethe states in $\mathfrak{su}_2$ as we have derived in Section \ref{sec2.3}

\begin{equation}
G_{ij}=\partial_{u_i}\Phi_{j},\qquad \mathrm{with}\ \Phi_{j}=-i\log\Big[\Big(\frac{u_j-\frac{i}{2}}{u_j+\frac{i}{2}}\Big)^L\prod_{n\ne j}\frac{u_n-u_j+i}{u_n-u_j-i}\Big],
\label{eqn:eq00013}
\end{equation}
$G_\pm$ are introduced because of the paired structure of the Bethe roots $\{u_i\}= \{-u_i\}$
\begin{equation}
(G_{\pm})_{ij}=\partial_{u_i}\Phi_{j}\pm\partial_{u_{i+\frac{M}{2}}}\Phi_{j},
\label{eqn:eq00014}
\end{equation}
$M$ is the number of the excitations.

\subsection{$\mathfrak{su}_3$ sector}
The $\mathfrak{su}_3$ sector is quiet similar to the $\mathfrak{su}_2$ case, but with the local Hilbert space is $\mathbb{C}^3$, with basis $\lvert 1\rangle, \lvert 2\rangle,\lvert 3\rangle$. The corresponding MPS is 
\begin{equation}
\lvert \mathrm{MPS}\rangle_k=\sum_{i_k=1}^3\mathrm{Tr}\, [t_{i_{1}}\dots t_{i_{L}}]\lvert i_1\dots i_L\rangle,
\label{eqn:eq00015}
\end{equation}
 $t_{i}$ is what we define before. We denote $M_1$ as the total number of excitations $\lvert 2\rangle$ and $\lvert 3\rangle$,$M_2$ as the number of excitations $\lvert 3\rangle$. The Bethe equations, see Appendix \ref{AppendixC}, \ref{AppendixD}, for $SU(3)$ fundamental spin chain $(L,M_1,M_2)$ are 
\begin{equation}
\begin{split}
1&=\Big(\frac{u_m-\frac{i}{2}}{u_m+\frac{i}{2}}\Big)^L\prod_{n\ne m}^{M_1}\frac{u_m-u_n+i}{u_m-u_n-i}\prod_{n=1}^{M_2}\frac{u_m-v_n-\frac{i}{2}}{u_m-v_n+\frac{i}{2}},\quad m=1,\dots,M_1\\
1&=\prod_{m=1}^{M_1}\frac{v_n-u_m-\frac{i}{2}}{v_n-u_m+\frac{i}{2}}\prod_{m\ne n}^{M_2}\frac{v_n-v_m+i}{v_n-v_m-i},\quad n=1,\dots M_2,
\end{split}
\label{eqn:eq00016}
\end{equation}
one can use the above equation to fully determine Bethe states $\lvert \mathbf{u},\mathbf{v}\rangle$. The corresponding Gaudin determinant $G=\det G_{IJ}$ are \cite{su3_norm} defined by 

\begin{equation}
G_{IJ}=\partial_I\Phi_J,
\label{eqn:eq00017}
\end{equation}
with $I,J=1,\dots,M_1,M_1+1,\dots,M_1+M_2$, and $\Phi$ are obtained by taking the logarithm of the right hand side of Bethe equation  

\begin{equation}
\begin{split}
\phi^{v}_{m}&=-i\log \Big[\Big(\frac{v_m-\frac{i}{2}}{v_m+\frac{i}{2}}\Big)^L\prod_{n\ne m}^{M_1}\frac{v_m-v_n+i}{v_m-v_n-i}\prod_{n=1}^{M_2}\frac{v_m-w_n-\frac{i}{2}}{v_m-w_n+\frac{i}{2}}\Big],\quad m=1,\dots,M_1,
\\
\phi^{w}_{n}&=-i\log \Big[\prod_{m=1}^{M_1}\frac{w_n-v_m-\frac{i}{2}}{w_n-v_m+\frac{i}{2}}\prod_{m\ne n}^{M_2}\frac{w_n-w_m+i}{w_n-w_m-i}\Big],\quad n=1,\dots M_2.
\end{split}
\label{eqn:eq00018}
\end{equation}
Noted that due to integrable properties of MPS, the Bethe roots again have to be paired 
\begin{equation}
\{u_i,v_j\}=\{-u_i,-v_j\}.
\label{eqn:eq00019}
\end{equation}
And we can also define the $G_+$ and $G_-$ follow the same rules as $\mathfrak{su}(2)$. Then as shown in \cite{de_Leeuw_2016}, the one point function can be determinant via
\begin{equation}
C_k^{SU(3)}=\sqrt{\frac{Q_1(0)Q_1(\frac{1}{2})}{\bar{Q}_2(0)\bar{Q}_2(\frac{1}{2})}}T_{k-1}(0)\sqrt{\frac{\det G_+}{\det G_-}},
\label{eqn:eq00020}
\end{equation}
where
\begin{equation}
T_n(x)=\sum_{a=-\frac{n}{2}}^{\frac{n}{2}}(-ix+a)^L\frac{Q_1(-ix+\frac{(n+1)}{2})Q_2(-ix+a)}{Q_1(-ix+(a+\frac{1}{2}))Q_1(-ix+(a-\frac{1}{2}))}.
\label{eqn:eq00021}
\end{equation}
Also we define the Baxter Q-functions 
\begin{equation}
Q_1(x)=\prod_{i=1}^{M_1}(ix-u_i), \quad Q_2(x)=\prod_{j=1}^{M_2}(ix-v_j),
\label{eqn:eq00022}
\end{equation}
and $\bar{Q}_2(x)=\prod_{j=1,v_j\ne=0}^{M_2}(x-v_j)$. 
\subsection{$\mathfrak{so}_6$ sector}
The rank $3$ algebra $\mathfrak{so}_6$ sector is quiet similar to the $\mathfrak{su}_2$ case, but with $3$ types of excitations. The corresponding MPS is 
\begin{equation}
\lvert \mathrm{MPS}\rangle_k=\sum_{i_k=1}^6\mathrm{Tr}\, [t_{i_{1}}\dots t_{i_{L}}]\lvert i_1\dots i_L\rangle,
\label{eqn:eq00023}
\end{equation}
 $t_{i}, i=1,2,3$ is what we define before, $t_4=t_5=t_6=0$ and $\phi$ are the scalar fields. We denote $M$ as the total number of excitations, $N_{1}$, $N_{2}$ as the number of excitations for specific type. Follow Eq. (\ref{eqn:eqc61}) the Bethe equations for $SO(6)$ fundamental spin chain $(L,M,N_{1},N_{2})$ are 
\begin{equation}
\begin{split}
1&=\Big(\frac{u_i-\frac{i}{2}}{u_i+\frac{i}{2}}\Big)^L\prod_{j\ne i}^{M}\frac{u_i-u_j+i}{u_i-u_j-i}\prod_{k=1}^{N_1}\frac{u_i-v_k^{(1)}-\frac{i}{2}}{u_i-v_k^{(1)}+\frac{i}{2}}\prod_{k=1}^{N_2}\frac{u_i-v_k^{(2)}-\frac{i}{2}}{u_i-v_k^{(2)}+\frac{i}{2}},\quad i=1,\dots,M\\
1&=\prod_{k=1}^{M}\frac{v_i^{(1)}-u_k-\frac{i}{2}}{v_i^{(1)}-u_k+\frac{i}{2}}\prod_{m\ne i}^{N_1}\frac{v_i^{(1)}-v_m^{(1)}+i}{v_i^{(1)}-v^{(1)}_m-i},\quad i=1,\dots N_1,\\
1&=\prod_{k=1}^{M}\frac{v_i^{(2)}-u_k-\frac{i}{2}}{v_i^{(2)}-u_k+\frac{i}{2}}\prod_{m\ne i}^{N_2}\frac{v_i^{(2)}-v_m^{(2)}+i}{v_i^{(2)}-v^{(2)}_m-i},\quad i=1,\dots N_2.
\end{split}
\label{eqn:eq00024}
\end{equation}
one can use the above equation to fully determine Bethe eigenstates $\lvert \mathbf{u},\mathbf{v^{(1)}},\mathbf{v^{(2)}}\rangle$. The corresponding Gaudin determinant $G=\det G_{IJ}$ \cite{phd_Escobed_Jorge} are defined by 
\begin{equation}
G_{IJ}=\partial_I \Phi_J,
\label{eqn:eq00025}
\end{equation}
with $I,J=1,\dots,M+N_1+N_2$, and $\Phi$ are obtained by taking the logarithm of the right hand side of Bethe equation. Noted that due to integrable properties of MPS, the Bethe roots again have to be paired 
\begin{equation}
\{u_i,v^{(1)}_j,v^{(2)}_j\}=\{-u_i,-v^{(1)}_j,-v^{(2)}_j\}.
\label{eqn:eq00026}
\end{equation}
And we can also define the $G_+$ and $G_-$ follow the same rules as $\mathfrak{su}(2)$. Then as shown in \cite{de_Leeuw_2018}, the one point function can be determinant via
\begin{equation}
C_k^{SO(6)}=\sqrt{\frac{Q_1(0)Q_1(\frac{1}{2})Q_1(\frac{k}{2})Q_1(\frac{k}{2})}{\bar{Q}_2(0)\bar{Q}_2(\frac{1}{2})\bar{Q}_3(0)\bar{Q}_3(\frac{1}{2})}}\mathbb{T}_{k-1}(0)\sqrt{\frac{\det G_+}{\det G_-}},
\label{eqn:eq00027}
\end{equation}
where
\begin{equation}
\mathbb{T}_n(x)=\sum_{a=-\frac{n}{2}}^{\frac{n}{2}}(-ix+a)^L\frac{Q_2(-ix+a)Q_3(-ix+a)}{Q_1(-ix+(a+\frac{1}{2}))Q_1(-ix+(a-\frac{1}{2}))}.
\label{eqn:eq00028}
\end{equation}
Also we define the Baxter Q-functions 
\begin{equation}
Q_1(x)=\prod_{i=1}^{M}(ix-u_i), \quad Q_{2}(x)=\prod_{j=1}^{N_{1}}(ix-v^{(1)}_j),\quad Q_{3}(x)=\prod_{j=1}^{N_{2}}(ix-v^{(2)}_j)
\label{eqn:eq00029}
\end{equation}
and $\bar{Q}_{2}(x)=\prod_{j=1,v^{(1)}_j\ne0}^{N_{1}}(ix-v^{(1)}_j)$, $\bar{Q}_{3}(x)=\prod_{j=1,v^{(2)}_j\ne0}^{N_{2}}(ix-v^{(2)}_j)$. 
% Chapter 4

\chapter{Matrix product states and twisted Boundary Yang-Baxter Equation} % Main chapter title

\label{Chapter4} % For referencing the chapter elsewhere, use \ref{Chapter4}
%----------------------------------------------------------------------------------------
In defected $\mathcal{N}=4$ super Yang-Mills, our one-point functions calculation involves matrix product states (MPS). Such states will be proved integrable \cite{de_Leeuw_2018}, such that its overlaps with the eigenstates can be expressed in a simple factorized form. Moreover, one can use the definition of such integrability to show that any such MPS will correspond to a solution to the $twisted\ Boundary\ Yang\mbox{-}Baxter\ equation$ (BYB). On the other hand, the twisted Boundary Yang-Baxter equation is deeply related to the $twisted\ Yangian\ algebra$, a specific solution to BYB is a representation of twisted Yangian. Due to the Hopf algebra structure of twisted Yangian, one can get higher representation by taking a coproduct in the lower representation. Such procedure is called $dressing$, that it corresponds to acting with a transfer matrix on the lower state, and the new state is also integrable by definition. We will focus on the two cases we pay attend to in the Chapter \ref{Chapter2}, $(SU(3),SO(3))$ and $(SO(6),SO(3)\times SO(3))$. Where the first entry suggest the symmetry of the spin chain, and the second entry suggest the symmetry of the MPS which will be defined later. 
\section{Integrable states}
As a reminder, the key object in integrable model is the so-called R-matrix, and such integrable model will be symmetric with respect to certain Lie group $\mathcal{G}$ if 
\begin{equation}
[G_{1}\otimes G_2, R_{12}]=0,
\label{eqn:eq00001}
\end{equation}
with $G_1,G_2\in \mathcal{G}$. There are two classes of such models we will focus on. First $\mathcal{G}=SU(N)$, the corresponding R-matrix will be 
\begin{equation}
R(u)=\mathbb{I}-\frac{\mathbb{P}}{u},
\label{eqn:eq00002}
\end{equation}
second class would be $\mathcal{G}=SO(N)$ 
\begin{equation}
R(u)=\mathbb{I}-\frac{\mathbb{P}}{u}+\frac{\mathbb{K}}{u-\kappa}.
\label{eqn:eq00003}
\end{equation}
Here $\mathbb{P}$ is the permutation operator with matrix elements $\mathbb{P}_{ab}^{cd}=\delta_a^d\delta_b^c$, the first column denotes the first space, and the second column denotes the second space. $\mathbb{K}$ is the trace operator with matrix elements $\mathbb{K}_{ab}^{cd}=\delta_{ab}\delta^{cd}$, and $\kappa=\frac{N}{2}-1$. The above R-matrices are symmetric under the exchange of the two space. Both of them satisfy the unitary condition $R_{12}(u)R_{21}(-u)=f(u)\mathbb{I}$,  and Yang-Baxter equation
\begin{equation}
R_{12}(v)R_{13}(u)R_{23}(u-v)=R_{23}(u-v)R_{13}(u)R_{12}(v).
\label{eqn:eq00004}
\end{equation}
Let us define the monodromy matrix as usual
\begin{equation}
T(u)=R_{0L}(u)\dots R_{02}(u)R_{01}(u),
\label{eqn:eq00005}
\end{equation}
the transfer matrix is $t(u)=\mathrm{Tr}_0\, T(u)$, and $0$ denotes auxiliary space, $1,\dots,L$ denote local physical space. We can also define a set of local conserved charges of the model as 
\begin{equation}
Q_r=-\frac{i}{(r-1)!}\frac{d}{du^{r-1}}\log t(u)\Big\vert_{u=0},\quad r\ge2.
\label{eqn:eq00006}
\end{equation}
The behavior of the charges under space reflection is 
\begin{equation}
\sigma Q_j\sigma=(-1)^{j}Q_j,
\label{eqn:eq00007}
\end{equation}
such property can be proved by using the boost operator to recursively generate all the conserved charges \cite{L_D_Faddeev_1983,Loebbert_2016} and the fact that Hamiltonian $H=Q_2$ is invariant under space reflection . The space reflection operator $\sigma$ acts on the Hilbert space as
\begin{equation}
\sigma: v_1\otimes\dots\otimes v_L\mapsto v_L\otimes\dots\otimes v_1.
\label{eqn:eq00008}
\end{equation}
Also we introduce the space reflected transfer matrix 
\begin{equation}
\tilde{t}(u)=\sigma t(u)\sigma=\mathrm{Tr}_0\, R_{01}(u)\dots R_{0L}(u).
\label{eqn:eq00009}
\end{equation}
Before discussing about the integrability of  $\mathrm{MPS}$, we'd better introduce what $\mathrm{MPS}$ is. Let's take $N$ $k$-dimensional matrices $\omega_j,\enspace j=1,\dots,N$ acting on a $k$ dimensional auxiliary color space $V_c$ (differ from $V_0$), and $N$ is number denoting $SU(N)$ or $SO(N)$. The $\mathrm{MPS}$ is an element of the Hilbert space of the spin chain defined as
\begin{equation}
\vert\mathrm{MPS}_{\omega}\rangle=\sum_{j_1,\dots,j_L=1}^N\mathrm{Tr}_c\,[\omega_{j_L}\dots\omega_{j_1}]\vert j_L,\dots,j_1\rangle.
\label{eqn:eq000010}
\end{equation}
An $\mathrm{MPS}$ is invariant under a subgroup  $\mathcal{G}^{\prime}\subset\mathcal{G}$ if for every $G\in \mathcal{G}^{\prime}$
\begin{equation}
[\otimes_{j=1}^{L}G_{j}]\vert \mathrm{MPS}_{\omega}\rangle=\vert \mathrm{MPS}_{\omega}\rangle,
\label{eqn:eq000011}
\end{equation}
Such state can be constructed by a group invariant $\omega$. Let us assume that there is a representation $\Lambda_{\omega}$ of $\mathcal{G}^{\prime}$ acting on the auxiliary space $V_c$. The building block $\omega$ is group invariant, if for all $G \in \mathcal{G}^{\prime}$
\begin{equation}
\Lambda_{\omega}(G^{-1})\omega_{j}\Lambda_{\omega}(G)=\sum_{k}G_{jk}\omega_k,\quad\quad j=1,\dots,N,
\label{eqn:eq000012}
\end{equation}
where $G_{jk}$ are the matrix elements of $G$ in the defining representation. And this equation put restrictions on $\Lambda_{\omega} $.  And we denote such case as $(\mathcal{G},\mathcal{G}^{\prime})$, where $\mathcal{G}$ is the symmetry of the spin chain, and $\mathcal{G}^{\prime}$ is the symmetry of $\mathrm{MPS}$ \cite{Pozsgay_2019}.

For our $SU(3)$ sector in Chapter \ref{Chapter3}, it corresponds to symmetric pair $(SU(3),SO(3))$. And our group invariant is chosen as the generators of $SU(2)$ algebra $\omega_i=S_i,\ i=1,2,3$ 
\begin{equation}
[S_i,S_j]=i\epsilon_{ijk}S_{k},
\label{eqn:eq000013}
\end{equation}
in the finite irreducible representation. Clearly they satisfies relation Eq. (\ref{eqn:eq000012}) by $(G_{i})_{jk}=\epsilon_{ijk}$ and $\Lambda_{\omega}(G)\sim1+iS_{i}$ infinitesimally. 

For our $SO(6)$ sector in Chapter, it corresponds to symmetric pair $(SO(6),SO(3)\times SO(3))$. Since we require one of the $SO(3)$ in the trivial scalar representation, then $\mathcal{G}^{\prime}\simeq SO(3)$. We can choose $\omega_i = S_i,\ i=1,2,3$ and $\omega_i=0,\ i=4,5,6$. Then due to the sum over $i=1,\dots, 6$, we are actually reduces to the case of $SU(3)$ sector, thus Eq. (\ref{eqn:eq000012}) is again satisfied. Hence prove the invariance property.\footnote{For $(SO(6),SO(5))$ case, Eq. (\ref{eqn:eq000012}) turns out to be $[G_{k},G_{ij}]=\sum_{\ell}G_{ij}G_{\ell}=\delta_{ik}G_{j}-\delta_{kj}G_{i}$, where I have used $(G_{ij})_{kl}=\delta_{ik}\delta_{jl}-\delta_{il}\delta_{kj}$ for defining representation of $SO(N)$.}

\subsection{Definition of integrable MPS}
Consider an integrable spin chain with transfer matrix $t(u)$. A MPS is integrable \cite{Piroli_2017} if it satisfies 
\begin{equation}
\tilde{t}(u)\lvert\mathrm{MPS}_{\omega}\rangle=t(u)\lvert\mathrm{MPS}_{\omega}\rangle.
\label{eqn:eq000014}
\end{equation}
Since the parity reverses rapidity, $u\rightarrow -u$. By Eq. (\ref{eqn:eq00007}), the definition of conserved charges, one can first take logarithm of both side of Eq. (12), then expand it out in terms of $u$ around 0. By matching the order of $u$, Eq. (\ref{eqn:eq000014}) implies 

\begin{equation}
Q_{2n+1}\lvert \mathrm{MPS}_{\omega}\rangle=0.
\label{eqn:eq000015}
\end{equation}
While the even order conserved Charges are trivially satisfied. For detail proof of the integrability of MPS showed in Chapter \ref{Chapter3}, one can be found in \cite{de_Leeuw_2018}.

\section{From integrabilty condition to square root relation}
In this section, we will link our integrability condition Eq. (\ref{eqn:eq000014}) with a relation called square root relation (sq.r.r.).
Before going deep into analyze, let us introduce some definition and properties for MPS. First, a MPS is irreducible, if there is no proper subspace (exclude identity) $V_c^{\prime}\subset V_c$ which is an invariant subspace for all $\omega_a,\ a=1,\dots, N$. 
\begin{thm}
If two sets of matrices $\{\alpha_j\}$ and $\{\beta_j\}$ are completely reducible, and they produce the same $\mathrm{MPS}$ for all $L$:
\begin{equation}
\vert \Psi_\alpha(L)\rangle=\vert \Psi_\beta(L)\rangle,
\label{eqn:eq000016}
\end{equation}
then there is a simultaneous similarity transformation $S$ connecting the two sets as
\begin{equation}
\alpha_j=S^{-1}\beta_jS,\quad j=1,\dots,N.
\label{eqn:eq000017}
\end{equation}
\label{thm:th001}
\end{thm}
Here the completely reducibility is to ensure there is no difference in the off-diagonal terms which will not appear in the trace. Let us now examine the integrability condition Eq. (\ref{eqn:eq000014}). Introduce the R matrix as 
\begin{equation}
R_{10}(u)=\sum^N_{a,b=1}E_{ab}\otimes \mathcal{L}_{ab}(u),
\label{eqn:eq000018}
\end{equation}
where $E_{ab}$ acts on the physical space $V_1$, while $\mathcal{L}_{ab}$ acts on the auxiliary space $V_0$. Apply the space reflection operator on the MPS
\begin{equation}
\begin{split}
\sigma\lvert MPS_{\omega}\rangle&=\sum_{j_1,\dots,j_L=1}^N\mathrm{Tr}_c[\omega_{j_1}\dots \omega_{j_L}]\vert j_L,\dots,j_1\rangle\\
&=\sum_{j_1,\dots,j_L=1}^N\mathrm{Tr}_c[\omega^{T}_{j_L}\dots \omega^T_{j_1}]\vert j_L,\dots,j_1\rangle.
\end{split}
\label{eqn:eq000019}
\end{equation}
where I have used $\mathrm{Tr}(A)=\mathrm{Tr}(A^T)$. By the definition of transfer matrix, one can see
\begin{equation}
\begin{split}
t(u)\sigma\lvert \mathrm{MPS}_{\omega}\rangle&=\sum^N_{a_i,b_i=1}\sum_{j_i=1}^NE_{a_Lb_L}\otimes\dots\otimes E_{a_1b_1}\mathrm{Tr}_0(\mathcal{L}_{a_Lb_L}\dots\mathcal{L}_{a_1b_1})\mathrm{Tr}_c[\omega^{T}_{j_L}\dots \omega^T_{j_1}]\vert j_L,\dots,j_1\rangle\\
&=\sum^N_{j_i=1}\mathrm{Tr}_{0,c}(\mathcal{L}_{a_Lj_L}\otimes\omega^{T}_{j_L}\dots\mathcal{L}_{a_1j_1}\otimes\omega^T_{j_1})\vert a_L,\dots,a_1\rangle,
\end{split}
\label{eqn:eq000020}
\end{equation}
where I have used $\mathrm{Tr}_a(A)\mathrm{Tr}_b(B)=\mathrm{Tr}_{ab}(A\otimes B)$, and $E_{ij}\lvert l\rangle=\delta_{jl}\lvert i\rangle$. One can see that by acting with a transfer matrix, we are somehow expanding our color space. By using Eq. (\ref{eqn:eq000019}) again, the left hand side of integrability condition yields 
\begin{equation}
\begin{split}
\tilde{t}(u)\lvert\mathrm{MPS}_{\omega}\rangle&=\sigma\sum^N_{j_i=1}\mathrm{Tr}_{0,c}(\mathcal{L}_{a_Lj_L}\otimes\omega^{T}_{j_L}\dots\mathcal{L}_{a_1j_1}\otimes\omega^T_{j_1})\vert a_L,\dots,a_1\rangle\\
&=\sum^N_{j_i=1}\mathrm{Tr}_{0,c}(\mathcal{L}^{T}_{a_Lj_L}\otimes\omega_{j_L}\dots\mathcal{L}^T_{a_1j_1}\otimes\omega_{j_1})\vert a_L,\dots,a_1\rangle.
\end{split}
\label{eqn:eq000021}
\end{equation}
Compare with the right hand side of Eq. (\ref{eqn:eq000014}), one would easily get that the integrability condition is equivalent to
\begin{equation}
\lvert\mathrm{MPS}_{\alpha}\rangle=\lvert\mathrm{MPS}_{\beta}\rangle,
\label{eqn:eq000022}
\end{equation}
with 
\begin{equation}
\alpha_{j}(u)=\sum_{k}^{N}\mathcal{L}_{jk}(u)\otimes \omega_{k},\quad \beta_{j}(u)=\sum_{k}^{N}\mathcal{L}^T_{jk}(u)\otimes \omega_{k}.
\label{eqn:eq000023}
\end{equation}
\begin{thm}
 If the $\lvert\mathrm{MPS}_A\rangle$ and $\lvert\mathrm{MPS}_B\rangle$ are completely reducible, then there exists a simultaneous similarity invertible transformation $K(u)$ such that
 \begin{equation}
\alpha_j(u)K(u)=K(u)\beta_{j}(u), \quad j=1,\dots, N,
\label{eqn:eq000024}
\end{equation}
the $K(u)$ matrix acts on big auxiliary space $V_0\otimes V_c$, we will denote it as 

\begin{equation}
K(u)=\sum_{a,b}E_{ab}\otimes \psi_{ab}(u).
\label{eqn:eq000025}
\end{equation}
\label{thm:th002}
\end{thm}

Where $E_{ab}$ are elementary matrices acting on auxiliary space $V_0$, and $\psi_{ab}(u), a,b=1,\dots, N$ are matrices acting on color space $V_c$. The set $\{\psi_{ab}(u)\}$ will be denoted as $\psi(u)$, called two-site block, the reason for this name will be clear soon. 

It is important to impose the reducibility for Eq. (\ref{eqn:eq000024}). If our MPS is irreducible, then we can determine $K(u)$ up to a scalar function of rapidity. Later we will see that this is actually a reminiscent of twisted Yangian algebra, and this happens in the $(SO(6),SO(5))$ case. However, if our MPS is completely reducible, then the $K(u)$ is not unique since we can pick different scalar functions for each blocks, and this happens in the $(SU(3),SO(3))$and $(SO(6),SO(3)\times SO(3))$ case when embedding them into big groups. 

Insert Eq. (\ref{eqn:eq000025}) into Eq. (\ref{eqn:eq000024}), one can obtain the following square root relation with Einstein summation rule implict,
\begin{equation}
R_{jd}^{kb}(u)\omega_k\psi_{bc}(u)=R_{jc}^{kb}(u) \psi_{db}(u)\omega_k,
\label{eqn:eq000026}
\end{equation}
$Proof$:
For every $j=1,\dots, N$, Eq. (\ref{eqn:eq000024}) is equal to 
\begin{equation}
\sum_{a,b,k=1}^{N}\Big(\mathcal{L}_{jk}(u)E_{ab}\Big)\otimes \Big(\omega_{k}\psi_{ab}(u)\Big)=\sum_{a,b,k=1}^{N}\Big(E_{ab}\mathcal{L}^T_{jk}(u)\Big)\otimes \Big(\psi_{ab}(u)\omega_{k}\Big),
\label{eqn:eq000027}
\end{equation}
multiply with $E_{cd}\otimes I$ in $V_0\otimes V_c$ space, $c,d$ is unspecified indices. And take the trace over $V_0$, one would get

\begin{equation}
\sum_{a,b,k=1}^{N}\mathrm{Tr}_0[E_{cd}\mathcal{L}_{jk}(u)E_{ab}]\otimes \Big(\omega_{k}\psi_{ab}(u)\Big)=\sum_{a,b,k=1}^{N}\mathrm{Tr}_0[E_{cd}E_{ab}\mathcal{L}^T_{jk}(u)]\otimes \Big(\psi_{ab}(u)\omega_{k}\Big),
\label{eqn:eq000028}
\end{equation}
using $E_{ab}E_{cd}=\delta_{bc}E_{ad}$, $\mathrm{Tr}\,AB^T=\mathrm{Tr}\, A^T B$ and $E^T_{ab}=E_{ba}$, Eq. (28) yields
\begin{equation}
\sum_{a,k}\mathrm{Tr}_0\,[\mathcal{L}_{jk}(u)E_{ad}]\Big(\omega_{k}\psi_{ac}(u)\Big)=\sum_{b,k}\mathrm{Tr}_0[\mathcal{L}_{jk}(u)E_{bc}]\Big(\psi_{db}(u)\omega_{k}\Big).
\label{eqn:eq000029}
\end{equation}
Recall our R matrices definition Eq. (\ref{eqn:eq000018}), in component form
\begin{equation}
R_{bc}^{ad}=\sum_{ij}(E_{ij})^a_b\otimes(\mathcal{L}_{ij})^d_c=(\mathcal{L}_{ab})^d_c=\mathrm{Tr}_0\, (E_{cd}\mathcal{L}_{ab}),
\label{eqn:eq000030}
\end{equation}
plug into Eq. (\ref{eqn:eq000029}), one would get the sq.r.r..

For invertible $\omega_j$ the sq.r.r. implies the initial condition
\begin{equation}
\psi_{jk}(0)=\omega_j\omega_k,
\label{eqn:eq000031}
\end{equation}
up to a scalar factor. One can prove this by using $R(0)=\mathbb{P}$ for both $SU(N)$ and $SO(N)$, also "Schur lemma" in $\omega_i$ \cite{Pozsgay_2019}. Thus our integrable condition is converted to find invertible  solution $K(u)$ of sq.r.r.. 
\section{From square root relation to BYB}
In this section we will explain that our $K(u)$ matrix obtain from sq.r.r. relation satisfying the twisted Boundary Yang-Baxter equation \cite{BYB_1984,BYB_1994}
\begin{equation}
K_2(v)R^T_{21}(u+v)K_1(u)R_{12}(v-u) = R_{21}(v-u)K_1(u)R_{12}^T(u+v)K_{2}(v),
\label{eqn:eq000032}
\end{equation}
where the R matrix is defined as before, note that since our R matrix is symmetric under swapping spaces, so
\begin{equation}
(R^T)_{ab}^{cd}=(R^{T_1})_{ab}^{cd}=(R^{T_2})_{ab}^{cd}=(R)_{cb}^{ad}=(R)_{ad}^{cb}.
\label{eqn:eq000033}
\end{equation}
Plug all the definition of $K(u)$ and $R(u)$, the left hand side of Eq. (\ref{eqn:eq000032}) is
\begin{equation}
\begin{split}
LHS&=\sum_{ab}\overset{2} {E_{ab}}\otimes\overset{c}{\psi_{ab}}(v)\sum_{cd}\overset{2} {E_{dc}}\otimes\overset{1}{\mathcal{L}_{cd}}(u+v)\sum_{ef}\overset{1} {E_{ef}}\otimes\overset{c}{\psi_{ef}}(u)\sum_{gh}\overset{1} {E_{gh}}\otimes\overset{2}{\mathcal{L}_{ab}}(v-u)\\
&=\sum_{abcdefgh}\overset{2}{E_{ab}E_{dc}\mathcal{L}_{gh}(v-u)}\otimes\overset{1}{\mathcal{L}_{cd}(u+v)E_{ef}E_{gh}}\otimes\overset{c}{\psi_{ab}(v)\psi_{ef}(u)}\\
&=\sum_{abcef}\overset{2}{E_{ac}\mathcal{L}_{fh}(v-u)}\otimes\overset{1}{\mathcal{L}_{cd}(u+v)E_{eh}}\otimes\overset{c}{\psi_{ab}(v)\psi_{ef}(u)},
\end{split}
\label{eqn:eq000034}
\end{equation}
multiply with $\overset{2}{E_{\alpha\beta}}\otimes\overset{1}{E_{\gamma\omega}}$ on both side of Eq. (\ref{eqn:eq000034}), then take trace over space $1$ and $2$, the above equation is equivalent to 
\begin{equation}
\begin{split}
\widetilde{LHS}&=\sum_{befc}\mathrm{Tr}_{2}\, [E_{\alpha c}\mathcal{L}_{f\gamma}(v-u)]\times\mathrm{Tr}_{1}\, [E_{e\omega}\mathcal{L}_{cd}(u+v)]\times\psi_{\beta b}\psi_{ef}\\
&=\sum_{befc}R_{\gamma\alpha}^{fc}(v-u)\times R_{be}^{c\omega}(v+u)\times\psi_{\beta b}(v)\psi_{ef}(u)
\end{split}
\label{eqn:eq000035}
\end{equation}
 while the right side after the multiplication yields
\begin{equation}
\widetilde{RHS}=\sum_{bcdg}R_{dg}^{\gamma b}(u+v)\times R_{bc}^{\beta \omega}(v-u)\times\psi_{cd}(u)\psi_{g\alpha}(v).
\label{eqn:eq000036}
\end{equation}
Put $\widetilde{LHS}=\widetilde{RHS}$, one would obtain
\begin{equation}
R_{\gamma\alpha}^{dc}(v-u) R_{ab}^{c\omega}(v+u)\psi_{\beta a}(v)\psi_{bd}(u)=R_{da}^{\gamma b}(u+v)R_{bc}^{\beta \omega}(v-u)\psi_{cd}(u)\psi_{a\alpha}(v).
\label{eqn:eq000037}
\end{equation}
The above indices explicit form can also be got from Eq. (\ref{eqn:eq000032}) directly by matrix multiplication of individual space and using Eq.(\ref{eqn:eq000033}). It can be seen that if our two-site block $\psi$ is a solution of BYB, we need double dressing it  compared to sq.r.r.. From the integrability condition and commutation relation between transfer matrix, given an integrable state $\lvert \mathrm{MPS} \rangle$ one can obtain 
\begin{equation}
Q_{2n+1}t(u)\lvert\mathrm{MPS}\rangle=t(u)Q_{2n+1}\lvert\mathrm{MPS}\rangle=0,
\label{eqn:eq000038}
\end{equation}
that is to say $t(u)\lvert\mathrm{MPS}\rangle$ is also integrable, hence
\begin{equation}
t(u)t(v)\lvert\mathrm{MPS}\rangle=\tilde{t}(u)\tilde{t}(v)\lvert\mathrm{MPS}\rangle.
\label{eqn:eq000039}
\end{equation}
As a matter of fact, one can get infinite integrable state by keeping act it with transfer matrices, and we call this procedure $dressing\ procedure$. Following the same procedure as dressing once, above equation is equivalent to say 
\begin{equation}
\lvert\mathrm{MPS}_{\alpha}\rangle=\lvert\mathrm{MPS}_{\beta}\rangle,
\label{eqn:eq000040}
\end{equation}
with
\begin{equation}
\alpha_{j}(u,v)=\sum_{k,l=1}^{N}\mathcal{L}_{jk}(u)\mathcal{L}_{kl}(v)\otimes \omega_{l},\quad \beta_{j}(u,v)=\sum_{k,l=1}^{N}\mathcal{L}^T_{jk}(u)\mathcal{L}^T_{kl}(v)\otimes \omega_{l}.
\label{eqn:eq000041}
\end{equation}
Assuming our double dressing MPS are irreducible, then according to Theorem \ref{thm:th001}, the similarity transformation matrices are unique up to a scalar function. As proved by \cite{Pozsgay_2019}, using Yang-Baxter relation and sq.r.r., there are two such transformation matrices

\begin{equation}
\begin{split}
M_1(u,v)=K_{2}(v)R^T_{21}(u+v)K_{1}(u)R_{12}(v-u),\\
M_2(u,v)=R_{21}(v-u)K_{1}(u)R^T_{12}(u+v)K_{2}(v),
\end{split}
\label{eqn:eq000042}
\end{equation}
meet the requirement $M(u,v)\alpha_{j}(u,v)=\beta_{j}M(u,v)$.  Thus 
\begin{equation}
M_{1}(u,v)=f(u,v)M_2(u,v)
\label{eqn:eq000043}
\end{equation}
with arbitrary function $f(u,v)$, compare with Eq. (\ref{eqn:eq000037}), we can set $f(u,v)=1$. 
Eq. (\ref{eqn:eq000032}) is closely related to the twisted Yangian algebra, and our dressing procedure is quite the same as take coproduct in this algebra. And RTT relation for the spin chain are related to Yangian algebra for $SU(N)$ or extend Yangian algebra for $SO(N)$. We will see this in next chapter. 
\section{Solutions for sq.r.r. and BYB}
\subsection{Scalar solutions}
As one can easily verify that $\psi_{ij}(u)=f(u)\delta_{ij},\ i,j=1,2,3$ is a solution to the BYB Eq. (\ref{eqn:eq000037}) for $(SU(3),SO(3))$ case with corresponding R matrix. And the corresponding MPS yields
\begin{equation}
\lvert\mathrm{MPS}_{\delta}\rangle=\otimes_{j=1}^{L/2}(\lvert 11\rangle+\lvert 22\rangle+\lvert 33\rangle).
\label{eqn:eq000044}
\end{equation}
here the $1,2,3$ are defined in Chapter \ref{Chapter2}.
The situation for $(SO(6),SO(3)\times SO(3))$ is a bit complicated, we need to make the following ansatz for the solution
\begin{equation}
\begin{split}
\psi_{ij}&=f(u)\delta_{ij},\\
\psi_{Ii}&=\psi_{iI}=0,\\
\psi_{IJ}&=h(u)\delta_{IJ},
\end{split}
\label{eqn:eq000045}
\end{equation}
with $i,j\in\{1,\dots,D\},I,J\in\{D+1,\dots,6\}, D=3$. Then it can be easily shown that $f(u)=-h(u)$ is a solution for BYB\footnote{For general $(SO(N),SO(D)\times SO(N-D))$, one can make a similar ansatz, the scalar solution takes the form $\begin{pmatrix}(N-D-1+2u)I_D&0\\0&(-D+1-2u)I_{N-D}\end{pmatrix}$ with $I$ denote the identity matrix.}. Hence the corresponding MPS yields
\begin{equation}
\lvert\mathrm{MPS}_{\delta_{\pm}}\rangle=\otimes_{j=1}^{L/2}(\lvert 11\rangle+\lvert 22\rangle+\lvert 33\rangle-\lvert 44\rangle-\lvert 55\rangle-\lvert 66\rangle).
\label{eqn:eq000046}
\end{equation}
Once we get the integrable scalar solutions, the higher dimensional integrable solution can be generated by the dressing procedure. But any way, we can also independently solve the sqr.r.r. to find the higher dimensional solutions. If our MPS is irreducible, then these two methods can give us the same solution of $K(u)$ up to a normalization. The case for completely reducible is a bit complicated, we will talk about it in next Chapter. 
\subsection{Higher dimensional solutions}
As we have seen from the derivation of sq.r.r., when we do the dressing procedure, the color space will be expand by the auxiliary space of the transfer matrix. Hence, it will give us a higher dimensional MPS. But in this section, a different approach will be presented.
For $(SU(3),SO(3))$, let us make the following ansatz 
\begin{equation}
\begin{split}
\psi_{ij}(u)&=(g(u)+\tilde{g}(u))S_iS_j-\tilde{g}(u)[S_i,S_j]+f(u)\delta_{ij}\\
            &=g(u)S_iS_j+\tilde{g}(u)S_jS_i+f(u)\delta_{ij},
\end{split}
\label{eqn:eq000047}
\end{equation}
with $\omega_{i}=S_i,\ i=1,2,3$, $k$ dimensional $SU(2)$ generators. Plug into sq.r.r. Eq. (\ref{eqn:eq000026}), one would get
\begin{equation}
\begin{split}
-i u^2 g(u)+2 i u g(u)-i u^2 \tilde{g}(u)+i u \tilde{g}(u)+2 i \tilde{g}(u)=0\\
-u f(u)+2 f(u)+u^2 \tilde{g}(u)-2 u \tilde{g}(u)=0,
\end{split}
\label{eqn:eq000048}
\end{equation}
solve it in terms of $g(u)$, one would obtain $\tilde{g}(u)=-\frac{ug(u)}{1+u}$, $f(u)=-\frac{u^2g(u)}{1+u}$. Choose $g(u)=1+u$ since we can only fix $\psi(u)$ up to a normalization, then one would get
\begin{equation}
\begin{split}
g(u)&=1+u,\\
\tilde{g}(u)&=-u,\\
f(u)&=-u^2.
\end{split}
\label{eqn:eq000049}
\end{equation}
So the corresponding MPS would look like
\begin{equation}
\begin{split}
\lvert \mathrm{MPS}_{k}\rangle&=\sum_{i_{\ell},j_{\ell}=1}^{3}\mathrm{Tr}_c\,[\psi_{j_{L/2},i_{L/2}}(0)\dots\psi_{j_{1},i_{1}}(0)]\lvert j_{L/2}i_{L/2}\dots j_1i_1\rangle\\
&=\sum_{i_{\ell},j_{\ell}=1}^{3}\mathrm{Tr}_c\,[S_{j_{L/2}}S_{i_{L/2}}\dots S_{j_1}S_{i_1}]\lvert j_{L/2}i_{L/2}\dots j_1i_1\rangle,
\end{split}
\label{eqn:eq000050}
\end{equation}
which is exactly the MPS in our D3-D5 brane $SU(3)$ sector.

The thing get a bit complicated for our $(SO(6),SO(3)\times SO(3))$ case. Let us make the following ansatz 
\begin{equation}
\begin{split}
\psi_{ij}(u)&=(g(u)+\tilde{g}(u))S_iS_j-\tilde{g}(u)[S_i,S_j]+f(u)\delta_{ij}\\
            &=g(u)S_iS_j+\tilde{g}(u)S_jS_i+f(u)\delta_{ij},\\
\psi_{iI}&=\psi_{Ii}=0,\\
\psi_{IJ}&=h(u)\delta_{IJ},
\end{split}
\label{eqn:eq000051}
\end{equation}
where $i,j\in \{1,2,3\}$ and $I,J\in \{4,5,6\}$. With $\omega_{i}=S_i,\ i=1,\dots,6$, and $S_{4,5,6}=0$. Furthermore, $S_1,S_2,S_3$ form a k-dimensional irreducible representation of $SU(2)$ with quadratic Casimir $C$
\begin{equation}
\begin{split}
[S_a,S_b]&=i \epsilon_{abc} S_c,\\
S_1^2+S_2^2+S_3^2=C\cdot I_k,&\quad C=\frac{k^2-1}{4}=s(s+1).
\end{split}
\label{eqn:eq000052}
\end{equation}
Plug the ansatz into the sq.r.r., one would obtain
\begin{equation}
\begin{split}
ug(u)+(u-1)\tilde{g}(u)&=0,\\
Cug(u)+u(1+C+u)\tilde{g}(u)+2(1+u)f(u)&=0,\\
-(2+u)h(u)-u(Cg(u)+(C-1)\tilde{g}(u)+f(u))&=0.
\end{split}
\label{eqn:eq000053}
\end{equation}
Solving these equation in terms of $g(u)$ we get 
\begin{equation}
\begin{split}
\tilde{g}(u)&=-\frac{ug(u)}{u-1},\\
f(u)&=\frac{1}{2}\frac{u(C+u+u^2)}{u^2-1}g(u),\\
h(u)&=-\frac{1}{2}\frac{u(-C+u+u^2)}{u^2-1}g(u),
\end{split}
\label{eqn:eq000054}
\end{equation}
So we can choose arbitrary $g(u)$ to cancel all the singularities, which yields
\begin{equation}
\begin{split}
g(u)&=(1-u^2),\\
\tilde{g}(u)&= u (u+1),\\
f(u)&=-\frac{1}{2} u \left(c+u^2+u\right),\\
h(u)&=\frac{1}{2} u \left(-c+u^2+u\right).
\end{split}
\label{eqn:eq000055}
\end{equation}
So the corresponding MPS would look like
\begin{equation}
\begin{split}
\lvert \mathrm{MPS}_{k}\rangle&=\sum_{i_{\ell},j_{\ell}=1}^{6}\mathrm{Tr}_c\,[\psi_{j_{L/2},i_{L/2}}(0)\dots\psi_{j_{1},i_{1}}(0)]\lvert j_{L/2}i_{L/2}\dots j_1i_1\rangle\\
&=\sum_{i_{\ell},j_{\ell}=1}^{6}\mathrm{Tr}_c\,[S_{j_{L/2}}S_{i_{L/2}}\dots S_{j_1}S_{i_1}]\lvert j_{L/2}i_{L/2}\dots j_1i_1\rangle,
\end{split}
\label{eqn:eq000056}
\end{equation}
which is exactly the MPS in our D3-D5 brane $SO(6)$ sector. % Chapter 5

\chapter{Spin Chain Overlaps and Twisted Yangian for $(SU(3),SO(3))$} % Main chapter title

\label{Chapter5} % For referencing the chapter elsewhere, use \ref{Chapter5}
%----------------------------------------------------------------------------------------
As explained in last Chapter, the integrable matrix product state (MPS) can be constructed by solving the square root relation (sq.r.r.), which is actually an easy version of the twisted Boundary Yang-Baxter equation (BYB). One can also generate a set of integrable MPS by the dressing procedure. So in order to get a higher color dimensional MPS, we can make it in two ways, by directly solving sq.r.r. or using dressing procedure. 

In this Chapter, we will see that BYB is deeply related to the twisted Yangian algebra, the solutions of BYB correspond to representations of twisted Yangian \cite{tyangian2}. Our dressing procedure is just a reminiscent of its Hopf algebra properties left coideal structure, we can get higher representation of twisted Yangian by taking coproduct. And our Lax operator will map to the operator matrix element of T-matrix in Yangian algebra on given condition. Moreover as one can show that if two representations differ from each other only up to a scalar function, then these two representations are identical. This can be achieved by matching the highest weight of these two representations.

The whole idea of this chapter is: First, we find the scalar solution $\tilde{K}(u)$ of BYB, then transform it to the solution of tBYB $K(u)$ using transformation matrix $P$. We do the same thing for the higher dimensional solution of BYB $\tilde{S}(u)$ to solution of tBYB $S(u)$. Then we dress $K(u)$ to higher dimensional solution of tBYB $T(u)K(u)T^{t}(-u)$ with the same dimensional as $S(u)$. After matching the highest weight of these two same dimension representation, we transform it back to BYB using charge conjugation $S(u)C$ \cite{tyangian}, which will give us the MPS that we want. The dressing transfer matrix will give us the information for the overlap formula. 
\section{Definitions and preliminaries}
\subsection{Notation}
Let $n\in \mathbb{N}$ and set $N=2n$ or $N=2n+1$. I will assume that $\mathfrak{g}=\mathfrak{gl}_N$, $\mathfrak{g}=\mathfrak{sl}_N$ or $\mathfrak{g}=\mathfrak{g}_N$, where $\mathfrak{g}_N$ is the orthogonal Lie algebra $\mathfrak{so}_N$. I label the rows and columns of matrices in $\mathfrak{gl}_N$ by the indices $\{−n,\dots,−1,1,\dots,n\}$ if 
$N = 2n$ and $\{−n, \dots , −1, 0, 1, \dots , n\}$ if $N = 2n + 1$ (sometimes I will use also the convention of indices as $\{1,\dots,N\}$, it doesn't make a difference). The corresponding Lie algebra $\mathfrak{gl}_N$ is defined by
\begin{equation}
[E_{ij},E_{kl}]=\delta_{jk}E_{il}-\delta_{il}E_{kj},
\label{eqn:eq000001}
\end{equation}
with indices running from $1$ to $N$, and $E_{ij}$ are the generators for $\mathfrak{gl}_N$. We use a different convention transpose $t$
\begin{equation}
A^t=C A^TC^{-1},
\label{eqn:eq000002}
\end{equation}
$T$ is the standard transposition, and $C_{ij}=\delta_{i,-j}$ with $C^2=1$, and we call it charge conjugation matrix. 

The Lie algebra $\mathfrak{so}_N$ is defined by the relations in the $t$ convention transpose: 
\begin{equation}
[F_{ij},F_{kl}]=\delta_{jk}F_{il}-\delta_{il}F_{kj}+\delta_{j,-l}F_{k,-i}-\delta_{i,-k}F_{-j,l},
\label{eqn:eq000003}
\end{equation}

\begin{equation}
F^{t}_{ij}=F_{-j,-i}=-F_{ij},\quad \mathrm{with\ }F_{ij}=E_{ij}-E_{ij}^t
\label{eqn:eq000004}
\end{equation}
We may identify $\mathfrak{g}_N$ with $\mathrm{span}_{\mathbb{C}}\{F_{ij}:-n\le i,j\le n\}$ and we will use $\mathfrak{h}_N=\mathrm{span}_{\mathbb{C}}\{F_{ii}:1\le i\le n\}$ as Cartan subalgebra.
\begin{definition}
A representation $L(\lambda_1,\dots, \lambda_N)$ of the $\mathfrak{gl}_N$ is called highest weigh if there exist a non zero vector $v\in L$ such that 
\begin{equation}
E_{ij}\cdot v=0\quad \mathrm{for}\quad 1\le i<j\le N
\label{eqn:eq000005}
\end{equation}
and 
\begin{equation}
E_{ii}\cdot v=\lambda_i\cdot v\quad \mathrm{for}\quad i=1,\dots,N.
\label{eqn:eq000006}
\end{equation}
\end{definition}
\begin{definition}
A representation $V(\lambda_1,\dots,\lambda_n)$ is a highest weight representation if there exist a non zero vector $v\in V$ and the following condition are satisfied 
\begin{equation}
\begin{alignedat}{2}
F_{ij}\cdot v&=0\quad &&\mathrm{for}\ -n\le i<j\le n,\qquad\mathrm{and}\\
F_{ii}\cdot v&=\lambda_i\cdot v\quad &&\mathrm{for}\ 1\le i<j\le n,
\end{alignedat}
\label{eqn:eq000007}
\end{equation}
\end{definition}
\subsection{R-matrices}
A R-matrix is called group $\mathcal{G}$ invariant if 
\begin{equation}
G_1G_2R_{12}(u)=R_{12}(u)G_2G_1,
\label{eqn:eq000008}
\end{equation}
with $G\in\mathcal{G}$. R-matrix also satisfies quantum integrable condition the Yang-Baxter equation with spectral parameter
\begin{equation}
R_{12}(u-v)R_{13}(u)R_{23}(v)=R_{23}(v)R_{13}(v)R_{12}(u-v).
\label{eqn:eq000009}
\end{equation}
In our $SU(N)$ case, the R-matrix takes the form
\begin{equation}
R(u)=\mathbb{I}-\frac{\mathbb{P}}{u}.
\label{eqn:eq0000010}
\end{equation}
Also the two spaces of R-matrix are symmetric 
\begin{equation}
R^{t}=R^{t_1}=R^{t_2}.
\label{eqn:eq0000011}
\end{equation}
The R-matrix in the standard transpose $T$ is the same as in $t$ for $SU(N)$ case
\begin{equation}
\tilde{R}(u)=\mathbb{I}-\frac{\mathbb{P}}{u},
\label{eqn:eq0000012}
\end{equation}
this R-matrix is also symmetric in spaces. 
\subsection{Versions of BYB in different transpositions}
The reason why we choose a different convention for the transposition will be revealed later, it is a good way to define the highest weight representations for Yangian algebra and its descendants. But anyway, let's compare different versions for twisted Boundary Yang-Baxter equations, and seek a way to link them.

In the normal transposition with R-matrix Eq. (\ref{eqn:eq0000012}), the BYB takes the form
\begin{equation}
\tilde{S}_{2}(v)\tilde{R}^{T}_{21}(-u-v)\tilde{S}_1(u)\tilde{R}_{12}(u-v)=\tilde{R}_{21}(u-v)\tilde{S}_1(u)\tilde{R}^{T}_{12}(-u-v)\tilde{S}_{2}(v).
\label{eqn:eq0000013}
\end{equation}
Where $\tilde{S}(u)=\sum_{i,j=1}^NE_{ij}\otimes\tilde{\psi}_{ij}(u)$. Here again we can solve this equation to give us two-site block $\tilde{\psi}(u)$ for MPS. This BYB seems to be different from the one we define in last chapter, but actually they are the same. After changing some signs of spectral parameters in Eq. (\ref{eqn:eq0000013}) and the sign of spectral parameter in the R-matrix, one would get the same kind of form of BYB. Thus solutions from last chapter will also valid for Eq. (\ref{eqn:eq0000013}). 

In the $t$ transposition, with R-matrix Eq. (\ref{eqn:eq0000010}), the BYB takes the form 
\begin{equation}
S_{2}(v)R^{t}_{21}(-u-v)S_1(u)R_{12}(u-v)=R_{21}(u-v)S_1(u)R^{t}_{12}(-u-v)S_{2}(v),
\label{eqn:eq0000014}
\end{equation}
with $S(u)=\sum_{i,j=-n}^{n}E_{ij}\otimes\psi_{ij}(u)$. To distinguish BYB in different transposition convention, we will call Eq. (\ref{eqn:eq0000014}) as tBYB for short. One can actually find similarity transformation to connect BYB with tBYB, but the way to prove such statement turns out to be different for different R matrix. Let's focus on the $SU(N)$ case first.

Consider a constant transformation matrix $P\in SU(N)$, which brings the solutions of BYB to tBYB $S(u)=P\tilde{S}(u)P^{-1}$. The transformation matrix also satisfies 

\begin{equation}
PP^T=C,
\label{eqn:eq0000015}
\end{equation}
where $C$ is charge conjugation matrix (just a name). The left hand side of Eq. (\ref{eqn:eq0000014}) yields once we insert the guess.

\begin{equation}
\begin{split}
LHS&=P_2S_2(v)P^{-1}_{2}C_2R^T_{21}(-u-v)C^{-1}_2P_1S_1(u)P^{-1}_{1}R_{12}(u-v)\\
&=P_2S_2(v)P^{T}_{2}\tilde{R}^T_{21}(-u-v)P_1S_1(u)P^{-1}_{1}C^{-1}_2\tilde{R}_{12}(u-v)\\
&=P_2S_2(v)P_1\tilde{R}^T_{21}(-u-v)P^{T}_{2}S_1(u)P^{-1}_{1}C^{-1}_2\tilde{R}_{12}(u-v)\\
&=P_2P_1S_2(v)\tilde{R}^T_{21}(-u-v)S_1(u)P^{-1}_{1}P^{-1}_{2}\tilde{R}_{12}(u-v)\\
&=P_1P_2S_2(v)\tilde{R}^T_{21}(-u-v)S_1(u)\tilde{R}_{12}(u-v)P^{-1}_{2}P^{-1}_{1}
\end{split}
\label{eqn:eq0000016}
\end{equation}
where I have used $[A_1,B_2]=0$ for arbitrary operator, the invariant property of R-matrix Eq. (\ref{eqn:eq000008}) and its transposed form. The RHS follows the same calculation but using $C_1$ to take care of the transpose (since the Eq. (\ref{eqn:eq0000011}), we can use $C_1$ or $C_2$ freely), yields
\begin{equation}
RHS=P_2P_1\tilde{R}_{12}(u-v)K_1(u)\tilde{R}^T_{12}(-u-v)K_2(v)P^{-1}_{2}P^{-1}_{1}.
\label{eqn:eq0000017}
\end{equation}
$LHS=RHS$ will give us Eq. (\ref{eqn:eq0000013}), so our $P$ matrix is good.

There is another transformation turns out to be useful for our $(SU(3),SO(3))$ case
\begin{equation}
\tilde{K}(u)=K(u)C,
\label{eqn:eq0000018}
\end{equation}
gives us the opposite effect that brings us from tBYB to BYB. 

The specific form of $P$ matrix, as we will show, depends on how we choose the basis for Yangian. This is because the tBYB is part of the definition of twisted Yangian, when one choose a specific basis for Yangian, we need to also choose a specific transformation that ensure our twisted Yangian is a sub-algebra of Yangian.
\section{Yangian and twisted Yangian}
Before giving the formal definition of what Yangian algebra and twisted Yangian are, we need to specify the transformation matrix $P$ for our $(SU(3),SO(3))$ case 
\begin{equation}
\begin{pmatrix}S_{+}\\S_0\\S_{-}\end{pmatrix}=P\begin{pmatrix}S_1\\S_2\\S_3\end{pmatrix}, \quad P=\begin{pmatrix}\frac{1}{\sqrt{2}}&\frac{i}{\sqrt{2}}&0\\0&0&1\\\frac{1}{\sqrt{2}}&-\frac{i}{\sqrt{2}}&0\end{pmatrix}
\label{eqn:eq0000019}
\end{equation}
then our scalar solution $\psi_{ij}=\delta_{ij}$ is the same in tBYB. Here for convenience, one can identify $\{+,0,-\}$ with $\{-1,0,1\}$ respectively.

Fix $N\in \mathbb{Z}_{\ge 2}$, we introduce elements $t^{(r)}_{ij}$ with $-n\le i,j\le n$ and $r\in \mathbb{Z}_{\ge 0}$ such that $t_{ij}^{(0)}=\delta_{ij}$. Combining these into formal power series $t_{ij}(u)=\sum_{r\ge 0}t^{(r)}_{ij}u^{-r}$, we can then form the generating matrix $T(u)=\sum_{-n\le i,j\le n} E_{ij}\otimes t_{ij}(u)$. Let $R(u)$ be given by Eq. (\ref{eqn:eq0000010}
).
\begin{definition}
The Yangian $Y(N)$ is the unital associative $\mathbb{C}$-algebra generated by elements $t^{(r)}_{ij}$ with $-n\le i,j\le n$ and $r\in \mathbb{Z}_{\ge 0}$ satisfying the relation \cite{tyangian}
\begin{equation}
R(u-v)T_1(u)T_2(v)=T_2(v)T_1(u)R(u-v), 
\label{eqn:eq0000020}
\end{equation}
where $T_1(u)$ and $T_2(u)$ are the elements of $\mathrm{End}\mathbb{C}^{N}\otimes\mathrm{End}\mathbb{C}^{N}\otimes Y(\mathfrak{g}[[u^{-1}]])$ given by

\begin{equation}
T_1(u)=\sum_{i,j=-n}^n E_{ij}\otimes \mathbb{I}\otimes t_{ij}(u),\quad T_2(u)=\sum_{i,j=-n}^n\mathbb{I} \otimes E_{ij}\otimes t_{ij}(u),
\label{eqn:eq0000021}
\end{equation}
In terms of the power series elements $t_{ij}(u)$, Eq. (\ref{eqn:eq0000020}) is equivalent to
\begin{equation}
\begin{split}
[t_{ij}(u),t_{kl}(v)]&=\frac{1}{u-v}\left(t_{kj}(u)t_{il}(v)-t_{kj}(v)t_{il}(u)\right)
\end{split}
\label{eqn:eq0000022}
\end{equation}
The Hopf algebra structure of $Y(N)$ is given by
\begin{equation}
\Delta:t_{ij}(u)\mapsto \sum_{-n\le a\le n}t_{ia}(u)\otimes t_{aj}(u),\quad S:T(u)\mapsto T(u)^{-1},\quad \epsilon T(u)\mapsto \mathbb{I}.
\label{eqn:eq0000023}
\end{equation}
We will call the generating matrix of the Yangian $Y(N)$ the T-matrix.
\end{definition}
Let $A\in GL(N)$ be such that $AA^{t}=\mathbb{I}$, let $a\in \mathbb{C}$ be a constant, also let $B\in GL(N)$ be an arbitrary matrix. Moreover, consider an arbitrary formal series $f(u)$ of the form

 \begin{equation}
f(u)=1+f_1u^{-1}+f_2u^{-2}+\dots\mathbb{C}[[u^{-1}]].
\label{eqn:eq0000024}
\end{equation}
The automorphisms:
\begin{equation}
\begin{split}
\mu_f:T(u)\mapsto f(u)T(u),\quad \tau_a: T(u)\mapsto T(u+a), \\ \alpha_A: T(u)\mapsto AT(u)A^{t},\quad \beta_B:T(u)\mapsto BT(u)B^{-1},
\end{split}
\label{eqn:eq0000025}
\end{equation}
where the first two maps is readily to see that it satisfies RTT relation Eq. (\ref{eqn:eq0000020}), for $\alpha_A$ one should used the invariance of $R$-matrix

\begin{equation}
R_{12}A_1A_2=A_1A_2R_{12},\quad R_{12}A_1^{t}A_2^{t}=A_1^{t}A_2^{t}R_{12}.
\label{eqn:eq0000026}
\end{equation}
For $\beta_B$ it follows that $B^{-1}\in GL(N)$. 

The above Yangian algebra is more or less the symmetry algebra for our spin chain $SU(3)$ sector. The Yangian algebra is an infinite algebra with infinite generators, but keep in mind that our spin chain has finite length with periodic boundary condition. By integrability, such symmetry is not possible for finite spin chain. Indeed, as one can show that the symmetry will break in the boundary \cite{Loebbert_2016}. The T-matrix is the monodromy matrix in our spin chain with operator entry $t_{ij}$. Taking a coproduct means extend the spin chain from $L$ to $L+1$ if we choose $t_{ij}$ acting on local physical space. Or it means expanding the auxiliary space if we choose $t_{ij}$ acting on auxiliary space, which is the case for our dressing procedure (one need to take a coproduct on each site, meaning number $L$ coproduct, to generate a transfer matrix acting on the original integrable state).
\begin{definition}
The extended twisted Yangian $Y^{+}(N)$ is the subalgebra of $Y(N)$ generated by the coefficients $s^{(r)}_{ij}$ with $-n\le i,j\le n$ and $r\in \mathbb{Z}_{\ge 0}$ of the matrix elements $s_{ij}(u)$ in $S(u)=\sum_{ij}E_{ij}\otimes s_{ij}(u)$ matrix \cite{tyangian}

\begin{equation}
S(u)=T(u)T^{t}(-u).
\label{eqn:eq0000027}
\end{equation}
\end{definition}
We have $s_{ij}(u)=\sum_{a=-n}^{n}t_{ia}(u)t_{-j,-a}(-u)$ and by Eq. (\ref{eqn:eq0000023}), we have this property
\begin{equation}
\Delta(s_{ij}(u))=\sum_{-n\le a,b\le n}t_{ia}(u)t_{-j,-b}\otimes s_{ab}(u)
\label{eqn:eq0000028}
\end{equation}
which is the same as $Y^+(N)$ is a left coideal subalgebra of $Y(N)$. There are two important properties that connect the twisted Yangian with tBYB as well as the RTT relation.

$\mathbf{Property\ 1:}$ Make use of Eq. (\ref{eqn:eq0000020}), one can show that the S-matrix $S(u)$ satisfies the tBYB equation
\begin{equation}
S_{2}(v)R^{t}_{21}(-u-v)S_1(u)R_{12}(u-v)=R_{21}(u-v)S_1(u)R^{t}_{12}(-u-v)S_{2}(v). 
\label{eqn:eq0000029}
\end{equation}
Proof:
\begin{equation}
\begin{split}
&T_2(v)T_2^{t}(-v)R_{21}^{t}(-u-v)T_1(u)T_1^{t}(-u)R_{12}(u-v)\\
&=R_{21}(u-v)T_1(u)T_1^{t}(-u)R^{t}_{12}(-u-v)T_2(v)T_2^{t}(-v)\\
\Leftrightarrow\ &T_2(v)T_1(u)R_{21}^{t}(-u-v)T_2^{t}(-v)T_1^{t}(-u)R_{12}(u-v)\\
&=R_{21}(u-v)T_1(u)T_2(v)R^{t}_{12}(-u-v)T_1^{t}(-u)T_2^{t}(-v)\\
\Leftrightarrow\ &T_2(v)T_1(u)R_{21}^{t}(-u-v)R_{12}(u-v)T_2^{t}(-v)T_1^{t}(-u)\\
&=T_2(v)T_1(u)R_{21}(u-v)R^{t}_{12}(-u-v)T_2^{t}(-v)T_1^{t}(-u)\\
\Leftrightarrow\ &R_{21}^{t}(-u-v)R_{12}(u-v)=R_{21}(u-v)R^{t}_{12}(-u-v).
\end{split}
\label{eqn:eq0000030}
\end{equation}
Since we have scalar solution $\tilde{\psi}_{ij}=\delta_{ij}$ up to a normalization in BYB, the transformed one is the same $\psi_{ij}=\delta_{ij}$ as one can see from the transformation matrix. So for our case $(SU(3),SO(3))$, the last equation holds. Actually the last equation holds for general $(SU(N),SO(N))$. Thus we have connected the twisted Yangian with our tBYB.

$\mathbf{Property\ 2:}$ 
The symmetry relation for $(SU(3),SO(3))$:
\begin{equation}
S^{t}(u)=S(-u)+\frac{S(u)-S(-u)}{2u}.
\label{eqn:eq0000031}
\end{equation}
This one can be easily prove when expanding out in terms of $t_{ij}$, then use the commutation relation Eq. (\ref{eqn:eq0000022}). Therefore one needs RTT-relation to get Eq. (\ref{eqn:eq0000031}), another interesting fact is that the symmetry relation fixes the representation up to an arbitrary  even order scalar function. 

Multiplication with an even formal series $g(u)=1+g^{(2)}u^{-2}+g^{(4)}u^{-4}+\dots$ defines an automorphism of $Y^{+}(N)$:
\begin{equation}
S(u)\rightarrow g(u)S(u). 
\label{eqn:eq0000032}
\end{equation}
As one can easily check that it satisfies the tBYB and symmetry relation. 
\subsection{Representation of Yangian and twisted Yangian}
\begin{definition}
A representation $L(\lambda_{1}(u),\dots,\lambda_N(u))$ of the Yangian $Y(N)$ is called a highest weight representation if there exists a nonzero vector $v\in L$ such that $L$ is generated by $v$ and following relation holds 
\begin{equation}
\begin{alignedat}{2}
t_{ij}(u)\cdot v&=0\quad &&\mathrm{for}\quad 1\le i<j\le N\quad\mathrm{and}\\
t_{ii}(u)\cdot v&=\lambda_i(u)\cdot v\quad &&\mathrm{for}\quad i=1,\dots,N.
\end{alignedat}
\label{eqn:eq0000033}
\end{equation}
\end{definition}
About definition is too abstract for us to do some calculation. However, there is an algebra homomorphism called evolution representation (ev): $Y(N)\rightarrow U(\mathfrak{gl}_N)$ such that \cite{tyangian}
\begin{equation}
\mathrm{ev}:t_{ij}\rightarrow \delta_{ij}+u^{-1}E_{ij},
\label{eqn:eq0000034}
\end{equation}
where $E_{ij}$ are the generators of $\mathfrak{gl}_N$. Using the evaluation homomorphism Eq. (\ref{eqn:eq0000034}), the $\mathfrak{gl}_N$ highest weight representation $L(\lambda_1,\dots,\lambda_N)$ ($\lambda_i$ is the corresponding highest weight) is also an h.w. representation of $Y(N)$ with highest weight $\lambda_i(u)=1+\lambda_i u^{-1}$. So we can some assume that this two algebra share the same vector space. This idea is through the whole chapter and next chapter that the quantum algebra share the same vector space with classical algebra. Let us use the following relation to denote such evolution representation
\begin{equation}
t_{ij}(u)\cdot v=\mathcal{L}^{(\lambda_1,\dots,\lambda_N)}_{ij}\cdot v,
\label{eqn:eq0000035}
\end{equation}
for $v\in L(\lambda_1,\dots,\lambda_N)$.
\begin{definition}
A representation $V$ of the twisted Yangian $Y^{+}(N)$ is called a highest weight representation if there exists a nonzero vector $v\in V$ such that $V$ is generated by $v$ and following relation holds
\begin{equation}
\begin{alignedat}{2}
s_{ij}(u)\cdot v&=0\quad &&\mathrm{for}\quad -n\le i<j\le n\quad\mathrm{and}\\
s_{ii}(u)\cdot v&=\mu_i(u)\cdot v\quad &&\mathrm{for}\quad i=0,\dots,n.
\end{alignedat}
\label{eqn:eq0000036}
\end{equation}
\end{definition}
Since $\mathfrak{so}_N$ is a sub-algebra of $\mathfrak{gl}_N$, and the embedding rules are actually defined in Eq. (\ref{eqn:eq000004}). One can choose the following convention for our $\mathfrak{so}_3$ case
\begin{equation}
\begin{split}
S_{z}&=F_{-1-1},\\
S_{+}&=F_{01},\\
S_{-}&=F_{10}.
\end{split}
\label{eqn:eq0000037}
\end{equation}
Then the $\mathfrak{gl}_3$ module $L(\lambda_1,\lambda_2,\lambda_3)$ with corresponding highest weight $\lambda_i,\  i=1,2,3$. Apply operator $S_z$ on the vector $\lvert l_1,l_2,l_3\rangle\in L(\lambda_1,\lambda_2,\lambda_3)$ 
\begin{equation}
S_z\lvert l_1,l_2,l_3\rangle=l_1-l_3.
\label{eqn:eq0000038}
\end{equation}
Using branching rules, one would get representations of $\mathfrak{so}_3$ as a subspace of $L(\lambda_1,\lambda_1,\lambda_2)$
\begin{equation}
\begin{alignedat}{2}
    [0,s]&=[0,s-2] + (s),\quad &&\mathrm{for}\  s=0,1,2,\dots,\\
    [0,s-\frac{1}{2}]\otimes (\frac{1}{2}) &=[0,s-\frac{3}{2}]\otimes (\frac{1}{2}) + (s),\quad &&\mathrm{for}\ s=\frac{3}{2},\frac{5}{2},\frac{7}{2},\dots.
\end{alignedat}
\label{eqn:eq0000039a}
\end{equation}
Above I have used the $[a,b]$ to denote the Dynkin labels of type $A_2$, and $(a)$ for representations of $SO(3)$ of dimension $2a+1$. For $\mathfrak{gl}_N$ with highest weight $(\lambda_1,\lambda_2,\dots,\lambda_N)$ has dimension \cite{glndimension}
\begin{equation}
dim[\lambda]=\prod_{i<j}\frac{(\ell_i-\ell_j)}{(\ell^0_i-\ell^0_j)},
\label{eqn:eq0000039}
\end{equation}
with $\ell^{0}_{j}=N-j,\ell_{j}=\lambda_j+\ell^{0}_{j}$. One can check, $\mathfrak{so}_3$ with $s=\lambda_1-\lambda_3$ doesn't have the same dimension as $L(\lambda_1,\lambda_1,\lambda_2)$, so we need to know what the quotient space is in the sense of Yangian.  Using Eq. (\ref{eqn:eq0000027}) and Eq. (\ref{eqn:eq0000035}), one would get the highest weight for $Y^+(3)$ using $L(\lambda_1,\lambda_1,\lambda_2)$ $\mathfrak{gl}_3$ module
\begin{equation}
\begin{split}
\mu_1(u)=(1+\lambda_2u^{-1})(1-\lambda_1 u^{-1}),\\
\mu_0(u)=(1+\lambda_1u^{-1})(1-\lambda_1 u^{-1}).
\end{split}
\label{eqn:eq0000040}
\end{equation}
Or using the Lax-operator
\begin{equation}
s_{ij}(u)\cdot v = \sum_{a}\mathcal{L}_{ia}^{(\lambda_1,\lambda_1,\lambda_2)}(u)\mathcal{L}_{-j,-a}^{(\lambda_1,\lambda_1,\lambda_2)}(-u)v,
\label{eqn:eq0000041}
\end{equation}
for all $v\in L(\lambda_1,\lambda_1,\lambda_2)$.
\section{Highest weight matching in $(SU(3),SO(3))$}
\subsection{MPS side of calculation}
Start with the two-site solution of higher representation, the $MPS$ can be built from the $S$-matrix $\tilde{S}(u)=\sum_{ab=1}^3 E_{ij}\otimes \tilde{\psi}_{ab}(u)$, where $\tilde{\psi}_{ab}$ is given by Eq. (\ref{eqn:eq000047}) from last chapter with $-\frac{1}{u^2}$ normalization
\begin{equation}
\tilde{\psi}_{ij}^{(s)}(u)=\delta_{ab}+u^{-1}[S_a,S_b]-u^{-2}S_aS_b,\quad \mathrm{with}\  a,b\in\{1,2,3\}.
\label{eqn:eq0000042}
\end{equation}
Transform the solution to the case of tBYB
\begin{equation}
S(u)=P\tilde{S}(u)P^{-1},
\label{eqn:eq0000043}
\end{equation}
the matrix elements of $S(u)=\sum_{i,j=-1}^1E_{ij}\otimes \psi_{ij}(u)$ yields 
\begin{equation}
\begin{split}
\psi^{(s)}_{1,1}&=\frac{1}{2}(\psi^{(s)}_{1,1}-i\psi^{(s)}_{2,1}+i\psi^{(s)}_{1,2}+\psi^{(s)}_{2,2})\\
&=1-u^{-1}S_z-u^{-2}S_{-}S_{+}\\
\psi^{(s)}_{0,0}&=1-u^{-2}S^2_z,\\
\psi^{(s)}_{-1,-1}&=1+u^{-1}S_z-u^{-2}S_{+}S_{-},\\
\psi^{(s)}_{0,1}&=u^{-1}S_+-u^{-2}S_zS_+,\\
\psi^{(s)}_{0,-1}&=-u^{-1}S_--u^{-2}S_zS_-,\\
\psi^{(s)}_{-1,0}&=-u^{-1}S_+-u^{-2}S_+S_z,\\
\psi^{(s)}_{-1,1}&=-u^{-2}S_{+}^2,\\
\psi^{(s)}_{1,0}&=u^{-1}S_--u^{-2}S_-S_z,\\
\psi^{(s)}_{1,-1}&=-u^{-2}S_{-}^2.
\end{split}
\label{eqn:eq0000044}
\end{equation}
Above representation is a $k = 2s+1$ dimensional irreducible representation of the twisted Yangian $Y^+(3)$ in presentation tBYB. Let us denote it $V(s)$. Obviously
\begin{equation}
-u^2\psi_{i,j}(u)\big\vert_{u=0}=S_{i}S_{-j}.
\label{eqn:eq0000045}
\end{equation}
We use the notation
\begin{equation}
\begin{split}
S_{\pm}&=\frac{1}{\sqrt{2}}(S_1\pm iS_2),\\
S_z&=S_3.
\end{split}
\label{eqn:eq0000046}
\end{equation}
And the corresponding commutation relation becomes
\begin{equation}
[S_+,S_-]=S_z,\quad [S_z,S_{\pm}]=\pm S_{\pm}.
\label{eqn:eq0000047}
\end{equation}
From Eq. (\ref{eqn:eq0000046}) we know that $S_+$ increases the weights, $S_-$ decreases the weights, therefore
\begin{equation}
S_{+}\cdot v=0.
\label{eqn:eq0000048}
\end{equation}
The $Y^+(3)$ highest weights of $V(s)$ are
\begin{equation}
\begin{split}
\mu_{1}(u)=1-su^{-1},\\
\mu_0(u)=1-s^2u^{-2}.
\end{split}
\label{eqn:eq0000049}
\end{equation}
\subsection{Scalar solution side of calculation}
From Eq. (\ref{eqn:eq0000049}) we can see that $V(s)$ can be embedded into $L(\lambda_1,\lambda_1,\lambda_2)$ if $\lambda_1 = s$ and $\lambda_2 = 0$ but $L(s,s,0)$ is finite dimensional iff $s \in \mathbb{N}$, therefore we only have chance to find connection between $\lvert\Psi_{\delta}\rangle$ and $\lvert \mathrm{MPS}_k\rangle$ when k is odd. And by Eq. (\ref{eqn:eq0000028}) we can achieve this by dressing the scalar solution $s=0$. 

For even $k$ we have to use coproduct properties Eq. (\ref{eqn:eq0000028}) that , and build higher representation by dressing $s=\frac{1}{2}$ solution to make the $\mathfrak{gl}_4$ module finite. The representation $L(\lambda_1,\lambda_1,\lambda_2)\otimes V(1/2)$ has highest weight
\begin{equation}
\begin{split}
&\mu_{1}(u)=(1+\lambda_2u^{-1})(1-\lambda_1 u^{-1})(1-\frac{1}{2}u^{-1}),\\
&\mu_0(u)=(1-\lambda_1^2u^{-2})(1-\frac{1}{4}u^{-2}),
\end{split}
\label{eqn:eq0000050}
\end{equation}
we can see that $V(s)$ can be embedded into $L(\lambda_1,\lambda_1,\lambda_2)\otimes V(1/2)$ if $\lambda_1=s$ and $\lambda_2=1/2$.
\section{Overlap Formula for $(SU(3),SO(3))$}
\subsection{Odd $k=2s+1$}
We have seen that the even and odd $k$ cases must be treated differently. Let us start with the odd case. We can show that $V(s)$ is embedded into $L(s,s,0)$ where $s\in \mathbb{Z}_+$ for small $s$. In the end, one can get for general $s$ the twisted Yangian $Y^{+}(3)$ acts on $L(s,s,0)\cong V(s)\oplus L(s,s,2)$ \cite{de_Leeuw_2020} as 
\begin{equation}
\begin{split}
s_{ij}\cdot (v_1\otimes v_2)&=\mathcal{L}^{(s,s,0)}_{ia}(u)\mathcal{L}^{(s,s,0)}_{-j,-b}(-u)v_1\otimes\delta_{ab}v_2\\
&=\begin{pmatrix}\psi^{(s)}_{i,j}(u)&X\\0&\mathcal{L}^{(s,s,2)}_{ia}(u)\mathcal{L}^{(s,s,2)}_{-j,-b}(-u)\otimes\delta_{ab}\end{pmatrix}\begin{pmatrix}w_1\\w_2\otimes v_3\end{pmatrix}
\end{split}
\label{eqn:eq0000051}
\end{equation}
for all $S\in Y^{+}(3)$, where $v_1\in L(s,s,0)$, $v_2,v_3\in V(0)$, $w_1\in V(s)$, $w_2\in L(s,s,2)$ and $s\in \mathbb{Z}_{>1}$. For later convenience, we use $0^{\prime}$ to denote space $L(s,s,0)$, $A_s$ to denote space $V(s)$, and $0$ to denote $L(s,s,2)$. And notice here we don't put any physics space (cause it's in fundamental representation), it is just a way of representation of auxiliary spaces, and when we take the trace over auxiliary space, it will all gone.
\begin{conj}
The $L(s,s,2)$ is an irrep of $Y^+(3)$ for all $s>1$.
\end{conj}
In the following we will show that how to build the overlap of $\langle \mathbf{u}\lvert\mathrm{MPS}\rangle$ state from the knowledge of $\langle \mathbf{u}\lvert\mathrm{MPS}_{\delta}\rangle$. 
First let use a different convention of R-matrix rather than Eq. (\ref{eqn:eq000003})
\begin{equation}
\begin{split}
\bar{R}(u)=u\mathbb{I}+i\mathbb{P},
\end{split}
\label{eqn:eq0000052}
\end{equation}
and hence use rescaled matrices:
\begin{equation}
\begin{split}
\bar{s}_{ij}(u)=u^2s_{ij}(iu),\\
\bar{t}_{ij}=ut_{ij}(iu).
\end{split}
\label{eqn:eq0000053}
\end{equation}
Let $\rho$ be a representation of the twisted Yangian and let us use an isomorphism Eq. (18) to transform tBYB back to BYB, which we use for our integrable state $\mathrm{MPS}$:
\begin{equation}
\phi_{\alpha\beta}=\rho(\bar{s}_{\alpha\gamma}(0))C_{\gamma\beta}=\rho(\bar{s}_{\alpha,-\beta}(0)),
\label{eqn:eq0000054}
\end{equation}
where $C$ is the charge conjugation matrix, and $\phi_{\alpha\beta}$ satisfies BYB. Furthermore, let us define the following state
\begin{equation}
\vert\mathrm{MPS}_{A}\rangle=\sum_{i,j\in\{-1,0,1\}}\mathrm{Tr}_A\,[\phi_{i_1j_1}\dots\phi_{i_{L/2}j_{L/2}}]\vert i_1,j_1,\dots,i_{L/2},j_{L/2}\rangle,
\label{eqn:eq0000055}
\end{equation}
where $A$ denotes all the auxiliary space other than physical space. For the trivial representation $V(0)$
\begin{equation}
\rho(\bar{s}_{ij}(0))=\delta_{i,j},
\label{eqn:eq0000056}
\end{equation}
When we exploit Eq. (\ref{eqn:eq0000055}) based on Eq. (\ref{eqn:eq000051}) and Eq. (\ref{eqn:eq0000054}), also use $\mathrm{Tr}\,(A\otimes B)=\mathrm{Tr}\,(A)\mathrm{Tr}\,(B)$, and $\mathrm{Tr}\,(A\oplus B)=\mathrm{Tr}\,(A)+\mathrm{Tr}\,(B)$, we would get
\begin{equation}
\begin{split}
\sum_{i,j\in\{-1,0,1\}}\sum_{a,b\in\{-1,0,1\}}\mathrm{Tr}_{0^{\prime}}\,&[\bar{\mathcal{L}}^{(s,s,0)}_{i_1a_1}(0)\bar{\mathcal{L}}^{(s,s,0)}_{j_1,b_1}(0)\dots\bar{\mathcal{L}}^{(s,s,0)}_{i_{L/2}a_{L/2}}(0)\bar{\mathcal{L}}^{(s,s,0)}_{j_{L/2},b_{L/2}}(0)]\\
&\mathrm{Tr}_{A_0}\,[\delta_{a_1,-b_1}\dots \delta_{a_{L/2},-b_{L/2}}]\vert i_1 j_1\dots i_{L/2}j_{L/2}\rangle\\
=\sum_{i,j\in\{-1,0,1\}} \mathrm{Tr}_{A_s}\,&[\psi^{(s)}_{i_1,-j_1}\dots \psi^{(s)}_{i_{L/2},-j_{L/2}}]\vert i_1 j_1\dots i_{L/2}j_{L/2}\rangle\\
+\sum_{i,j\in\{-1,0,1\}}\sum_{a,b\in\{-1,0,1\}}\mathrm{Tr}_{0}\,&[\bar{\mathcal{L}}^{(s,s,2)}_{i_1a_1}(0)\bar{\mathcal{L}}^{(s,s,2)}_{j_1,b_1}(0)\dots\bar{\mathcal{L}}^{(s,s,2)}_{i_{L/2}a_{L/2}}(0)\bar{\mathcal{L}}^{(s,s,2)}_{j_{L/2},b_{L/2}}(0)]\\
&\mathrm{Tr}_{A_0}\,[\delta_{a_1,-b_1}\dots \delta_{a_{L/2},-b_{L/2}}]\vert i_1 j_1\dots i_{L/2}j_{L/2}\rangle.
\end{split}
\label{eqn:eq0000057}
\end{equation}
The MPS in $\{-1,0,1\}$ basis takes the form
\begin{equation}
\lvert \mathrm{MPS}_{2s+1}\rangle=\sum_{i,j\in\{-1,0,1\}}\mathrm{Tr}_{A_s}\,[\psi^{(s)}_{i_1,-j_1}\dots \psi^{(s)}_{i_{L/2},-j_{L/2}}]\lvert i_1,j_1,\dots,i_{L/2},j_{L/2}\rangle,
\label{eqn:eq0000058}
\end{equation}
where we have made use of
\begin{equation}
\begin{split}
\sum_{a,b\in\{1,2,3\}}\tilde{\psi}_{ab}\lvert a\rangle\lvert b\rangle&=\sum_{a,b\in\{1,2,3\}}\sum_{i,j\in\{-1,0,1\}}\sum_{i^{\prime},j^{\prime}\in\{-1,0,1\}}P_{ai}^{-1}\psi_{ij}P_{jb}\times P^{-1}_{ai^{\prime}}\lvert i^{\prime}\rangle P^{-1}_{bj^{\prime}}\lvert j^{\prime}\rangle\\
&=\sum_{i,j\in\{-1,0,1\}}\sum_{i^{\prime},j^{\prime}\in\{-1,0,1\}}\sum_{a,b\in\{1,2,3\}}P^{*}_{i^{\prime}a}P_{ai}^{-1}P_{jb}P^{-1}_{bj^{\prime}}\psi_{ij}\lvert i^{\prime}\rangle \lvert j^{\prime}\rangle\\
&=\sum_{i,j\in\{-1,0,1\}}\sum_{i^{\prime},j^{\prime}\in\{-1,0,1\}}\delta_{i,-i^{\prime}}\delta_{j,j^{\prime}}\psi_{ij}\lvert i^{\prime}\rangle \lvert j^{\prime}\rangle\\
&=\sum_{i,j\in\{-1,0,1\}}\psi_{-i,j}\lvert i\rangle \lvert j\rangle,
\end{split}
\label{eqn:eq0000059}
\end{equation}
Also there exist an automorphism of generators of $SU(2)$ which acts as follows
\begin{equation}
V S_i V^{-1}=-S_{-i}, \quad i\in \{-,0,+\},
\label{eqn:eq0000060}
\end{equation}
then one can change the sign in Eq.(\ref{eqn:eq0000059}), hence gives us Eq. (\ref{eqn:eq0000058}). The above procedure also works for $\vert\mathrm{MPS}_{\delta}\rangle$ state.

Then Eq. (\ref{eqn:eq0000057}) is equivalent to

\begin{equation}
\vert \mathrm{MPS}_{2s+1}\rangle=\left(\bar{T}^{(s,s,0)}(0)-\bar{T}^{(s,s,2)}(0)\right)\vert\mathrm{MPS}_{\delta}\rangle
\label{eqn:eq0000061}
\end{equation}
where
\begin{equation}
\bar{T}^{(s,s,m)}(u)=\mathrm{Tr}_0\,[\bar{\mathcal{L}}_{01}^{(s,s,m)}(u)\dots\bar{\mathcal{L}}_{0L}^{(s,s,m)}(u)],
\label{eqn:eq0000062}
\end{equation}
and
\begin{equation}
\begin{split}
\bar{L}^{(s,s,m)}(u)&=\bar{\mathcal{L}}_{i,j}^{(s,s,m)}(u)\otimes \lvert e_{i}\rangle\langle e_{j}\lvert=u\mathbb{I}-iE_{i,j}^{(s,s,m)}\otimes e_{ij}\\
&=u\mathbb{I}+iE_{ij}^{(-m,-s,-s)}\otimes e_{ji}\\
&=u\mathbb{I}-is\mathbb{I}+iE_{ij}^{(-m+s,0,0)}\otimes e_{ji},
\end{split}
\label{eqn:eq0000063}
\end{equation}
where the second line we have used that $E_{ij}\rightarrow -E_{ji}$ is a Lie-algebra automorphism and it connects a representation to its dual one, and the third line we take an identity matrix out. One can see that given our $\vert\mathrm{MPS}_{\delta}\rangle$ is integrable, then by acting with transfer matrix we can generate higher dimensional integrable $\vert \mathrm{MPS}\rangle$. Here one may be puzzled about how the physical space of $\vert\mathrm{MPS}_{2s+1}\rangle$ is the same as $\vert\mathrm{MPS}_{\delta}\rangle$, it is the same because both of them are in the fundamental representation. One can see this clearly when taking the inner product between them and Bethe state $\vert \mathbf{u}\rangle$, which is in the fundamental representation in our model.

Let us use another notation of Eq. (\ref{eqn:eq0000063}):
\begin{equation}
\bar{\mathcal{L}}^{(s)}(u)=u-i\frac{s-1}{2}+iE_{i,j}^{(s,0,0)}\otimes e_{ji},
\label{eqn:eq0000064}
\end{equation}

\begin{equation}
\bar{\mathcal{T}}^{(s)}(u)=\mathrm{Tr}_0\,[\bar{\mathcal{L}}_{01}^{(s)}(u)\dots \bar{\mathcal{L}}_{0L}^{(s)}(u)].
\label{eqn:eq0000065}
\end{equation}
From Eq. (\ref{eqn:eq0000063}) and Eq. (\ref{eqn:eq0000064}), one can see that 
\begin{equation}
\begin{split}
    \bar{L}^{(s,s,0)}(u)=\bar{\mathcal{L}}^{(s)}(u-i\frac{s+1}{2}),\\
\bar{L}^{(s,s,2)}(u)=\bar{\mathcal{L}}^{(s-2)}(u-i\frac{s+3}{2}).
\end{split}
\label{eqn:eq0000066}
\end{equation}
Therefore
\begin{equation}
\vert \mathrm{MPS}_{2s+1}\rangle=\left(\bar{\mathcal{T}}^{(s)}(-i\frac{s+1}{2})-\bar{\mathcal{T}}^{(s-2)}(-i\frac{s+3}{2})\right)\vert \mathrm{MPS}_{\delta}\rangle,
\label{eqn:eq0000067}
\end{equation}
and the ratio of the overlaps is equal to the difference of eigenvalues of the transfer matrices
\begin{equation}
\frac{\langle \mathrm{MPS}_{2s+1}\vert\mathbf{u}\rangle}{\langle \mathrm{MPS}_{\delta}\vert\mathbf{u}\rangle}=\bar{\mathcal{T}}^{(s)}(-i\frac{s+1}{2})-\bar{\mathcal{T}}^{(s-2)}(-i\frac{s+3}{2}).
\label{eqn:eq0000068}
\end{equation}
The eigenvalues of the transfer matrices can be written as, 
\begin{equation}
\begin{split}
\bar{\mathcal{T}}^{(s)}(u)=Q_1(-iu-\frac{s}{2})Q_{2}(-iu+\frac{s+3}{2})\sum_{k=0}^s\frac{(u+i\frac{s+1}{2}-ik)^LQ_2(-iu+\frac{s+1}{2}-k)}{Q_1(-iu+\frac{s}{2}-k)Q_1(-iu+\frac{s+2}{2}-k)}\\ \times \sum_{l=0}^k\frac{Q_1(-iu+\frac{s+2}{2}-l)}{Q_2(-iu+\frac{s+1}{2}-l)Q_2(-iu+\frac{s+3}{2}-l)}.
\end{split}
\label{eqn:eq0000069}
\end{equation}
I will derive this by Tableau sum in Appendix \ref{AppendixE}. Let us assume that $L,M_1,M_2$ (where $M_1$ and $M_2$ are defined as numbers of excitation stated in chapter 3) are even, then 
\begin{equation}
\begin{split}
\bar{\mathcal{T}}&^{(s)}(-i\frac{s+1}{2})-\bar{\mathcal{T}}^{(s-2)}(-i\frac{s+3}{2})\\
=&Q_1(s+\frac{1}{2})Q_2(1)\sum_{k=0}^s(ik)^L\frac{Q_2(k)}{Q_1(k+\frac{1}{2})Q_1(k-\frac{1}{2})}\sum_{l=0}^k\frac{Q_1(l-\frac{1}{2})}{Q_2(l)Q_2(l-1)}\\
-&Q_1(s+\frac{1}{2})Q_2(1)\sum_{k=0}^{s-2}(ik+i2)^L\frac{Q_2(k+2)}{Q_1(k+\frac{5}{2})Q_1(k+\frac{3}{2})}\sum_{l=0}^k\frac{Q_1(l+\frac{3}{2})}{Q_2(l+2)Q_2(l+1)}\\
=&\frac{Q_1(\frac{1}{2})}{Q_2(0)}Q_1(s+\frac{1}{2})\sum_{k=1}^s2(ik)^L\frac{Q_2(k)}{Q_1(k+\frac{1}{2})Q_1(k-\frac{1}{2})}.
\end{split}
\label{eqn:eq0000070}
\end{equation}
Where I have defined two type Baxter polynomials as 
\begin{equation}
Q_1(a)=\prod_{i=1}^{M_1} (ia-u_i),\quad Q_{2}(a)=\prod_{j=1}^{M_2}(ia-v_{j}).
\label{eqn:eq0000070a}
\end{equation}

The derivations] is as following. 
Eq. (\ref{eqn:eq0000070}) is equivalent to
\begin{equation}
\begin{split}
Eq.(70)&=Q_1(s+\frac{1}{2})Q_2(1)\Big[\sum_{k=0}^1f(k)\sum_{l=0}^k g(l)+\sum_{k=2}^sf(k)(\sum_{l=0}^1g(l)+\sum_{l=2}^kg(l))\Big]\\
&\qquad -Q_1(s+\frac{1}{2})Q_2(1)\sum_{m=2}^sf(m)\sum_{n=2}^m
g(n)\\
&=Q_1(s+\frac{1}{2})Q_2(1)\Big[\sum_{k=0}^1f(k)\sum_{l=0}^kg(l)+\sum_{k=2}^sf(k)\sum_{l=0}^1g(l)\Big]\\
&=Q_1(s+\frac{1}{2})Q_2(1)\Big[f(0)g(0)+\sum_{k=1}^sf(k)\sum_{l=0}^1g(l)\Big]\\
&=\frac{Q_1(s+\frac{1}{2})Q_1(\frac{1}{2})}{Q_{2}(0)}\Big(\frac{Q_2(1)Q_1(-1/2)}{Q_2(-1)Q_1(1/2)}+1\Big)\sum_{k=1}^sf(k)\\
&=2\frac{Q_1(s+\frac{1}{2})Q_1(\frac{1}{2})}{Q_{2}(0)}\sum_{k=1}^sf(k)
\end{split}
\label{eqn:eq0000071}
\end{equation}
we denote
\begin{equation}
f(k)=(ik)^L\frac{Q_2(k)}{Q_{1}(k+\frac{1}{2})Q_{1}(k-\frac{1}{2})},\quad g(l)=\frac{Q_1(l-\frac{1}{2})}{Q_{2}(l)Q_{2}(l-1)}.
\label{eqn:eq0000072}
\end{equation}
In the last line of Eq. (\ref{eqn:eq0000071}) we have used the fact that the paired momentum $Q(x)=Q(-x)$ for all excitations. Then
\begin{equation}
\bar{\mathcal{T}}^{(s)}(-i\frac{s+1}{2})-\bar{\mathcal{T}}^{(s-2)}(-i\frac{s+3}{2})=\frac{Q_1(\frac{1}{2})}{Q_2(0)}\mathbb{T}_{2s}(0).
\label{eqn:eq0000073}
\end{equation}
From \cite{de_Leeuw_2020}, we know the scalar solution overlap can be obtained by Thermodynamic Bethe Ansatz
\begin{equation}
\frac{\langle \mathrm{MPS}_{\delta}\vert\mathbf{u}\rangle}{\sqrt{\langle \mathbf{u}\lvert \mathbf{u}\rangle}}=\frac{Q_2(0)}{Q_2(\frac{1}{2})}\sqrt{\frac{Q_1(0)Q_1(\frac{1}{2})}{\bar{Q}_2(0)\bar{Q}_2(\frac{1}{2})}}\sqrt{\frac{\det G_+}{\det G_-}},
\label{eqn:eq0000074}
\end{equation}
hence
\begin{equation}
\frac{\langle \mathrm{MPS}_{2s+1}\lvert\mathbf{u}\rangle}{{\sqrt{\langle \mathbf{u}\lvert \mathbf{u}\rangle}}}=\mathbb{T}_{2s}(0)\sqrt{\frac{Q_1(0)Q_1(\frac{1}{2})}{\bar{Q}_2(0)\bar{Q}_2(\frac{1}{2})}}\sqrt{\frac{\det G_+}{\det G_-}}.
\label{eqn:eq0000075}
\end{equation}
where

\begin{equation}
\mathbb{T}_{2s}(0)=Q_1(s+\frac{1}{2})\sum_{k=1/2}^s2(ik)^L\frac{Q_2(k)}{Q_1(k+\frac{1}{2})Q_1(k-\frac{1}{2})}
\label{eqn:eq0000076}
\end{equation}
This agree with the overlap formula from Chapter \ref{Chapter3}.
\subsection{Even k=2s+1}
Let us now concentrate on even case. Follow Eq. (\ref{eqn:eq0000050}), one can see that $V(s)$ can be embedded into $L(s,s,1/2)\otimes V(1/2)$ where $s=\frac{3}{2},\frac{5}{2},\dots$. For general $s$, the twisted Yangian $Y^{+}(3)$ acts on $L(s,s,1/2)\otimes V(1/2)\cong V(s)\oplus (L(s,s,2)\otimes V(1/2))$ \cite{de_Leeuw_2020} as 
\begin{equation}
\begin{split}
s_{ij}(u)&\cdot(v_1\otimes v_2)=\mathcal{L}_{ia}^{(s,s,1/2)}(u)\mathcal{L}_{-j,-b}^{(s,s,1/2)}(-u)v_1\otimes\psi_{a,b}^{(1/2)}(u)v_2\\
&=\begin{pmatrix}(1-\frac{1}{4}u^{-2})\psi^{(s)}_{i,j}(u)&X\\0&\mathcal{L}^{(s,s,3/2)}_{ia}(u)\mathcal{L}^{(s,s,3/2)}_{-j,-b}(-u)\otimes\psi^{(1/2)}_{ab}\end{pmatrix}\begin{pmatrix}w_1\\w_2\otimes w_3\end{pmatrix}
\end{split}
\label{eqn:eq0000077}
\end{equation}
for all $v_1\in L(s,s,1/2),\ v_2,w_3 \in V(1/2),\ w_1\in V(s), w_2\in L(s,s,3/2)$ and $s\in \mathbb{Z}_{+}+1/2$.
We used the following conjecture:
\begin{conj}
The $L(s,s,3/2) \otimes V(1/2)$ is an irrep of $Y^{+}(3)$ for all $s\in\mathbb{Z}_{+}+1/2$. 
\end{conj}
Use the same argument as odd case, one would find Eq. (\ref{eqn:eq0000077}) connects $\vert \mathrm{MPS}_{2s+1}\rangle$ with $\vert \mathrm{MPS}_{2}\rangle$ as
\begin{equation}
\lvert\mathrm{MPS}_{2s+1}\rangle=(\frac{2}{i})^{L}\left(\bar{T}^{(s,s,1/2)}(0)-\bar{T}^{(s,s,3/2)}(0)\right)\vert \mathrm{MPS}_{2}\rangle.
\label{eqn:eq0000078}
\end{equation}
From Eq. (\ref{eqn:eq0000063}) one can obtain
\begin{equation}
\begin{split}
&\bar{\mathcal{L}}^{(s,s,1/2)}(u)=\bar{\mathcal{L}}^{(m)}(u-i\frac{m+2}{2}),\\
&\bar{\mathcal{L}}^{(s,s,3/2)}(u)=\bar{\mathcal{L}}^{(m-1)}(u-i\frac{m+3}{2}),
\end{split}
\label{eqn:eq0000079}
\end{equation}
where 
\begin{equation}
m=s-1/2.
\label{eqn:eq0000080}
\end{equation}
Thus
\begin{equation}
\lvert\mathrm{MPS}_{2s+1}\rangle=(\frac{2}{i})^{L}\left(\bar{T}^{(m)}(-i\frac{m+2}{2})-\bar{T}^{(m-1)}(-i\frac{m+3}{2})\right)\vert \mathrm{MPS}_{2}\rangle,
\label{eqn:eq0000081}
\end{equation}
Use Eq. (\ref{eqn:eq0000069}), one would find 
\begin{equation}
\begin{split}
\bar{T}&^{(m)}(-i\frac{m+2}{2})-\bar{T}^{(m-1)}(-i\frac{m+3}{2})\\
=&Q_1(m+1)Q_2(\frac{1}{2})\sum_{k=0}^m(ik+\frac{i}{2})^L\frac{Q_2(k+\frac{1}{2})}{Q_1(k+1)Q_1(k)}\sum_{l=0}^k\frac{Q_1(l)}{Q_2(l+\frac{1}{2})Q_2(l-\frac{1}{2})}\\
-&Q_1(m+1)Q_2(\frac{1}{2})\sum_{k=0}^{m-1}(ik+\frac{3i}{2})^L\frac{Q_2(k+\frac{3}{2})}{Q_1(k+2)Q_1(k+1)}\sum_{l=0}^k\frac{Q_1(l+1)}{Q_2(l+\frac{3}{2})Q_2(l+\frac{1}{2})}\\
=&\frac{Q_1(0)}{Q_2(\frac{1}{2})}Q_1(s+\frac{1}{2})\sum_{k=1/2}^s(ik)^L\frac{Q_2(k)}{Q_1(k+\frac{1}{2})Q_1(k-\frac{1}{2})}.
\end{split}
\label{eqn:eq0000082}
\end{equation}
Use the explicit form of $\mathbb{T}_{2s}(0)$ and $\mathbb{T}_{1}(0)$ in $s=\mathbb{Z}_{+}+1/2$ case
\begin{equation}
\begin{split}
&\mathbb{T}_{2s}(0)=Q_1(s+\frac{1}{2})\sum_{k=1/2}^s2(ik)^L\frac{Q_2(k)}{Q_1(k+\frac{1}{2})Q_1(k-\frac{1}{2})},\\
&\mathbb{T}_{1}(0)=2(\frac{i}{2})^{L}\frac{Q_{2}(\frac{1}{2})}{Q_1(0)},
\end{split}
\label{eqn:eq0000083}
\end{equation}
hence
\begin{equation}
(\frac{2}{i})^{L}\left(\bar{T}^{(m)}(-i\frac{m+2}{2})-\bar{T}^{(m-1)}(-i\frac{m+3}{2})\right)=\frac{\mathbb{T}_{2s}(0)}{\mathbb{T}_{1}(0)}.
\label{eqn:eq0000084}
\end{equation}
Finally, the connection between $\vert \mathrm{MPS}_{2s+1}\rangle$ with $\vert \mathrm{MPS}_{2}\rangle$ is 
\begin{equation}
\frac{\langle\mathrm{MPS}_{2s+1}\vert \mathbf{u}\rangle}{\langle\mathrm{MPS}_{2}\vert \mathbf{u}\rangle}=\frac{\mathbb{T}_{2s}(0)}{\mathbb{T}_{1}(0)}.
\label{eqn:eq0000085}
\end{equation}
 
% Chapter 6

\chapter{Spin Chain Overlaps and extended twisted Yangian for $(SO(6),SO(3)\times SO(3))$} % Main chapter title

\label{Chapter6} % For referencing the chapter elsewhere, use \ref{Chapter6}
%----------------------------------------------------------------------------------------
\section{Definitions and preliminaries}
The notions of Lie algebras follow the same convention as last chapter, also the transposition rules.

\subsection{R-matrices}
For $\mathfrak{so}_N$ models one can build the trace operator
\begin{equation}
\mathbb{K}=\mathbb{P}^{t_1}=\mathbb{P}^{t_2}=\sum_{ij}E_{ij}\otimes(E_{ji})^{t}=\sum_{ij}E_{ij}\otimes E_{-i,-j},
\label{eqn:eq0000001}
\end{equation}
where $\mathbb{P}$ is the permutation operator $\mathbb{P}=\sum_{ij}E_{ij}\otimes E_{ji}$. The corresponding $\mathfrak{so}_N$ R-matrices
\begin{equation}
R(u)=\mathbb{I}-\frac{\mathbb{P}}{u}+\frac{\mathbb{K}}{u-\kappa},
\label{eqn:eq0000002}
\end{equation}
where $\kappa=N/2-1$, it is a solution of the quantum Yang-Baxter equation with spectral parameter
\begin{equation}
R_{12}(u)R_{13}(u+v)R_{23}(v)=R_{23}(v)R_{13}(u+v)R_{12}(u).
\label{eqn:eq0000003}
\end{equation}
The $\mathfrak{so}_N$ R-matrix satisfies the crossing equation
\begin{equation}
R^{t}(u)=R^{t_1}=R^{t_2}=R(\kappa-u).
\label{eqn:eq0000004}
\end{equation}
In the normal convention of transpose, the corresponding trace operator is defined as 
\begin{equation}
\tilde{\mathbb{K}}=\mathbb{P}^{T_1}=\mathbb{P}^{T_2}=\sum_{ij}E_{ij}\otimes(E_{ji})^{T}=\sum_{ij}E_{ij}\otimes E_{i,j},
\label{eqn:eq0000005}
\end{equation}
and the corresponding R-matrix up to a normalization is 
\begin{equation}
\tilde{R}(u)=\mathbb{I}-\frac{\mathbb{P}}{u}+\frac{\tilde{\mathbb{K}}}{u-\kappa}.
\label{eqn:eq0000006}
\end{equation}
And the crossing equation becomes
\begin{equation}
\tilde{R}^{T}(u)=\tilde{R}^{T_1}(u)=\tilde{R}^{T_2}(u)=\tilde{R}(\kappa-u).
\label{eqn:eq0000007}
\end{equation}
\subsection{Versions of BYB in different transpositions}
Recall we have used a different convention of twisted Boundary Yang-Baxter equations for these two separated case
\begin{equation}
S_{2}(v)R^{t}_{21}(-u-v)S_1(u)R_{12}(u-v)=R_{21}(u-v)S_1(u)R^{t}_{12}(-u-v)S_{2}(v),
\label{eqn:eq0000008}
\end{equation}

\begin{equation}
\tilde{S}_{2}(v)\tilde{R}^{T}_{21}(-u-v)\tilde{S}_1(u)\tilde{R}_{12}(u-v)=\tilde{R}_{21}(u-v)\tilde{S}_1(u)\tilde{R}^{T}_{12}(-u-v)\tilde{S}_{2}(v).
\label{eqn:eq0000009}
\end{equation}
Same as before for twisted Yangian case, we typically solve Eq. (\ref{eqn:eq0000009}) in the usual basis based on the symmetry, then we transform the solution to tBYB case Eq. (\ref{eqn:eq0000008}), which is actually Highest weight friendly way for our algebra. 

Now, let's turn to the case for $SO(N)$. Consider a constant transformation matrix $P$, which brings the solutions of BYB to tBYB $K(u)=P\tilde{K}(u)P^{-1}$. Insert it into Eq. (\ref{eqn:eq0000008}) with the help of crossing relation Eq. (\ref{eqn:eq0000007}), one would get
\begin{equation}
\begin{split}
&P_2\tilde{K}_2(v)P_2^{-1}R_{21}(2+u+v)P_1\tilde{K}_1(u)P_1^{-1}R_{12}(u-v)\\&=R_{21}(u-v)P_1\tilde{K}_1(u)P_1^{-1}R_{12}(2+u+v)P_2\tilde{K}_2(v)P_2^{-1}\\
\Leftrightarrow&\tilde{K}_2(v)P_2^{-1}R_{21}(2+u+v)P_1\tilde{K}_1(u)P_1^{-1}R_{12}(u-v)P_2\\&=P_2^{-1}R_{21}(u-v)P_1\tilde{K}_1(u)P_1^{-1}R_{12}(2+u+v)P_2\tilde{K}_2(v),
\end{split}
\label{eqn:eq00000010}
\end{equation}
compare with Eq. (\ref{eqn:eq0000008}), for $\tilde{K}(u)$ is a solution of BYB, the sufficient but unnecessary condition is 
\begin{equation}
\tilde{R}_{12}(u)=P_{2}^{-1}P_{1}^{-1}R_{12}(u)P_{2}P_{1}.
\label{eqn:eq00000011}
\end{equation}
Use the Eq. (\ref{eqn:eq0000004}) and Eq. (\ref{eqn:eq0000007}), one can show $P$ matrix should satisfy the following relation 
\begin{equation}
PP^{T}=C,
\label{eqn:eq00000012}
\end{equation}
$C$ is the charge conjugation operator  $C_{ij}=\delta_{i,-j}$, with $C^2=1$ . Here is detailed proof

\begin{equation}
\begin{split}
&C_1R^{T_1}_{12}(u)C_1=C_1\left(P_{2}P_{1}\tilde{R}_{12}(u)P_{2}^{-1}P_{1}^{-1}\right)^{T_1}C_1\\
\Leftrightarrow &R^{t_1}_{12}(u)=C_1(P_{1}^{-1})^{T_1}P_2\tilde{R}^{T_1}_{12}(u)(P_{1})^{T_1}P_{2}^{-1}C_1\\
\Leftrightarrow &R_{12}(\kappa-u)=C_1(P_{1}^{T_1})^{-1}P_2\tilde{R}_{12}(\kappa-u)P_{2}^{-1}(P_{1})^{T_1}C_1\\
\Leftrightarrow &P_{2}P_{1}\tilde{R}_{12}(\kappa-u)P_{2}^{-1}P_{1}^{-1}=C_1(P_{1}^{T_1})^{-1}P_2\tilde{R}_{12}(\kappa-u)P_{2}^{-1}(P_{1})^{T_1}C_1,
\end{split}
\label{eqn:eq00000013}
\end{equation}
then we will get $P_{2}P_{1}=C_1(P_{1}^{T_1})^{-1}P_2,\  P_{2}^{-1}P_{1}^{-1} = P_{2}^{-1}(P_{1})^{T_1}C_1$, then one can easily verify that Eq. (\ref{eqn:eq00000012}) satisfies this relation with $C^2=1$. 
\subsection{Extended Yangian and extended twisted Yangian}
Before giving the formal definition of what extended Yangian algebra and extended twisted Yangian are, we need to specify the transformation matrix $P$ for our $(SO(6),SO(3)\times SO(3))$ case. We'd like to transform the scalar solution from BYB to tBYB, the transformation matrix for $(SO(6),SO(3)\times SO(3))$ takes the form
\begin{equation}
\begin{pmatrix}\lvert -3\rangle\\\lvert -2\rangle\\\lvert -1\rangle\\\lvert 1\rangle\\\lvert 2\rangle\\\lvert 3\rangle\end{pmatrix}=P\begin{pmatrix}\lvert 1\rangle\\\lvert 2\rangle\\\lvert 3\rangle\\\lvert 4\rangle\\\lvert 5\rangle\\\lvert 6\rangle\end{pmatrix}, \quad P_{SO(3)\times SO(3)}=\frac{1}{\sqrt{2}}\begin{pmatrix}1&i&0&0&0&0\\0&0&0&0&1&i\\0&0&1&i&0&0\\0&0&1&-i&0&0\\0&0&0&0&1&-i\\1&-i&0&0&0&0\end{pmatrix},
\label{eqn:eq00000014}
\end{equation}
the reason for why we choose such transformation will be clear later. 

Up to different normalization of $\tilde{K}(u)$, one can find the scalar solution with above transformation matrix becomes
\begin{equation}
K_{SO(3)\times SO(3)}=\begin{pmatrix}
 1 & 0 & 0 & 0 & 0 & 0 \\
 0 & -1 & 0 & 0 & 0 & 0 \\
 0 & 0 & 0 & 1 & 0 & 0 \\
 0 & 0 & 1 & 0 & 0 & 0 \\
 0 & 0 & 0 & 0 & -1 & 0 \\
 0 & 0 & 0 & 0 & 0 & 1 \\
\end{pmatrix}.
\label{eqn:eq00000015}
\end{equation}
Correspondingly, one can see the following symmetry relation
\begin{equation}
K^{t}_{SO(3)\times SO(3)}=K.
\label{eqn:eq00000016}
\end{equation}
Fix $N\in \mathbb{Z}_{\ge 3}$, we introduce elements $t^{(r)}_{ij}$ with $-n\le i,j\le n$ and $r\in \mathbb{Z}_{\ge 0}$ such that $t_{ij}^{(0)}=\delta_{ij}$. Combining these into formal power series $t_{ij}(u)=\sum_{r\ge 0}t^{(r)}_{ij}u^{-r}$, we can then form the generating matrix $T(u)=\sum_{-n\le i,j\le n} E_{ij}\otimes t_{ij}(u)$. Let $R(u)$ be given by Eq. (\ref{eqn:eq0000002}).

\begin{definition}
The extended Yangian $X(\mathfrak{g}_N)$ is the unital associative $\mathbb{C}$-algebra generated by elements $t^{(r)}_{ij}$ with $-n\le i,j\le n$ and $r\in \mathbb{Z}_{\ge 0}$ satisfying the relation \cite{e_t_yangian}
 \begin{equation}
R(u-v)T_1(u)T_2(v)=T_2(v)T_1(u)R(u-v), 
\label{eqn:eq00000017}
\end{equation}
where $T_1(u)$ and $T_2(u)$ are the elements of $\mathrm{End}\mathbb{C}^{N}\otimes\mathrm{End}\mathbb{C}^{N}\otimes X$($\mathfrak{g}[[u^{-1}]]$) given by 
\begin{equation}
T_1(u)=\sum_{i,j=-n}^n E_{ij}\otimes \mathbb{I}\otimes t_{ij}(u),\quad T_2(u)=\sum_{i,j=-n}^n\mathbb{I} \otimes E_{ij}\otimes t_{ij}(u),
\label{eqn:eq00000018}
\end{equation}
 In terms of the power series elements $t_{ij}(u)$, Eq. (\ref{eqn:eq00000017}) is equivalent to
\begin{equation}
\begin{split}
[t_{ij}(u),t_{kl}(v)]&=\frac{1}{u-v}\left(t_{kj}(u)t_{il}(v)-t_{kj}(v)t_{il}(u)\right)\\
&-\frac{1}{u-v-\kappa}\sum_{-n\le a \le n}\left(\delta_{k,-i}t_{aj}(u)t_{-a,l}(v)-\delta_{l,-j}t_{k,-a}(v)t_{i,a}(u)\right)
\end{split}
\label{eqn:eq00000019}
\end{equation}
The Hopf algebra structure of $X(\mathfrak{g}_N)$ is given by
\begin{equation}
\Delta:t_{ij}(u)\mapsto \sum_{-n\le a\le n}t_{ia}(u)\otimes t_{aj}(u),\quad S:T(u)\mapsto T(u)^{-1},\quad \epsilon T(u)\mapsto \mathbb{I}.
\label{eqn:eq00000020}
\end{equation}
We will call the generating matrix of the extended Yangian $X(\mathfrak{g}_N)$ the $T$-matrix.
\end{definition}
Let $A\in GL(N)$ be such that $AA^{t}=\mathbb{I}$ and let $a\in \mathbb{C}$ be a constant. Moreover, consider an arbitrary formal series $f(u)$ of the form
\begin{equation}
f(u)=1+f_1u^{-1}+f_2u^{-2}+\dots\mathbb{C}[[u^{-1}]].
\label{eqn:eq00000021}
\end{equation}
The automorphisms:
\begin{equation}
\mu_f:T(u)\mapsto f(u)T(u),\quad \tau_a: T(u)\mapsto T(u+a), \quad \alpha_A: T(u)\mapsto AT(u)A^{t},
\label{eqn:eq00000022}
\end{equation}
where the first two maps is readily to see that it satisfies RTT relation Eq. (\ref{eqn:eq00000017}), for $\alpha_A$ one should used the invariance of $R$-matrix
\begin{equation}
R_{12}A_1A_2=A_1A_2R_{12},\quad R_{12}A_1^{t}A_2^{t}=A_1^{t}A_2^{t}R_{12}.
\label{eqn:eq00000023}
\end{equation}
The above Extended Yangian algebra is more or less the symmetry algebra for our spin chain $SO(6)$ sector. The semi-infinite size of spin chain enforce such symmetry will be broken on the boundary. Again the physical meaning of the coproduct depends on how you associate $t_{ij}$ with what space. 
\begin{definition}
The extended twisted Yangian $X(\mathfrak{g}_N,\mathcal{G})$ is the subalgebra of $X(\mathfrak{g}_N)$ generated by the coefficients  $s^{(r)}_{ij}$ with $-n\le i,j\le n$ and $r\in \mathbb{Z}_{\ge 0}$ of the matrix elements $s_{ij}(u)$ in $S(u)=\sum_{ij}E_{ij}\otimes s_{ij}(u)$ matrix \cite{e_t_yangian}
\begin{equation}
S(u)=T(u)K(u)T^{t}(-u),
\label{eqn:eq00000024}
\end{equation}
where $K(u)$ is scalar solution of tBYB.
\end{definition}
The algebra $X(\mathfrak{g}_N,\mathcal{G})$ is a left coideal subalgebra of $X(\mathfrak{g}_N): \Delta(X(\mathfrak{g}_N,\mathcal{G}))\subset X(\mathfrak{g}_N)\otimes X(\mathfrak{g}_N,\mathcal{G})$, or in terms of matrix entries $s_{ij}(u)=\sum_{a,b=-n}^{n}t_{ia}(u)k_{ab}(u)t_{-j,-b}(-u)$, where $k_{ab}(u)$ are the matrix entries of $K(u)$. And by Eq. (\ref{eqn:eq00000019}), one can prove this property 
\begin{equation}
\Delta(s_{ij}(u))=\sum_{-n\le a,b\le n}t_{ia}(u)t_{-j,-b}\otimes s_{ab}(u)
\label{eqn:eq00000025}
\end{equation}
There are two important properties that connect the Extend twisted Yangian $X(\mathfrak{g}_N,\mathcal{G})$ with the tBYB equation as well as the RTT relation. 

$\mathbf{Property\ 1.}$ Make use of Eq. (\ref{eqn:eq00000016}), one can show that the S-matrix $S(u)$ satisfies the tBYB equation
\begin{equation}
S_{2}(v)R^{t}_{21}(-u-v)S_1(u)R_{12}(u-v)=R_{21}(u-v)S_1(u)R^{t}_{12}(-u-v)S_{2}(v). 
\label{eqn:eq00000026}
\end{equation}
$Proof:$
\begin{equation}
\begin{split}
&T_2(v)K_2(v)T_2^{t}(-v)R_{21}^{t}(-u-v)T_1(u)K_1(u)T_1^{t}(-u)R_{12}(u-v)\\
&=R_{21}(u-v)T_1(u)K_1(u)T_1^{t}(-u)R^{t}_{12}(-u-v)T_2(v)K_2(v)T_2^{t}(-v)\\
\Leftrightarrow\ &T_2(v)K_2(v)T_1(u)R_{21}^{t}(-u-v)T_2^{t}(-v)K_1(u)T_1^{t}(-u)R_{12}(u-v)\\
&=R_{21}(u-v)T_1(u)K_1(u)T_2(v)R^{t}_{12}(-u-v)T_1^{t}(-u)K_2(v)T_2^{t}(-v)\\
\Leftrightarrow\ &T_2(v)T_1(u)K_2(v)R_{21}^{t}(-u-v)K_1(u)R_{12}(u-v)T_2^{t}(-v)T_1^{t}(-u)\\
&=T_2(v)T_1(u)R_{21}(u-v)K_1(u)R^{t}_{12}(-u-v)K_2(v)T_2^{t}(-v)T_1^{t}(-u)\\
\Leftrightarrow\ &K_2(v)R_{21}^{t}(-u-v)K_1(u)R_{12}(u-v)=R_{21}(u-v)K_1(u)R^{t}_{12}(-u-v)K_2(v).
\end{split}
\label{eqn:eq00000027}
\end{equation}
$\mathbf{Property\ 2}$
The symmetry relation for $(SO(6),SO(3)\times SO(3))$:
\begin{equation}
S^{t}(u)=S(-u)+\frac{1}{2u}(S(u)-S(-u))-\frac{\mathrm{tr}(S(u))}{2u-2}\mathbb{I}.
\label{eqn:eq00000028}
\end{equation}
$Proof:$
\begin{equation}
\begin{split}
&[S^{t}(u)]_{ij}\\&=S_{-j,-i}(u)=\sum_{a,b}k_{a,b}t_{-j,a}(u)t_{i,-b}(-u)\\
&=\sum_{a,b}k_{a,b}(u)t_{i,-b}(-u)t_{-j,a}(u)+\frac{1}{2u}\sum_{a,b}k_{a,b}(u)[t_{i,a}(u)t_{-j,-b}(-u)-t_{ia}(-u)t_{-j,-b}(u)]\\&+\frac{1}{2u-2}\sum_{a,b,c}k_{a,b}(u)(\delta_{a,b}t_{i,-c}(-u)t_{-j,c}(u)-\delta_{i,j}t_{c,a}(u)t_{-c,-b}(u))\\
&=\sum_{a,b}k_{-b,-a}(u)t_{i,-b}(-u)t_{-j,a}(u)+\frac{1}{2u}\sum_{a,b}k_{a,b}(u)[t_{i,a}(u)t_{-j,-b}(-u)-t_{ia}(-u)t_{-j,-b}(u)]\\&+\frac{1}{2u-2}\Big(0-\delta_{i,j}\mathrm{Tr}\,S(u)\Big),\\
&=(S(-u))_{ij}+\frac{1}{2u}(S(u)-S(-u))_{ij}-\frac{1}{2u-2}\delta_{i,j}\mathrm{Tr}\,S(u)
\end{split}
\label{eqn:eq00000029}
\end{equation}
where I have used the commutation relation Eq. (\ref{eqn:eq00000019}), also $K(u)=K(-u)=K$ and $\mathrm{Tr}\,(K)=0$. 

Hence for our physical system $(SO(6),SO(3)\times SO(3))$, the corresponding Boundary Yangian Baxter equation combined with RTT relation are equivalent to Extend Twisted Yangian of type $(SO(6),SO(3)\times SO(3))$. 

The automorphism of twisted Yangian Eq. (\ref{eqn:eq0000032}) is not an automorphism of  $X(\mathfrak{g}_N,\mathcal{G})$ since the symmetry relation is not satisfied.

\subsection{Representations of extended Yangian and extended twisted Yangian}
\begin{definition}
A representation $V$ of $X(\mathfrak{g}_N)$ is a highest weight representation if there exists a nonzero vector $v\in V$ such that $V=X(\mathfrak{g}_N)\cdot v$ and satisfies the following conditions \cite{rep_e_t_yangian_I,rep_e_t_yangian_II}
\begin{equation}
\begin{alignedat}{2}
t_{ij}(u)\cdot v&=0\qquad &&\mathrm{for}\ -n\le i<j\le n,\quad \mathrm{and}\\
t_{ii}(u)\cdot v&=\lambda_i(u) v\qquad &&\mathrm{for}\ -n\le i\le n,
\end{alignedat}
\label{eqn:eq00000031}
\end{equation}
and for each $i$, $\lambda_{i}(u)$ is a formal power series in $u^{-1}$ 
\begin{equation}
\lambda_{i}=1+\sum_{r=1}^{\infty}\lambda^{(r)}_{i}u^{-r},\quad \lambda^{(r)}_{i}\in \mathbb{C}
\label{eqn:eq00000032}
\end{equation}
the vector $v$ is called the highest weight vector of $V$, and $\lambda(u)=(\lambda_{-n}(u),\dots,\lambda_{n}(u))$ is the highest weight. 
\end{definition}
In contrast to the Yangian $Y(N)$, there is no surjective homeomorphism for $X(\mathfrak{so}_N)$ onto the algebra $U(\mathfrak{so}_N)$, so we cannot use $\mathfrak{so}_N$ modules as $X(\mathfrak{so}_N)$ modules in general. But there is a way out of this if one only consider $X(\mathfrak{so}_6)$, because there is an isomorphism between $\mathfrak{so}_6$ and $\mathfrak{gl}_4$, so we can find a way to connect $X(\mathfrak{so}_6)$ and $Y(\mathfrak{gl}_4)$. 

Consider the vector space $\mathbb{C}^4$ with its canonical basis $e_1, e_2,e_3, e_4$, and denote the subspace of $\mathbb{C}^4\times\mathbb{C}^4$ by $V$ with  $\mathbb{C}^6$ regarding $v_{-3},v_{-2},v_{-1},v_{1},v_{2},v_{3}$ as its canonical basis. Let's identify
\begin{equation}
\begin{split}
v_{-3}=e_1\otimes e_2-e_2\otimes e_1,\quad v_{3}=e_3\otimes e_4-e_4\otimes e_3,\\
v_{-2}=e_3\otimes e_1-e_1\otimes e_3,\quad v_{2}=e_2\otimes e_4-e_4\otimes e_2,\\
v_{-1}=e_1\otimes e_4-e_4\otimes e_1,\quad v_{1}=e_2\otimes e_3-e_3\otimes e_2.
\end{split}
\label{eqn:eq00000033}
\end{equation}
Such mapping can be generated by isomorphism between $\mathfrak{so}_6$ and $\mathfrak{gl}_4$, for detail see Appendix \ref{AppendixF}.
The mapping
\begin{equation}
T^{\mathfrak{so}_6}(u)\mapsto \frac{1}{2}(1-P_{12})\cdot T_1^{\mathfrak{gl}_4}(u)T_2^{\mathfrak{gl}_4}(u-1)
\label{eqn:eq00000034}
\end{equation}
defines an isomorphism $\phi: X(\mathfrak{so}_6)\rightarrow Y(\mathfrak{gl}_4)$. More explicitly, the images of the generators under the isomorphism are given by the formulas
\begin{equation}
t^{\mathfrak{so}_6}_{ij}(u)=t^{\mathfrak{gl}_4}_{a_1[i],a_1[j]}(u)t^{\mathfrak{gl}_4}_{a_2[i],a_2[j]}(u-1)-t^{\mathfrak{gl}_4}_{a_2[i],a_1[j]}(u)t^{\mathfrak{gl}_4}_{a_1[i],a_2[j]}(u-1),
\label{eqn:eq00000035}
\end{equation}
where
\begin{equation}
\begin{matrix}a_1[-3]=1,& a_1[-2]=3,& a_1[-1]=1,& a_1[1]=2, & a_1[2]=2, & a_1[3]=3;\\
a_2[-3]=2,
& a_2[-2]=1,
& a_2[-1]=4,
 & a_2[1]=3,
 & a_2[2]=4,& a_2[3]=4;
\end{matrix}
\label{eqn:eq00000036}
\end{equation}
is a mapping of indices from $\mathfrak{so}_6$ $\{-3,-2,-1,1,2,3\}$ to $\mathfrak{gl}_4$ $\{1,2,3,4\}$. We will prove Eq. (\ref{eqn:eq00000034}) using the so-called fusion procedure in Appendix \ref{AppendixF}.

For extended twisted Yangian, there do exist a formal definition of it. However, such definition turns out doesn't work for our $X(\mathfrak{so}_6,\mathfrak{so}_3\otimes \mathfrak{so}_3)$ case. We are going to "invent" one for $X(\mathfrak{so}_6,\mathfrak{so}_3\otimes \mathfrak{so}_3)$ by inspecting the classical algebra.

The tBYB Eq. (\ref{eqn:eq0000008}) and the symmetry relation Eq. (\ref{eqn:eq00000028}) form reflection algebra $\mathcal{B}(\mathcal{G})$, where $\mathcal{G}=\mathfrak{so}_3\oplus \mathfrak{so}_3$ , with the R-matrix given by Eq. (2), $\kappa=2$. As show in \cite{e_t_yangian}, $\mathcal{B}(\mathcal{G})$ are isomorphic to $X(\mathfrak{so}_6,\mathfrak{so}_3\otimes \mathfrak{so}_3)$. Due to the algebra isomorphism Eq. (\ref{eqn:eq00000034}), and the evaluation homomorphism from last Chapter, one can conjecture, that the $X(\mathfrak{so}_6,\mathfrak{so}_3\otimes \mathfrak{so}_3)$ module $V$ is a highest weight if there exists a nonzero vector $v\in V$ such that $V$ is generated by $v$ and \cite{de_Leeuw_2020}
\begin{equation}
\begin{split}
s_{ij}\cdot v&=0,\quad \mathrm{for\ all\ i<j\ where\ (i,j)\ne (-1,1)},\\
s_{ii}\cdot v&=\mu_i(u) v,\\
s_{1,-1}\cdot v&=\mu^{+}(u) v,\\
s_{-1,1}\cdot v&=\mu^{-}(u) v.
\end{split}
\label{eqn:eq00000037}
\end{equation}
Note that if we only focus on the "Cartan" subalgebra part of $X(\mathfrak{so}_6,\mathfrak{so}_3\otimes \mathfrak{so}_3)$, these choices of "Cartan" generators are due to the choice of the transformation matrix Eq. (\ref{eqn:eq00000014}), the corresponding matrix entries, i.e. $s_{1,-1}$, are made up of Cartan generator of $\mathfrak{so}_3\otimes \mathfrak{so}_3$. The same logic goes for raising and lowering operators. Actually the choice of transformation matrix Eq. (\ref{eqn:eq00000014}) is coming from the choice of generators Eq. (\ref{eqn:eq00000035}) and basis Eq. (\ref{eqn:eq00000033}) of $X(\mathfrak{so}_6)$. Since we need to embed $X(\mathfrak{so}_6,\mathfrak{so}_3\otimes \mathfrak{so}_3)$) as a subalgebra of $X(\mathfrak{so}_6)$, an arbitrary transformation cannot do so.  As one can show that the highest weights of higher dimensional two site solutions from BYB and the highest weights of the corresponding dressing scalar solutions from BYB  will only differ by a pure function of rapidity parameter $u$, if we choose the transformation matrix Eq. (\ref{eqn:eq00000014}). By the automorphism $\mu_f$ in Eq. (\ref{eqn:eq00000022}), the representations from higher two-site solutions and dressing scalar solutions are in fact the same representation of $X(\mathfrak{so}_6,\mathfrak{so}_3\otimes \mathfrak{so}_3)$\footnote{Here we don't put the symmetry relation constraint on higher dimensional solutions, so automorphism $\mu_f$ may differ.}, this is all we want to connect MPS and scalar state, consequently the overlap formula. 
\section{Highest weight matching between dressing scalar solution and MPS}
\subsection{MPS side of calculation}
Start with the two-site solution of higher representation, the $MPS$ can be built from the $S$-matrix $\tilde{S}(u)=\sum_{a,b=1}^6 e_{ab}\otimes \tilde{\psi}_{ab}(u)$, where $\tilde{\psi}_{ab}$ is given in Chapter \ref{Chapter4},
\begin{equation}
\begin{split}
\psi_{ij}(u)&=(-u^2+1)S_iS_j+u(u+1)S_jS_i-\frac{1}{2}u(u^2+u+C)\delta_{ij},\\
\psi_{iI}&=\psi_{Ii}=0,\\
\psi_{IJ}&=\frac{1}{2}u(u^2+u-C)\delta_{IJ},
\end{split}
\label{eqn:eq00000038}
\end{equation}
where $i,j\in \{1,2,3\}$ and $I,J\in \{4,5,6\}$. Transform the solution to the case of tBYB
\begin{equation}
S(u)=P_{SO(3)\times SO(3)}\tilde{S}(u)P^{-1}_{SO(3)\times SO(3)},
\label{eqn:eq00000039}
\end{equation}
for simplicity, I will only show the relevant matrix elements
\begin{equation}
\begin{split}
s_{33}(u)&=\frac{1}{2}(s+1)s-\frac{1}{2}u^3-\frac{1}{2}u^2-\frac{1}{2}S_z \left(-2 u^2-u+1\right)-\frac{1}{2}(u+1)S_z^2 ,\\
s_{22}(u)&=-\frac{1}{2}(s+1)s u +\frac{1}{2} u^2 +\frac{1}{2} u^3,\\
s_{11}(u)&=-\frac{1}{2}(s+1)s u + \frac{1}{2}(1 + u) S_z^2,\\
s_{1-1}(u)&=s_{-11}(u)=-\frac{1}{2}u^2 -\frac{1}{2} u^3 +\frac{1}{2} (1 + u) S_z^2.
\end{split}
\label{eqn:eq00000040}
\end{equation}
We use the notation
\begin{equation}
\begin{split}
S_{\pm}&=\frac{1}{\sqrt{2}}(S_1\pm iS_2),\\
S_z&=S_3.
\end{split}
\label{eqn:eq00000041}
\end{equation}
And the corresponding commutation relation becomes
\begin{equation}
[S_+,S_-]=S_z,\quad [S_z,S_{\pm}]=\pm S_{\pm}.
\label{eqn:eq00000042}
\end{equation}
We will denote the states by $(J_z,K_z)$. Since one of the $\mathfrak{so}_3$ is in the trivial representation, the states can be characterized by $(S_z,0)$ or $(0,S_z)$. The $S_i$ form a h.w. representation of $\mathfrak{so}_3\oplus \mathfrak{so}_3$ i.e. there exist a lowest weight vector $v$ for which $(\lambda_1,\lambda_2)=(s,0)$ or $(\lambda_1,\lambda_2)=(0,s)$. From Eq. (\ref{eqn:eq00000042}) we know that $S_+$ increases the weights, $S_-$ decreases the weights, therefore
\begin{equation}
S_{+}\cdot v=0.
\label{eqn:eq00000043}
\end{equation}
Using this we can calculate highest weights as
\begin{equation}
s_{33}(u)\cdot v=-\frac{1}{2}u (s-u-1) (s-u)\cdot v,
\label{eqn:eq00000044}
\end{equation}

\begin{equation}
s_{22}(u)\cdot v=\frac{1}{2}u(u-s)(s+u+1)\cdot v,
\label{eqn:eq00000045}
\end{equation}

\begin{equation}
s_{11}\cdot v=\frac{1}{2}s(s-u)\cdot v,
\label{eqn:eq00000046}
\end{equation}

\begin{equation}
s_{1-1}(u)\cdot v= s_{-11}(u)\cdot v=-\frac{1}{2}(u+1) (u-s) (s+u)\cdot v.
\label{eqn:eq00000047}
\end{equation}
Combine all the elements together
\begin{equation}
\begin{split}
\mu_3(u)&=-\frac{1}{2}u (s-u-1) (s-u),\\
\mu_2(u)&=\frac{1}{2}u (u-s) (s+u+1)\\
\mu_1(u)&=\frac{1}{2}s (s-u),\\
\mu^{(+)}(u)&=\mu^{(-)}(u)=-\frac{1}{2}(u+1) (u-s) (s+u).
\end{split}
\label{eqn:eq00000048}
\end{equation}
\subsection{Scalar solution side of calculation}
Now let try to use the dressing scalar solution to calculate the highest weight, recall that the higher dimensional representation of $X(\mathfrak{so}_6,\mathfrak{so}_3\oplus\mathfrak{so}_3)$ can be obtained by 
\begin{equation}
S^{D}(u)=T^{\mathfrak{so}_6}(u)K(u)(T^{\mathfrak{so}_6})^t(-u),
\label{eqn:eq00000049}
\end{equation}
acting in the $\mathfrak{so}_3\oplus\mathfrak{so}_3$ module. Or in the matrix elements form
\begin{equation}
s^{D}_{ij}(u)=\sum_{ab}t^{\mathfrak{so}_6}_{ia}(u)K_{ab}(u)t^{\mathfrak{so}_6}_{-j,-b}(-u),
\label{eqn:eq00000050}
\end{equation}
the $X(\mathfrak{so}_6)$ generators $t^{\mathfrak{so}_6}_{ij}(u)$ can be mapped to $Y(4)$ generators $t^{\mathfrak{gl}_4}_{ij}(u)$ as stated in Eq. (\ref{eqn:eq00000039}). However, since we don't know how to connect $X(\mathfrak{so}_6)$ with the  algebra $U(\mathfrak{so}_6)$, we cannot just embed $\mathfrak{so}_3\oplus\mathfrak{so}_3$ into $\mathfrak{so}_6$ and do all the calculation in the $\mathfrak{so}_6$ module. The mapping of basis Eq. (\ref{eqn:eq00000033}) and isomorphism Eq. (\ref{eqn:eq00000034}) hint us that we can do all the calculation in the $\mathfrak{gl}_4$ module $L(\lambda_1,\lambda_2,\lambda_3,\lambda_4)$, so let us first try to directly associate $\mathfrak{so}_3\oplus\mathfrak{so}_3$ with just one representation of $\mathfrak{gl}_4$. 

Recall $\mathfrak{so}_4$ Lie algebra is a direct sum of two $\mathfrak{so}_3$ Lie algebras, $\mathfrak{so}_4=\mathfrak{so}_3\oplus\mathfrak{so}_3$. One can easily show that the following identification
\begin{equation}
\begin{split}
&J_{+}=F_{-1,2}=-F_{-2,1},\ J_{-}=F_{2,-1}=-F_{1,-2},J_z=-\frac{1}{2}(F_{11}+F_{22});\\
&K_{+}=F_{1,2}=-F_{-2,-1},\ K_{-}=F_{2,1}=-F_{-1,-2},K_z=\frac{1}{2}(F_{11}-F_{22});
\end{split}
\label{eqn:eq00000051}
\end{equation}
will give such relation. Then the definition of $F_{ij}=E_{ij}-E_{-j-i}$ gives us the embedding rules, we choose the Cartan subalgebra of $\mathfrak{so}_3\oplus\mathfrak{so}_3$ as $\mathfrak{h}_5=\{-\frac{1}{2}(F_{11}+F_{22}),\frac{1}{2}(F_{11}-F_{22})\}$, then $L(\lambda_1,\lambda_2,\lambda_3,\lambda_4)$ will have highest weight $(-\frac{\lambda_3-\lambda_2+\lambda_4-\lambda_1}{2},\frac{\lambda_3-\lambda_2-\lambda_4+\lambda_1}{2})$ as $\mathfrak{so}(3)\oplus \mathfrak{so}(3)$ module. We know that our $\mathfrak{so}_3\oplus\mathfrak{so}_3$ is in $(s,0)$ or $(0,s)$ representation. Therefore the $\mathfrak{gl}_4$ highest weight has to be $(a+s,b+s,b,a)$ or $(a+s,b,b+s,a)$. Let's deal with the first case $(a+s,b+s,b,a)$. Eq. (\ref{eqn:eq00000048}) tells us that we need to impose $\mu^{D(+)}(u)=\mu^{D(-)}(u)$, 
\begin{equation}
\begin{split}
s^{D}_{-11}\cdot v&=\sum_{a,b}t^{\mathfrak{so}_6}_{-1a}(u)K_{ab}t^{\mathfrak{so}_6}_{-1-b}(-u)\cdot v\\
&=\Big(t^{\mathfrak{so}_6}_{-1-1}(u)t^{\mathfrak{so}_6}_{-1-1}(-u)-t^{\mathfrak{so}_6}_{-12}(u)t^{\mathfrak{so}_6}_{-1-2}(-u)+t^{\mathfrak{so}_6}_{-13}(u)t^{\mathfrak{so}_6}_{-1-3}(-u)\Big)\cdot v\\
&=\Big(t^{\mathfrak{gl}_4}_{11}(u)t^{\mathfrak{gl}_4}_{44}(u-1)t^{\mathfrak{gl}_4}_{11}(u)t^{\mathfrak{gl}_4}_{44}(u-1)+\dots \Big)\cdot v\\
&=\frac{(a-u-1) (a+u-1) (a+s-u) (a+s+u)}{(u-1) u^2 (u+1)}\cdot v\\
&=\mu^{D(+)}(u)\cdot v.
\end{split}
\label{eqn:eq00000052}
\end{equation}
Similarly one can get
\begin{equation}
\begin{split}
s^{D}_{1-1}\cdot v&=\sum_{a,b}t^{\mathfrak{so}_6}_{-1a}(u)K_{ab}t^{\mathfrak{so}_6}_{-1-b}(-u)\cdot v\\
&=\frac{(b-u-1) (b+u-1) (b+s-u) (b+s+u)}{(u-1) u^2 (u+1)}\cdot v\\
&=\mu^{D(-)}(u)\cdot v.
\end{split}
\label{eqn:eq00000053}
\end{equation}
So $\mu^{D(+)}(u)=\mu^{D(-)}(u)$ will give us $a=b$. Matching the dimensions of the representation, for $\mathfrak{gl}_4$, we use the dimensional calculation formula in Chapter \ref{Chapter5}.  $(a+s,a+s,a,a)$ will be
\begin{equation}
d_{\mathfrak{gl}_4}=\frac{1}{12} (s+2)^2 (s+1) (s+3),
\label{eqn:eq00000054}
\end{equation}
the dimension for our $\mathfrak{so}(3)\oplus \mathfrak{so}(3)$ should be $d=2s+1$, then the embedding is not possible in this case. For the case $(0,s)$, $\mathfrak{gl}_4$ representation $(a+s,b,b+s,a)$ is not a valid, so we need to choose a different kind of Cartan subalgebra. One can choose $\mathfrak{h}_5=\{\frac{1}{2}(F_{11}-F_{22}),-\frac{1}{2}(F_{11}+F_{22})\}$ in this case, then the rest is exactly the same, the embedding is also not possible.

So if we cannot identify one representation of $\mathfrak{gl}_4$, like the case of $\mathfrak{so}_5$, with $\mathfrak{so}(3)\oplus \mathfrak{so}(3)$,  then maybe we can embed $\mathfrak{so}(3)\oplus \mathfrak{so}(3)$ as a subspace of a representation of $\mathfrak{gl}_4$ using branching rules, like the case of $(\mathfrak{su}_3,\mathfrak{so}_3)$. And we identify the quotient space with a direct sum of several representations of $\mathfrak{gl}_4$. Indeed, we can embed our $(s,0)$ or $(0,s)$ into $L(s,s,0,0)$ for integer spin $s$, and $L(s,s,\frac{1}{2},\frac{1}{2})\otimes (\frac{1}{2},0)$ for half-integer spin $s$. 
\subsubsection{$\mathbf{L(s,s,0,0)}$ for integer spin $\mathbf{s=0,1,2,\dots}$ case}
Use the same calculation as before, the corresponding highest weights are
\begin{equation}
s^{D}_{33}(u)\cdot v=\frac{(s-u-1) (s-u)}{u (u+1)}\cdot v,
\label{eqn:eq00000055}
\end{equation}
\begin{equation}
s^{D}_{22}(u)\cdot v=\frac{(s-u) (s+u+1)}{u (u+1)}\cdot v,
\label{eqn:eq00000056}
\end{equation}

\begin{equation}
s^{D}_{11}(u)\cdot v=\frac{s (u-s)}{u^2 (u+1)}\cdot v,
\label{eqn:eq00000057}
\end{equation}

\begin{equation}
s^{D}_{1-1}(u)\cdot v=s^{D}_{-11}(u)\cdot v=(1-\frac{s^2}{u^2})\cdot v,
\label{eqn:eq00000058}
\end{equation}
hence,
\begin{equation}
\begin{split}
\mu^{D}_3(u)&=\frac{(s-u-1) (s-u)}{u (u+1)},\\
\mu^{D}_2(u)&=\frac{(s-u) (s+u+1)}{u (u+1)},\\
\mu^{D}_1(u)&=\frac{s (u-s)}{u^2 (u+1)},\\
\mu^{D(+)}(u)&=\mu^{D(-)}(u)=1-\frac{s^2}{u^2}.
\end{split}
\label{eqn:eq00000059}
\end{equation}
Compare with Eq. (\ref{eqn:eq00000048}), one could claim
\begin{equation}
S(u)=-\frac{1}{2}u^2 (u+1)S^{D}(u),
\label{eqn:eq00000060}
\end{equation}
for integer spin case.
\subsubsection{$\mathbf{L(s,s,\frac{1}{2},\frac{1}{2})\otimes (\frac{1}{2},0)}$ for half integer spin $\mathbf{s=\frac{1}{2},\frac{3}{2},\dots}$ case}
In this case, since we are dressing $(\frac{1}{2},0)$ representation $S^{s=\frac{1}{2}}(u)$, the corresponding $S^{D}(u)$ will look different
\begin{equation}
S^{D}(u)=T^{\mathfrak{so}_6}(u)S^{s=\frac{1}{2}}(u)(T^{\mathfrak{so}_6})^t(-u),
\label{eqn:eq00000061}
\end{equation}
where $S^{s=\frac{1}{2}}(u)$ in components are
\begin{equation}
\begin{split}
\psi^{(1/2)}_{-3,-3}(u)&=-\frac{1}{2}(u+1) \left(S_z (2 u-1)+u^2\right)-\frac{1}{2}S_z^2 (u+1)+\frac{3}{8},\\
\psi^{(1/2)}_{-2,-2}(u)&=\psi^{(1/2)}_{2,2}(u)=-\frac{1}{8} u (3-4 u (u+1)),\\
\psi^{(1/2)}_{-1,-1}(u)&=\psi^{(1/2)}_{1,1}(u)=-\frac{3 u}{8}+\frac{1}{2}S_z^2 (u+1),\\
\psi^{(1/2)}_{3,3}(u)&=-\frac{1}{2}(u+1) \left( S_z (1-2u)u^2\right)-\frac{1}{2}S_z^2 (u+1)+\frac{3}{8},\\
\psi^{(1/2)}_{1,-1}(u)&=\psi^{(1/2)}_{-1,1}(u)=-\frac{1}{2}(u+1) \left(u^2-S_z^2\right),\\
\psi^{(1/2)}_{-3,-1}(u)&=\psi^{(1/2)}_{-3,1}(u)=-\frac{1}{2}\sqrt{2} (u+1) ((u-1) S_+S_z-u S_zS_+),\\
\psi^{(1/2)}_{-3,3}(u)&= S_+^2 (u+1),\\
\psi^{(1/2)}_{-1,-3}(u)&=\psi^{(1/2)}_{1,-3}(u)=\frac{\sqrt{2}}{2} (u+1) (u S_-S_z-(u-1) S_zS_-),\\
\psi^{(1/2)}_{-1,3}(u)&=\psi^{(1/2)}_{1,3}(u)= \frac{\sqrt{2}}{2} (u+1) (u S_+S_z-(u-1) S_zS_+),\\
\psi^{(1/2)}_{3,-3}(u)&=S_-^2 (u+1),\\
\psi^{(1/2)}_{3,-1}(u)&=\psi^{(1/2)}_{3,1}(u)=-\frac{\sqrt{2}}{2} (u+1) ((u-1) S_-S_z-u S_zS_-).
\end{split}
\label{eqn:eq00000062}
\end{equation}
Eq. (\ref{eqn:eq00000061}) acting rules in components would be 
\begin{equation}
s^{D}_{ij}(u)\cdot (v_1\otimes v_2)=t^{\mathfrak{so}_6}_{ia}(u)t^{\mathfrak{so}_6}_{-j,-b}(-u)\cdot v_1\otimes \psi^{(1/2)}_{a,b}(u)\cdot v_2,
\label{eqn:eq00000063}
\end{equation}
where $v_1$ is a vector in $\mathfrak{so}_6$ module, and $v_2$ is a vector in $\mathfrak{so}(3)\oplus \mathfrak{so}(3)$ module. Use the same calculation as before, for instance
\begin{equation}
\begin{split}
&s^{D}_{33}(u)\cdot (v_1\otimes v_2)\\&=t^{\mathfrak{so}_6}_{33}(u)t^{\mathfrak{so}_6}_{-3,-3}(-u)\cdot v_1\otimes \psi^{(1/2)}_{3,3}(u)\cdot v_2,\\
&=t^{\mathfrak{gl}_4}_{33}(u)t^{\mathfrak{gl}_4}_{44}(u-1)t^{\mathfrak{gl}_4}_{11}(-u)t^{\mathfrak{gl}_4}_{22}(-u-1)\cdot v_1\\&\quad\otimes (-\frac{1}{2}(u+1) \left( S_z (1-2u)+u^2\right)-\frac{1}{2}S_z^2 (u+1)+\frac{3}{8})\cdot v_2\\
&=-\frac{\left(1-4 u^2\right)^2 (s-u-1) (s-u)}{32 u \left(u^2-1\right)}(v_1\otimes v_2).
\end{split}
\label{eqn:eq00000064}
\end{equation}
Use the same procedure
\begin{equation}
s^{D}_{22}(u)\cdot v=-\frac{\left(1-4 u^2\right)^2 (s-u) (s+u+1)}{32 u \left(u^2-1\right)}\cdot v,
\label{eqn:eq00000065}
\end{equation}

\begin{equation}
s^{D}_{11}(u)\cdot v=\frac{s \left(1-4 u^2\right)^2 (s-u)}{32 u^2 \left(u^2-1\right)}\cdot v,
\label{eqn:eq00000066}
\end{equation}
 
\begin{equation}
s^{D}_{1-1}(u)\cdot v=s^{D}_{-11}(u)\cdot v=-\frac{\left(1-4 u^2\right)^2 (u-s) (s+u)}{32 (u-1) u^2}\cdot v,
\label{eqn:eq00000067}
\end{equation}
hence,
\begin{equation}
\begin{split}
\mu^{D}_3(u)&=-\frac{\left(1-4 u^2\right)^2 (s-u-1) (s-u)}{32 u \left(u^2-1\right)},\\
\mu^{D}_2(u)&=-\frac{\left(1-4 u^2\right)^2 (s-u) (s+u+1)}{32 u \left(u^2-1\right)},\\
\mu^{D}_1(u)&=\frac{s \left(1-4 u^2\right)^2 (s-u)}{32 u^2 \left(u^2-1\right)},\\
\mu^{D(+)}(u)&=\mu^{D(-)}(u)=-\frac{\left(1-4 u^2\right)^2 (u-s) (s+u)}{32 (u-1) u^2}.
\end{split}
\label{eqn:eq00000068}
\end{equation}
Compare with Eq. (\ref{eqn:eq00000048}), one could claim
\begin{equation}
S(u)=\frac{16 (u-1) u^2 (u+1)}{(2 u-1)^2 (2 u+1)^2}S^{D}(u),
\label{eqn:eq00000069}
\end{equation}
for half integer spin case. 
\section{Overlap Formula for $(SO(6),SO(3)\times SO(3))$}
As we have seen in last section that we need to treat the integer spin and half integer spin differently, so I will discuss them separately. Since $V_L(s)=(s,0)$ and $V_R(s)=(0,s)$ are related by a global rotation we assume that 
\begin{equation}
\langle MPS_L\lvert \mathbf{u}\rangle = A\langle MPS_R\lvert \mathbf{u}\rangle, 
\label{eqn:eq00000070}
\end{equation}
with $A$ a constant.
\subsection{Integer spin case $\mathbf{s = 0,1,2,\dots}$}
Using the branching rules, we have found the following relation
\begin{equation}
\begin{split}
[0,s,0]\oplus[2,s-3,0]=[2,s-2,0]\oplus[0,s-3,0]
\oplus(0,s)\oplus(s,0),\quad \mathrm{for}\ s> 0.
\end{split}
\label{eqn:eq00000071}
\end{equation}
Above I have used the $[a,b,c]$ to denote the Dynkin labels of type $A_3$, and $(a, b)$ denote Dynkin labels of $SO(3)\times SO(3)$ of dimension $(2a+1)\times(2b+1)$. So one would expect the overlap formula as
\begin{equation}
\langle MPS_3\lvert \mathbf{u}\rangle=\lim_{u\rightarrow 0}\Big( T^{(0,1,0)}(a)\langle MPS_{\delta_{\pm}}\lvert\mathbf{u}\rangle\Big),
\label{eqn:eq00000072}
\end{equation}
where $a$ is due to automorphism $\tau_a$, one can fix it by compare the eigenvalue of the transfer matrix. Again from \cite{de_Leeuw_2020}, we know the scalar solution overlap can be obtained by Thermodynamic Bethe Ansatz
\begin{equation}
\frac{\langle \mathrm{MPS}_{\delta_{\pm}}\vert\mathbf{u}\rangle}{\sqrt{\langle \mathbf{u}\lvert \mathbf{u}\rangle}}=\sqrt{\frac{Q_1(0)Q_2(0)Q_3(0)}{Q_1(\frac{1}{2})Q_2(\frac{1}{2})Q_3(\frac{1}{2})}}\sqrt{\frac{\det G_+}{\det G_-}},
\label{eqn:eq00000072a}
\end{equation}
where three type of Baxter polynomials are defined as
\begin{equation}
    Q_1(a)=\prod_{i=1}^{M} (ia-u_i),\quad Q_{2}(a)=\prod_{j=1}^{N_{2}}(ia-v^{(2)}_{j})\quad Q_{3}(a)=\prod_{j=1}^{N_{3}}(ia-v^{(3)}_{j}).
    \label{eqn:eq00000072b}
\end{equation}
Hence for higher spin
\begin{equation}
\begin{split}
\langle MPS_{2s+1}\lvert \mathbf{u}\rangle=& \lim_{u\rightarrow 0} \Big(e(u)T^{(0,s,0)}(a)+f(u)T^{(2,s-3,0)}(b)\\
&-g(u)T^{(2,s-2,0)}(c)-h(u)T^{(0,s-3,0)}(d)\Big)\langle\mathrm{MPS}_{\delta_{\pm}}\lvert\mathbf{u}\rangle,
\end{split}
\label{eqn:eq00000073}
\end{equation}
again the function $e(u),\ f(u),\ g(u),\  h(u)$ in front of the transfer matrices are due to the automorphisms in Eq. (\ref{eqn:eq00000022}). One can only fix one of the pre-factor (i.e. $T^{(0,s,0)}$), the rest probably can be fixed by compare the highest weight. And we need to fix $a,b,c,d$ by the eigenvalue of the transfer matrix. Or one could use the same method by acting on the states to determine the whole parameters. 

\subsection{Half integer spin case $\mathbf{s=\frac{1}{2},\frac{3}{2},\dots}$}
Using the branching rules, we have found the following relation
\begin{equation}
\begin{split}
\big([0,s-\frac{1}{2},0]\otimes (\frac{1}{2},0)\big)\oplus\big([1,s-\frac{5}{2},0]\otimes (0,\frac{1}{2})\big)=\big([1,s-\frac{3}{2},0]\otimes (0,\frac{1}{2})\big)\\\oplus\big([0,s-\frac{5}{2},0]\otimes (\frac{1}{2},0)\big)\oplus(s,0),\quad \mathrm{for}\ s\ge \frac{1}{2}.
\end{split}
\label{eqn:eq00000074}
\end{equation}
So one would expect the overlap formula as
\begin{equation}
\langle MPS_4\lvert \mathbf{u}\rangle= \lim_{u\rightarrow 0}\Big( \alpha(u)T^{(0,1,0)}(p)-\beta(u)T^{(1,0,0)}(q)\Big)\langle MPS_2\lvert\mathbf{u}\rangle,
\label{eqn:eq00000075}
\end{equation}
where $a$ is due to automorphism $\tau_a$, one maybe can fix it by compare the eigenvalue of the transfer matrix.
And for higher spin
\begin{equation}
\begin{split}
\langle MPS_{2s+1}\lvert \mathbf{u}\rangle=\lim_{u\rightarrow 0} \Big(\alpha(u)T^{(0,s-\frac{1}{2},0)}(p)-\beta(u)T^{(1,s-\frac{3}{2},0)}(q)-\gamma(u)T^{(0,s-\frac{5}{2},0)}(r)\\+\delta(u)T^{(1,s-\frac{5}{2},0)}(x)\Big)\langle MPS_2\lvert\mathbf{u}\rangle,
\end{split}
\label{eqn:eq00000076}
\end{equation}
again the function $\alpha(u),\ \beta(u),\ \gamma(u),\delta(u)$ in front of the transfer matrices are due to the automorphisms in Eq. (\ref{eqn:eq00000022}). One can only fix one of the pre-factor (i.e. $T^{(0,s,0)}$) using a different choose of states, the rest probably can be fixed by compare the highest weight. And we need to fix $p,q,r,x$ by the eigenvalue of the transfer matrix. Or one could use the same method by acting on the states to determine the whole parameters. 
%----------------------------------------------------------------------------------------
%	THESIS CONTENT - APPENDICES
%----------------------------------------------------------------------------------------

\appendix % Cue to tell LaTeX that the following "chapters" are Appendices

% Include the appendices of the thesis as separate files from the Appendices folder
% Uncomment the lines as you write the Appendices

% Appendix A

\chapter{$SU(2)$ Spin 1 scattering matrix and Bethe equation}  % Main appendix title

\label{AppendixA} % For referencing this appendix elsewhere, use \ref{AppendixA}

\section{Hamiltonian kernel setup}

A naïve gauss of the Hamiltonian kernel for spin 1 $SU(2)$ would just rewrite the spin $\frac{1}{2}$ kernel 
\begin{equation}
H_{n,n+1}= \frac{1}{2}I_4-\frac{1}{2}\sigma^a\otimes\sigma^a
\label{eqn:eqa1}
\end{equation}
to the form
\begin{equation}
H_{n,n+1}=I_9-t^a\otimes t^a
\label{eqn:eqa2}
\end{equation}
is a map from $\mathbb{C}^3\otimes\mathbb{C}^3$. Where we have chosen $\lvert 0\dots0\rangle $ as vacuum state to fixed. And $\sigma$ are the Pauli matrices, $t$ are the $su(2)$ generators in spin one representation, take the form

\begin{equation}
t_1=\frac{1}{\sqrt2}\begin{pmatrix}0&1&0\\1&0&1\\0&1&0\end{pmatrix},t_2=\frac{1}{\sqrt2}\begin{pmatrix}0&-i&0\\i&0&-i\\0&i&0\end{pmatrix},t_3=\begin{pmatrix}1&0&0\\0&0&0\\0&0&-1\end{pmatrix}
\label{eqn:eqa3}
\end{equation}
and satisfy $su(2)$ Lie algebra. But as proved in \cite{Faddeev_aba} by constructing the Lax operator in spin $s$ case, one can show that the above Hamiltonian doesn't give rise to an integrable model.

The correct form of Hamiltonian kernel for spin $1$ is derived by Faddeev 
\begin{equation}
H_{n,n+1}=t_n^it_{n+1}^i-(t_n^it_{n+1}^i)^2,
\label{eqn:eqa4}
\end{equation}
where we have chosen $\lvert 0\dots0\rangle $ as vacuum state to fixed. Denote spin 1 as $\vert 2\rangle$, spin 0 as $\lvert 1\rangle$, and spin -1 as $\lvert 0\rangle$
\begin{equation}
\vert 2\rangle=\begin{pmatrix}1\\0\\0\end{pmatrix},\vert 1\rangle=\begin{pmatrix}0\\1\\0\end{pmatrix},\vert 0\rangle=\begin{pmatrix}0\\0\\1\end{pmatrix}
\label{eqn:eqa5}
\end{equation}
The Hamiltonian kernel in basis $\{\vert 22\rangle,\vert 21\rangle,\vert 20\rangle,\vert 12\rangle,\vert 11\rangle,\vert 10\rangle,\vert 02\rangle,\vert 01\rangle,\vert 00\rangle\}$ yields
\begin{equation}
H_{n,n+1}=\begin{pmatrix}0&0&0&0&0&0&0&0&0\\0&-1&0&1&0&0&0&0&0\\0&0&-3&0&2&0&-1&0&0\\0&1&0&-1&0&0&0&0&0\\0&0&2&0&-2&0&2&0&0\\0&0&0&0&0&-1&0&1&0\\0&0&-1&0&2&0&-3&0&0\\0&0&0&0&0&1&0&-1&0\\0&0&0&0&0&0&0&0&0\end{pmatrix}
\label{eqn:eqa6}
\end{equation}
\section{One-magnon state}
With one magnon state defined as 
\begin{equation}
\lvert k\rangle=\lvert 0 0\dots0 \overset{k}1 0\dots0\rangle,
\label{eqn:eqa7}
\end{equation}
one can form the following eigenstate with one excitation, also fixed momentum 
\begin{equation}
\lvert p\rangle=\sum_k^L\mathrm{e}^{ipk}\lvert k\rangle.
\label{eqn:eqa8}
\end{equation}
Act with $\mathcal{H}$ on $\lvert p\rangle$ and obtain
\begin{equation}
\begin{split}
\mathcal{H}\lvert p\rangle&=\sum_k\mathrm{e}^{ipk}\left(\overbrace{-\lvert k\rangle+\lvert k-1\rangle}^{H_{k-1,k}}+\overbrace{-\lvert k\rangle+\lvert k+1\rangle}^{H_{k,k+1}}\right)\\
&=\sum_k\mathrm{e}^{ipk}(-1+\mathrm{e}^{ip}-1+\mathrm{e}^{-ip})\\
&=E(p)\lvert p\rangle
\end{split}
\label{eqn:eqa9}
\end{equation}
with eigenenergy
\begin{equation}
E(p)=-2+\mathrm{e}^{ip}+\mathrm{e}^{-ip}=-4\sin^2(\frac{1}{2}p)
\label{eqn:eqa10}
\end{equation}
\section{Two-magnon state and scattering phase}
For two excitation states, there are two different states 
\begin{equation}
\lvert k_1<k_2\rangle=\lvert 0 0\dots0 \overset{k_1}1 0\dots 0 \overset{k_2}10\dots0\rangle,
\label{eqn:eqa11}
\end{equation}
and
\begin{equation}
\lvert \bar{k}\rangle=\lvert 0 0\dots0 \overset{\bar{k}}2 0\dots0\rangle.
\label{eqn:eqa12}
\end{equation}
Here, analogous with one magnon state, we make a plane wave eigenstate ansatz
\begin{equation}
\lvert p,q\rangle=\lvert p<q\rangle+S(p,q)\lvert q<p\rangle+C(p,q)\lvert p+q;2\rangle,
\label{eqn:eqa13}
\end{equation}
where 
\begin{equation}
\lvert p<q\rangle=\sum_{k<l} \mathrm{e}^{ipk+iql}\lvert k<l\rangle.
\label{eqn:eqa14}
\end{equation}
and denote states with Eq. (\ref{eqn:eqa12}) excitation as
\begin{equation}
\lvert p;2\rangle=\sum_{k} \mathrm{e}^{ipk}\lvert \bar{k}\rangle.
\label{eqn:eqa15}
\end{equation}
Act with $\mathcal{H} $ on $\lvert p<q\rangle$,
\begin{equation}
\begin{split}
\mathcal{H}\lvert p<q\rangle
=\sum_{k<l}&\mathrm{e}^{ipk+iql}[\overbrace{-\lvert k,l\rangle+\lvert k-1,l}^{H_{k-1,k}}\rangle\overbrace{-\lvert k,l\rangle+\lvert k+1,l}^{H_{k,k+1}}\rangle\\
&\overbrace{-\lvert k,l\rangle+\lvert k,l-1}^{H_{l-1,l}}\rangle\overbrace{-\lvert k,l\rangle+\lvert k,l+1}^{H_{l,l+1}}\rangle]\\
+\sum_{k=l-1}& \mathrm{e}^{ipk+iql}[\overbrace{-2\lvert k,l\rangle+2\lvert \bar{k},0\rangle+2\lvert 0,\overline{k+1}}^{H_{k,k+1}}\rangle\\
&\overbrace{-\lvert k,k+1\rangle+\lvert k-1,k+1}^{H_{k-1,k}}\rangle\overbrace{-\lvert k,k+1\rangle+\lvert k,k+2}^{H_{k+1,k+2}}\rangle]\\
=(-4+&\mathrm{e}^{-ip}+\mathrm{e}^{ip}+\mathrm{e}^{-iq}+\mathrm{e}^{iq})\lvert p<q\rangle\\
-(\mathrm{e}^{ip}&+\mathrm{e}^{-iq})\sum_{k=l-1}\mathrm{e}^{ipk+iql}\lvert k,k+1\rangle+2(\mathrm{e}^{-ip}+\mathrm{e}^{iq})\lvert p+q;2\rangle.
\end{split}
\label{eqn:eqa16}
\end{equation}
So 
\begin{equation}
\begin{split}
\ [&\mathcal{H}-E(p)-E(q)]\lvert p<q\rangle \\
=&-(\mathrm{e}^{i(p+q)}+1)\sum_{k}\mathrm{e}^{i(p+q)k}\lvert k,k+1\rangle+2(\mathrm{e}^{-ip}+\mathrm{e}^{iq})\lvert p+q;2\rangle.
\end{split}
\label{eqn:eqa17}
\end{equation}
Similarly, one can swap $p$ and $q$ to get
\begin{equation}
\begin{split}
\ [&\mathcal{H}-E(p)-E(q)]\lvert p>q\rangle \\
=&-(\mathrm{e}^{i(p+q)}+1)\sum_{k}\mathrm{e}^{i(p+q)k}\lvert k,k+1\rangle+2(\mathrm{e}^{-iq}+\mathrm{e}^{ip})\lvert p+q;2\rangle
\end{split}
\label{eqn:eqa18}
\end{equation}
Using the same procedure, act $\mathcal{H}$ on Eq. (\ref{eqn:eqa14})
\begin{equation}
\begin{split}
\mathcal{H}\lvert p+q;2\rangle
=&\sum_{k<l}\mathrm{e}^{i(p+q)k}[\overbrace{-3\lvert \bar{k},0\rangle+2\lvert k,k+1\rangle-\lvert 0,\overline{k+1}\rangle}^{H_{k,k+1}}\\
&\overbrace{-3\lvert 0,\bar{k}\rangle+2\lvert k-1,k\rangle-\lvert \overline{k-1},0\rangle}^{H_{k-1,k}}].
\end{split}
\label{eqn:eqa19}
\end{equation}
then
\begin{equation}
\begin{split}
\ [\mathcal{H}-E(p)-E(q)]\lvert p+q;2\rangle=&-(2+\mathrm{e}^{-ip}+\mathrm{e}^{ip}+\mathrm{e}^{-iq}+\mathrm{e}^{iq})\lvert p+q;2\rangle \\
&-(\mathrm{e}^{i(p+q)}+\mathrm{e}^{-i(p+q)})\lvert p+q;2\rangle\\
&+2(1+\mathrm{e}^{i(p+q)})\sum_{k}\mathrm{e}^{i(p+q)k}\lvert k,k+1\rangle.
\end{split}
\label{eqn:eqa20}
\end{equation}
So combine Eq. (\ref{eqn:eqa13}), Eq. (\ref{eqn:eqa17}), Eq. (\ref{eqn:eqa18}) and Eq. (\ref{eqn:eqa20}), 
\begin{equation}
([\mathcal{H}-E(p)-E(q)])\lvert p,q\rangle=0,
\label{eqn:eqa21}
\end{equation}
one can obtain
\begin{equation}
\begin{cases}
-(1+\mathrm{e}^{i(p+q)})-(1+\mathrm{e}^{i(p+q)})S+2(1+\mathrm{e}^{i(p+q)})C=0,\\
\begin{split} 2&(\mathrm{e}^{iq}+\mathrm{e}^{-ip})+2S(\mathrm{e}^{-iq}+\mathrm{e}^{ip})\\&-C(2+\mathrm{e}^{iq}+\mathrm{e}^{ip}+\mathrm{e}^{-iq}+\mathrm{e}^{-ip}+\mathrm{e}^{-i(p+q)}+\mathrm{e}^{i(p+q)})=0.
\end{split}\end{cases}
\label{eqn:eqa22}
\end{equation}
Solve this equation, the solution is 

\begin{equation}
S=-\frac{1+\mathrm{e}^{ip}+\mathrm{e}^{i(p+q)}-3\mathrm{e}^{iq}}{1-3\mathrm{e}^{ip}+\mathrm{e}^{i(p+q)}+\mathrm{e}^{iq}},\ C=-\frac{2(\mathrm{e}^{ip}-\mathrm{e}^{iq})}{1-3\mathrm{e}^{ip}+\mathrm{e}^{i(p+q)}+\mathrm{e}^{iq}}.
\label{eqn:eqa23}
\end{equation}
Set
\begin{equation}
\mathrm{e}^{ip}=\frac{u+i}{u-i},\ \mathrm{e}^{iq}=\frac{v+i}{v-i},
\label{eqn:eqa24}
\end{equation}
then Eq. (\ref{eqn:eqa23}) read 
\begin{equation}
S(u,v)=\frac{u-v-i}{u-v+i},C(u,v)=\frac{u-v}{u-v+i}.
\label{eqn:eqa25}
\end{equation}
With periodic boundary condition, the Bethe equation yields 

\begin{equation}
(\frac{u_k+i}{u_k-i})^L=\prod_{j=1,\ j\ne k}^M \frac{u_k-u_j+i}{u_k-u_j-i}.
\label{eqn:eqa26}
\end{equation}
For general spin $s$ case, Hamiltonian Eq. (\ref{eqn:eqa4}) is not applicable anymore, a better way to do it is using the algebraic Bethe ansatz. The corresponding Bethe equation takes the form

\begin{equation}
(\frac{u_k+is}{u_k-is})^L=\prod_{j=1,j\ne k}^M \frac{u_k-u_j+i}{u_k-u_j-i}.
\label{eqn:eqa27}
\end{equation}

% Appendix B

\chapter{$SU(3)$ and $SU(N)$ fundamental spin chain scattering matrix}  % Main appendix title

\label{AppendixB} % For referencing this appendix elsewhere, use \ref{AppendixB}

\section{$su(3)$ fundamental representation}
The generators, $t$, of the Lie algebra $su(3)$ in fundamental representation, are 
\begin{equation}
t_a=\frac{\lambda_a}{2},\quad \mathrm{with}\ a=1,\dots,8,
\label{eqn:eqb1}
\end{equation}
where $\lambda$ are the famous Gell-Mann matrices.
\begin{equation}
H_1=t_3,\quad H_2=t_8,
\label{eqn:eqb2}
\end{equation}
are the Cartan subalgebra.
A convenient basis of states for the fundamental representation of $SU(3)$ is simply
\begin{equation}
\lvert 2\rangle =V_1=\begin{pmatrix}1\\0\\0\end{pmatrix},\ \lvert 1\rangle=V_2=\begin{pmatrix}0\\1\\0\end{pmatrix},\ \lvert0\rangle=V_3=\begin{pmatrix}0\\0\\1\end{pmatrix}.
\label{eqn:eqb3}
\end{equation}
Their eigenvalues under $H_1$ and $H_2$ can be read off by inspection, since $H_1$ and $H_2$ are diagonal. Thus the eigenvalues under $(H_1,H_2)$ for the vectors $V_1$, $V_2$ and $V_3$ are 
\begin{equation}
V_1:(\frac{1}{2},\frac{1}{2\sqrt{3}}), \quad V_2:(-\frac{1}{2},\frac{1}{2\sqrt{3}})\quad V_3:(0,-\frac{1}{\sqrt{3}}).
\label{eqn:eqb4}
\end{equation}
\subsection{Hamiltonian kernel setup}
Analogous with $su(2)$ fundamental spin chain, the Hamiltonian kernel takes the form
\begin{equation}
H_{n,n+1}=\frac{2}{3}I_9-2t_n^i\otimes t_{n+1}^i.
\label{eqn:eqb5}
\end{equation}
Where the Einstein summation rule is explicit, and $i=1,\dots, 8$. We have chosen $\lvert 0\dots0\rangle $ as vacuum state to fix the ratio between identity matrix and tensor product matrix. And $t$ are the $su(3)$ generators in fundamental representation. Using the structure constant $f_{abc}$ in commutation relation $[t^a,t^b]=if_{abc}t^c$ is totally antisymmetric, one can show that $\sum_{i=1}^8[t^{j}_{n}\otimes I+I\otimes t^{j}_{n+1},t^{i}_{n}\otimes t^{i}_{n+1}]=0$, hence
\begin{equation}
[Q^{i},H_{n,n+1}]=0,\quad \mathrm{with}\ i=1,\dots,8,
\label{eqn:eqb6}
\end{equation}
which justifies our spin chain Hamiltonian is $SU(N)$ invariant. 
 In $\mathbb{C}^3\otimes \mathbb{C}^3$ space, the permutation operator can be rewrote as 
\begin{equation}
P_{n,n+1}=2t_n^i\otimes t_{n+1}^i+\frac{1}{3}I_9,
\label{eqn:eqb7}
\end{equation}
then $H_{n,n+1}$ reads
\begin{equation}
H_{n,n+1}=I_9-P_{n,n+1}.
\label{eqn:eqb8}
\end{equation}
\subsection{One-magnon state}
In SU(3), we have two kinds of excitations $\lvert 1 \rangle$ and $\lvert 2 \rangle$, one need to consider both of them. Actually, since the Hamiltonian kernel is symmetric under the exchange of these two excitations, we only need to calculate one. 

With one magnon state $\lvert 1 \rangle$ defined as 
\begin{equation}
\lvert k\rangle=\lvert 0 0\dots0 \overset{k}1 0\dots0\rangle,
\label{eqn:eqb9}
\end{equation}
and one magnon state $\lvert 2 \rangle$ defined as
\begin{equation}
\lvert \bar{k}\rangle=\lvert 0 0\dots0 \overset{\bar{k}}2 0\dots0\rangle.
\label{eqn:eqb10}
\end{equation}
One can form the following eigenstate with one excitation, also fixed momentum
\begin{equation}
\lvert p\rangle=\sum_k^L\mathrm{e}^{ipk}\lvert k\rangle, \quad 
\lvert \bar{p}\rangle=\sum_k^L\mathrm{e}^{ipk}\lvert \bar{k}\rangle
\label{eqn:eqb11}
\end{equation}
Act with $\mathcal{H}$ on $\lvert p\rangle$ and obtain
\begin{equation}
\begin{split}
\mathcal{H}\lvert p\rangle&=\sum_k\mathrm{e}^{ipk}\left(\overbrace{\lvert k\rangle-\lvert k-1\rangle}^{H_{k-1,k}}+\overbrace{\lvert k\rangle-\lvert k+1\rangle}^{H_{k,k+1}}\right)\\
&=\sum_k\mathrm{e}^{ipk}(1-\mathrm{e}^{ip}+1-\mathrm{e}^{-ip})\\
&=E(p)\lvert p\rangle,
\end{split}
\label{eqn:eqb12}
\end{equation}
with eigenenergy
\begin{equation}
E(p)=2-\mathrm{e}^{ip}-\mathrm{e}^{-ip}=4\sin^2(\frac{1}{2}p),
\label{eqn:eqb13}
\end{equation}
The same case for $\lvert \bar{p}\rangle$.
\subsection{Two-magnon state and scattering phase}
For two excitation states, there are four different states
\begin{equation}
\begin{split}
&\lvert k_1<k_2\rangle=\lvert 0 0\dots0 \overset{k_1}1 0\dots 0 \overset{k_2}10\dots0\rangle,\\
&\lvert k_1<\bar{k}_2\rangle=\lvert 0 0\dots0 \overset{k_1}1 0\dots 0 \overset{\bar{k}_2}20\dots0\rangle,\\
&\lvert \bar{k}_1<\bar{k}_2\rangle=\lvert 0 0\dots0 \overset{\bar{k}}2 0\dots 0 \overset{\bar{k}_2}20\dots0\rangle,\\
&\lvert\bar{k}_1< k_2\rangle=\lvert 0 0\dots0 \overset{\bar{k}_1}2 0\dots 0 \overset{k_2}10\dots0\rangle,
\end{split}
\label{eqn:eqb14}
\end{equation}
but again since the symmetry of the $H_{n,n+1}$, we only need to consider the first two of them.

Let us consider the first case with two same color excitation, it is completely the same as in $su(2)$ case. Here, analogous with one magnon state, we make a plane wave eigenstate ansatz
\begin{equation}
\lvert p,q\rangle=\lvert p<q\rangle+S_{11}^{11}(p,q)\lvert q<p\rangle,
\label{eqn:eqb15}
\end{equation}
where 
\begin{equation}
\lvert p<q\rangle=\sum_{k<l} \mathrm{e}^{ipk+iql}\lvert k<l\rangle.
\label{eqn:eqb16}
\end{equation}
Act with $\mathcal{H} $ on $\lvert p<q\rangle$,
\begin{equation}
\begin{split}
\mathcal{H}\lvert p<q\rangle
=\sum_{k<l}&\mathrm{e}^{ipk+iql}[\overbrace{\lvert k,l\rangle-\lvert k-1,l}^{H_{k-1,k}}\rangle+\overbrace{\lvert k,l\rangle-\lvert k+1,l}^{H_{k,k+1}}\rangle\\
&+\overbrace{\lvert k,l\rangle-\lvert k,l-1}^{H_{l-1,l}}\rangle+\overbrace{\lvert k,l\rangle-\lvert k,l+1}^{H_{l,l+1}}\rangle]\\
+\sum_{k=l-1}& \mathrm{e}^{ipk+iql}[\overbrace{\lvert k,k+1\rangle-\lvert k-1,k+1\rangle}^{H_{k-1,k}}\\
&+\overbrace{0}^{H_{k,k+1}}+\overbrace{\lvert k,k+1\rangle-\lvert k,k+2}^{H_{k+1,k+2}}\rangle]\\
=(4-&\mathrm{e}^{-ip}-\mathrm{e}^{ip}-\mathrm{e}^{-iq}-\mathrm{e}^{iq})\lvert p<q\rangle\\
+(-&2+\mathrm{e}^{ip}+\mathrm{e}^{-iq})\sum_{k=l-1}\mathrm{e}^{ipk+iql}\lvert k,k+1\rangle.
\end{split}
\label{eqn:eqb17}
\end{equation}
So
\begin{equation}
\begin{split}
\ [&\mathcal{H}-E(p)-E(q)]\lvert p<q\rangle \\
=&(\mathrm{e}^{i(p+q)}+1-2\mathrm{e}^{ip})\sum_{k}\mathrm{e}^{i(p+q)k}\lvert k,k+1\rangle.
\end{split}
\label{eqn:eqb18}
\end{equation}
Similarly, one can swap $p$ and $q$ to get
\begin{equation}
\begin{split}
\ [&\mathcal{H}-E(p)-E(q)]\lvert p>q\rangle \\
=&(\mathrm{e}^{i(p+q)}+1-2\mathrm{e}^{iq})\sum_{k}\mathrm{e}^{i(p+q)k}\lvert k,k+1\rangle.
\end{split}
\label{eqn:eqb19}
\end{equation}
Then from Eq. (\ref{eqn:eqb15}), one can get the scattering phase 
\begin{equation}
S_{11}^{11}(p,q)=-\frac{\mathrm{e}^{i(p+q)}+1-2\mathrm{e}^{iq}}{\mathrm{e}^{i(p+q)}+1-2\mathrm{e}^{ip}},\quad S_{22}^{22}(p,q)=S_{11}^{11}(p,q).
\label{eqn:eqb20}
\end{equation}
Now let's consider the second case with two different excitation. Due to the state might change color after scattering, the scattering phase is a bit hard in this circumstance. Assume the eigenstate of the Hamiltonian as
\begin{equation}
\lvert p,\bar{q}\rangle=\lvert p<\bar{q}\rangle+S_{12}^{12}\lvert p>\bar{q}\rangle+S_{12}^{21}\lvert \bar{p}>q\rangle.
\label{eqn:eqb21}
\end{equation}
Using the same procedure, act $\mathcal{H}$ on $\lvert p<\bar{q}\rangle$
\begin{equation}
\begin{split}
\mathcal{H}\lvert p<\bar{q}\rangle
=\sum_{k<l}&\mathrm{e}^{ipk+iql}[\overbrace{\lvert k,\bar{l}\rangle-\lvert k-1,\bar{l}}^{H_{k-1,k}}\rangle+\overbrace{\lvert k,\bar{l}\rangle-\lvert k+1,\bar{l}}^{H_{k,k+1}}\rangle\\
&+\overbrace{\lvert k,\bar{l}\rangle-\lvert k,\overline{l-1}}^{H_{l-1,l}}\rangle+\overbrace{\lvert k,\bar{l}\rangle-\lvert k,\overline{l+1}}^{H_{l,l+1}}\rangle]\\
+\sum_{k=l-1}& \mathrm{e}^{ipk+iql}[\overbrace{\lvert k,\overline{k+1}\rangle-\lvert k-1,\overline{k+1}\rangle}^{H_{k-1,k}}\\
&+\overbrace{\lvert k,\overline{k+1}\rangle-\lvert \bar{k},k+1\rangle}^{H_{k,k+1}}+\overbrace{\lvert k,\overline{k+1}\rangle-\lvert k,\overline{k+2}}^{H_{k+1,k+2}}\rangle]\\
=(E(p&)+E(q))\lvert p<\bar{q}\rangle-\mathrm{e}^{iq}\sum_{k}\mathrm{e}^{i(p+q)k}\lvert \bar{k},k+1\rangle\\
+(&-\mathrm{e}^{iq}+1+\mathrm{e}^{i(p+q)})\sum_{k}\mathrm{e}^{i(p+q)k}\lvert k,\overline{k+1}\rangle,
\end{split}
\label{eqn:eqb22}
\end{equation}
then

\begin{equation}
\begin{split}
\ [\mathcal{H}-E(p)-E(q)]\lvert p<\bar{q}\rangle=&-\mathrm{e}^{iq}\sum_{k}\mathrm{e}^{i(p+q)k}\lvert \bar{k},k+1\rangle\\
+(-&\mathrm{e}^{iq}+1+\mathrm{e}^{i(p+q)})\sum_{k}\mathrm{e}^{i(p+q)k}\lvert k,\overline{k+1}\rangle.
\end{split}
\label{eqn:eqb23}
\end{equation}
so $\mathcal{H}\lvert \bar{q}<p\rangle$ can be easily obtained by swapping $p$ with $q$ and color, and  $\mathcal{H}\lvert q<\bar{p}\rangle$ can be obtained by swapping $p$ with $q$ 
\begin{equation}
\begin{split}
\ [\mathcal{H}-E(p)-E(q)]\lvert \bar{q}<p\rangle=&-\mathrm{e}^{ip}\sum_{k}\mathrm{e}^{i(p+q)k}\lvert k,\overline{k+1}\rangle\\
+(-&\mathrm{e}^{ip}+1+\mathrm{e}^{i(p+q)})\sum_{k}\mathrm{e}^{i(p+q)k}\lvert \bar{k},k+1\rangle,
\end{split}
\label{eqn:eqb24}
\end{equation}

\begin{equation}
\begin{split}
\ [\mathcal{H}-E(p)-E(q)]\lvert q<\bar{p}\rangle=&-\mathrm{e}^{ip}\sum_{k}\mathrm{e}^{i(p+q)k}\lvert \bar{k},k+1\rangle\\
+(-&\mathrm{e}^{ip}+1+\mathrm{e}^{i(p+q)})\sum_{k}\mathrm{e}^{i(p+q)k}\lvert k,\overline{k+1}\rangle.
\end{split}
\label{eqn:eqb25}
\end{equation}
Combine Eq. (\ref{eqn:eqb21}) with Eq. (\ref{eqn:eqb23}), Eq. (\ref{eqn:eqb24}) and Eq. (\ref{eqn:eqb25}), one can obtain

\begin{equation}
\begin{cases}
-\mathrm{e}^{iq}+S_{12}^{12}(1+\mathrm{e}^{i(p+q)}-\mathrm{e}^{ip})+S_{12}^{21}(-\mathrm{e}^{ip})=0,\\
(1+\mathrm{e}^{i(p+q)}-\mathrm{e}^{iq})+S_{12}^{12}(-\mathrm{e}^{ip})+S_{12}^{21}(1+\mathrm{e}^{i(p+q)}-\mathrm{e}^{ip})=0\end{cases}
\label{eqn:eqb26}
\end{equation}
Solve this equation, the solution is
\begin{equation}
S_{12}^{12}=-\frac{\mathrm{e}^{ip}-{e}^{iq}}{1-2\mathrm{e}^{ip}+\mathrm{e}^{i(p+q)}},
\ S_{12}^{21}=-\frac{(-1+\mathrm{e}^{ip})(-1+\mathrm{e}^{iq})}{1-2\mathrm{e}^{ip}+\mathrm{e}^{i(p+q)}}.
\label{eqn:eqb27}
\end{equation}
Set
\begin{equation}
\mathrm{e}^{ip}=\frac{u+i/2}{u-i/2},\ \mathrm{e}^{iq}=\frac{v+i/2}{v-i/2},
\label{eqn:eqb28}
\end{equation}
then the scattering matrix can be written as a compact form 
\begin{equation}
S_{ab}^{cd}(u,v)=\frac{(u-v)\delta_a^c\delta_b^d-i\delta_a^d\delta_b^c}{u-v+i},
\label{eqn:eqb29}
\end{equation}
where $a,b,c,d$ is the color index of the excitations, $\{1,2\}$.  Above scattering matrix satisfied Yang-Baxter equation, and we will prove it in $SU(N)$ case.
\begin{figure}[h]
    \centering
    \includegraphics[scale=0.8]{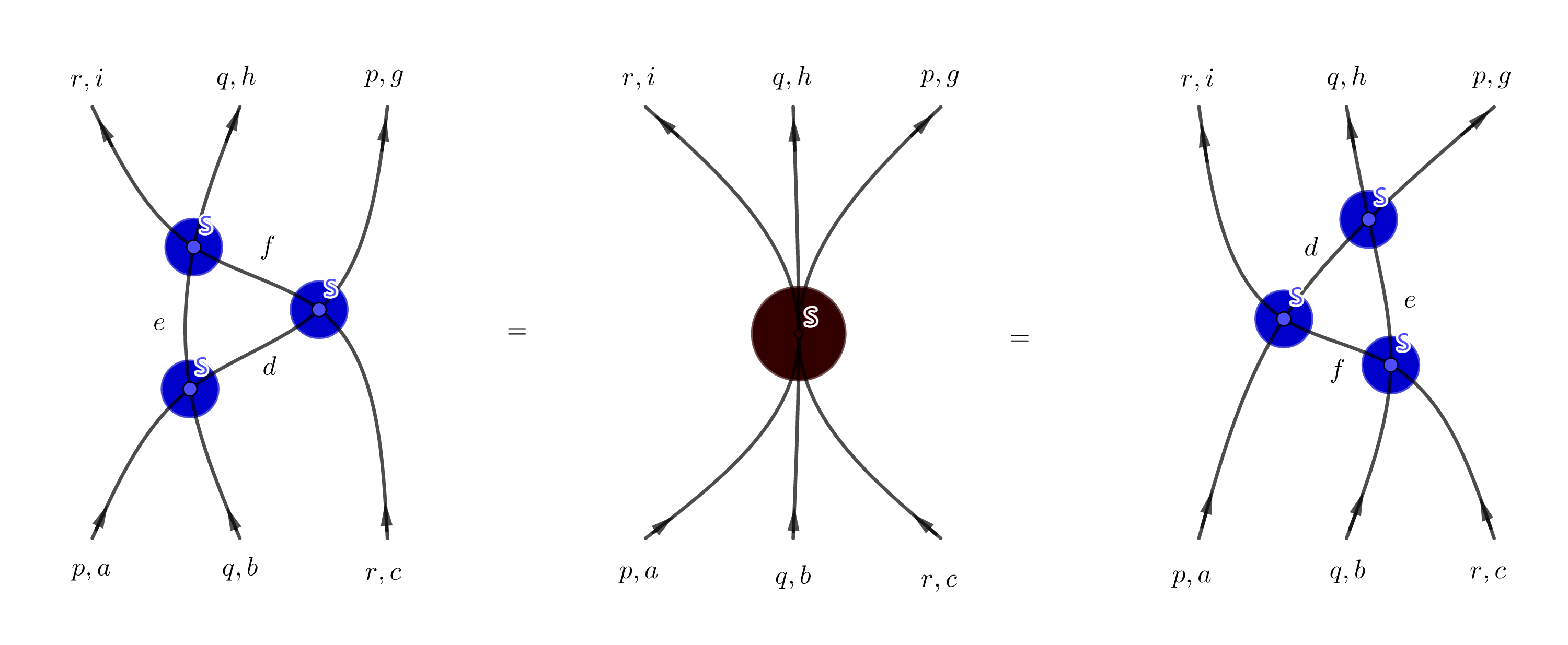}
    \caption{$SU(N)$ Yang-Baxter equation for scattering matrices \cite{intro_int_NB}}
    \label{fig:B1}
\end{figure}
\section{$SU(N)$ fundamental scattering matrix}
For $SU(N)$ the building block for Hamiltonian would be
\begin{equation}
t_n^{a}\otimes t_{n+1}^{a},
\label{eqn:eqb30}
\end{equation}
where the $t$ are the generators of $SU(N)$, the corresponding invariant property can be prove in the same as in $SU(3)$. In the space $(\mathbb{C}^{N})_n\otimes(\mathbb{C}^{N})_{n+1}$, using the properties of tensor product, one can say that 
\begin{equation}
\begin{split}
(t_n^{a}\otimes t_{n+1}^{a})_{iN+j,kN+l}&=(t_n^a)_{ik}\times(t_n^a)_{jl}\\
&=\frac{1}{2}\Big(\delta_{il}\delta_{kj}-\frac{1}{N}\delta_{ik}\delta_{jl}\Big).
\end{split}
\label{eqn:eqb31}
\end{equation}
 On the other hand, since the acting rule of permutation operator on state $P\lvert i,j \rangle=\lvert j,i \rangle$, the components of $P$ matrix reads
\begin{equation}
P_{aN+b,cN+d}=\delta_{ad}\delta_{dc},
\label{eqn:eqb32}
\end{equation}
same for identity matrix $I_{aN+b,cN+d}=\delta_{ac}\delta_{bd}$. So the building block can be written as 
\begin{equation}
(t_n^{a}\otimes t_{n+1}^{a})_{iN+j,kN+l}=\frac{1}{2}\Big(P_{iN+j,kN+j}-\frac{1}{N}I_{iN+j,kN+j}).
\label{eqn:eqb33}
\end{equation}
So we can write the permutation operator $P_{n,n+1}$ in terms of $t_n^{a}\otimes t_{n+1}^{a}$. Correspondingly the following Hamiltonian is $SU(N)$ invariant
\begin{equation}
\begin{split}H&=\sum_{n}H_{n,n+1}=\sum_{n}\Big(I_{n,n+1}-P_{n,n+1}\Big)\\
&=\sum_n \Big[(1-\frac{1}{N})I_{n,n+1}-2t_n^{a}\otimes t_{n+1}^{a}\Big].
\end{split}
\label{eqn:eqb34}
\end{equation}
 Here we have fixed the pre-factors by choosing the vacuum $\lvert 0,0,\dots,0\rangle$. Apply the Hamiltonian on a general state, 
\begin{equation}
H_{k,k+1}\lvert a,b\rangle=\lvert a,b\rangle-\lvert b,a\rangle,
\label{eqn:eqb35}
\end{equation}
so the states $\lvert 1,1,\dots, 1\rangle,\dots,\lvert N-1,N-1,\dots, N-1\rangle$ are automatically vacuums. Due to Eq. (\ref{eqn:eqb35}), the scattering matrix for $SU(N)$ is of the same format as $SU(3)$, but with more components,
\begin{equation}
S_{ab}^{cd}(u,v)=\frac{(u-v)\delta_a^c\delta_b^d-i\delta_a^d\delta_b^c}{u-v+i},
\label{eqn:eqb36}
\end{equation}
where $a,b,c,d$ is the color index of the excitations, $\{1,2,\dots,N-1\}$. Insert the above equation, one can easily check the Yang-Baxter equation is satisfied
\begin{equation}
S_{ab}^{de}(u,v)S_{dc}^{gf}(u,w)S_{ef}^{hi}(v,w)=S_{bc}^{ef}(v,w)S_{af}^{di}(u,w)S_{de}^{gh}(u,v),
\label{eqn:eqb37}
\end{equation}
or in Fig. \ref{fig:B1}.

% Appendix C

\chapter{$SU(3)$, $SU(N)$ Nested Bethe equations and generalizations}  % Main appendix title

\label{AppendixC} % For referencing this appendix elsewhere, use \ref{AppendixC}

\section{$SU(3)$}
Using CBA, we have seen that the scattering matrix is a rank $4$ tensor with off-diagonal terms. Since there is no universal way to diagonalize a rank $4$ tensor, we will use a brilliant way to diagonalize the scattering matrix, so called $Nested\ Bethe\ Ansatz$. Write out the scattering matrix in the basis $\lvert 11\rangle$, $\lvert 21\rangle$, $\lvert 12\rangle$, $\lvert 22\rangle$ 
\begin{equation}
S_{k,j}(u_k,u_j)=\begin{pmatrix}S_{11}^{11}(u_k,u_j)&&&\\&S_{21}^{21}(u_k,u_j)&S_{12}^{21}(u_k,u_j)&\\&S_{21}^{12}(u_k,u_j)&S_{12}^{12}(u_k,u_j)&\\&&&S_{22}^{22}(u_k,u_j)\end{pmatrix},
\label{eqn:eqc1}
\end{equation}
where the empty elements is zero. Let's consider a spin chain of length $L$ with $ M_1$ excitations consist of  $ M_1-M_2$ color $1$ particle, and $M_2$ color $2$ particles. The Bethe equations can be rewritten as 
\begin{equation}
e^{-ip_k L}\lvert\Psi\rangle=S_{k,k+1}\dots S_{k,M_1}S_{k,1}\dots S_{k,k-1}\lvert\Psi\rangle.
\label{eqn:eqc2}
\end{equation}
However, unlike in $SU(2)$ case, these are matrix Bethe equations. In this 2 components basis, one should think the state $\lvert \Psi\rangle$ as a "short spin chain", and $S$ as some kind of Hamiltonian. Since the vacuum state $\lvert 00\rangle$ gives us identity, we might as well "cut" it. Let's pick a vacuum on this short chain, say the color $1$ component, then one can say that we have a length $M_1$ spin chain with $M_2$ excitations color 2. We achieve this by the reduced two-body scattering operator \cite{su_3_bethe_ansatz}
\begin{equation}
s_{k,j}=\begin{pmatrix}1&&&\\&t_{21}^{21}(u_k,u_j)&r_{12}^{21}(u_k,u_j)&\\&r_{21}^{12}(u_k,u_j)&t_{12}^{12}(u_k,u_j)&\\&&&1\end{pmatrix},
\label{eqn:eqc3}
\end{equation}
where
\begin{equation}
S_{k,j}=S_{11}^{11}(u_k,u_j)s_{k,j}.
\label{eqn:eqc4}
\end{equation}
So due to the indistinguishable property of excitations kind, define 
\begin{equation}
t_{k,j}=\frac{u_k-u_j}{u_k-u_j-i},\ r_{k,j}=\frac{-i}{u_k-u_j-i},
\label{eqn:eqc5}
\end{equation}
then 
\begin{equation}
t^{21}_{21}(u_k,u_j)=t^{12}_{12}(u_k,u_j)=t_{k,j}, \quad r^{12}_{21}(u_k,u_j)=r^{21}_{12}(u_k,u_j)=r_{k,j}P_{k,j}
\label{eqn:eqc6}
\end{equation}
where P is the permutation operator. We can rewrite Eq. (\ref{eqn:eqc2}) as
\begin{equation}
\begin{split}
\lambda_k\lvert \Psi\rangle=s_{k,k+1},\dots s_{k,N_1}s_{k,1}\dots s_{k,k-1}\lvert \Psi\rangle,
\end{split}
\label{eqn:eqc7}
\end{equation}
where $\lambda_k$ is the common eigenvalue of the reduced many-body scattering operator
\begin{equation}
\lambda_k=e^{-ip_k L}\prod_{j\ne k}^{N_1}(S_{11}^{11})^{-1}(p_k,p_j),
\label{eqn:eqc8}
\end{equation}
On the vacuum of the short spin chain the reduced operators $s_{k,j}$ act trivially and we can find all values of $k$

\begin{equation}
\lambda_k\lvert 11\dots11\rangle=\lvert 11\dots11\rangle,
\label{eqn:eqc9}
\end{equation}
i.e. $\lambda_k=1$, which is our $su(2)$ Bethe equation.

\subsection{One-magnon problem}
Next we solve the one-magnon problem of the short chain. For one-magnon, Eq. (\ref{eqn:eqc7}) reduces to
\begin{equation}
\lambda_k\lvert \Psi\rangle=(t_{k,k+1}+r_{k,k+1}P_{k,k+1}),\dots,(t_{k,k-1}+r_{k,k-1}P_{k,k-1})\lvert \Psi\rangle,
\label{eqn:eqc10}
\end{equation}
and $P_{i,j}$ is the permutation operator because the color has changed when. Since there is no scattering between color 2 excitations. We cannot use the plane wave anymore since our short spin chain is not translationally invariant. Introduce a coordinate-space wave function $\psi_l$ 
\begin{equation}
\lvert \Psi\rangle=\sum_{1\le l\le N_1}\psi_l \lvert 1\dots1\overset{l}21\dots1\rangle=\sum_{1\le l\le N_1}\psi_l \lvert l\rangle.
\label{eqn:eqc11}
\end{equation}
Then apply the reduced scattering matrix $s_{k,j}$ on the state as Eq. (\ref{eqn:eqc10}), one can get recursion relations for the wave function $\psi_l$,
\begin{equation}
\psi_k^{(j)}=\psi_k^{(j-1)}t_{k,k-j}+\psi_{k-j}r_{k,k-j}
\label{eqn:eqc12}
\end{equation}
where $\psi_k^{(j)}$ is the wave function of $\lvert k\rangle$ after apply reduced scattering matrix $s_{k,j}$ $j$ times. And the state $\lvert k-j\rangle$ will not change anymore, gives us the eigenvalue equation 
\begin{equation}
\sigma_k\psi_{k-j}=\psi_{k-j}t_{k,k-j}+\psi_k^{(j-1)}r_{k,k-j},
\label{eqn:eqc13}
\end{equation}
Here $\sigma_k$ is the eigenvalue for one excitation on short chain. Combine Eq. (\ref{eqn:eqc12}) and Eq. (\ref{eqn:eqc13}), one can find that
\begin{equation}
\begin{split}
\frac{\psi_{k-j-1}}{\psi_{k-j}}&=\frac{r^2_{k,k-j}-t^2_{k,k-j}+\sigma_kt_{k,k-j}}{\sigma_k-t_{k,k-j-1}}\frac{r_{k,k-j-1}}{r_{k,k-j}}\\
&=\frac{u_{k-j}-(u_k+\frac{i}{2}\frac{1+\sigma_k}{1-\sigma_k})-\frac{i}{2}}{u_{k-j-1}-(u_k+\frac{i}{2}\frac{1+\sigma_k}{1-\sigma_k})+\frac{i}{2}}.
\end{split}
\label{eqn:eqc14}
\end{equation}
The L.H.S. of this equation only depends on the difference $k-j$, thus this equation can only be consistent if the right hand side does not depend on $k$, so
\begin{equation}
v=u_k+\frac{i}{2}\frac{1+\sigma_k}{1-\sigma_k}
\label{eqn:eqc15}
\end{equation}
must be a constant. We can interpret it as an auxiliary Bethe root for color $2$. For later use, we can invert above equation
\begin{equation}
\sigma_k(v)=\frac{u_k-v+\frac{i}{2}}{u_k-v-\frac{i}{2}},
\label{eqn:eqc16}
\end{equation}
it plays the same role as $e^{ip}$, moving excitation on the short spin chain. By choose $\psi_1=1$, the recursion relation Eq. (\ref{eqn:eqc14}) gives us
\begin{equation}
\psi_k(v)=\prod_{j=1}^{k-1}\frac{u_j-v+\frac{i}{2}}{u_{j+1}-v-\frac{i}{2}}=\frac{u_1-v-\frac{i}{2}}{u_k-v-\frac{i}{2}}\sigma_1\dots\sigma_{k-1}.
\label{eqn:eqc17}
\end{equation}
Similarly, by combining Eq. (\ref{eqn:eqc13}) and Eq. (\ref{eqn:eqc15}), one can get the following expression
\begin{equation}
\psi_k^{(j)}(v)=\frac{u_{k-j-1}-v+\frac{i}{2}}{u_k-v-\frac{i}{2}}\psi_{k-j-1}(v)=\frac{\psi_k(v)}{\sigma_{k-1}\dots\sigma_{k-j}}.
\label{eqn:eqc18}
\end{equation}
But remember our iteration process only works for $j=1$ to $j=k-1$. One can also get another iteration relation when $j>k-1$, but there is no need for this. We can just set $k=M_1$, then this iteration works for all $s_{k,j}$. When we are through, all the wave function satisfy the eigenvalue equation but the final one, the final equation is 
\begin{equation}
\sigma_{M_1}(v)\sum_{j=1}^{M_1-1}\psi_j\lvert j\rangle+\psi_{M_1}^{(M_1-1)}\lvert M_1\rangle.
\label{eqn:eqc19}
\end{equation}
Demanding this equation satisfying the eigenvalue equation yields
\begin{equation}
\psi_{M_1}^{(M_1-1)}=\sigma_{M_1}\psi_{M_1},
\label{eqn:eqc20}
\end{equation}
insert Eq. (\ref{eqn:eqc18}), one can show that
\begin{equation}
1=\prod_{k}^{M_1}\sigma_k=\prod_k^{M_1} e^{-i\tilde{p}_kM_1}=e^{-i\tilde{P}M_1},
\label{eqn:eqc21}
\end{equation}
which is similar in $su(2)$ case that an overall shift of $M_1$ site will change none, where $\tilde{p}_k$ is the pseudo-momentum of the excitations in the short spin chain, $\tilde{P}$ is the overall pseudo-momentum.

Combine Eq. (\ref{eqn:eqc16}) with Eq. (\ref{eqn:eqc6}), we have find the first Bethe equation with $\lambda_k=\sigma_k$ for one excitation:
\begin{equation}
\frac{u_k-v+\frac{i}{2}}{u_k-v-\frac{i}{2}}=\left(\frac{u_k-\frac{i}{2}}{u_k+\frac{i}{2}}\right)^L\prod_{j\ne k}^{M_1}\frac{u_k-u_j+i}{u_k-u_j-i}
\label{eqn:eqc22}
\end{equation}
This completely solves the one-magnon problem $M_2=1$. 

\subsection{Two and more magnon states}
Next we will consider the matrix eigenvalue equation Eq. (\ref{eqn:eqc2}) for two magnons problem $M_2=2$. The states now are describe by
\begin{equation}
\begin{split}
\lvert\Psi\rangle&=\sum_{1\le l_1\le l_2\le M_1}\psi_{l_1,l_2} (v_1,v_2)\lvert 1\dots1\overset{l_1}21\dots1\overset{l_2}21\dots1\rangle\\
&=\sum_{1\le l_1\le l_2\le M_1}\psi_{l_1,l_2}(v_1,v_2) \lvert l_1,l_2\rangle.
\end{split}
\label{eqn:eqc23}
\end{equation}
Use the same procedure as before, we act the reduce scattering matrix one by one, the only difference is that now we have to consider the scattering between color $2$ excitations. Just like in $su(2)$ case, here the two-body wave function should be the product of two one-body wave functions. Since we need to consider the scattering, so the wave equation by Bethe Ansatz is 
\begin{equation}
\psi_{l_1,l_2}(v_1,v_2)=B\psi_{l_1}(v_1)\psi_{l_2}(v_2)+B^{\prime}\psi_{l_1}(v_2)\psi_{l_2}(v_1).
\label{eqn:eqc24}
\end{equation}
Denote $\psi_{l}(v_1)$, $\psi_{l}(v_2)$ by $\psi_{l}$, $\psi_{l}^{\prime}$ and $\sigma_{l}(v_1)$, $\sigma_{l}(v_2)$ by $\sigma_{l}$, $\sigma_{l}^{\prime}$. Substitute this ansatz into eigenequation Eq. (\ref{eqn:eqc2}), one can find that
\begin{equation}
\lambda_k=\sigma_k(v_1)\sigma_k(v_2),
\label{eqn:eqc25}
\end{equation}
and 

\begin{equation}
\frac{B}{B^{\prime}}=-\frac{\psi_j^{\prime}\psi_{j-k}\sigma_j/(\sigma^{\prime}_{j-1}\dots\sigma^{\prime}_{j-k})-\psi_j\psi_{j-k}^{\prime}/(\sigma_{j-1}\dots\sigma_{j-k+1})}{\psi_j\psi_{j-k}^{\prime}\sigma_j^{\prime}/(\sigma_{j-1}\dots\sigma_{j-k})-\psi_j^{\prime}\psi_{j-k}/(\sigma_{j-1}^{\prime}\dots\sigma_{j-k+1}^{\prime})}.
\label{eqn:eqc26}
\end{equation}
Substitute Eq. (\ref{eqn:eqc16}) and Eq. (\ref{eqn:eqc17}) into Eq. (\ref{eqn:eqc26}), one would find out 
\begin{equation}
S(v_1,v_2)=\frac{B^{\prime}}{B}=\frac{v_1-v_2-i}{v_1-v_2+i},
\label{eqn:eqc27}
\end{equation}
which is nothing but the scattering matrix on this short spin chain, and also the same expression of $su(2)$ scattering matrix. This completely solves two-magnon problem.

For $M_2$-magnon problem, one can just use the Bethe ansatz for the wavefunction, and the corresponding Bethe equations are 
\begin{equation}
\prod_{j=1}^{M_1}\sigma_{j}(v_l)=\prod_{m\ne l}^{M_2}S(v_l,v_m),\quad l=1,\dots,M_2.
\label{eqn:eqc28}
\end{equation}
And also the eigenvalue equations
\begin{equation}
\lambda_k=\prod_{j=1}^{M_2}\sigma_k(v_j)=e^{-ip_k L}\prod_{j\ne k}^{M_1}(S_{11}^{11})^{-1}(u_k,u_j),\quad k=1,\dots,M_1.
\label{eqn:eqc29}
\end{equation}
Write out Eq. (\ref{eqn:eqc28}) and Eq. (\ref{eqn:eqc29})

\begin{equation}
\prod_{j=1}^{M_1}\frac{u_j-v_l+\frac{i}{2}}{u_j-v_l-\frac{i}{2}}=\prod_{m\ne l}^{M_2}\frac{v_l-v_m-i}{v_l-v_m+i},\quad l=1,\dots,M_2,
\label{eqn:eqc30}
\end{equation}

\begin{equation}
\prod_{j=1}^{M_2}\frac{u_k-v_j+\frac{i}{2}}{u_k-v_j-\frac{i}{2}}=\left(\frac{u_k-\frac{i}{2}}{u_k+\frac{i}{2}}\right)^L\prod_{j\ne k}^{M_1}\frac{u_k-u_j+i}{u_k-u_j-i},\quad k=1,\dots,M_1.
\label{eqn:eqc31}
\end{equation}
 These equations completely determine all the Bethe roots and the auxiliary Bethe roots. 
 \section{SU(N)}
 Just the same as in $su(3)$ Nested Bethe equations, we can reproduce a whole set of Bethe equations layer by layer. Write the scattering matrix in the basis $\lvert ab\rangle$, with $a,b=1,\dots,N-1$
\begin{equation}
S^{(N)}_{k,j}(u_k,u_j)=\begin{pmatrix}S_{11}^{11}(u_k,u_j)&&&\\&S_{21}^{21}(u_k,u_j)&S_{12}^{21}(u_k,u_j)&\\&S_{21}^{12}(u_k,u_j)&S_{12}^{12}(u_k,u_j)&\\&&&S^{(N-1)}(u_k,u_j)\end{pmatrix},
\label{eqn:eqc32}
\end{equation}
where $S^{(m)}$ denote the scattering matrix for $su(m)$, the scattering matrix take the form of block diagonal form is due to the conservation of charge. Let's consider a chain with excitations denoted as $\{M_1,M_2,\dots,M_{N-1}\}$, which means that we first create $M_1$ number  of color $1$ excitations out of the vacuum chain of site $L$, then we create $M_2$ color $2$ excitations on top of the short chain $M_1$ sites, we iterate it until we recover the whole spin chain configuration. The Bethe equations from periodicity of first level short spin chain $\lvert \Psi\rangle^{1}$ can be rewritten as \footnote{ One can also say the vacuum level Bethe equations are 
$e^{-ip_k L}\lvert\Psi\rangle^{0}=\lvert\Psi\rangle^{0}$.}
\begin{equation}
e^{-ip_k L}\lvert\Psi\rangle^{1}=S^{(N)}_{k,k+1}\dots S^{(N)}_{k,M_1}S^{(N)}_{k,1}\dots S^{(N)}_{k,k-1}\lvert\Psi\rangle^{1}.
\label{eqn:eqc33}
\end{equation}
\subsection{One-magnon problem}
Let's pick a vacuum on the "first level spin chain" by considering the reduced scattering matrix
\begin{equation}
s^{(N)}_{k,j}=\begin{pmatrix}1&&&\\&t_{21}^{21}(u_k,u_j)&r_{12}^{21}(u_k,u_j)&\\&r_{21}^{12}(u_k,u_j)&t_{12}^{12}(u_k,u_j)&\\&&&s^{(N-1)}_{k,j}\end{pmatrix},
\label{eqn:eqc34}
\end{equation}
 where 
\begin{equation}
S^{(N)}_{k,j}=S_{11}^{11}(u_k,u_j)s^{(N)}_{k,j}.
\label{eqn:eqc35}
\end{equation}
We can rewrite Eq. (\ref{eqn:eqc33}) as
\begin{equation}
\begin{split}
\lambda^{(1)}_k\lvert \Psi\rangle^{1}=s^{(N)}_{k,k+1},\dots s^{(N)}_{k,M_1}s^{(N)}_{k,1}\dots s^{(N)}_{k,k-1}\lvert \Psi\rangle^{1},
\end{split}
\label{eqn:eqc36}
\end{equation}
where $\lambda^{(1)}_k$ is the common eigenvalue of the reduced many-body scattering operator
\begin{equation}
\lambda^{(1)}_k=e^{-ip_k L}\prod_{j\ne k}^{M_1}(S_{11}^{11})^{-1}(u_k,u_j).
\label{eqn:eqc37}
\end{equation}
On the vacuum of the short spin chain the reduced operators $s^{N}_{k,j}$ act trivially and we can find all values of $k$

\begin{equation}
\lambda^{(1)}_k\lvert 11\dots11\rangle=\lvert 11\dots11\rangle,
\label{eqn:eqc38}
\end{equation}
i.e. $\lambda^{(1)}_k=1$. Now we can create excitation on top of "first level spin chain", for example one-magnon state:
\begin{equation}
\lvert \Psi\rangle^{1}=\sum_{1\le l\le N_1}\psi_l(u_j^{(a)}) \lvert 1\dots1\overset{l}a1\dots1\rangle=\sum_{1\le l\le N_1}\psi_l(u_j^{(a)}) \lvert (l,a)\rangle,
\label{eqn:eqc39}
\end{equation}
with $a=2,\dots,N-1$. Demanding this state as the solution of Eq. (\ref{eqn:eqc2}), we can get
\begin{equation}
\sigma^{(1)}_k(u^{(a)})=\frac{u^{(1)}_k-u_j^{(a)}+\frac{i}{2}}{u^{(1)}_k-u_j^{(a)}-\frac{i}{2}},
\label{eqn:eqc40}
\end{equation}
$\sigma^{(1)}_k(u_j^{(a)})$ is the eigenvalue of one-magnon state, which is exactly the same as in $su(3)$ case, just the type of excitations becomes more. This is all the consequence of the conservation of charge, each subsector with only one type of excitation reduces to the $su(3)$ case, and this subsector must fulfill Eq. (\ref{eqn:eqc36}), leads to Eq. (\ref{eqn:eqc40}). More generally, one can get 

\begin{equation}
\sigma_{k}^{(a)}(u_j^{(b)})=\frac{u^{(a)}_k-u_j^{(b)}+\frac{i}{2}}{u^{(a)}_k-u_j^{(b)}-\frac{i}{2}},\quad \mathrm{with}\quad  a<b,
\label{eqn:eqc41}
\end{equation}
due to the indistinguishable property of excitations and equivalence of vacuums. For $M_2$ color $2$ excitation, we can conclude that
\begin{equation}
\lambda^{(1)}_{k}=\prod_{j=1}^{M_2}\sigma^{(1)}_k(u_j^{(2)})=e^{-ip_k L}\prod_{j\ne k}^{M_1}(S_{11}^{11})^{-1}(u_k,u_j),\quad k=1,\dots,M_1.
\label{eqn:eqc42}
\end{equation}

\subsection{Two and more magnon problem}
The scattering matrix on "first level spin chain" actually reduces to the $su(n-1)$ case, to prove this, we should again use the same procedure as in $su(3)$ case. There are two kinds of state that we need to pay attention to, the first kind is the same as in $su(3)$ case, 
\begin{equation}
\begin{split}
\lvert\Psi\rangle&=\sum_{1\le l_1\le l_2\le M_1}\psi_{l_1,l_2} (v_1,v_2)\lvert 1\dots1\overset{l_1}a1\dots1\overset{l_2}a1\dots1\rangle\\
&=\sum_{1\le l_1\le l_2\le M_1}\psi_{l_1,l_2}(v_1,v_2) \lvert l_1,l_2\rangle.
\end{split}
\label{eqn:eqc43}
\end{equation}
This state gives us the scattering between the same color, 

\begin{equation}
S_{aa}^{aa}(v_1,v_2)=\frac{v_1-v_2-i}{v_1-v_2+i}.
\label{eqn:eqc44}
\end{equation}
There are another state which contains two different colors excitations
\begin{equation}
\begin{split}
\lvert\Psi\rangle&=B_1\sum_{1\le l_1\le \bar{l}_2\le M_1}\psi_{l_1} (v_1)\psi_{\bar{l}_2} (v_2)\lvert 1\dots1\overset{l_1}a1\dots1\overset{\bar{l}_2}b1\dots1\rangle\\
&+B_2\sum_{1\le \bar{l}_1\le l_2\le M_1}\psi_{\bar{l}_1} (v_2)\psi_{l_2} (v_1)\lvert 1\dots1\overset{\bar{l}_1}b1\dots1\overset{l_2}a1\dots1\rangle\\
&+B_3\sum_{1\le \bar{l}_1\le l_2\le M_1}\psi_{\bar{l}_1} (v_1)\psi_{l_2} (v_2)\lvert 1\dots1\overset{\bar{l}_1}b1\dots1\overset{l_2}a1\dots1\rangle\\
&=B_1\lvert v_1<\bar{v}_2\rangle+B_2\lvert\bar{v}_2 <v_1\rangle+B_3\lvert \bar{v}_1<v_2\rangle.
\end{split}
\label{eqn:eqc45}
\end{equation}
Plug this into Eq. (\ref{eqn:eqc36}), use the reduced scattering $r$ and $t$ between color $a$ and color $b$, one would obtain 
\begin{equation}
S_{ab}^{ab}(v_1,v_2)=\frac{B_2}{B_1}=\frac{v_1-v_2}{v_1-v_2+i}, \quad S_{ab}^{ba}(v_1,v_2)=\frac{B_3}{B_1}=\frac{-i}{v_1-v_2+i}
\label{eqn:eqc46}
\end{equation}
This is nothing but the scattering matrix element we need for $su(n-1)$, thus proves what we want. The Bethe equations from periodicity of second level short spin chain $\lvert \Psi\rangle^{2}$  can be written as
\begin{equation}
\prod_{j=1}^{M_1}\sigma^{(1)}_j(u_k^{(2)})\lvert\Psi\rangle^{2}=S^{(N-1)}_{k,k+1}\dots S^{(N-1)}_{k,M_2}S^{(N-1)}_{k,1}\dots S^{(N-1)}_{k,k-1}\lvert\Psi\rangle^{2}.
\label{eqn:eqc47}
\end{equation}
\subsection{Nesting for higher excitations}
Let's pick a vacuum on the "second level spin chain" by considering the reduced scattering matrix 
\begin{equation}
s^{(N-1)}_{k,j}=\begin{pmatrix}1&&&\\&t_{32}^{32}(u_k,u_j)&r_{23}^{32}(u_k,u_j)&\\&r_{32}^{23}(u_k,u_j)&t_{23}^{23}(u_k,u_j)&\\&&&s^{(N-2)}_{k,j}\end{pmatrix},
\label{eqn:eqc48}
\end{equation}
where
\begin{equation}
S^{(N-1)}_{k,j}=S_{22}^{22}(u_k,u_j)s^{(N-1)}_{k,j}.
\label{eqn:eqc49}
\end{equation}
We can rewrite Eq. (\ref{eqn:eqc47}) as
\begin{equation}
\lambda^{(2)}_{k}\lvert \Psi\rangle^{2}=s^{(N-1)}_{k,k+1},\dots s^{(N-1)}_{k,M_2}s^{(N-1)}_{k,1}\dots s^{(N-1)}_{k,k-1}\lvert \Psi\rangle^{2},
\label{eqn:eqc50}
\end{equation}
where $\lambda^{(2)}_{kl}$ is the common eigenvalue of the reduced many-body scattering operator
\begin{equation}
\lambda^{(2)}_{k}=\prod_{j=1}^{M_1}\sigma^{(1)}_j(u_k^{(2)})\prod_{j\ne k}^{M_2}(S_{22}^{22})^{-1}(u_k,u_j).
\label{eqn:eqc51}
\end{equation}
Again, as before, for $M_3$ number of color $3$ excitation on the "second level spin chain", one should obtain
\begin{equation}
\lambda^{(2)}_{k}=\prod_{j=1}^{M_3}\sigma^{(2)}_k(u_j^{(3)}).
\label{eqn:eqc52}
\end{equation}
Combine with the Bethe equation Eq. (\ref{eqn:eqc47}), the second set of equations comes out naturally
\begin{equation}
\lambda^{(2)}_{k}=\prod_{j=1}^{M_3}\sigma^{(2)}_k(u_l^{(3)})=\prod_{j=1}^{M_1}\sigma^{(1)}_j(u_k^{(2)})\prod_{j\ne k}^{M_2}(S_{22}^{22})^{-1}(u_k,u_j),\quad k=1,\dots,M_2.
\label{eqn:eqc53}
\end{equation}
So for "level $a$ spin chain", with $a=1,\dots,N-2$, one can get the eigenvalue equations
\begin{equation}
\lambda^{(a)}_{k}=\prod_{l=1}^{M_{a+1}}\sigma^{(a)}_k(u_l^{(a+1)})=\prod_{j=1}^{M_{a-1}}\sigma^{(a-1)}_j(u_k^{(a)})\prod_{j\ne k}^{M_a}(S_{aa}^{aa})^{-1}(u_k,u_j),\quad k=1,\dots,M_a.
\label{eqn:eqc54}
\end{equation}
On the top level spin chain $a=N-1$, the reduced scattering matrix becomes
\begin{equation}
s^{(2)}_{k,j}=S^{(2)}_{k,j}/S_{N-1,N-1}^{N-1,N-1}(u_k,u_j)=1,
\label{eqn:eqc55}
\end{equation}
then the corresponding eigenvalue $\lambda_{k}^{(N-1)}=1$. Then the eigenvalue equations is 
\begin{equation}
\lambda_{k}^{(N-1)}=1=\prod_{j=1}^{M_{N-2}}\sigma^{(N-2)}_j(u_k^{(N-1)})\prod_{j\ne k}^{M_{N-1}}(S_{N-1,N-1}^{N-1,N-1})^{-1}(u_k,u_j),\quad k=1,\dots,M_{N-1}.
\label{eqn:eqc56}
\end{equation}
Eq.(\ref{eqn:eqc42}), Eq. (\ref{eqn:eqc54}) and Eq. (\ref{eqn:eqc56}) give all Bethe equations that we want, at first level
\begin{equation}
\prod_{l=1}^{M_2}\frac{u_k^{(1)}-u_l^{(2)}+\frac{i}{2}}{u_k^{(1)}-u_l^{(2)}-\frac{i}{2}}=\left(\frac{0-u_k^{(1)}+\frac{i}{2}}{0-u_k^{(1)}-\frac{i}{2}}\right)^L\prod_{j\ne k}^{M_1}\frac{u_k^{(1)}-u_j^{(1)}+i}{u_k^{(1)}-u_j^{(1)}-i},
\label{eqn:eqc57}
\end{equation}
and at $a$ level
\begin{equation}
\prod_{j=1}^{M_{a+1}}\frac{u_k^{(a)}-u_j^{(a+1)}+\frac{i}{2}}{u_k^{(a)}-u_j^{(a+1)}-\frac{i}{2}}=\prod_{j=1}^{M_{a-1}}\frac{u_j^{(a-1)}-u_k^{(a)}+\frac{i}{2}}{u_j^{(a-1)}-u_k^{(a)}-\frac{i}{2}}\prod_{j\ne k}^{M_a}\frac{u_k^{(a)}-u_j^{(a)}+i}{u_k^{(a)}-u_j^{(a)}-i},
\label{eqn:eqc58}
\end{equation}
and at top level
\begin{equation}
1=\prod_{j=1}^{M_{N-2}}\frac{u_j^{(N-2)}-u_k^{(N-1)}+\frac{i}{2}}{u_j^{(N-2)}-u_k^{(N-1)}-\frac{i}{2}}\prod_{j\ne k}^{M_{N-1}}\frac{u_k^{(N-1)}-u_j^{(N-1)}+i}{u_k^{(N-1)}-u_j^{(N-1)}-i}
\label{eqn:eqc59}
\end{equation}
As one can easily see that above Bethe equations are just eigenvalue problem. More specifically, we can get such set of equations by imposing periodic boundary condition on each level of spin chain. 
One can actually construct the wave function of such system \cite{phd_Escobed_Jorge},the Bethe equations are obtained in this way.

\section{Generalizations}

Let us generalize our spin chain of length $L$ with Lie (super) group of rank $r$ with Cartan matrix $M_{a,b}$. The physical space at each site sits at the representation denoted by Dynkin label $V_a$. Thus a general Bethe state can be identified by a set of Bethe roots $\{u_{j}^{(a)}\}$, where $a=1,\dots,r$ denotes the Dynkin node and $j=1,\dots,N_a$ denotes the number of excitations of type $a$. Let's also introduce the Baxter polynomials
\begin{equation}
Q_a(u)=\prod_{j=1}^{N_a}(u-u^{(a)}_j)
\label{eqn:eqc60}
\end{equation}
The general Bethe equations then are \cite{phd_Escobed_Jorge}
\begin{equation}
\Big(\frac{u^{(a)}_{j}+V_a\frac{i}{2}}{u^{(a)}_{j}-V_a\frac{i}{2}}\Big)^L=s_a\prod_{b=1}^r\frac{Q_b(u^{(a)}_j+M_{a,b}\frac{i}{2})}{Q_b(u^{(a)}_j-M_{a,b}\frac{i}{2})},
\label{eqn:eqc61}
\end{equation}
where $s_a=-1,1$ if node $a$ is bosonic or fermionic respectively. 
% Appendix D

\chapter{$GL(3)$ and $GL(N)$ nested algebraic Bethe ansatz}  % Main appendix title

\label{AppendixD} % For referencing this appendix elsewhere, use \ref{AppendixD}
\section{$GL(3)$ nested algebraic Bethe ansatz}
\subsection{Notion setup}
In algebraic Bethe ansatz (ABA), we need to find the $R$-matrix acting in $\mathbb{C}^N\otimes\mathbb{C}^N$ that satisfy the Yang-Baxter equation
\begin{equation}
R_{12}(u_1,u_2)R_{13}(u_1,u_3)R_{23}(u_2,u_3)=R_{23}(u_2,u_3)R_{13}(u_1,u_3)R_{12}(u_1,u_2),
\label{eqn:eqd1}
\end{equation}
in order to provide compatibility of the $RTT$-relation
\begin{equation}
R_{12}(u,v)T_1(u)T_2(v)=T_2(v)T_1(u)R_{12}(u,v).
\label{eqn:eqd2}
\end{equation}
Actually the non-trivial $R$-matrix has exactly the same form as in the case of $\mathbb{C}^2$ auxiliary space:
\begin{equation}
R(u,v)=I+g(u,v)P,\quad g(u,v)=\frac{i}{u-v}.
\label{eqn:eqd3}
\end{equation}
Here $\mathbb{I}$ is the identity operator in $\mathbb{C}^N\otimes\mathbb{C}^N$, $P$ is the permutation operator. The permutation operator has the form
\begin{equation}
P=\sum_{i,j=1}^{N}E_{ij}\otimes E_{ji},
\label{eqn:eqd4}
\end{equation}
where $(E_{i,j})_{lk}=\delta_{il}\delta_{jk}$, $i,j,l,k=1,\dots,N$. The $R$-matrix Eq. (\ref{eqn:eqd3}) is called $\mathfrak{gl}_N$-invariant due to the property
\begin{equation}
[R_{12}(u,v),g_1\otimes \mathbb{I}+\mathbb{I}\otimes g_2]=0,\quad g_1,g_2\in\mathfrak{gl}_N,
\label{eqn:eqd5}
\end{equation}
 this has been proven in the coordinate Bethe ansatz, or in finite version
\begin{equation}
[R_{12}(u,v),G_1\otimes G_2]=0,\quad g_1,g_2\in GL_N.
\label{eqn:eqd6}
\end{equation}
For $SU(N)$ $XXX$ fundamental $L$ sites spin chain, the corresponding monodromy matrix is similar to $SU(2)$ case
\begin{equation}
T(u)=R_{0L}(u)\dots R_{01}(u),
\label{eqn:eqd7}
\end{equation}
the $R_{0i}$ is in the space $\mathbb{C}^N\otimes \mathbb{C}^N$. The auxiliary space of the monodromy matrix is $V_0$, the corresponding physical space is 
\begin{equation}
\mathcal{H}=V_1\otimes\dots \otimes V_L=\mathbb{C}^N\otimes\dots\otimes\mathbb{C}^N.
\label{eqn:eqd8}
\end{equation}
And we pick the a vacuum for this quantum space as 
\begin{equation}
\lvert 0 \rangle=\begin{pmatrix}1\\0\\\vdots\\0\end{pmatrix}\otimes\dots\begin{pmatrix}1\\0\\\vdots\\0\end{pmatrix}.
\label{eqn:eqd9}
\end{equation}

The $RTT$-relation Eq. (\ref{eqn:eqd2}) with $R$-matrix Eq. (\ref{eqn:eqd3}) implies
\begin{equation}
\begin{split}
\ [T_{ij}(u),T_{kl}(v)]&=g(u,v)(T_{kj}(v)T_{il}(u)-T_{kj}(u)T_{il}(v))\\
&=g(u,v)(T_{il}(u)T_{kj}(v)-T_{il}(v)T_{kj}(u)),
\end{split}
\label{eqn:eqd10}
\end{equation}
where $T_{i,j}$ are matrix elements of the monodromy matrix expanding in the $V_0$ as $T=\sum_{i,j}E_{ij}\otimes T_{ij}$, and they act on the physical space $\mathcal{H}$.\footnote{Here commutation relation Eq. (\ref{eqn:eqd10}) is sometimes denoted as the defining relation for Yangian. $T_{ij}$ is choosing to act on physical space, but we choose it to act on auxiliary space in Chapter \ref{Chapter4},\ref{Chapter5} \ref{Chapter6}.}

To simplify our tedious writings, here we introduce some notion for convenience.
\begin{itemize}
  \item Two rational functions, $f(x,y)$ and $g(x,y)$
  \begin{equation}
    g(x,y)=\frac{i}{x-y}, \quad f(x,y)=1+g(x,y).
    \label{eqn:eqd11}
  \end{equation}
  \item Sets of variables. We denote sets of variables by a bar: $\bar{x},\bar{u},\bar{v}$ etc. Individual elements of the sets are denoted by the subscripts: $v_j, u_k$ etc. A notation $\bar{u}_i$ means the set of $u$ without element $u_i$.
  \item Shorthand notation for product. If the functions $\lambda_i$, $g$, $f$, as well as the operators $T_{i,j}$ depend on sets of variables, this means that one should take the product over the corresponding set. For example
  \begin{equation}
    T_{i,j}(\bar{u})=\prod_{u_k\in \bar{u}}T_{i,j}(u_k);\quad g(z,\bar{w}_i)=\prod_{w_j\ne w_i}g(z,w_j);\quad f(\bar{u},\bar{v})=\prod_{u_j\in \bar{u}}\prod_{v_k\in \bar{v}}f(u_j,v_k).
   \label{eqn:eqd12}
  \end{equation}
  By definition, any product over empty set is equal to 1.
\end{itemize}
Since we have already construct the $su(N)$-invariant $R$-matrix as in Eq. (\ref{eqn:eqd3}), all we need to do is let $N=3$
\begin{equation}
R(u,v)=I+g(u,v)P,
\label{eqn:eqd13}
\end{equation}
with identity matrix $I$ and the permutation matrix $P$ are $9\times 9$ matrices:
\begin{equation}
I_{ij}^{kl}=\delta_{ij}\delta_{kl},\quad P_{ij}^{kl}=\delta_{il}\delta_{jk}, \quad \ i,j,k,l=1,2,3,
\label{eqn:eqd14}
\end{equation}
where we have used the lower two indices to denote the first space, and the upper two indices to denote the second space. And also the $su(2)$-invariant $R$-matrix, which we denote by $r(u,v)$:
\begin{equation}
r(u,v)=\mathbf{1}+g(u,v)\mathbf{p}.
\label{eqn:eqd15}
\end{equation}
where the identity matrix $\mathbf{1}$ and the permutation matrix $\mathbf{p}$ have the same form as in Eq. (\ref{eqn:eqd14}), but with less components, are $4\times 4$ matrices. The monodromy matrix is the same as in Eq. (\ref{eqn:eqd7}), we can rewrite it in the expansion of the auxiliary space $V_0$

\begin{equation}
\begin{split}
T(u)&=\begin{pmatrix}T_{11}(u)&T_{12}(u)&T_{13}(u)\\T_{21}(u)&T_{22}(u)&T_{23}(u)\\T_{31}(u)&T_{32}(u)&T_{33}(u)\end{pmatrix}\\
&=\begin{pmatrix}A(u)&\mathbb{B}(u)\\\mathbb{C}(u)&\mathbb{D}(u)\end{pmatrix}
=\begin{pmatrix}A(u)&B_{1}(u)&B_{2}(u)\\C_{1}(u)&D_{11}(u)&D_{12}(u)\\C_{2}(u)&D_{21}(u)&D_{22}(u)\end{pmatrix}.
\end{split}
\label{eqn:eqd16}
\end{equation}
For a general representation of $su(N)$ group, we cannot use Eq. (\ref{eqn:eqd7}) to build the monodromy matrix anymore, we need to construct the corresponding Lax operator $L_{0i}$ such that 

\begin{equation}
R_{12}(u,v)L_1(u)L_2(v)=L_2(v)L_1(u)R_{12}(u,v),
\label{eqn:eqd17}
\end{equation}
and also for monodromy matrix
\begin{equation}
T(u)=L_{0L}(u)\dots L_{01}(u).
\label{eqn:eqd18}
\end{equation}
But this is irrelevant for now, we will focus on fundamental representation. 

Generally speaking, that the operator $T_{ij}$ with $i<j$ are the creation operator, but for our case, the operator $T_{23}$ will annihilates the vacuum vector: $T_{23}\lvert 0\rangle=0$. This can be proved by induction \cite{intro_ABA}.\footnote{For one site spin chain $T(u)=R(u,0)$, then one can get $T_{23}(u)=E_{23}+g(u,0)E_{32}$. Recall that we choose vacuum as $\lvert 0\rangle=\begin{pmatrix}1\\0\\0\end{pmatrix}$, so $T_{23}(u)\lvert 0\rangle=0$. One can easily generalize this to $L$ site by $T_{23}^L(u)=T_{21}^{L-1}(u)T_{13}^{(1)}(u)+T_{22}^{L-1}(u)T_{23}^{(1)}(u)+T_{23}^{L-1}(u)T_{33}^{(1)}(u)$. This actually is true for general model if we are focusing on the defining representation.}

Before going deep into the calculations, we first want to use the commutation relations Eq. (\ref{eqn:eqd10}) give us some intuition. Denote $T_{ii}(u)\lvert 0\rangle=\lambda_i\lvert0\rangle$, and we hit a excitation state, say $T_{13}\lvert 0\rangle$, by the transfer matrix, the commutation relation gives us 

\begin{equation}
\begin{split}
\mathrm{Tr}\, T(z)T_{13}(u)\lvert 0\rangle&=\left(f(u,z)\lambda_1(z)+\lambda_2(z)+f(z,u)\lambda_3(z)\right)T_{13}\lvert 0\rangle\\
&\quad \quad +g(z,u)(\lambda_1(u)-\lambda_3(u))T_{13}\lvert 0\rangle\\
&\quad \quad +g(z,u)\left(T_{12}(u)T_{23}(z)-T_{12}(z)T_{23}(u)\right)\lvert 0\rangle.
\end{split}
\label{eqn:eqd19}
\end{equation}
we call the second line \textit{the unwanted term of the first type}, and the third line \textit{the unwanted term of the second type}. Just like in $su(2)$ case, when we impose the state to be the eigenstate of the transfer matrix, we can get a set of Bethe equations by killing the  unwanted term of the first type, but now we will get two sets of Bethe equations by killing these two unwanted terms. Here the unwanted term of the second type automatically vanishes because $T_{23}\lvert 0\rangle=0$, but it is not the case for neither $\mathrm{Tr}\,  T(z)T_{12}(u)\lvert 0\rangle$ nor a general state. 
The condition that $T_{23}(u)\lvert 0\rangle=0$ will make $\lambda_2(u)=\kappa \lambda_3(u)$, where $\kappa$ is a constant. Follows Eq. (\ref{eqn:eqd10}) 
\begin{equation}
[T_{23}(u),T_{32}(v)]=g(u,v)(T_{22}(u)T_{33}(v)-T_{22}(v)T_{33}(u)).
\label{eqn:eqd20}
\end{equation}
Apply this equation on $\lvert 0\rangle$ we obtain
\begin{equation}
0=g(u,v)(\lambda_2(u)\lambda_3(v)-\lambda_2(v)\lambda_3(u))\lvert 0\rangle
\label{eqn:eqd21}
\end{equation}
Therefore, the ratio $\lambda_2(u)/\lambda_3(u)$ does not depend on $u$. Normalize the monodromy matrix such that $\lambda_3(u)=1$, and we assume $\lambda_2(u)=1$ for simplicity. Also from Eq. (\ref{eqn:eqd7}) one can see that $\lambda_1(u)=f(u,\bar{\xi})$ because 

\begin{equation}
\begin{split}T^{(L)}_{11}(u)\lvert 0\rangle &= T_{11}^{(L-1)}(u)T_{11}^{(1)}(u)\lvert 0\rangle +T_{12}^{L-1}(u)T_{21}^{(1)}(u)\lvert 0\rangle +T_{13}^{L-1}(u)T_{31}^{(1)}(u)\lvert 0\rangle \\ &=(T_{11}^{(1)}(u))^L\lvert 0\rangle+0\\
&=f(u,\bar\xi)\lvert 0\rangle,
\end{split}
\label{eqn:eqd22}
\end{equation}
put all the eigenvalues of transfer matrix together
\begin{equation}
\lambda_1(u)=f(u,\bar{\xi}),\quad\lambda_2(u)=\lambda_3(u)=1.
\label{eqn:eqd23}
\end{equation}
\subsection{Action of the operator $D_{\xi\alpha}$}
In our circumstance, there are only two creation operator: $T_{12}(u)=B_1(u)$ and $T_{13}=B_2(u)$. We can try as usual that 
\begin{equation}
B_{\beta_1}(u_1)\dots B_{\beta_a}(u_a)\lvert 0\rangle,\quad a=0,1,\dots,
\label{eqn:eqd24}
\end{equation}
as the eigenstate for the transfer matrix. Here $\beta_i=1$ or $\ 2$. Let act the transfer matrix 

\begin{equation}
\mathrm{Tr}\, T(z)=A(z)+D_{11}(z)+D_{11}(z)
\label{eqn:eqd25}
\end{equation}
on this vector. 
Let's start with the action of the operator $D_{\alpha\alpha}(z)$. From Eq. (\ref{eqn:eqd10}), one can obtain
\begin{equation}
D_{\xi\alpha}(z)B_{\beta}(u)=B_{\beta}(u)D_{\xi\alpha}(z)+g(z,u)B_{\alpha}(u)D_{\xi\beta}(z)+g(u,z)B_{\alpha}(z)D_{\xi\beta}(u),
\label{eqn:eqd26}
\end{equation}
we can see that we may get the the unwanted terms of the second type. Indeed, the operator $D_{11}$ acting on $B_2$ gives contributions with $B_1$, and the operator $D_{22}$ acting on $B_1$ gives contributions with $B_2$. Thus, vector Eq. (\ref{eqn:eqd24}) is not invariant under the action of $D_{11}(z)+D_{22}(z)$. Meanwhile, as we will show later that the action of $A(z)$ does not generate the unwanted term of the second type. So Eq. (\ref{eqn:eqd24}) definite can not be the eigenvector of the transfer matrix. 

A explicit way to solve the problem is to construct a state as some kind superposition of state \cite{su3_norm} like Eq. (\ref{eqn:eqd24}), 

\begin{equation}
\vert\Psi_a(\bar{u})\rangle=\sum_{\beta_1,\dots,\beta_a}B_{\beta_1}(u_1)\dots B_{\beta_a}(u_a)F_{\beta_1\dots\beta_a}\vert 0\rangle,\quad a=0,1,\dots
\label{eqn:eqd27}
\end{equation}
with $\#a$ momentum carry excitations. Then when acting with $D_{\xi \alpha}$, the state will close. Here $F_{\beta_1\dots\beta_a}$ are some numerical coefficients. The sum is taken over every $\beta_{i}$, each $\beta_i$ takes value $1$ or $2$.

We can rewrite Eq. (\ref{eqn:eqd27}) in the form of a scalar product. For this we introduce a two component vector $\mathbb{B}(u)=(B_{1}(u),B_{1}(u))$ with operator-value components. Then we can write down the vector $\lvert \Psi_a(\bar{u})\rangle$ as the scalar product 

\begin{equation}
\lvert \Psi_a(\bar{u})\rangle=\mathbb{B}_1(u_1)\mathbb{B}_2(u_2)\dots\mathbb{B}_a(u_a)\mathbb{F}(\bar{u})\lvert 0\rangle,
\label{eqn:eqd28}
\end{equation}
where $\mathbb{F}(\bar{u})$ is a vector belonging to the space
\begin{equation}
\overbrace{\mathbb{C}^2\otimes\dots\otimes\mathbb{C}^2}^{a\quad times}.
\label{eqn:eqd29}
\end{equation}
The commutation relations Eq. (\ref{eqn:eqd26}) can be written as follows 
\begin{equation}
\mathbb{D}_0(z)\mathbb{B}_1(u)=\mathbb{B}_1(u)\mathbb{D}_0(z)r_{01}(z,u)+g(u,z)\mathbb{B}_1(z)\mathbb{D}_0(u)p_{01},
\label{eqn:eqd30}
\end{equation}
where $p_{01}$ is a $4\times 4$ permutation matrix, and $r_{01}$ is a $R$-matrix for $su(2)$. We call the first term on the R.H.S. of Eq. (\ref{eqn:eqd30}) $the\ first\ scheme\ of \ commutation$. The second term in the R.H.S. of Eq. (\ref{eqn:eqd30}) is called $the\ second\ scheme\ of \ commutation$. Also from the commutation relations Eq. (\ref{eqn:eqd10}), the commutation relation of $D$ operators reduce to $su(2)$ case

\begin{equation}
[D_{\alpha\beta}(u),D_{\gamma\delta}(v)]=g(u,v)(D_{\gamma\beta}(v)D_{\alpha\delta}(u)-D_{\gamma\beta}(u)D_{\alpha\delta}(v)).
\label{eqn:eqd31}
\end{equation}
 Hence, the matrix $\mathbb{D}(u)$ satisfies the $RTT$-relation with the $R$-matrix $r(u,v)$
\begin{equation}
r_{12}(u,v)\mathbb{D}_1(u)\mathbb{D}_2(v)=\mathbb{D}_2(v)\mathbb{D}_1(u)r_{12}(u,v).
\label{eqn:eqd32}
\end{equation}
So $\mathbb{D}(u)$ can be treated as the monodromy matrix of a model with $su(2)$-invariant $R$-matrix.
Act $\mathrm{tr}\, \mathbb{D}(z)$ on $\lvert \Psi_a(\bar{u})\rangle$, we have
\begin{equation}
\begin{split}
\mathrm{tr}_0\,\mathbb{D}_0(z)\lvert \Psi_a(\bar{u})\rangle&=\mathrm{tr}_0\,\mathbb{D}_0(z)\mathbb{B}_1(u_1)\mathbb{B}_2(u_2)\dots\mathbb{B}_a(u_a)\mathbb{F}(\bar{u})\lvert 0\rangle
\\&=\mathrm{tr}_0\,\left(\mathbb{B}_1(u_1)\mathbb{D}_0(z)r_{01}(z,u_1)+g(u_1,z)\mathbb{B}_1(z)\mathbb{D}_0(u_1)\right)\\
&\qquad \qquad \qquad \times\mathbb{B}_2(u_2)\dots\mathbb{B}_a(u_a)\mathbb{F}(\bar{u})\lvert 0\rangle,
\end{split}
\label{eqn:eqd33}
\end{equation}
upon using the commutation relation,  we have two contributions. The second one is not what we want, since it changes the spectral parameters. As in $su(2)$ ABA case, we can deal with the first term only, leave the second term for now. So we have 
\begin{equation}
\mathrm{tr}_0\,\mathbb{D}_0(z)\lvert \Psi_a(\bar{u})\rangle=\mathrm{tr}_0\,\mathbb{B}_1(u_1)\mathbb{D}_0(z)r_{01}(z,u_1)\mathbb{B}_2(u_2)\dots\mathbb{B}_a(u_a)\mathbb{F}(\bar{u})\lvert 0\rangle+\mathcal{Z},
\label{eqn:eqd34}
\end{equation}
where $\mathcal{Z}$ denote the unwanted term, and it will have the same meaning for the rest of the note, but maybe represents different terms. Finally we will end up with

\begin{equation}
\begin{split}
\mathrm{tr}_0\,\mathbb{D}_0(z)\lvert \Psi_a(\bar{u})\rangle=&\mathrm{tr}_0\,\mathbb{B}_1(u_1)\mathbb{B}_2(u_2)\dots\mathbb{B}_a(u_a)\mathbb{D}_0(z)r_{0a}(z,u_a)\dots r_{01}(z,u_1)\mathbb{F}(\bar{u})\lvert 0\rangle\\&+\mathcal{Z}.
\end{split}
\label{eqn:eqd35}
\end{equation}
We have obtained a matrix $\hat{\mathcal{T}}^{(a)}(z)$

\begin{equation}
\hat{\mathcal{T}}_0^{(a)}(z)=\mathbb{D}_0(z)\mathcal{T}_0^{(a)}(z),\quad \mathrm{where}\quad \mathcal{T}_0^{(a)}(z)=r_{0a}(z,u_a)\dots r_{01}(z,u_1).
\label{eqn:eqd36}
\end{equation}
Recall that $\mathbb{D}_0(z)$ can be treated the $su(2)$-invariant monodromy matrix. Its matrix element act in the original Hilbert space $\mathcal{H}$ as follows

\begin{equation}
\begin{split}
D_{11}(z)\lvert0\rangle=T_{22}(z)\lvert0\rangle=1,\quad D_{22}(z)\lvert0\rangle=T_{33}(z)\lvert0\rangle=1,\\
D_{12}(z)\lvert0\rangle=T_{23}(z)\lvert0\rangle=0, \quad D_{21}(z)\lvert0\rangle=T_{32}(z)\lvert0\rangle=0.
\end{split}
\label{eqn:eqd37}
\end{equation}
The matrix $\mathcal{T}_0^{(a)}(z)$ is the new monodromy matrix of the $su(2)$-invariant $XXX$ chain of the length $a$,

\begin{equation}
\mathcal{T}^{(a)}(z)=\begin{pmatrix}\mathcal{A}^{(a)}(z)&\mathcal{B}^{(a)}(z)\\\mathcal{C}^{(a)}(z)&\mathcal{D}^{(a)}(z)\end{pmatrix}.
\label{eqn:eqd38}
\end{equation}
And the physical space is $\mathcal{H}^{(a)}$ with the vacuum chosen to be 

\begin{equation}
\lvert\Omega^{(a)}\rangle=\overbrace{\begin{pmatrix}1\\0\end{pmatrix}\otimes\dots\begin{pmatrix}1\\0\end{pmatrix}}^{a\ \mathrm{times}},
\label{eqn:eqd39}
\end{equation}
consequently 

\begin{equation}
\mathcal{A}^{(a)}(z)\lvert\Omega^{(a)}\rangle=f(z,\bar{u})\lvert\Omega^{(a)}\rangle,\quad \mathcal{D}^{(a)}(z)\lvert\Omega^{(a)}\rangle=\lvert\Omega^{(a)}\rangle.
\label{eqn:eqd40}
\end{equation}
One can check the above relation by using the explicit expression of $\mathcal{T}_0^{(a)}(z)$. So $\hat{\mathcal{T}}_0^{(a)}(z)$ is the product of two monodromy matrices whose entries act in different spaces. So

\begin{equation}
\begin{split}
\mathrm{tr}_0\,\hat{\mathcal{T}}_0^{(a)}(z)&\mathbb{F}(\bar{u})\lvert 0\rangle\\
&=\left(D_{11}(z)\mathcal{A}^{(a)}(z)+D_{12}(z)\mathcal{C}^{(a)}(z)+D_{21}(z)\mathcal{B}^{(a)}(z)+D_{22}(z)\mathcal{D}^{(a)}(z)\right)\mathbb{F}(\bar{u})\lvert 0\rangle\\
&=\left(\mathcal{A}^{(a)}(z)+\mathcal{D}^{(a)}(z)\right)\mathbb{F}(\bar{u})\lvert 0\rangle=\mathrm{tr}\,\mathcal{T}^{(a)}(z)\mathbb{F}(\bar{u})\lvert 0\rangle.
\end{split}
\label{eqn:eqd41}
\end{equation}
Thus, if we do not want to have unwanted terms of the second type in the action of $\mathrm{tr}\, \mathbb{D}(z)$, we should demand that $\mathbb{F}(\bar{u})$ be an eigenvector of the transfer matrix $\mathrm{tr}\,\mathcal{T}^{(a)}(z)$. Hence, it have the form

\begin{equation}
\mathbb{F}(\bar{u})=\mathcal{B}^{(a)}(v_1)\dots\mathcal{B}^{(a)}(v_b)\lvert\Omega^{(a)}\rangle
\label{eqn:eqd42}
\end{equation}
and the set $\bar{v}=\{v_1,\dots,v_b\}$ satisfies the Bethe equations from Eq. (38)

\begin{equation}
f(v_j,\bar{u})=\frac{f(v_j,\bar{v}_j)}{f(\bar{v}_j,v_j)}, \quad j=1,\dots,b,
\label{eqn:eqd43}
\end{equation}
where $b\le a$ as the number of non-momentum carry excitations, since this excitations is built upon the short chain. Written Eq. (\ref{eqn:eqd43}) out, 
\begin{equation}
\begin{split}
\prod_{l=1}^{a}(1+\frac{i}{v_j-u_l})&=\prod_{m\ne j}^{b}\frac{1+\frac{i}{v_j-v_m}}{1+\frac{i}{v_m-v_j}}\quad j=1,\dots,b\\
\Leftrightarrow\prod_{l=1}^{a}(\frac{v_j-u_l+\frac{i}{2}}{v_j-u_l-\frac{i}{2}})&=\prod_{m\ne j}^{b}\frac{v_j-v_m+i}{v_j-v_m-i}\quad j=1,\dots,b,
\end{split}
\label{eqn:eqd44}
\end{equation}
where in the second line I have shift the parameter $u_l\rightarrow u_l+\frac{i}{2}$. Also one can immediately get the eigenvalue for Eq. (\ref{eqn:eqd41})

\begin{equation}
\mathrm{tr}_0\,\mathbb{D}_0(z)\lvert \Psi_{a,b}(\bar{u},\bar{v})\rangle=\tau_{D}(z\lvert\bar{u},\bar{v})\lvert \Psi_{a,b}(\bar{u},\bar{v})\rangle+\mathcal{Z}
\label{eqn:eqd45}
\end{equation}
with 
\begin{equation}
\tau_{D}(z\lvert\bar{u},\bar{v})=f(z,\bar{u})f(\bar{v},z)+f(z,\bar{v}).
\label{eqn:eqd46}
\end{equation}
And we follow the same procedure as before, and use the properties that $\lvert \Psi_{a,b}(\bar{u},\bar{v})\rangle$ is totally symmetric under $u$, finally we will end up with

\begin{equation}
\mathrm{tr}_0\,\mathbb{D}_0(z)\lvert \Psi_{a,b}(\bar{u},\bar{v})\rangle=\tau_{D}(z\lvert\bar{u},\bar{v})\lvert \Psi_{a,b}(\bar{u},\bar{v})\rangle+\sum_{k=1}^ag(u_k,z)f(u_k,\bar{u}_k)f(\bar{v},u_k)\lvert \Phi_{a,b}(z,u_k,\bar{u},\bar{v})\rangle
\label{eqn:eqd47}
\end{equation}
where

\begin{equation}
\lvert \Phi_{a,b}(z,u_k,\bar{u},\bar{v})\rangle=\mathbb{B}_1(u_1)\mathbb{B}_2(u_2)\dots\mathbb{B}_k(z)\dots\mathbb{B}_a(u_a)\mathbb{F}(\bar{u})\lvert 0\rangle,
\label{eqn:eqd48}
\end{equation}
\subsection{Action of the operator A(z)}
The commutation relation between $A(z)$ and $B_{\beta}(u)$ is 
\begin{equation}
A(z)B_{\beta}(u)=f(u,z)B_{\beta}A(z)+g(z,u)B_{\beta}(z)A(u).
\label{eqn:eqd49}
\end{equation}
Again as before, repeat all the moving $A(z)$ from left to the right of the state
\begin{equation}
A(z)\vert \Psi_{a,b}(\bar{u},\bar{v})\rangle=\tau_A(z\vert\bar{u},\bar{v})\vert \Psi_{a,b}(\bar{u},\bar{v})\rangle+\sum_{k=1}^a
\Lambda_k\vert \Phi_{a,b}(z,u_k,\bar{u},\bar{v})\rangle
\label{eqn:eqd50}
\end{equation}
with 
\begin{equation}
\tau_A(z\vert\bar{u},\bar{v})=\lambda_1(z)f(\bar{u},z),\quad \Lambda_k=\lambda_1(u_k)g(z,u_k)f(\bar{u}_k,u_k).
\label{eqn:eqd51}
\end{equation}
Combine Eq. (\ref{eqn:eqd47}) and Eq. (\ref{eqn:eqd50}), 

\begin{equation}
\mathrm{Tr}\, T(z) \vert \Psi_{a,b}(\bar{u},\bar{v})\rangle=\tau(z\vert\bar{u},\bar{v})\vert \Psi_{a,b}(\bar{u},\bar{v})\rangle+\sum_{k=1}^aM_k\vert\Phi_{a,b}(z,u_k,\bar{u},\bar{v})\rangle,
\label{eqn:eqd52}
\end{equation}
with
\begin{equation}
\tau(z\vert \bar{u},\bar{v})=\tau_D(z\vert\bar{u},\bar{v})+\tau_A(z\vert\bar{u},\bar{v}),\quad M_k=\Lambda_k+g(u_k,z)f(u_k,\bar{u}_k)f(\bar{v},u_k).
\label{eqn:eqd53}
\end{equation}
Set $M_k=0$, we will get another set of Bethe equations
\begin{equation}
\lambda_1(u_k)=\frac{f(u_k,\bar{u}_k)}{f(\bar{u}_k,u_k)}f(\bar{v},u_k), \quad k=1,\dots, a
\label{eqn:eqd54}
\end{equation}
Written Eq. (\ref{eqn:eqd54}) out, and remember we have shift our momentum carrying Bethe roots $u_l\rightarrow u_l+\frac{i}{2}$, so 

\begin{equation}
\begin{split}
\prod_{l=1}^{L}\frac{u_k-\xi_l+\frac{3i}{2}}{u_k-\xi_l+\frac{i}{2}}&=\prod_{j\ne k}^{a}\frac{u_k-u_j+i}{u_k-u_j-i}\prod_{m=1}^{b}\frac{u_k-v_m-\frac{i}{2}}{u_k-v_m+\frac{i}{2}},\quad k=1,\dots, a,\\
\Leftrightarrow
\Big(\frac{u_k+\frac{i}{2}}{u_k-\frac{i}{2}}\Big)^L&=\prod_{j\ne k}^{a}\frac{u_k-u_j+i}{u_k-u_j-i}\prod_{m=1}^{b}\frac{u_k-v_m-\frac{i}{2}}{u_k-v_m+\frac{i}{2}},\quad k=1,\dots, a,
\end{split}
\label{eqn:eqd55}
\end{equation}
where in the second line I have set $\xi=i$ for all sites. Eq. (\ref{eqn:eqd44}) and Eq. (\ref{eqn:eqd55}) give us the Nested Bethe equations for $su(3)$ system as we want.

\section{GL(N)}
For general $GL(N)$\footnote{We typically do not distinguish $GL(N)$, $SL(N)$ and $SU(N)$. The $GL(N)$ system and $SL(N)$ system differ only up to a center or quantum determinant \cite{tyangian}, while on the complex field, $SL(N)$ and $SU(N)$ are basical the same algebra.}-invariant $XXX$ spin chain, the R-matrix takes the same form as in Eq. (3). The $N\times N$ monodromy matrix takes the form
\begin{equation}
T(u)=\begin{pmatrix}A(u)&\mathbb{B}(u)\\\mathbb{C}(u)&\mathbb{D}(u)\end{pmatrix}.
\label{eqn:eqd56}
\end{equation}

Notice we set $\mathbb{D}$ as the matrix of size $(N-1)\times (N-1)$. Using the same argument, one can see that $T_{ij},\ i=2,\dots,N;\ i<j$ will annihilate the vacuum state Eq. (\ref{eqn:eqd9}). The Bethe states look similar to Eq. (\ref{eqn:eqd28})
\begin{equation}
\lvert\Psi\rangle=\mathbb{B}_1(u_1)\dots\mathbb{B}_{a}(u_a)\mathbb{F}\lvert0\rangle.
\label{eqn:eqd57}
\end{equation}
Act with transfer matrix $t(u)=A(u)+\mathrm{Tr}_a\, D_a(u)$, for the unwanted term to cancel, we need to impose Bethe equation for $N-1$ level excitations. In the meantime, we have reduced $GL(N-1)$- invariant $XXX$ spin chain. See \cite{Kulish1983GL_N} for details.
% Appendix E

\chapter{Fusion procedure and Tableau sum}  % Main appendix title

\label{AppendixE} % For referencing this appendix elsewhere, use \ref{AppendixE}
We give a short introduction for $gl(N)$ spin chain fusion procedure. Start with the fundamental R-matrix in $\mathbb{C}^N\otimes \mathbb{C}^N$, we can build the R-matrix in $V^{\lambda}\otimes \mathbf{C}^N$ with physical space in the representation $\lambda$. The consistency of the fusion procedure depends on the Yang-Baxter equation for R-matrix, consequently the RTT relation. Consequently, one would get a set of commuting transfer matrix, these transfer matrices are functionally dependent. They satisfy a number of functional relation called fusion relations, which looks quite similar to the relation for characters of general linear groups. Let us start the fusion. 

\section{Quantum fusion relations}
\subsection{Notion}
The fundamental R-matrix acting in $\mathbb{C}^N\otimes \mathbb{C}^N$ has the form
\begin{equation}
R(u)=u+2\mathcal{P},
\label{eqn:eqe1}
\end{equation}
also the R-matrix satisfies the Yang-Baxter equation

\begin{equation}
R_{12}(u_1-u_2)R_{13}(u_1-u_3)R_{23}(u_2-u_3)=R_{23}(u_2-u_3)R_{13}(u_1-u_3)R_{12}(u_1-u_2).
\label{eqn:eqe2}
\end{equation}
The monodromy matrix ($\mathcal{T}\mbox{-}$matrix) is the following product of R-matrix in the auxiliary space $V_0=\mathbb{C}^N$:

\begin{equation}
\mathcal{T}(u)=R_{0L}(u-y_L)\dots R_{01}(u-y_1),
\label{eqn:eqe3}
\end{equation}
it acts on the quantum space $\otimes_{i=1}^{L} V_i,\ V_i=\mathbb{C}^N$. Then the RTT relation is a direct consequence of Eq. (\ref{eqn:eqe2})

\begin{equation}
R_{12}(u-v)\mathcal{T}_{13}(u)\mathcal{T}_{23}(v)=\mathcal{T}_{23}(v)\mathcal{T}_{13}(u)R_{12}(u-v),
\label{eqn:eqe4}
\end{equation}
where the two auxiliary spaces are $\mathcal{V}_1=\mathcal{V}_2=\mathbb{C}^N$ and the quantum space is $\mathcal{V}_3=\otimes_{i=1}^{L} V_i$. 

\subsection{Fusion procedure}
Using the R-matrix Eq. (\ref{eqn:eqe1}) as a building block, it is possible to construct more complicated solutions to the Yang-Baxter equation. Notice that the special points $\pm2$ of the fundamental R-matrix are symmetric projector and anti-symmetric projector respectively. Thus one can project our fundamental space to arbitrary subspace by such projector.  
Let $\lambda=(\lambda_1,\lambda_2,\dots,\lambda_m)$ be a Young diagram with n boxes and m lines $\lambda_1\ge \lambda_2\ge \dots \ge\lambda_m$, $m\le k$, $\sum_{i=1}^{m}\lambda_i=n$. Let 
\begin{equation}
P_{\lambda}:\otimes_{i=1}^{n} V_i\rightarrow V^{(\lambda)}
\label{eqn:eqe5}
\end{equation}
be the projection operator on the space of the irreducible representation of $GL(N)$ corresponding to $\lambda$. Assign the box with coordinates $(i,j)$ ($i$-th row and $j$-th column) with the number

\begin{equation}
s_{(ij)}=u-2(i-j).
\label{eqn:eqe6}
\end{equation}
\begin{figure}
    \centering
    \ytableausetup{centertableaux}
\begin{ytableau}
\scriptstyle u & \scriptstyle u+2 & \scriptstyle u+4 &\scriptstyle u+6 \\
\scriptstyle u-2 & \scriptstyle u & \scriptstyle u+2 \\
\scriptstyle u-4
\end{ytableau}
    \caption{Young diagram for $\lambda=(4,3,1)$ }
    \label{fig:young diagram}
\end{figure}

For example $\lambda=(4,3,1)$, see Fig. \ref{fig:young diagram}. 
Label all the boxes in the order that start from left to right in the first row, then continue for second row and so on, one gets a set of numbers $s_1, s_2,\dots, s_n$ (in our example $s_1=u,\ s_2=u+2,\ s_3=u+4,\ s_4=u+6,$$\ s_5=u-2,\ s_6=u,\ s_7=u+2,\ s_8=u-4 $).
The R-matrix acting in $V^{(\lambda)}\otimes \mathbb{C}^N$ is 

\begin{equation}
R^{(\lambda)}(u)=P_{\lambda}R_{n0}(s_n)\otimes\dots\otimes R_{20}(s_2)\otimes R_{10}(s_1)P_{\lambda}
\label{eqn:eqe7}
\end{equation}
with 

\begin{equation}
P_{\lambda}\propto \prod_{i<j}^{n}R_{ij}(s_i-s_j)
\label{eqn:eqe8}
\end{equation}
The fundamental R-matrix $R_{i0}(s_i)$ are tensor multiplied in the auxiliary spaces $V_i=\mathbf{C}^k$, their entries being multiplied in the common quantum space $V_0=\mathbf{C}^k$ in the indicated order. The $R^{(\lambda)}(u)$ satisfies the Yang-Baxter equation in $\mathcal{V}_1\otimes\mathcal{V}_2\otimes\mathcal{V}_3$ space 

\begin{equation}
R_{12}^{(\lambda)}(u_1-u_2)R_{13}^{(\lambda)}(u_1-u_3)R_{23}(u_2-u_3)=R_{23}(u_2-u_3)R_{13}^{(\lambda)}(u_1-u_3)R_{12}^{(\lambda)}(u_1-u_2),
\label{eqn:eqe9}
\end{equation}
where $\mathcal{V}_1=V^{(\lambda)},\ \mathcal{V}_2=\mathcal{V}_3=\mathbb{C}^N$.
The proof is simple: one can replace the projector $P_{\lambda}$ as a product of fundamental R-matrix, and then use the Yang-Baxter equation (\ref{eqn:eqe2})\footnote{Due to the naive assignment Eq. (9), one would meet $R(0)$ in the product, then the complementary projectors $P^{+}$ and $P^{-}$ meet together, our $P_{\lambda}=0$ in this case. So one needs to do some regularization,  see \cite{fusion_procedure} for more detail.}.
The fusion procedure in the quantum space is made in a similar way. This procedure allows one to define the R-matrix $R^{(\lambda)(\mu)}(u)$ acting in $V(\lambda)\otimes V(\mu)$ for two arbitrary Young diagrams $\lambda, \mu$. The Yang-Baxter Equation then reads
\begin{equation}
R_{12}^{(\lambda)(\mu)}(u_1-u_2)R_{13}^{(\lambda)}(u_1-u_3)R_{23}^{(\mu)}(u_2-u_3)=R_{23}^{(\mu)}(u_2-u_3)R_{13}^{(\lambda)}(u_1-u_3)R_{12}^{(\lambda)(\mu)}(u_1-u_2),
\label{eqn:eqe10}
\end{equation}
where $\mathcal{V}_1 = V^{(\lambda)}$, $\mathcal{V}_1 = V^{(\mu)}$, $\mathcal{V}_3 = \mathbb{C}^N$. One can use the same procedure as before to prove this equation by using Eq. (\ref{eqn:eqe9}). The monodromy matrix in this case would be 

\begin{equation}
\mathcal{T}^{(\lambda)}(u)=R_{0L}^{(\lambda)}(u-y_L)\dots R_{02}^{(\lambda)}(u-y_2)R_{01}^{(\lambda)}(u-y_1).
\label{eqn:eqe11}
\end{equation}
where the auxiliary space is $V_0=V^{(\lambda)}$ and quantum space is $\otimes_{i=1}^{L}V_i$. Follow Eq. (\ref{eqn:eqe10}) and Eq. (\ref{eqn:eqe11}), the RTT relation would be

\begin{equation}
R_{12}^{(\lambda)(\mu)}(u-v)\mathcal{T}_{13}^{(\lambda)}(u)\mathcal{T}_{23}^{(\mu)}(v)=\mathcal{T}_{23}^{(\mu)}(v)\mathcal{T}_{13}^{(\lambda)}(u)R_{12}^{(\lambda)(\mu)}(u-v)
\label{eqn:eqe12}
\end{equation}
where the two auxiliary spaces are $\mathcal{V}_1=V^{(\lambda)},\mathcal{V}_1=V^{(\mu)}$, and the quantum space is $\mathcal{V}_3=\otimes_{i=1}^{L}V_i$. 
We proceed by introducing the transfer matrix, obtained by taking the trace of $\mathcal{T}^{(\lambda)}(u)$ in the auxiliary space:
\begin{equation}
T^{(\lambda)}(u)=\mathrm{Tr}_{aux}\mathcal{T}^{\lambda}(u-\lambda_1+\lambda_1^{\prime})
\label{eqn:eqe13}
\end{equation}
where $\lambda_i^{\prime}$ denotes the height of the $i\mbox{-}th$ column of $\lambda$ , and $\lambda_i$ are the corresponding Dynkin label. The shift of the spectral parameter is introduce for later convenience. Using Eq. (\ref{eqn:eqe1}), one can easily prove that the transfer matrices are commute regardless of spectral parameter $u$ and representation $\lambda$
\begin{equation}
[T^{(\lambda)}(u),T^{(\mu)}(v)]=0.
\label{eqn:eqe14}
\end{equation}
For the case of rectangular diagrams $a\times m$ (height $a$ and length $m$) we introduce the special notation
\begin{equation}
T_{m}^{a}(u)\equiv T^{(a\times m)}(u).
\label{eqn:eqe15}
\end{equation}
\subsection{Functional relations (Fusion rules)}
A combination of the fusion procedure and the Yang-Baxter equation results in numerous functional relations (fusion rules) for quantum transfer matrices.

We illustrate the origin of these relations on the simplest example,
\begin{equation}
T_{1}^{1}(u+1)T_{1}^{1}(u-1)=T_{1}^{2}(u)+T_{2}^{1}(u).
\label{eqn:eqe16}
\end{equation}
Consider the product
\begin{equation}
T_{1}^{1}(u+1)T_{1}^{1}(u-1)=\mathrm{Tr}_{V_1\otimes V_2}(\mathcal{T}_{20}(u+1)\otimes \mathcal{T}_{10}(u-1)),
\label{eqn:eqe17}
\end{equation}
where $V_1=V_2=\mathbb{C}^N$ are auxiliary spaces and the quantum space $V_0$ is in fundamental representation . Using the cyclicity of the trace, the property $(P^{\pm})^{2}=P^{\pm},\ P^{+}P^{-}=0$ and the Yang-Baxter equation for $R(-2)=P^{-}$,

\begin{equation}
R_{12}(-2)R_{13}(u-1)R_{23}(u+1)=R_{23}(u+1)R_{13}(u-1)R_{12}(-2),
\label{eqn:eqe18}
\end{equation}
then

\begin{equation}
\begin{split}
T_{1}^{1}&(u-1)T_{1}^{1}(u+1)=\mathrm{Tr}_{V_1\otimes V_2}(\mathcal{T}_{20}(u-1)\otimes \mathcal{T}_{10}(u+1))\\
&=\mathrm{Tr}_{V_1\otimes V_2}[(P_{21}^{+}+P_{21}^{-})R_{2,\alpha_N}(u-1)R_{1,\alpha_N}(u+1)(P_{21}^{+}+P_{21}^{-})\\&\dots(P_{21}^{+}+P_{21}^{-})R_{2,\alpha_1}(u-1)R_{1,\alpha_1}(u+1)(P_{21}^{+}+P_{21}^{-})]\\
&=\mathrm{Tr}_{V_1\otimes V_2}[(P_{21}^{+})R_{2,\alpha_N}(u-1)R_{1,\alpha_N}(u+1)(P_{21}^{+})\dots(P_{21}^{+})R_{2,\alpha_1}(u-1)R_{1,\alpha_1}(u+1)(P_{21}^{+})]\\
&\quad +\mathrm{Tr}_{V_1\otimes V_2}[(P_{12}^{-})R_{1,\alpha_N}(u+1)R_{2,\alpha_N}(u-1)(P_{12}^{-})\dots(P_{12}^{-})R_{1,\alpha_1}(u+1)R_{2,\alpha_2}(u-1)(P_{12}^{-})]\\
&=\mathrm{Tr}_{V_1\otimes V_2}(\mathcal{T}^{1\times 2}(u-1))+\mathrm{Tr}_{V_1\otimes V_2}(\mathcal{T}^{2\times 1}(u+1))\\
&=T_{2}^{1}(u)+T_{1}^{2}(u)
\end{split}
\label{eqn:eqe19}
\end{equation}
where the second equation I have used $I = P^{+} + P^{-}$, and the third equation follows Eq. (\ref{eqn:eqe17}), and the last one from the definition Eq. (\ref{eqn:eqe13}). In terms of group theory, one would get a similar relation
\begin{equation}
\ydiagram{1} \otimes \ydiagram{1} = \ydiagram{1,1} \oplus \ydiagram{2}
\label{eqn:eqe20}
\end{equation}
A more general relation derived in a similar way is 

\begin{equation}
T_{m}^{1}(u+1)T_{1}^{1}(u-m)=T^{(\lambda=(m,1))}(u)+T_{m+1}^{1}(u).
\label{eqn:eqe21}
\end{equation}
The Clebsch-Gordan decomposition of the corresponding tensor product of Young diagram yields the same relation. Proceeding further, it is possible to show that $T^{\lambda}(u)$ can be expressed through either $T^{1}_m(u)$ or $T^{a}_1(u)$ only.

To get a general transfer matrix out of $T^{1}_m(u)$ or $T^{a}_1(u)$, we introduce the famous Bazhanov-Reshetikhin determinant formula (BR formula) \cite{BR_formula} for $GL(N)$,
\begin{equation}
\begin{split}
T^{(\lambda)}(u)=\det_{1\le i,j\le \lambda_1^{\prime}}(T^{1}_{\lambda_i-i+j}(u-\lambda_1^{\prime}+\lambda_1-\lambda_i+i+j-1)),\\
T^{(\lambda)}(u)=\det_{1\le i,j\le \lambda_1}(T_{1}^{\lambda_i^{\prime}-i+j}(u-\lambda_1+\lambda_1^{\prime}-\lambda_i^{\prime}+i+j-1))    
\end{split}
\label{eqn:eqe22}
\end{equation}
where $T^{\emptyset}(u)=1$, $\emptyset$ denotes the empty diagram.

For the rectangular diagram $T_{m}^{a}(u)$ for $A_r$, corresponding $\lambda_i=m,\ \lambda_j^{\prime}=a$, Eq. (\ref{eqn:eqe22}) reads
\begin{equation}
\begin{split}
T_{m}^{a}(u)=\det (T^{1}_{m-i+j}(u+i+j-a-1)),\quad i,j=1,\dots,a,\\
T_{m}^{a}(u)=\det (T_{1}^{a-i+j}(u+i+j-m-1)),\quad i,j=1,\dots,m ,
\end{split}
\label{eqn:eqe23}
\end{equation}
where $\ T_1^{(b)}(u)=T^1_{(b)}(u)=0$, unless $0\le b\le r+1$, and $\ T_1^{(0)}(u)=T_0^{(1)}(u)=T_1^{(r+1)}(u)=1$. These determinant formulas exhibits a very nice structure. However, they give only a partial solution because $T^1_m(u)$ or $T^a_1(u)$ (entering as ”input”) are still to be determined. 
\subsection{Bilinear form of the fusion rules}
It follows from Eq. (\ref{eqn:eqe23}) that transfer matrices for rectangular Young diagrams obey a closed set of relations among themselves. Using the Jacobi identity for determinants,
\begin{equation}
D\begin{bmatrix}m+1\\m+1\end{bmatrix}D\begin{bmatrix}1\\1\end{bmatrix}=D\begin{bmatrix}1,m+1\\1,m+1\end{bmatrix}D+D\begin{bmatrix}1\\m+1\end{bmatrix}D\begin{bmatrix}m+1\\1\end{bmatrix},
\label{eqn:eqe24}
\end{equation}
where $D\begin{bmatrix}i_1,i_2,\dots\\j_1,j_2,\dots\end{bmatrix}$is the minor of $D$ removing $i_k$'s rows and $j_k$'s columns. and they can be represented in the following model-independent bilinear form:

\begin{equation}
T_{m}^{a}(u+1)T_{m}^{a}(u-1)-T_{m+1}^{a}(u)T_{m-1}^{a}(u)=T_{m}^{a+1}(u)T_{m}^{a-1}(u).
\label{eqn:eqe25}
\end{equation}
Since $T^a_m(u)$ commute at different $u, a, m$, the same equation holds for all eigenvalues of the transfer matrix, so we can treat $T^a_m(u)$ as a number-valued function.
\section{Tableau sum for Type $A_r$}
Let $\fbox{1}_u,\dots,\fbox{r+1}_u$ be variables depending on $u$. If we set $T_{1}^{1}(u)=\sum_{a=1}^{r+1}\fbox{a}_u$, then 
\begin{equation}
T_{1}^{1}(u-1)T_{1}^{1}(u+1)=\sum_{a\le b}\fbox{a}_{u-1}\fbox{b}_{u+1}+\sum_{a>b}\begin{matrix}\fbox{b}_{u+1}\\\fbox{a}_{u-1}\end{matrix},
\label{eqn:eqe26}
\end{equation}
where the both arrays of the boxes stand for the product. Comparing this with the T-system relation Eq. (\ref{eqn:eqe16}), one may identify $T_{1}^{2}(u)$ and $T_{2}^{1}(u)$ individually with the two terms in Eq. (\ref{eqn:eqe26}) and try to further establish similar formulas for higher $T_m^a(u)$. Such a procedure leads to a solution of the T-system expressed as a sum of tableaux. In fact, if one forgets the spectral parameter $u$ in Eq. (\ref{eqn:eqe26}), it can be viewed as the identity among Schur functions corresponding to the irreducible decomposition of the $A_r$-modules Eq. (\ref{eqn:eqe20}).
In this sense the result presented in what follows for $A_r$ is a deformation of the classical tableau sum formula for the Schur functions.

Consider the Young diagram ($m^a$) of $a \times m$ rectangular shape. Let Tab($m^a$) be the set of semistandard tableaux on ($m^a$) with numbers $\{1, 2,\dots, r+1\}$. The inscribed numbers are strictly increasing to the bottom and non-decreasing to the right. 

Note that Tab($m^a$) is empty for $a>r+1$. We define \cite{T_system_Y_system}
\begin{equation}
T_u=\prod_{i=1}^a\prod_{j=1}^m\fbox{$t_{ij}$}_{u+a-m-2i+2j}\quad \mathrm{for}\ T=(t_{ij})\in \mathrm{Tab}(m^a),
\label{eqn:eqe27}
\end{equation}
where $t_{ij}$ denotes the entry of the box in the $i$-th row and $j$-th column from the top left. 
\begin{thm*}
\begin{equation}
T_{m}^{a}(u)=\sum_{T\in \mathrm{Tab}(m^a)}T_u \quad (1\le a\le r+1)
\label{eqn:eqe28}
\end{equation}
is a solution of the T-system for $A_r$ Eq.(\ref{eqn:eqe25})
\end{thm*}
We note that $T_m^{r+1}(u)$ here is not just 1 but non-trivially chosen as Eq. (\ref{eqn:eqe28}) as opposed to the original definition of the T-system. However, Tab($m^{r+1}$) consists of only one tableau. Therefore, Eq. (\ref{eqn:eqe28}) states that
\begin{equation}
T_{m}^{r+1}(u)=\prod_{j=1}^{m}T_1^{r+1}(u-m-1+2j),\quad T_1^{r+1}(u)=\prod_{i=1}^{r+1}\fbox{i}_{u+r+2-2i}.
\label{eqn:eqe29}
\end{equation}
Thus $T_{m}^{r+1}(u)=1$ can be restored if the variables $\fbox{1}_u,\dots,\fbox{r+1}_u$ are chosen so as to satisfy the simple relation $T_1^{r+1}(u)=1$. 

In the $A_r$ case, one can use Thermodynamic Bethe Ansatz to construct the corresponding building block $\fbox{a}_u$. In our model, the nested Bethe Ansatz involves $r$ sets of Bethe rapidities, which will be denoted as $\{\{\lambda_j^a\}_{j=1,\dots,N_a}\}_{a=1,\dots,r}$. The $Q$-functions are 

\begin{equation}
Q_0(u)=u^L,\quad Q_a(u)=\prod_{j=1}^{N_a} (u-\lambda_j^{(a)}),\quad Q_{r+1}(u)=1.
\label{eqn:eqe30}
\end{equation}
The eigenvalue of the fundamental transfer matrix is 

\begin{equation}
T(u)=\sum_{j=1}^{r+1}z^{(j)}(u),
\label{eqn:eqe31}
\end{equation}
where 

\begin{equation}
\fbox{$\ell$}_u=z^{(\ell)}(u)=\frac{Q_0(u)}{Q_0(u+i)}\frac{Q_{\ell-1}(u+i\frac{\ell+1}{2})Q_{\ell}(u+i\frac{\ell-2}{2})}{Q_{\ell-1}(u+i\frac{\ell-1}{2})Q_{\ell}(u+i\frac{\ell}{2})},\quad \ell=1, \dots,N.
\label{eqn:eqe32}
\end{equation}
The Bethe equations are

\begin{equation}
\begin{split}
    \frac{Q_{\ell-1}(\lambda_j^{(\ell)}+i\frac{1}{2})}{Q_{\ell-1}(\lambda_j^{(\ell)}-i\frac{1}{2})}\frac{Q_{\ell}(\lambda_j^{(\ell)}-i)}{Q_{\ell}(\lambda_j^{(\ell)}+i)}\frac{Q_{\ell+1}(\lambda_j^{(\ell)}+i\frac{1}{2})}{Q_{\ell+1}(\lambda_j^{(\ell)}-i\frac{1}{2})}=-1,
\\ j=1,\dots,N_{\ell},\quad \ell=2,\dots,r.
\end{split}
\label{eqn:eqe33}
\end{equation}
In this convention, the Hirota equation Eq. (\ref{eqn:eqe25}) turns to 

\begin{equation}
\begin{split}
    T_{m}^{a}(u+\frac{i}{2})T_{m}^{a}(u-\frac{i}{2})-T_{m+1}^{a}(u)T_{m-1}^{a}(u)=T_{m}^{a+1}(u)T_{m}^{a-1}(u)\\ a=1,\dots,r,\quad m=1,2,\dots.
\end{split}
\label{eqn:eqe34}
\end{equation}
And we specify the boundary condition to this system as 

\begin{equation}
T_{0}^{a}=1,\quad T^{0}_{m}=\frac{Q_0(u-i\frac{m}{2})}{Q_0(u+i\frac{m}{2})},\quad T_{m}^{r+1}(u)=1.
\label{eqn:eqe35}
\end{equation}
\section{Calculation of Eq. (\ref{eqn:eq0000069})}
\begin{figure}
    \centering
    \ytableausetup{centertableaux}
\begin{ytableau}
\scriptstyle 1 & \scriptstyle $\dots$ & \scriptstyle 1 &\scriptstyle 2 &\scriptstyle $\dots$ &\scriptstyle 2 &\scriptstyle 3 &\scriptstyle $\dots$ &\scriptstyle 3
\end{ytableau}
    \caption{Tableau sum for $T^{(1)}_{m}$ }
    \label{fig:young diagram2}
\end{figure}
The sum rule for $a=1$ is the following
\begin{equation}
T_{m}^{(1)}(u)=\sum_{\tau}\prod_{l=1,\dots,m}z^{(\tau_{l})}(u+i\frac{-1-m+2l}{2})
\label{eqn:eqe36}
\end{equation}
for $\mathfrak{su}_{3}$ the tableaux looks like Fig. \ref{fig:young diagram2}, where $\#1 =k_1$, $\#2=k_2$ and $\#3=k_3$ with $\sum k_{i}=m$. 
Then the tableau yields 
\begin{equation}
T^{(1)}_{m}(u)=\sum_{k_1,k_2}\prod_{l_1=1}^{k_1}z^{(1)}(u+i\frac{-1-m+2l_1}{2})\prod_{l_2=k_2+1}^{k_1+k_2}z^{(2)}(\dots)\prod_{l_{3}=k_1+k_2+1}^{m},
\label{eqn:eqe37}
\end{equation}
where the sum is a short hand of 

\begin{equation}
\sum_{k_1,k_2}=\sum_{k_1=0}^{m}\sum_{k_2=0}^{m-k_1},
\label{eqn:eqe38}
\end{equation}
also the $z$-functions

\begin{equation}
\begin{split}
z^{(1)}(u)&=\frac{Q_{1}(u-\frac{i}{2})}{Q_{1}(u+\frac{i}{2})};\\
z^{(2)}(u)&=\frac{Q_0(u)}{Q_0(u+i)}\frac{Q_{1}(u+\frac{3i}{2})}{Q_{1}(u+\frac{i}{2})}\frac{Q_2(u)}{Q_2(u+i)};\\
z^{(3)}(u)&=\frac{Q_0(u)}{Q_0(u+i)}\frac{Q_2(u+2i)}{Q_2(u+i)}.
\end{split}
\label{eqn:eqe39}
\end{equation}
The sum is quite straight forward in separate cases 

\begin{equation}
\prod_{l_1=1}^{k_1}z^{(1)}(u+i\frac{-1-m+2l_1}{2})=\frac{Q_{1}(-\frac{i}{2}+u+i\frac{-1-m+2}{2})}{Q_{1}(\frac{i}{2}+u+i\frac{-1-m+2k_1}{2})}
\label{eqn:eqe40}
\end{equation}

\begin{equation}
\begin{split}
\prod_{l_2=k_1+1}^{k_2}z^{(2)}(u+i\frac{-1-m+2l_2}{2})=&\frac{Q_{0}(u+i\frac{-1-m+2(k_1+1)}{2})}{Q_{0}(u+i\frac{-1-m+2(k_1+k_2)}{2}+i)}\frac{Q_{1}(\frac{3i}{2}+u+i\frac{-1-m+2(k_1+k_2)}{2})}{Q_{1}(\frac{i}{2}+u+i\frac{-1-m+2(k_1+1)}{2})}\\
&\times \frac{Q_{2}(u+i\frac{-1-m+2(k_1+1)}{2})}{Q_{2}(i+u+i\frac{-1-m+2(k_1+k_2)}{2})}
\end{split}
\label{eqn:eqe41}
\end{equation}

\begin{equation}
\begin{split}
\prod_{l_3=k_1+k_2+1}^{k_2}z^{(3)}(u+i\frac{-1-m+2l_3}{2})=&\frac{Q_{0}(u+i\frac{-1-m+2(k_1+k_2+1)}{2})}{Q_{0}(u+i\frac{-1-m+2m}{2}+i)}\frac{Q_{2}(2i+u+i\frac{-1-m+2m}{2})}{Q_{2}(i+u+i\frac{-1-m+2(k_1+k_2+1)}{2})}
\end{split}
\label{eqn:eqe42}
\end{equation}
then the sum can be written as

\begin{equation}
\begin{split}
T^{(1)}_{m}=&\sum^{m}_{k_1=0}\sum^{m-k_1}_{k_2=0}\frac{Q_{0}(u+i\frac{-1-m+2(k_1+1)}{2})Q_{2}(u+i\frac{-1-m+2(k_1+1)}{2})}{Q_{1}(\frac{i}{2}+u+i\frac{-1-m+2k_1}{2})Q_{1}(\frac{i}{2}+u+i\frac{-1-m+2(k_1+1)}{2})}\\
&\times \frac{Q_{1}(\frac{3i}{2}+u+i\frac{-1-m+2(k_1+k_2)}{2})}{Q_{2}(i+u+i\frac{-1-m+2(k_1+k_2)}{2})Q_{2}(i+u+i\frac{-1-m+2(k_1+k_2+1)}{2})}
\end{split}
\label{eqn:eqe43}
\end{equation}

make the following change of variables
\begin{equation}
p=m-k_1,q=m-k_1-k_2,
\label{eqn:eqe44}
\end{equation}
then Eq. (\ref{eqn:eqe43}) yields

\begin{equation}
\begin{split}
T^{(1)}_{m}=&\frac{Q_{1}(u-i\frac{m}{2})Q_{2}(u+i\frac{m+3}{2})}{Q_{0}(u+i\frac{m+1}{2})}\sum_{p=0}^{m}\frac{Q_{0}(u+i\frac{1+m}{2}-ip)Q_{2}(u+i\frac{1+m}{2}-ip)}
{Q_{1}(u+i\frac{m}{2}-ip)Q_{1}(u+i\frac{2+m}{2}-ip)}\\
&\times \sum_{q=0}^{p}\frac{Q_{1}(u+i\frac{2+m}{2}-iq)}{Q_{2}(u+i\frac{1+m}{2}-iq)Q_{2}(u+i\frac{3+m}{2}-ip)}
\end{split}
\label{eqn:eqe45}
\end{equation}
Due to different choice of normalization \cite{tableausum_Pozsgay}, we can neglect $Q_{0}(u+i\frac{m+1}{2})$ in the denominator. So we have recovered Eq. (\ref{eqn:eq0000069}).
% Appendix F

\chapter{Homomorphism between $Y(4)$ and $X(SO(6))$}  % Main appendix title

\label{AppendixF} % For referencing this appendix elsewhere, use \ref{AppendixF}
As stated in Chapter \ref{Chapter6}, in contrast to the Yangian $Y(N)$, there is no surjective homeomorphism for $X(\mathfrak{so}_N)$ onto the algebra $U(\mathfrak{so}_N)$, so we cannot use $\mathfrak{so}_N$ modules as $X(\mathfrak{so}_N)$ modules in general. But there is a way out of this if one only consider $X(\mathfrak{so}_6)$, because there is an isomorphism between $\mathfrak{so}_6$ and $\mathfrak{gl}_4$, so we can find a way to connect $X(\mathfrak{so}_6)$ with $Y(\mathfrak{gl}_4)$, then we can use the evaluation homomorphism.
The mapping
\begin{equation}
T^{\mathfrak{so}_6}(u)\mapsto \frac{1}{2}(1-P_{12})\cdot T_1^{\mathfrak{gl}_4}(u)T_2^{\mathfrak{gl}_4}(u-1)
\label{eqn:eqf1}
\end{equation}
defines an homomorphism $\phi: X(\mathfrak{so}_6)\rightarrow Y(\mathfrak{gl}_4)$. More explicitly, the images of the generators under the isomorphism are given by the formulas
\begin{equation}
t^{\mathfrak{so}_6}_{ij}(u)=t^{\mathfrak{gl}_4}_{a_1[i],a_1[j]}(u)t^{\mathfrak{gl}_4}_{a_2[i],a_2[j]}(u-1)-t^{\mathfrak{gl}_4}_{a_2[i],a_1[j]}(u)t^{\mathfrak{gl}_4}_{a_1[i],a_2[j]}(u-1),
\label{eqn:eqf2}
\end{equation}
where
\begin{equation}
\begin{matrix}a_1[-3]=1,& a_1[-2]=3,& a_1[-1]=1,& a_1[1]=2, & a_1[2]=2, & a_1[3]=3;\\
a_2[-3]=2,
& a_2[-2]=1,
& a_2[-1]=4,
 & a_2[1]=3,
 & a_2[2]=4,& a_2[3]=4;
\end{matrix}
\label{eqn:eqf3}
\end{equation}
is a mapping of indices from $\mathfrak{so}_6$ $\{-3,-2,-1,1,2,3\}$ to $\mathfrak{gl}_4$ $\{1,2,3,4\}$. We will prove Eq. (\ref{eqn:eqf1}) using the so-called fusion procedure \cite{extend_yangian_A_Molev}, which is explained in Appendix \ref{AppendixE}. First let take a look at the relation between the classical group $\mathfrak{so}_6$ and $\mathfrak{gl}_4$ before jumping to the quantum world. 
\section{Isomorphism between $\mathfrak{so}_6$ and $\mathfrak{gl}_4$}
Instead of specifying the mapping for all the 15 generators from $\mathfrak{so}_6$ to $\mathfrak{gl}_4$, we can just identify the mapping $\psi$ for the simple roots (one can see this by inspect the possible distance between two lower indices) between these two groups
\begin{equation}
\psi(F_{2,-1})=E_{2,1},\quad \psi(F_{3,2})=E_{3,2},\quad \psi(F_{2,1})=E_{4,3},
\label{eqn:eqf4}
\end{equation}
and we need to show that the commutation relations are preserved 
\begin{equation}
\begin{split}
[F_{ij},F_{kl}]=\delta_{jk}F_{il}-\delta_{il}F_{kj}+\delta_{j,-l}F_{k,-i}-\delta_{i,-k}F_{-j,l}, \quad F_{ij}\in \mathfrak{so}_6\\
[E_{ij},E_{kl}]=\delta_{jk}E_{il}-\delta_{il}E_{kj},\quad E_{ij}\in \mathfrak{gl}_4.
\end{split}
\label{eqn:eqf5}
\end{equation}
For instance, we have
\begin{equation}
\begin{split}
&F_{3,-2}=[F_{31},F_{1,-2}]=-[[F_{32},F_{21}],F_{2,-1}],\\
\mapsto\quad & E_{41}=-[[E_{32},E_{43}],E_{2,1}].
\end{split}
\label{eqn:eqf6}
\end{equation}
Then it is oblivious to show that
\begin{equation}
\begin{split}
&[F_{3,-2},F_{32}]=F_{3,-3}=0;\\
\mapsto\quad &[E_{41},E_{32}]=0,
\end{split}
\label{eqn:eqf7}
\end{equation}
One can use the simple roots to generate all the 15 generators and verify that the commutation relation by the same calculation.

Consider the vector space $\mathbb{C}^4$ with its canonical basis $e_1, e_2,e_3, e_4$ and denote by $V$ the six-dimensional subspace of $\mathbb{C}^4\times\mathbb{C}^4$. Let identify 
\begin{equation}
v_{-3}=e_1\otimes e_2-e_2\otimes e_1,
\label{eqn:eqf8}
\end{equation}
then one can generating all the rest states by
\begin{equation}
\begin{split}
v_{-2}=&\psi(F_{-2,-3}).v_{-3}=(-E_{3,2}\otimes I-I\otimes E_{3,2})(e_1\otimes e_2-e_2\otimes e_1)\\=&e_3\otimes e_1-e_1\otimes e_3,\\
v_{-1}=&\psi(F_{-1,-2}).v_{-2}=e_3\otimes e_1-e_1\otimes e_3,\\
v_{1}=&\psi(F_{1,-2}).v_{-2}=e_2\otimes e_3-e_3\otimes e_2,\\
v_{2}=&\psi(F_{2,1}).v_{1}=e_2\otimes e_4-e_4\otimes e_2,\\
v_{3}=&\psi(F_{3,2}).v_{2}=e_3\otimes e_4-e_4\otimes e_3.
\end{split}
\label{eqn:eqf9}
\end{equation}
 So we identify $V$ with $\mathbb{C}^6$ regarding $v_{-3},v_{-2},v_{-1},v_{1},v_{2},v_{3}$ as its canonical basis
\begin{equation}
\begin{split}
v_{-3}=e_1\otimes e_2-e_2\otimes e_1,\quad v_{3}=e_3\otimes e_4-e_4\otimes e_3,\\
v_{-2}=e_3\otimes e_1-e_1\otimes e_3,\quad v_{2}=e_2\otimes e_4-e_4\otimes e_2,\\
v_{-1}=e_1\otimes e_4-e_4\otimes e_1,\quad v_{1}=e_2\otimes e_3-e_3\otimes e_2.
\end{split}
\label{eqn:eqf10}
\end{equation}

\section{Homomorphism between $X(\mathfrak{so}_6)$ and $Y(4)$}
Now we are prepared to show the homomorphism mapping Eq. (\ref{eqn:eqf1}). We start by showing that the mapping defines an algebra homomorphism for R-matrices. We use a version of the well known fusion procedure for R-matrices. Then we will show the mapping preserved the "commutation relation" RTT relation.  Consider the tensor product space $(\mathbb{C}^4)^{\otimes4}$. In the following we consider $V\otimes V$ as a natural subspace of $(\mathbb{C}^4)^{\otimes2}\otimes(\mathbb{C}^4)^{\otimes2}$. Obviously, the operator
  \begin{equation}
    (1-P_{12})(1-P_{34})=R^{\circ}_{12}(+1)R^{\circ}_{34}(+1)
    \label{eqn:eqf11}
  \end{equation}
is a projection of $(\mathbb{C}^4)^{\otimes2}\otimes(\mathbb{C}^4)^{\otimes2}$ to the subspace $V\otimes V$, here I have used $R^{\circ}$ to denote the R-matrix for $GL(4)$. Let us set 
  \begin{equation}
    \begin{split}
R_{V}(u)&=\frac{1}{4}(1-P_{12})(1-P_{34})R^{\circ}_{14}(u+1)R^{\circ}_{13}(u)R^{\circ}_{24}(u)R^{\circ}_{23}(u-1)\\
&=\frac{1}{4}R^{\circ}_{23}(u-1)R^{\circ}_{24}(u)R^{\circ}_{13}(u)R^{\circ}_{14}(u+1)(1-P_{12})(1-P_{34}),
\end{split}
   \label{eqn:eqf12}
  \end{equation}
where the second line follows the Yangian-Baxter equation. From Eq. (\ref{eqn:eq0000002}) we have the following equality of operators in $V\otimes V$,
\begin{equation}
R_{V}(u)=\frac{u-2}{u-1}(1-\frac{P_V}{u}+\frac{Q_V}{u-2}).
\label{eqn:eqf13}
\end{equation}
where the pre-factor is determined later by matching with its action on $(\mathbb{C}^4)^{\otimes4}$ space. Note that from Eq. (\ref{eqn:eqf12}) one can write 
\begin{equation}
\begin{split}
    R_V=\frac{1}{4}(1-P_{12})(1-P_{34})R^{\circ}_{14}(u+1)R^{\circ}_{13}(u)R^{\circ}_{24}(u)R^{\circ}_{23}(u-1)\\
=\frac{1}{4}(1-P_{12})(1-P_{34})(1-\frac{P_{14}}{u+1})(1-\frac{P_{13}}{u})(1-\frac{P_{24}}{u})(1-\frac{P_{23}}{u-1})
\end{split}
\label{eqn:eqf14}
\end{equation}
The proof is explicit by tedious, one can complete the proof by the application of Eq. (\ref{eqn:eqf13}) to all basis $v_i\otimes v_i$ and compare it with Eq. (12) applied on $(\mathbb{C}^4)^{\otimes4}$. We will show one example $v_{-1}\otimes v_{1}$, on one hand
\begin{equation}
\begin{split}
&R_V(u)(v_{-1}\otimes v_{1})\\&=\frac{4(u-2)}{u-1}(1-\frac{P_V}{u}+\frac{Q_V}{u-2})(v_{-1}\otimes v_{1})\\
&=\frac{u-2}{u-1}\Big[v_{-1}\otimes v_{1}-\frac{1}{u}(v_{1}\otimes v_{-1})
\\&\quad+\frac{1}{u-2}(v_{-1}\otimes v_{1}+v_{1}\otimes v_{-1}+v_{2}\otimes v_{-2}+v_{-2}\otimes v_{2}+v_{3}\otimes v_{-3}+v_{-3}\otimes v_{3})\Big]\\
&=v_{-1}\otimes v_{1}+\frac{2}{u(u-1)}(v_{1}\otimes v_{-1})\\&\quad+\frac{1}{u-1}(v_{2}\otimes v_{-2}+v_{-2}\otimes v_{2}+v_{3}\otimes v_{-3}+v_{-3}\otimes v_{3}),
\end{split}
\label{eqn:eqf15}
\end{equation}
on the other hand

\begin{equation}
\begin{split}
&R_V(u)(v_{-1}\otimes v_{1})\\&=\frac{1}{4}(1-P_{12})(1-P_{34})(1-\frac{P_{14}}{u+1})(1-\frac{P_{13}}{u})(1-\frac{P_{24}}{u})(1-\frac{P_{23}}{u-1})\\&\quad (e_1\otimes e_4\otimes e_1\otimes e_4-e_4\otimes e_1\otimes e_1\otimes e_4-e_1\otimes e_4\otimes e_4\otimes e_1+e_4\otimes e_1\otimes e_4\otimes e_1)\\
&=(v_{-1}\otimes v_{1})+\frac{2}{u(u-1)}(v_{1}\otimes v_{-1})\\&\quad+\frac{1}{u-1}(v_{2}\otimes v_{-2}+v_{-2}\otimes v_{2}+v_{3}\otimes v_{-3}+v_{-3}\otimes v_{3}).
\end{split}
\label{eqn:eqf16}
\end{equation}
The remaining cases are verified by the same calculation. Therefore, the element $R_V(u)$ coincides with the R-matrix for $\mathfrak{so}_6$ up to a scalar factor, which doesn't make a difference. 

So, in order to verify that the mapping Eq. (\ref{eqn:eqf1}) defines a homomorphism $X(\mathfrak{so}_6)\rightarrow Y(\mathfrak{gl}_4)$ we need to show that the "commutation relation" RTT relation
\begin{equation}
R_V(u-v)T_{1^{\prime}}(u)T_{2^{\prime}}(v)=T_{2^{\prime}}(v)T_{1^{\prime}}(u)R_V(u-v)
\label{eqn:eqf17}
\end{equation}
remains valid when $T(u)$ is replaced by its image. Here we use primed indices to indicate the copies of the space $V$ in the tensor product $V\otimes V$. We reserve unprimed indices for the copies of $\mathbb{C}^4$ in the tensor product $(\mathbb{C}^4)^{\otimes4}$. The left hand side of Eq. (\ref{eqn:eqf17}) reads
\begin{equation}
\begin{split}
    \frac{1}{4}\times\frac{1}{4}(1-P_{12})(1-P_{34})R^{\circ}_{14}(u-v+1)R^{\circ}_{13}(u-v)R^{\circ}_{24}(u-v)R^{\circ}_{23}(u-v-1)
\\ \times(1-P_{12})\cdot T_1^{\mathfrak{gl}_4}(u)T_2^{\mathfrak{gl}_4}(u-1)(1-P_{34})\cdot T_3^{\mathfrak{gl}_4}(v)T_4^{\mathfrak{gl}_4}(v-1).
\end{split}
\label{eqn:eqf18}
\end{equation}

Writing the product of R-matrices in the equivalent form Eq. (\ref{eqn:eqf14}), we simplify this using
\begin{equation}
P^2=I,\quad (I-P)^2=2(I-P),\quad [I-P_{12},I-P_{34}]=0,
\label{eqn:eqf19}
\end{equation}
obtain
\begin{equation}
\begin{split}
    \frac{1}{16}&R^{\circ}_{23}(u-v-1)R^{\circ}_{24}(u-v)R^{\circ}_{13}(u-v)R^{\circ}_{14}(u-v+1)(1-P_{12})\\&\times(1-P_{34})(1-P_{12})(1-P_{34})\times T_1^{\mathfrak{gl}_4}(u)T_2^{\mathfrak{gl}_4}(u-1)\cdot T_3^{\mathfrak{gl}_4}(v)T_4^{\mathfrak{gl}_4}(v-1)\\
=\frac{1}{4}&(1-P_{12})(1-P_{34})R^{\circ}_{14}(u-v+1)R^{\circ}_{13}(u-v)R^{\circ}_{24}(u-v)R^{\circ}_{23}(u-v-1)
\\& \times T_1^{\mathfrak{gl}_4}(u)T_2^{\mathfrak{gl}_4}(u-1)\cdot T_3^{\mathfrak{gl}_4}(v)T_4^{\mathfrak{gl}_4}(v-1).
\end{split}
\label{eqn:eqf20}
\end{equation}
Now apply the $RTT\mbox{-}$relation repeatedly to bring this expression to the form
\begin{equation}
\begin{split}
\frac{1}{4}(1-P_{12})(1-P_{34})T_3^{\mathfrak{gl}_4}(v)T_4^{\mathfrak{gl}_4}(v-1)T_1^{\mathfrak{gl}_4}(u)T_2^{\mathfrak{gl}_4}(u-1)\\\times  R^{\circ}_{14}(u-v+1)R^{\circ}_{13}(u-v)R^{\circ}_{24}(u-v)R^{\circ}_{23}(u-v-1).
\end{split}
\label{eqn:eqf21}
\end{equation}
Finally, since $R^{\circ}(+1)$ is a projection, from $RTT\mbox{-}$relation 
\begin{equation}
\frac{1}{2}(1-P_{12})T_1^{\mathfrak{gl}_4}(u)T_2^{\mathfrak{gl}_4}(u-1)=\frac{1}{4}(1-P_{12})T_1^{\mathfrak{gl}_4}(u)T_2^{\mathfrak{gl}_4}(u-1)(1-P_{12}),
\label{eqn:eqf22}
\end{equation}
one can recover the right hand side of Eq. (\ref{eqn:eqf17}).
The explicit images of the generators of $X(\mathfrak{so}_6)$ are found by taking the matrix
elements in Eq. (\ref{eqn:eqf1}). Indeed, the application of $T^{\mathfrak{so}_6}(u)$ to the basis vector $v_{−1}$ of $V$ gives
\begin{equation}
\begin{split}
T^{\mathfrak{so}_6}(u)(v_{-1})=&t_{-3,-1}(u)v_{-3}\\&+t_{-2,-1}(u)v_{-2}+t_{-1,-1}(u)v_{-1}+t_{1,-1}(u)v_{1}+t_{2,-1}(u)v_{2}+t_{3,-1}(u)v_{3},    
\end{split}
\label{eqn:eqf23}
\end{equation}
while
\begin{equation}
\begin{split}
&\frac{1}{2}(1-P_{12})\cdot T^{\mathfrak{gl}_4}_1(u)T^{\mathfrak{gl}_4}_2(u-1)(v_{-1})=\frac{1}{2}(1-P_{12})\cdot T^{\mathfrak{gl}_4}_1(u)T^{\mathfrak{gl}_4}_2(u-1)(e_1\otimes e_4-e_4\otimes e_1)\\
&=\frac{1}{2}(1-P_{12})\Big[\sum_{a,b=1}^{4}T_{a1}(u)T_{b4}(u-1)e_a\otimes e_b-\sum_{a,b=1}^{4}T_{a4}(u)T_{b1}(u-1)e_a\otimes e_b\Big]\\
&=\frac{1}{2}\sum_{a,b=1}^4\Big[T_{a1}(u)T_{b4}(u-1)-T_{a4}(u)T_{b1}(u-1)\Big](e_a\otimes e_b-e_b\otimes e_a),
\end{split}
\label{eqn:eqf24}
\end{equation}
one can get

\begin{equation}
\begin{split}
t_{-3,-1}(u)&=\frac{1}{2}\Big[T_{11}(u)T_{24}(u-1)-T_{14}(u)T_{21}(u-1)-T_{21}(u)T_{14}(u-1)+T_{24}(u)T_{11}(u-1)\Big],\\
&=T_{11}(u)T_{24}(u-1) - T_{14}(u)T_{21}(u-1)\\
t_{-2,-1}(u)&=\frac{1}{2}\Big[T_{31}(u)T_{14}(u-1)-T_{34}(u)T_{11}(u-1)-T_{11}(u)T_{34}(u-1)+T_{14}(u)T_{31}(u-1),\\
&=T_{31}(u)T_{14}(u-1)-T_{34}(u)T_{11}(u-1)\\
t_{-1,-1}(u)&=\frac{1}{2}\Big[T_{11}(u)T_{44}(u-1)-T_{14}(u)T_{41}(u-1)-T_{41}(u)T_{14}(u-1)+T_{44}(u)T_{11}(u-1)\Big],\\
&=T_{11}(u)T_{44}(u-1)-T_{14}(u)T_{41}(u-1)\\
t_{1,-1}(u)&=\frac{1}{2}\Big[T_{21}(u)T_{34}(u-1)-T_{24}(u)T_{31}(u-1)-T_{31}(u)T_{24}(u-1)+T_{34}(u)T_{21}(u-1)\Big],\\
&=T_{21}(u)T_{34}(u-1)-T_{24}(u)T_{31}(u-1)\\
t_{2,-1}(u)&=\frac{1}{2}\Big[T_{21}(u)T_{44}(u-1)-T_{24}(u)T_{41}(u-1)-T_{41}(u)T_{24}(u-1)+T_{44}(u)T_{21}(u-1)\Big],\\
&=T_{21}(u)T_{44}(u-1)-T_{24}(u)T_{41}(u-1)\\
t_{3,-1}(u)&=\frac{1}{2}\Big[T_{31}(u)T_{44}(u-1)-T_{34}(u)T_{41}(u-1)-T_{41}(u)T_{34}(u-1)+T_{44}(u)T_{31}(u-1)\Big]\\
&=T_{31}(u)T_{44}(u-1)-T_{34}(u)T_{41}(u-1).
\end{split}
\label{eqn:eqf25}
\end{equation}
Where we have used the antisymmetric property of the basis vector. After find the link between the generators of $X(\mathfrak{so}_6)$ and $Y(\mathfrak{gl}_4)$, we can use the evaluation homomorphism, the $\mathfrak{gl}_4$ module $L(\lambda_1,\lambda_2,\lambda_3,\lambda_4)$ is an $X(\mathfrak{so}_6)$ module. The rest entries can be obtain in the same way. %----------------------------------------------------------------------------------------
%	BIBLIOGRAPHY
%----------------------------------------------------------------------------------------
%----------------------------------------------------------------------------------------
\bibliography{main.bib}

\end{document}